\documentclass[12pt,preprint]{aastex}

\def\lessim{\mathrel{\hbox{\rlap{\hbox{\lower4pt\hbox{$\sim$}}}\hbox{$<$}}}}
\def\grtsim{\mathrel{\hbox{\rlap{\hbox{\lower4pt\hbox{$\sim$}}}\hbox{$>$}}}}

\shorttitle{M31 Novae}
\shortauthors{Shafter et al.}

\begin{document}


\title{A Spectroscopic and Photometric Survey of Novae in M31}


\author{A. W. Shafter\altaffilmark{1}, M. J. Darnley\altaffilmark{2}, K. Hornoch\altaffilmark{3}, A. V. Filippenko\altaffilmark{4}, M. F. Bode\altaffilmark{2}, R. Ciardullo\altaffilmark{5}, K. A. Misselt\altaffilmark{6}, R. A. Hounsell\altaffilmark{2}, R. Chornock\altaffilmark{4,7}, and T. Matheson\altaffilmark{8}}
\altaffiltext{1}{Department of Astronomy, San Diego State University, San Diego, CA 92182, USA}
\altaffiltext{2}{Astrophysics Research Institute, Liverpool John Moores University, Birkenhead CH41 1LD, UK}
\altaffiltext{3}{Astronomical Institute, Academy of Sciences, CZ-251~65~Ond\v{r}ejov, Czech Republic}
\altaffiltext{4}{Department of Astronomy, University of California, Berkeley, CA 94720-3411, USA}
\altaffiltext{5}{Department of Astronomy and Astrophysics, The Pennsylvania State University, 525 Davey Lab, University Park, PA 16802, USA}
\altaffiltext{6}{Steward Observatory, University of Arizona, Tucson, AZ, 85721, USA}
\altaffiltext{7}{Harvard-Smithsonian Center for Astrophysics, 60 Garden Street, Cambridge, MA 02138, USA}
\altaffiltext{8}{National Optical Astronomy Observatory, 950 North Cherry Avenue, Tucson, AZ 85719-4933, USA}



\begin{abstract}
We report the results of a multi-year spectroscopic and
photometric survey of novae in M31
that resulted in a total of 53 spectra of 48 individual
nova candidates. Two of these, M31N~1995-11e and
M31N~2007-11g, were revealed to be long-period Mira variables, not novae.
These data double the number of
spectra extant for novae in M31 through the end of 2009 and bring to 91
the number of M31 novae with known spectroscopic classifications.
We find that
75 novae (82\%) are confirmed or likely members of the
Fe~II spectroscopic class, with the remaining
16 novae (18\%) belonging to the He/N (and related) classes.
These numbers are consistent with those found for Galactic novae.
We find no compelling evidence that spectroscopic
class depends sensitively on spatial position or population
within M31 (i.e., bulge vs. disk), although the distribution for
He/N systems appears slightly more extended than that for the Fe~II class.
We confirm the existence of a correlation between speed class and
ejection velocity (based on line width), as in the case of Galactic novae.
Follow-up photometry allowed us to determine
light-curve parameters
for a total of 47 of the 91 novae with known spectroscopic class.
We confirm that more luminous novae generally fade the fastest, and
that He/N novae are typically faster and brighter than their Fe~II counterparts.
In addition, we find a weak dependence of nova speed class
on position in M31, with the spatial distribution of
the fastest novae being slightly more extended than that of slower novae.

\end{abstract}




\section{Introduction}

Classical novae
form a subclass of the cataclysmic variable stars.
They are semidetached binary star systems
where a late-type Roche-lobe-filling
star transfers
mass to a white dwarf companion~\citep[e.g.,][]{war95, war08}.
If the mass accretion rate onto the white dwarf is sufficiently low
to allow the accreted gas to become degenerate,
a thermonuclear runaway (TNR) will eventually ensue in the accreted envelope,
leading to a nova eruption. These eruptions can reach an
absolute magnitude as bright as M$_V \approx -10$~\citep[e.g.,][]{sta08},
making them among the most luminous explosions in the Universe.
Their high luminosities and rates~\citep[
$\sim50$~yr$^{-1}$ in a galaxy like M31;][]{sha01, dar06}
make novae powerful
probes of the properties of
close binaries in different (extragalactic) stellar
populations.
The most thoroughly studied extragalactic system is M31, where
more than 800 novae have been discovered over the past century
\citep[e.g., see][and references therein]{pie07e, sha08}.

Despite this large number of novae discovered,
very few follow-up studies of their photometric, or particularly
their spectroscopic, properties have been attempted.
Most recent M31 surveys have been undertaken
through narrow-band filters in order to take advantage of the
fact that novae fade more slowly in H$\alpha$ than they do in the continuum
\citep[e.g.,][]{cia87, sha01}. Such observations are ideal
for determining the rate and spatial distribution of novae within a galaxy,
but not for characterizing the nova light curves.
As shown by \citet{cia90b} the H$\alpha$ light curves
are not simply correlated with the peak luminosity as are the
broad-band light curves.
Moreover, most of the broad-band
light-curve data for M31 novae come from the early photographic studies
of~\citet{arp56} and~\citet{ros64, ros73},
as summarized by~\citet{cap89}, and from observations in Crimea and Latvia
during the period between 1977 and 1990~\citep{sha92}.

Similarly, spectroscopic data for M31 novae have, until recently,
also been limited. The dearth of available spectra
is not surprising given that novae in M31 are by definition transient,
and relatively faint,
reaching peak brightnesses in the range $m_V \approx 18$ to $m_V \approx 
15$ mag before fading back to quiescence. Furthermore,
since their eruptions are not predictable in advance, spectroscopic
observations require not only timely access to large telescopes, but
coordination with a photometric survey dedicated
to discovering suitable targets.
\citet{hum32}
was the first to report spectroscopic observations of classical
novae in M31, and it was not until more than half a century later
that
\citet{cia83}
published the spectra of four H$\alpha$ emission-line
sources, which they identified as classical novae in eruption. The
number of nova spectra has increased dramatically in recent years
thanks to greater access to
queue scheduling on large telescopes, such as the Hobby-Eberly Telescope (HET).

In order to better understand the spectroscopic properties of novae in M31,
and to study any variation with
spatial position in the galaxy,
we began a multi-year M31 nova survey in the Fall of 2006 using the HET.
The program was motivated in large part by the work of \citet{wil92},
who realized that the spectra of Galactic novae
(taken within a few weeks of eruption)
can be divided into one of two principal spectroscopic types: Fe~II and He/N.
These types are believed to be related to fundamental properties
of the progenitor binary such as the white dwarf mass. 
As part of our HET program,
we have measured the spectra of
26 M31 novae (representing $\sim 1/3$ of the total for which
spectra are available), making it the most significant
spectroscopic follow-up study of Local Group novae since the pioneering work
of~\citet{tom92}. To supplement these data, we have included
21 M31 nova spectra obtained over the years
with the Lick Observatory 3-m Shane reflector. Finally, in order
to characterize the light curves, we have employed a variety
of telescopes to acquire
broad-band photometric measurements of many of the novae
in our spectroscopic sample.
Here, we report the results of our survey.

\section{Observations}

\subsection{Spectroscopy}

During the early part of our survey, optical spectra were obtained
primarily with the Lick 3-m Shane reflector using the Kast double
spectrograph \citep{mil93}, although
the observation of M31N 1990-10b was taken with the older UV Schmidt
spectrograph \citep{mil87}.  Most spectra were acquired in one of two
basic instrument configurations.  One used the D55 dichroic beamsplitter to
split the spectrum over $\sim$5200--5500~\AA, with blue light being passed
through the 600/4310 grism and red light being reflected off a 300/7500 (or
600/5000) grating.  The resulting combined spectra cover the range
$\sim$3300--10000~\AA\ ($\sim$3300--7900~\AA), with the $2.0''$-wide 
slit giving a spectral resolution of $\sim$6~\AA\ in the blue 
and $\sim$11~\AA\ (6~\AA) in the
red.  The other setup removed the beamsplitter and all light was sent to a
600/5000 grating on the red side, providing $\sim$6~\AA\ resolution over the
range $\sim$4300--7000~\AA.

   The data were reduced using standard techniques
\citep[e.g.,][]{fol03}.  Routine CCD processing and spectrum
extraction were completed with IRAF\footnote{ IRAF is distributed by
  the National Optical Astronomy Observatory, which is operated by the
  Association for Research in Astronomy, Inc.  under cooperative
  agreement with the National Science Foundation.}, and the data were
extracted with the optimal algorithm of \citet{Horne86}.  We obtained
the wavelength scale from low-order polynomial fits to
calibration-lamp spectra.  Small wavelength shifts were then applied
to the data after cross-correlating a template sky to the night-sky
lines that were extracted with the nova.  Using our own IDL routines, we
fit spectrophotometric standard-star spectra to the data in order to
flux calibrate our spectra and to remove telluric lines \citep{Wade88,
  Matheson00}.  A summary of the Lick observations is given in
Table~\ref{lickdata}.

Most of the spectra in this survey, and all of the ones obtained
in the past few years, were acquired
with the Marcario Low-Resolution Spectrograph
\citep[LRS;][]{hil98} on the HET. Initially, we employed the 
$g2$ grating with a 2.0$''$ slit and the GG385 order-blocking filter,
covering 4275--7250\,\AA\ at a resolution of $R \approx 650$. Later, to
obtain more coverage at longer wavelengths, we opted to use
the lower resolution $g1$ grating with a
1.0$''$ slit and the GG385 filter. This choice increased our
wavelength coverage
to 4150--11000\,\AA\ while yielding a resolution of $R \approx 600$.
In practice, the useful
spectral range of the $g1$ grating is limited to
$\lambda\lessim9000$\,\AA\ where the effects of order overlap are minimal.
All HET spectra were reduced using standard IRAF routines to flat-field
the data and to optimally extract the spectra.
These were placed on relative flux scales through comparison with
observations of spectrophotometric standards routinely used at the HET.
The data were not corrected for absorption by the Earth's atmosphere, and
telluric absorption features are visible in some of the spectra.
A summary of the HET observations is given in Table~\ref{hetdata}.

\begin{figure}
\includegraphics[angle=0,scale=.60]{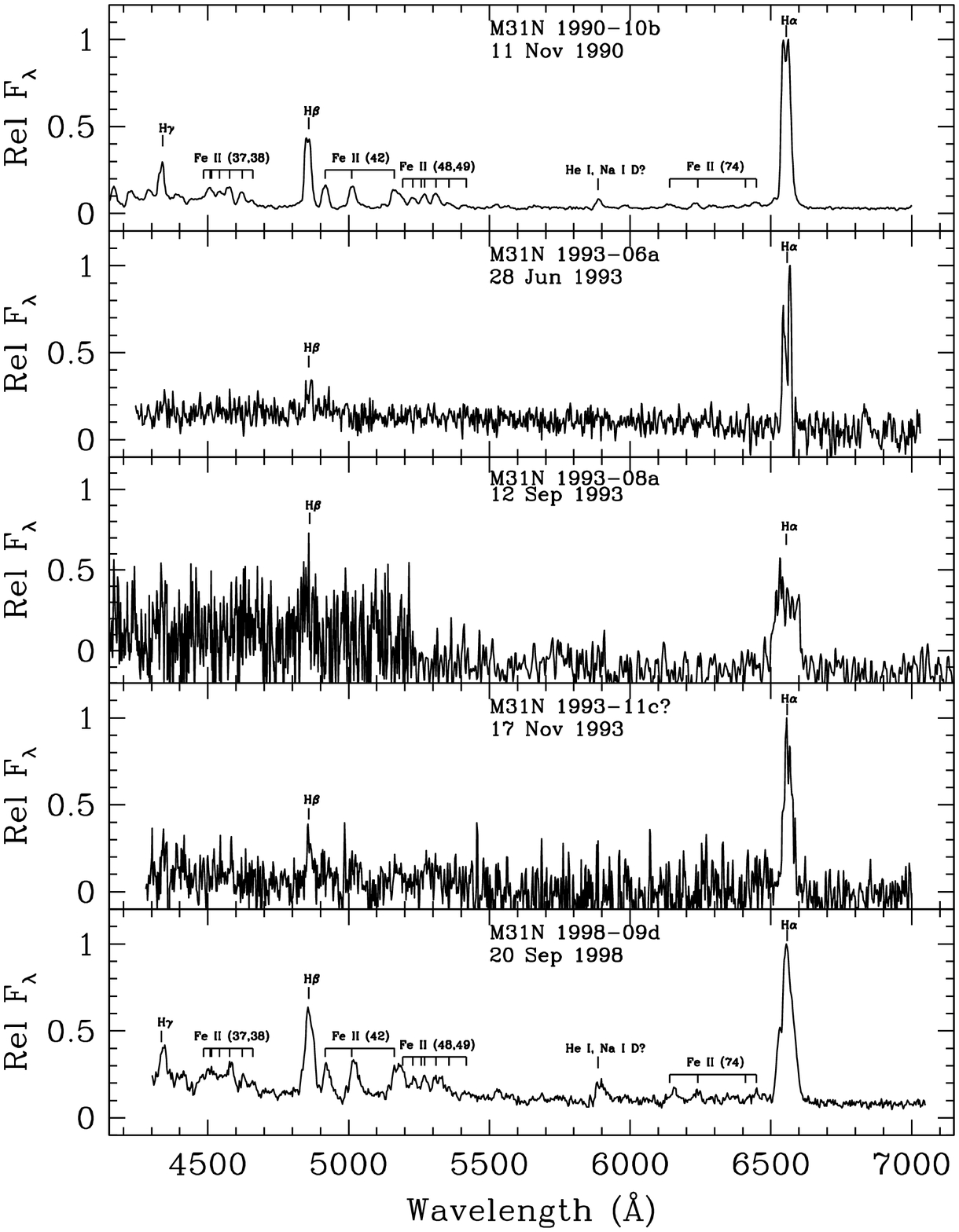}
\caption{Spectra of the M31 novae M31N~1990-10b, 1993-06a, 1993-08a, 1993-11c (ID uncertain; see text), and 1998-09d, taken 19, 25, 37, 9, and 9 days post discovery, respectively. All are Fe~II systems with the possible exception of M31N~1993-08a, which has a broad H$\alpha$ emission line characteristic of the He/N novae.
\label{fig1}}
\end{figure}

\begin{figure}
\includegraphics[angle=0,scale=.60]{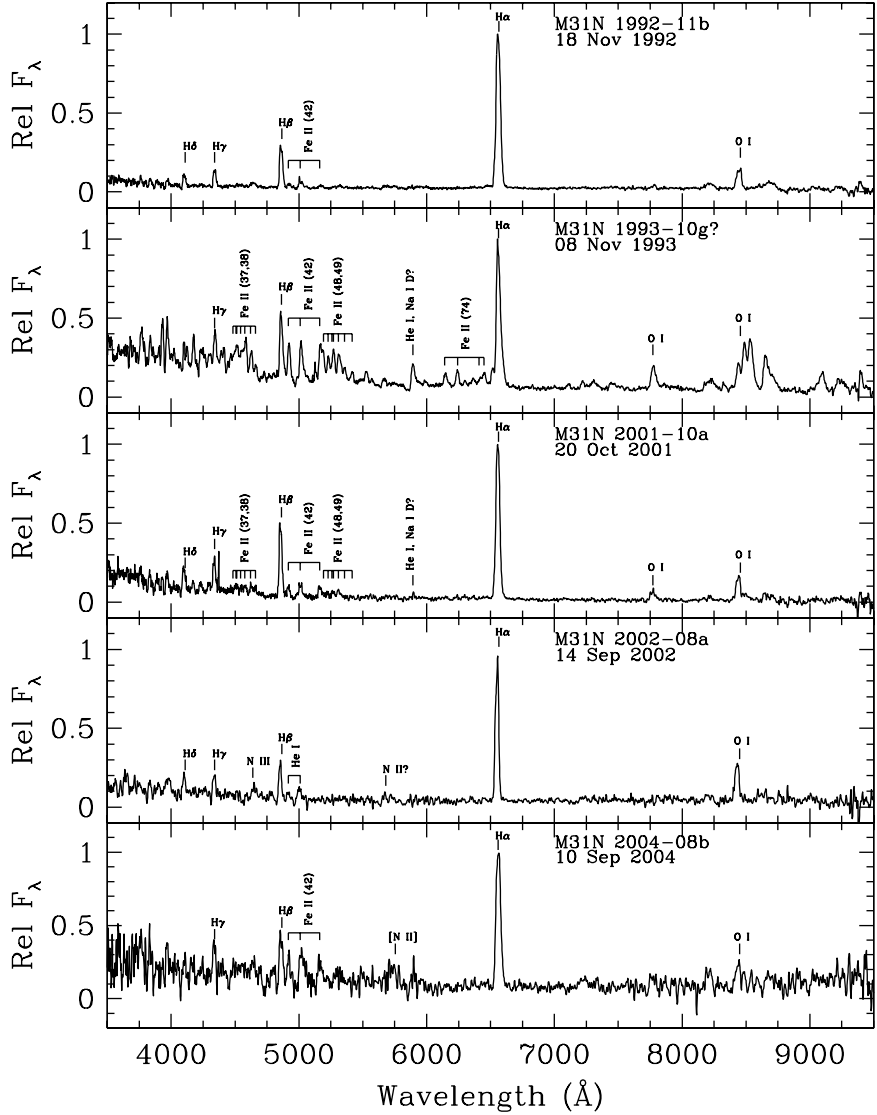}
\caption{Spectra of the M31 novae M31N~1992-11b, 1993-10g (ID uncertain, see text), 2001-10a, 2002-08a, and 2004-08b obtained 10, 21, 17, 40, and 34 days post discovery, respectively. All are Fe~II systems with the possible exception of 2002-08a, which was observed well past maximum
light and has a spectrum similar to that of an He/Nn nova.
\label{fig2}}
\end{figure}

\begin{figure}
\includegraphics[angle=0,scale=.60]{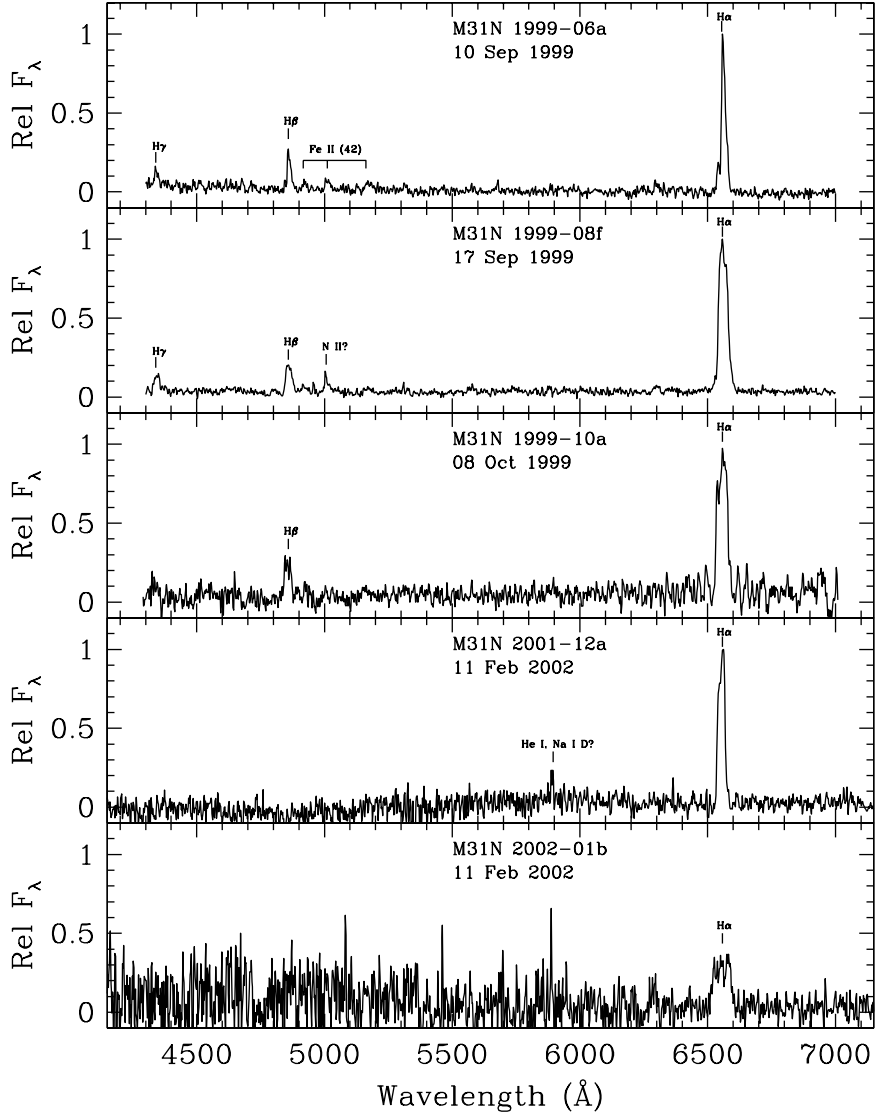}
\caption{Spectra of the M31 novae M31N~1999-06a, 1999-08f, 1999-10a, 2001-12a, and 2002-01b, taken 75, 19, 6, 58, and 35 days post discovery, respectively. All are Fe~II novae with the exception of 1999-08f, where the type is uncertain, and 2002-01b, where the broad H$\alpha$ line suggests that the object may be a He/N system.
\label{fig3}}
\end{figure}

\begin{figure}
\includegraphics[angle=0,scale=.60]{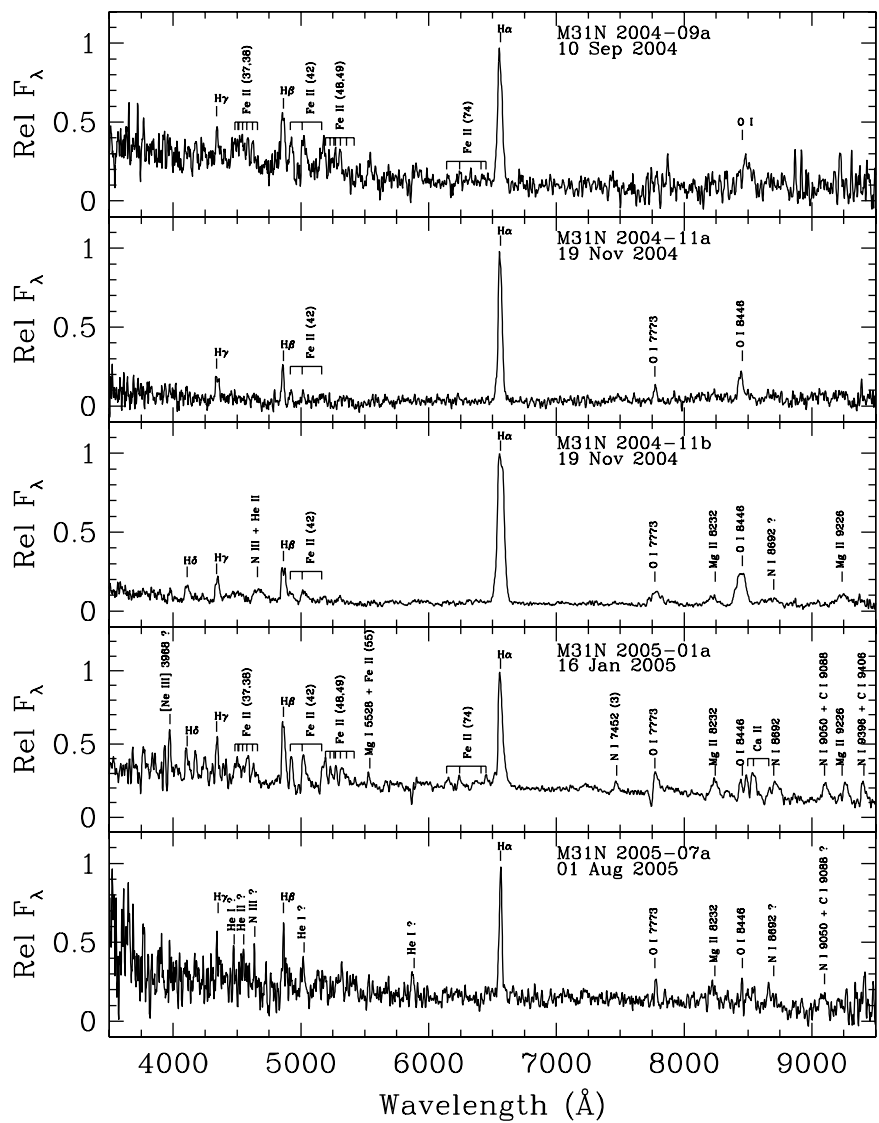}
\caption{Spectra of the M31 novae M31N~2004-09a, 2004-11a, 2004-11b, 2005-01a, and 2005-07a taken 8, 14, 14, 8, and 2 days post discovery, respectively. All are Fe~II novae with the possible exception of 2004-11b, which has a spectrum characterized by relatively broad Balmer and N III emission similar to that of an Fe~IIb, or hybrid nova, and M31N~2005-07a, which might be a He/Nn system.
\label{fig4}}
\end{figure}

\begin{figure}
\includegraphics[angle=0,scale=.60]{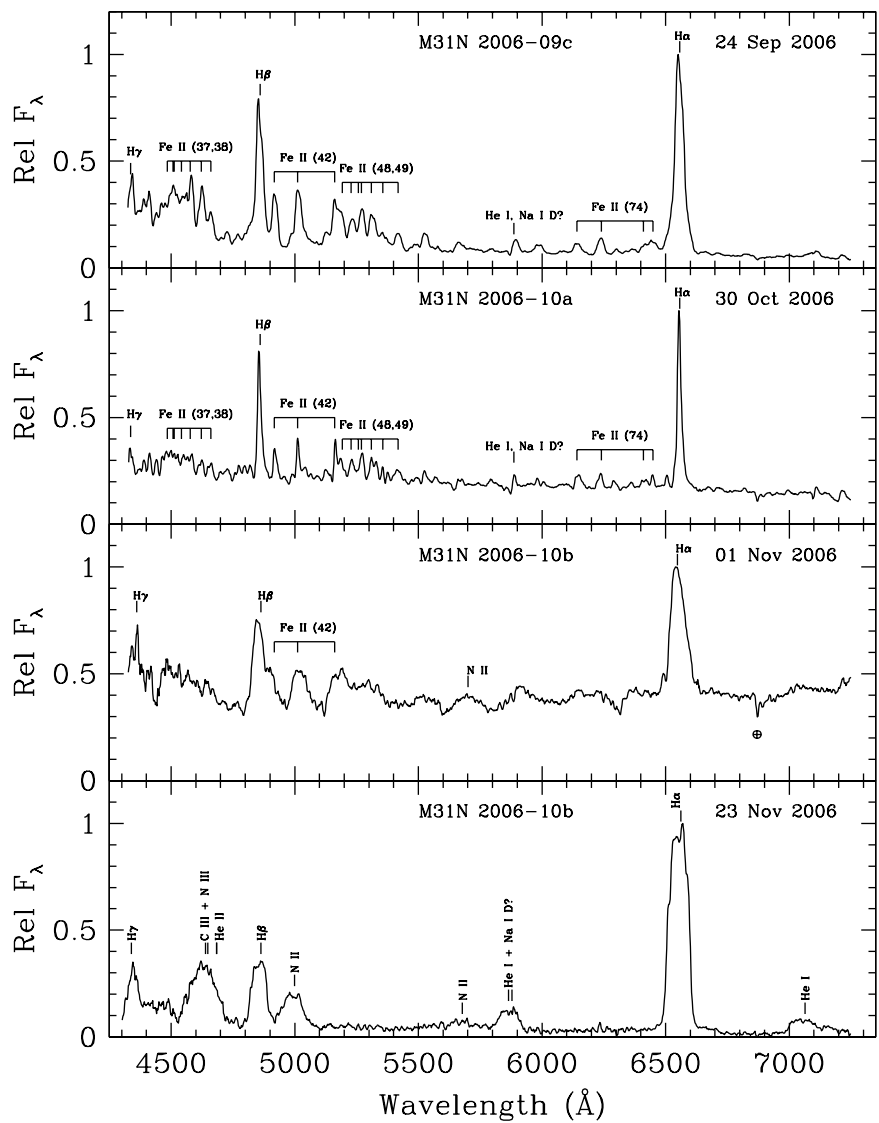}
\caption{Spectra of the M31 novae M31N~2006-09c, 2006-10a, and
two observations of 2006-10b taken 7, 8, 2, and 24 days
post discovery, respectively.
Both M31N~2006-09c and M31N~2006-10a are typical Fe~II novae,
while M31N~2006-10b (observed twice)
is an example of a hybrid or Fe~IIb nova. By the time of the second
spectrum, M31N~2006-10b had evolved into a classic He/N nova.
\label{fig5}}
\end{figure}

\begin{figure}
\includegraphics[angle=0,scale=.60]{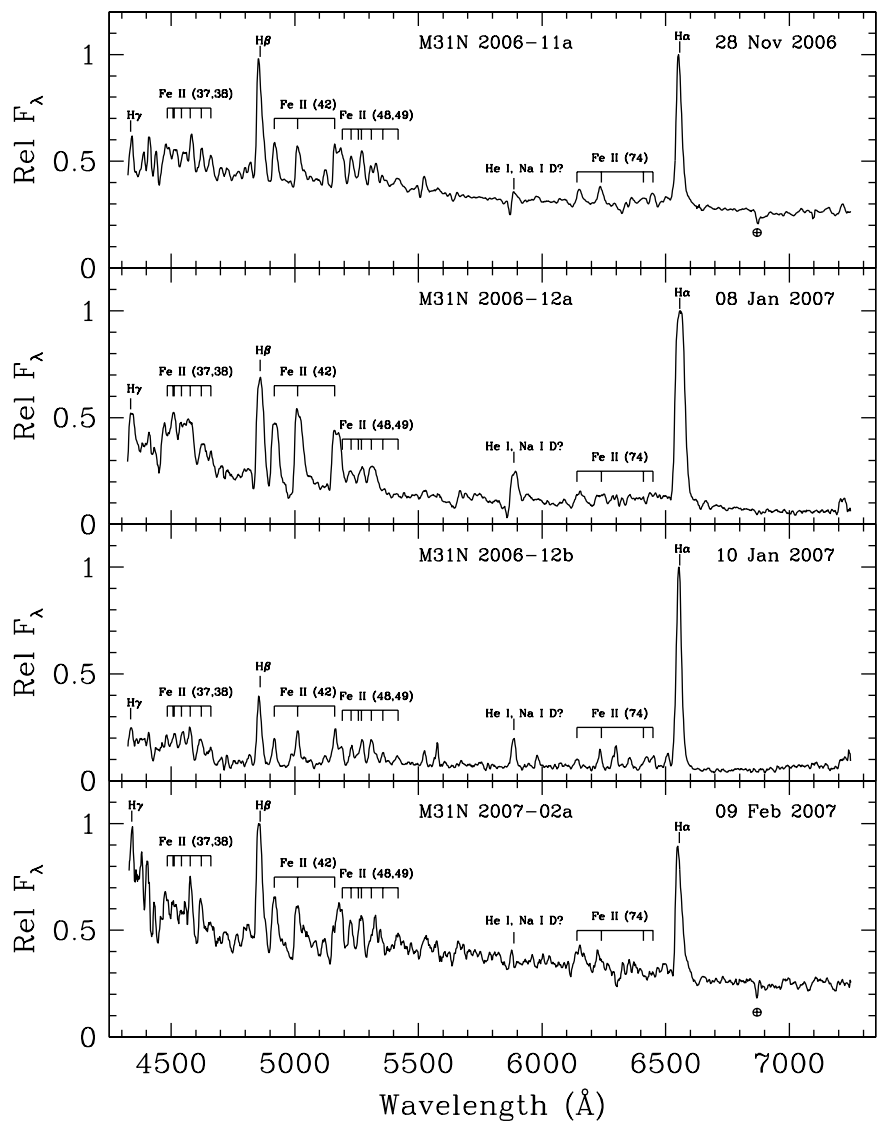}
\caption{Spectra of the M31 novae M31N~2006-11a, 2006-12a,
2006-12b, and 2007-02a, taken 4, 24, 18, and 3~days post discovery, respectively. All four novae are typical members of the Fe~II class.
\label{fig6}}
\end{figure}

\begin{figure}
\includegraphics[angle=0,scale=.60]{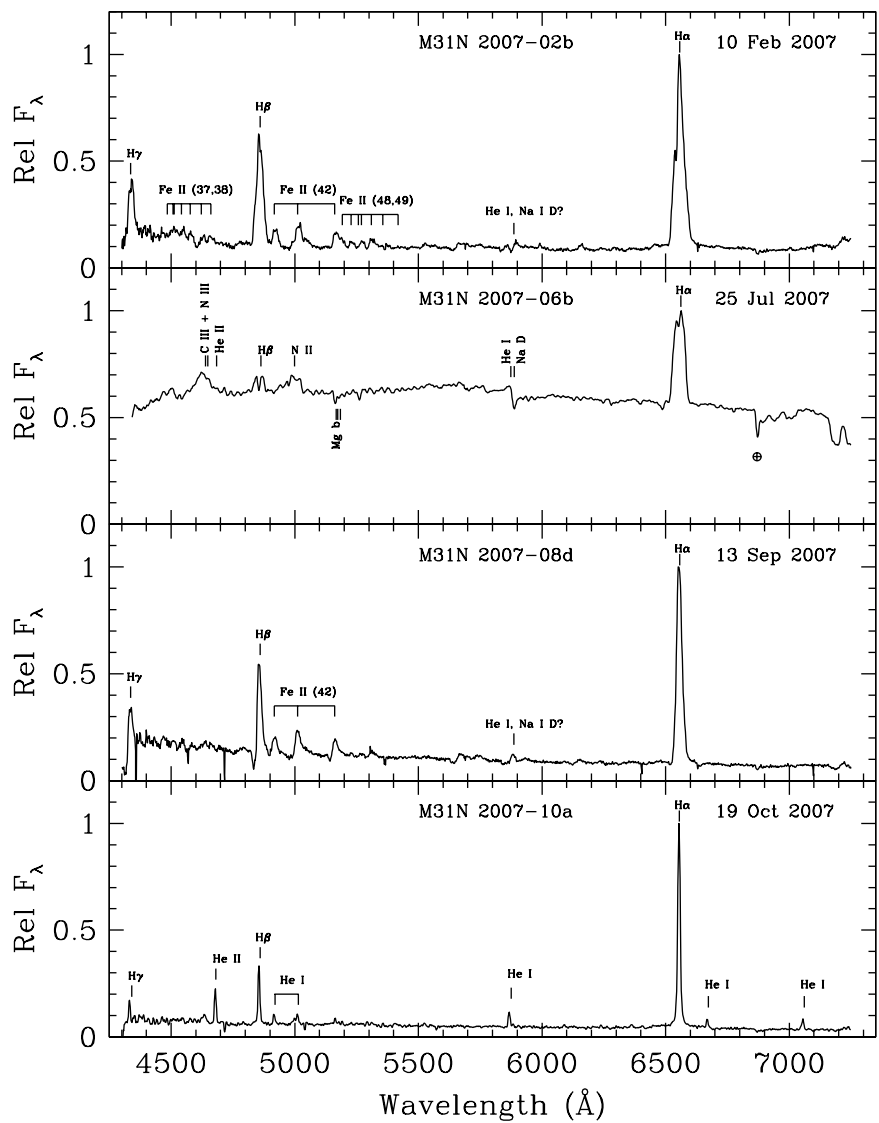}
\caption{Spectra of the M31 novae M31N~2007-02b, 2007-06b, 2007-08d
and
2007-10a, taken 7, 37, 21, and 14 days post discovery, respectively.
M31N~2007-02b is likely to be an Fe~II system, although the
broad component in the Balmer lines is often seen in hybrid novae.
M31N~2007-06b is a He/N nova that originated in the M31 globular cluster
Bol~111~\citep{sha07b}. M31N 2007-08d is a Fe~II system.
M31N~2007-10a is an unusual nova
displaying prominent Balmer and He~I lines. The nova is not typical
of either the Fe~II class (no Fe~II lines) or the He/N class (the lines
are narrow and there is no sign of nitrogen lines).
This is the prototype of our
new class of the narrow-lined He novae (He/Nn).
\label{fig7}}
\end{figure}

\begin{figure}
\includegraphics[angle=0,scale=.60]{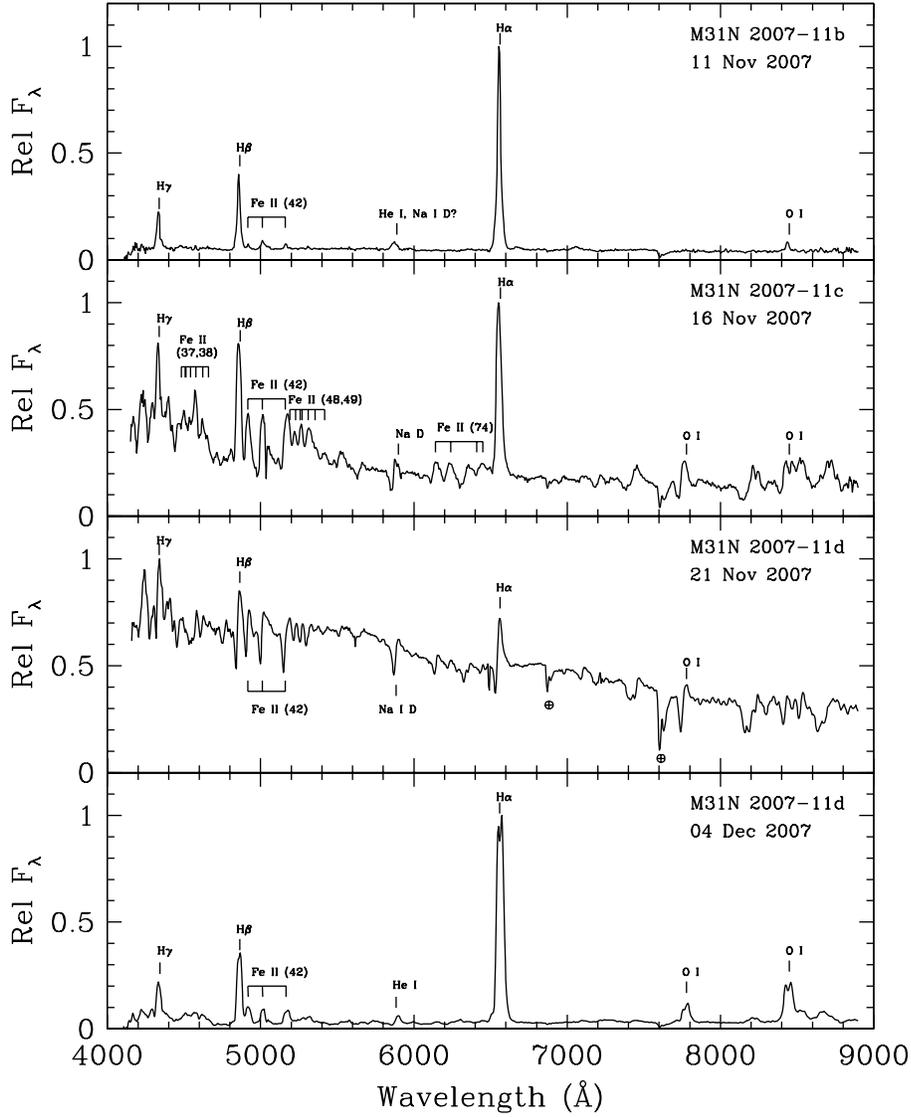}
\caption{Spectra of the M31 novae M31N~2007-11b, 2007-11c, and
2007-11d (two spectra), taken 11, 5, 5, and 18 days post discovery, respectively. All are Fe~II novae. M31N~2007-11d was observed twice,
shortly after eruption when the P~Cyg line profiles were clearly evident,
and roughly two weeks later after the continuum had faded considerably.
\label{fig8}}
\end{figure}

\begin{figure}
\includegraphics[angle=0,scale=.60]{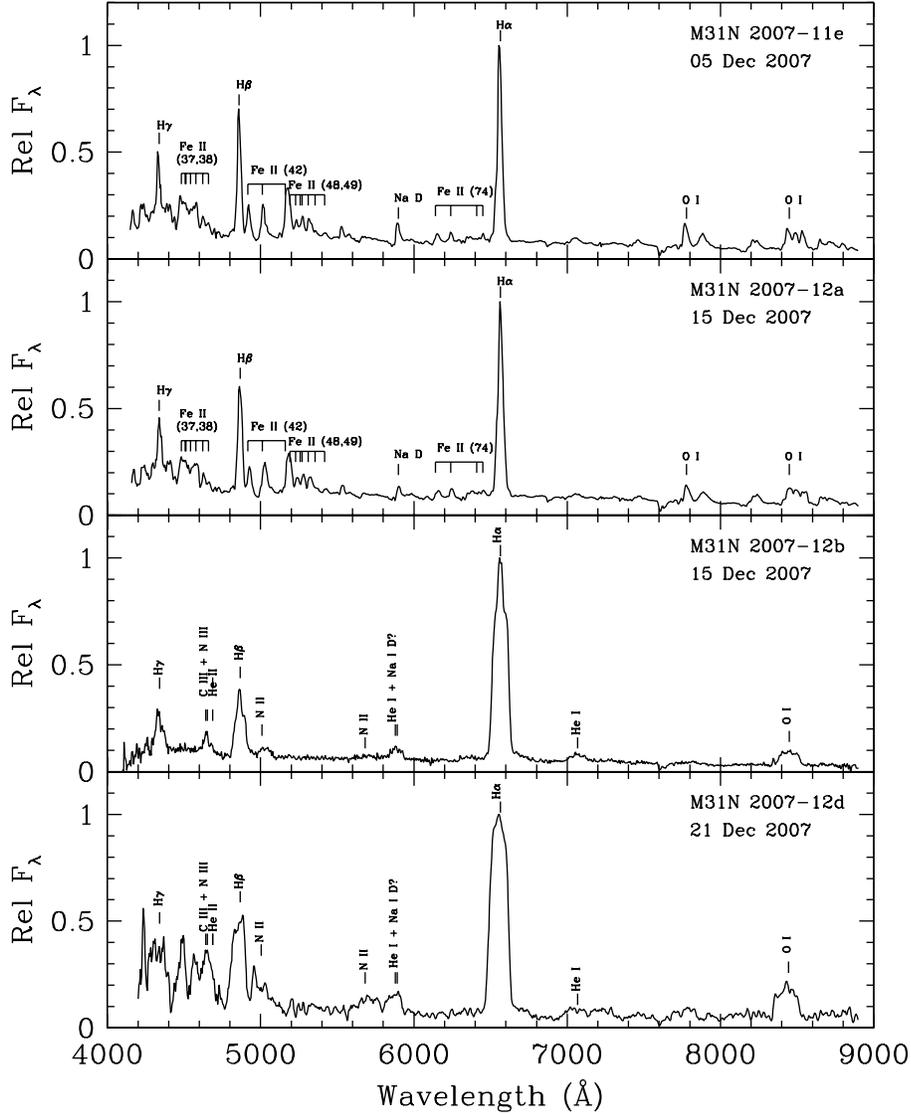}
\caption{Spectra of the M31 novae M31N~2007-11e, 2007-12a,
2007-12b, and
2007-12d, taken 8, 10, 6, and 4 days post discovery, respectively.
M31N~2007-11e and M31N~2007-11a are typical Fe~II novae,
while M31N~2007-12b and M31N~2007-12d are both examples of He/N novae.
\label{fig9}}
\end{figure}

\begin{figure}
\includegraphics[angle=0,scale=.60]{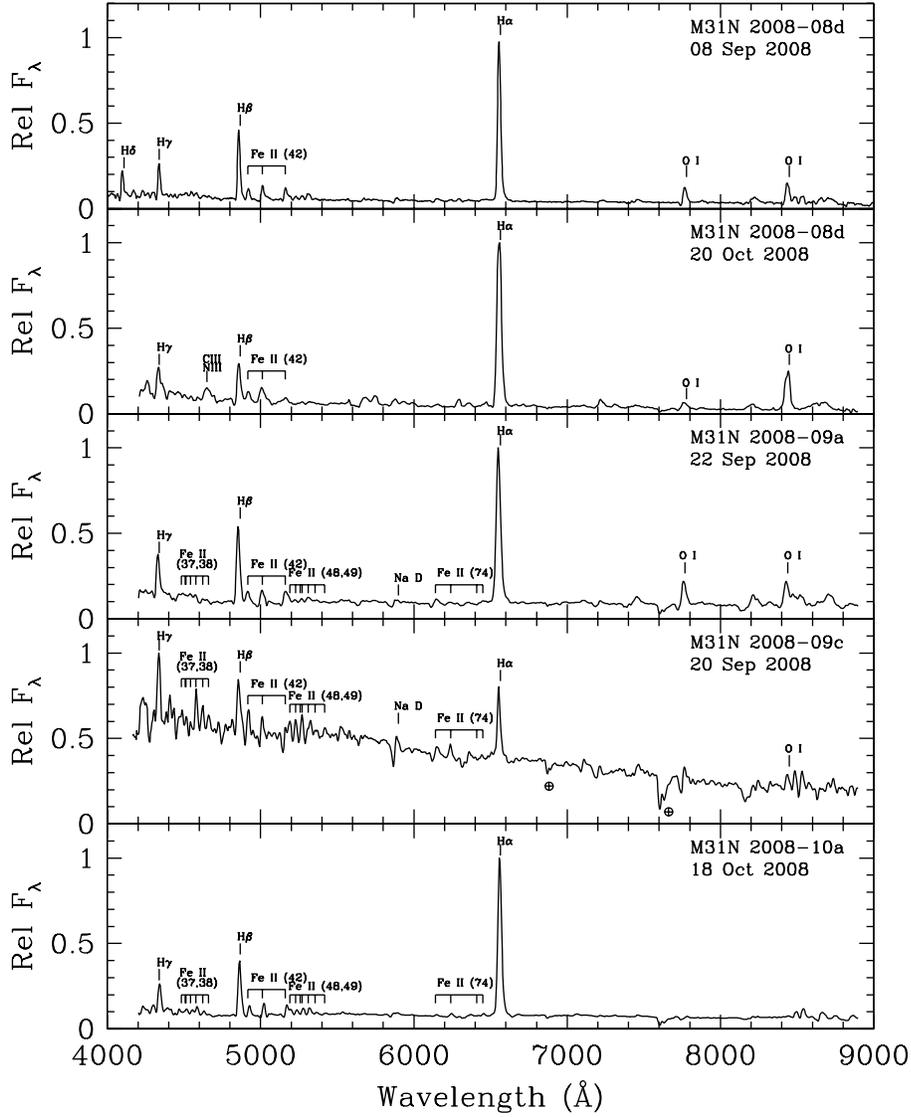}
\caption{Spectra of the M31 novae M31N~2008-08d (two spectra), 2008-09a,
2008-09c, and 2008-10a, taken 14, 57, 10, 6, and 11 days post discovery, respectively. All four novae are of the
Fe~II spectral type.
\label{fig10}}
\end{figure}

\begin{figure}
\includegraphics[angle=0,scale=.60]{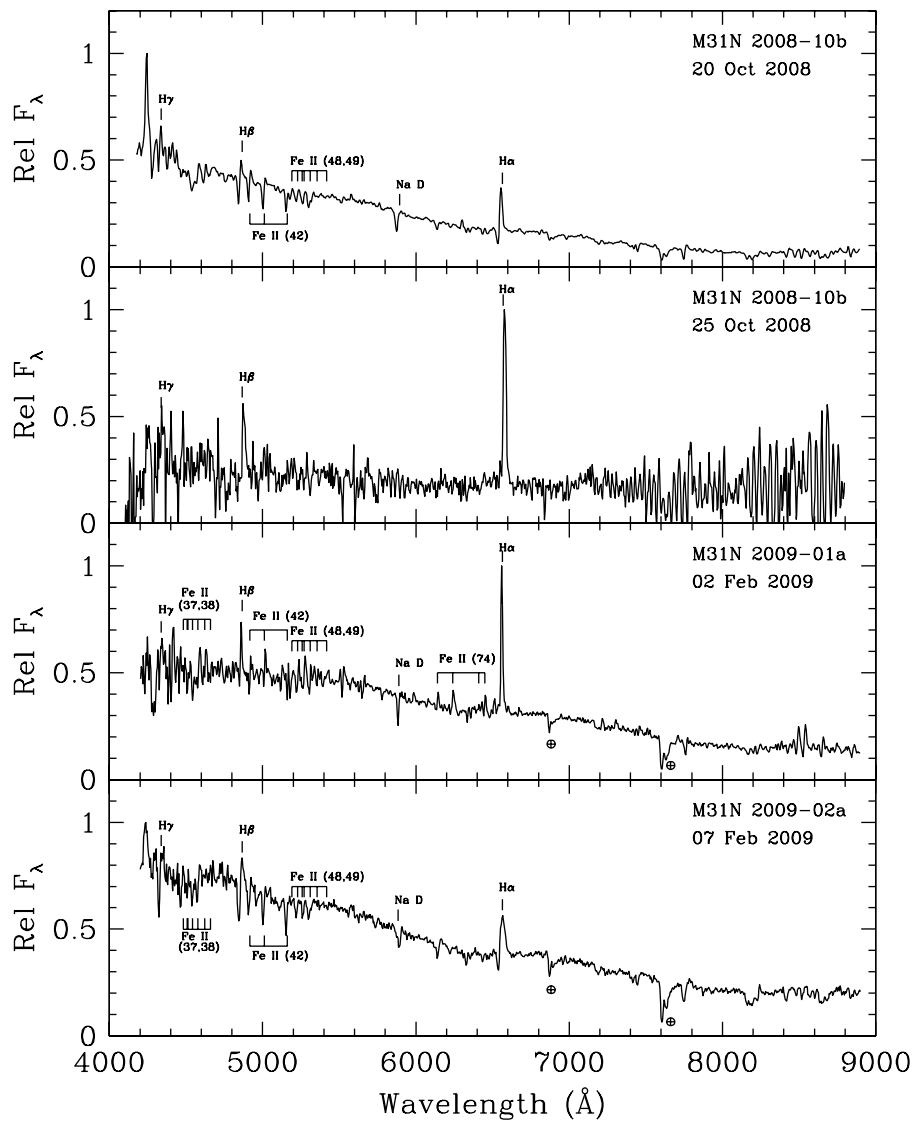}
\caption{Spectra of the M31 novae M31N~2008-10b (two spectra), 2009-01a, and
2009-02a, taken 15, 20, 6, and 2 days, respectively. M31N~2008-10b is a Fe~II nova that was observed twice.
The first spectrum of M31N~2008-10b and the spectra
of and M31N~2009-01a and
M31N~2009-02a display P~Cyg profiles indicating that they were
observed shortly after eruption.
\label{fig11}}
\end{figure}

\begin{figure}
\includegraphics[angle=0,scale=.60]{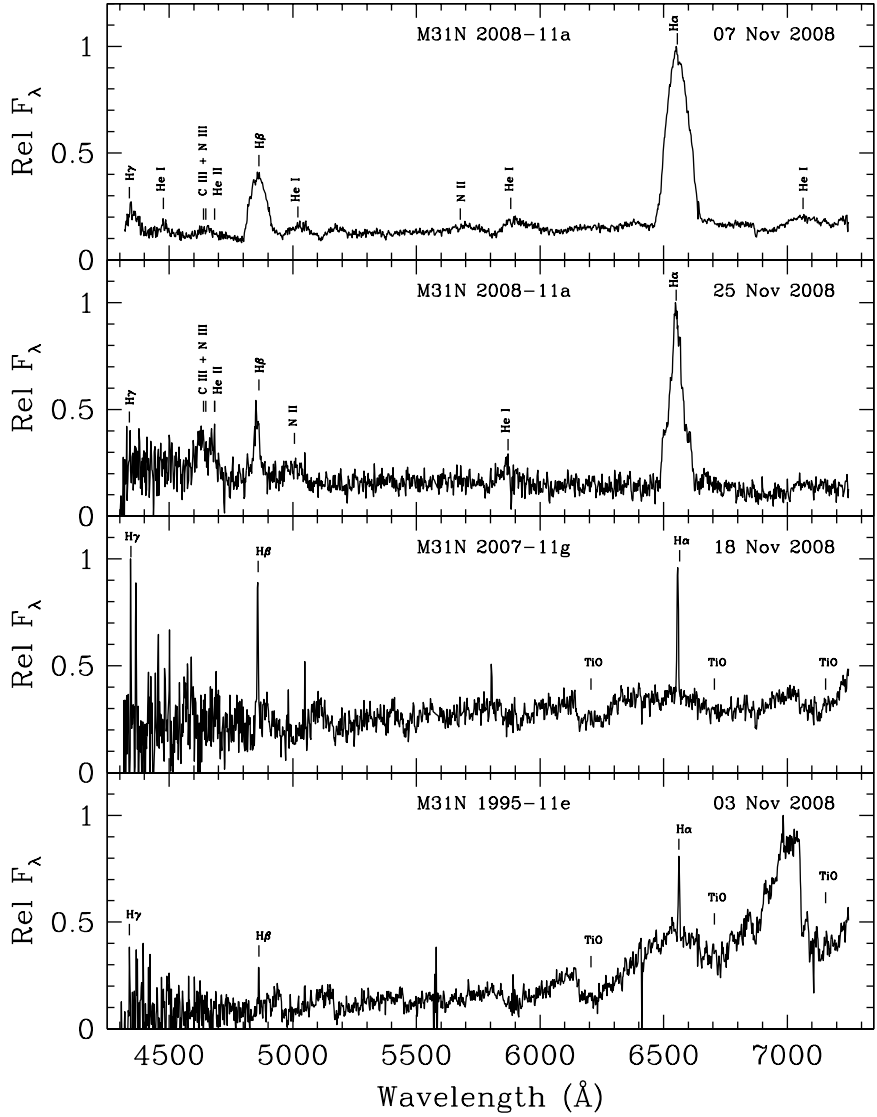}
\caption{Spectra of the M31 nova M31N~2008-11a, a classic He/N nova,
taken 4 and 22 days post discovery.
The final two objects, M31N~2007-11g and M31N~1995-11e,
are examples of long-period variable stars that were mistakenly
classified as novae.
\label{fig12}}
\end{figure}

Between the Lick and HET observations,
we obtained a total of 53 spectra of M31 nova candidates
(21 with the Shane and 32 with the HET). These
data, shown in Figures~\ref{fig1}--\ref{fig12}, include
46 M31 novae (5 with two spectra each),
and two long-period Mira variables originally thought to be novae.
Because the observations were made under a variety of atmospheric
conditions with the stellar image typically overfilling the spectrograph 
slit, our data can not be considered spectrophotometric. Thus, all 
spectra have been displayed on a relative flux scale.
In the caption for each figure
we have indicated the time elapsed between discovery of the nova
and the date of our spectroscopy.
Although the date of discovery does not necessarily reflect the time of maximum
light, M31 has been monitored much more frequently over the past decade, and
we estimate that
the discovery date is likely within a few days of peak brightness
in the majority of recent cases.

\subsection{Photometry}

To complement our spectroscopic survey, we were able to amass sufficient
photometric observations to produce light curves
for many of the novae in our survey.
Our primary motivation was to measure nova fade rates ($t_2$) that could
then be correlated with other properties, such as spectroscopic class.
The photometric data consist both
of targeted (mostly $B$ and $V\/$-band) observations,
which were obtained primarily with the
Liverpool Telescope \citep[LT;][]{ste04}
and the Faulkes Telescope North \citep[FTN;][]{bug07},
and survey images
(mainly $R$ band), taken over the years with a variety of
telescopes.
The LT and FTN data were reduced using a combination of IRAF
and Starlink software, calibrated using standard stars from
\citet{lan92}, and checked against secondary standards
from \citet{mag92}, \citet{hai94}, and \citet{mas06}.

Our extensive $R\/$-band observations were taken largely from the
photometric database compiled by one of us (K.H.) as part
of an ongoing program to monitor nova light curves in M31.
These data include both
survey and targeted images taken with various telescopes with diameters of
0.26--6~m. Most of the images come from the 0.65-m telescope of
the Ond\v{r}ejov Observatory (operated partly by the Charles University,
Prague) and the 0.35-m telescope in the private observatory of K.H. at
Lelekovice.
Standard reduction procedures for raw CCD images were applied
(bias and dark-frame subtraction and flat-field correction) using
the SIMS\footnote{\tt http://ccd.mii.cz/} and Munipack\footnote
{\tt http://munipack.astronomy.cz/} programs.
Reduced images of the same series were coadded to improve
the signal-to-noise ratio; the total exposure time of these series
varied from a few minutes
up to about one hour. To facilitate nova detection,
the gradient of the galaxy background
was flattened by the spatial median filter
using SIMS.
Photometric and astrometric measurements
of the novae were then performed using
``Optimal Photometry'' (based on fitting of point-spread function 
profiles) in GAIA\footnote{\tt http://www.starlink.rl.ac.uk/gaia} and
APHOT (a synthetic aperture photometry and astrometry software package
developed by M. Velen and P. Pravec at the Ond\v{r}ejov Observatory;
see Pravec et al. [1994]), respectively.
Five to ten comparison stars were used for individual brightness
measurements.
$B$, $V$, and $R$ magnitudes for comparison stars located in the M31
field were taken from~\citet{mas06}.
Finally,
in the case of images taken using the Sloan Digital Sky Survey
(SDSS) $g'$ and $r'$ filters, we computed $g'$ and $r'$ magnitudes for
comparison stars from $BVRI$ magnitudes taken from \citet{mas06}
using empirical color transformations between the SDSS $ugriz$ system
and the Johnson-Cousins $UBVRI$ system published by \citet{jor06}.
A summary of all of our photometric observations is presented in Table~\ref{photobs}.

\begin{figure}
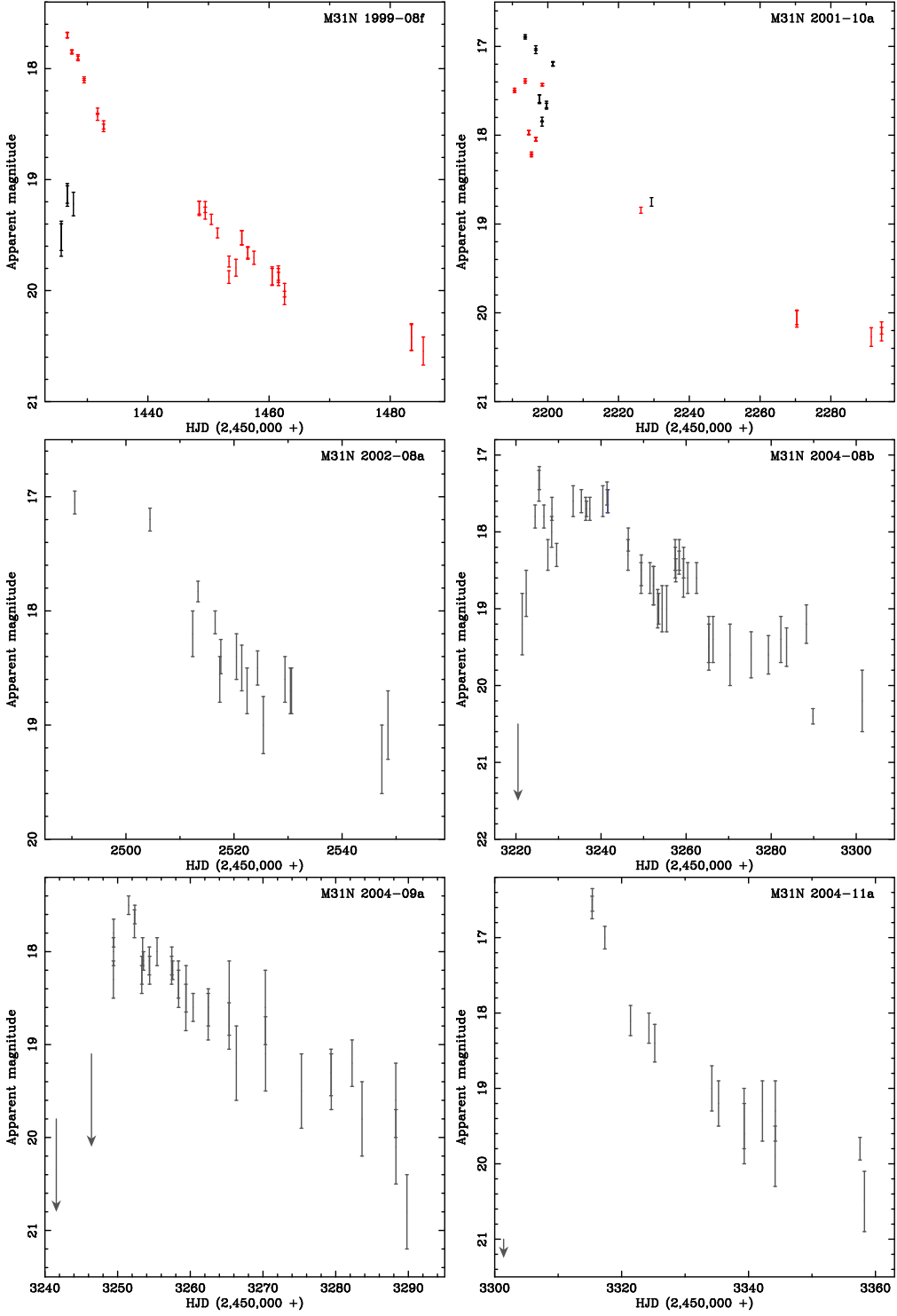

\includegraphics[width=0.40\textwidth]{f13a.eps}
\includegraphics[width=0.40\textwidth]{f13b.eps}
\newline
\includegraphics[width=0.40\textwidth]{f13c.eps}
\includegraphics[width=0.40\textwidth]{f13d.eps}
\newline
\includegraphics[width=0.40\textwidth]{f13e.eps}
\includegraphics[width=0.40\textwidth]{f13f.eps}
\caption{Nova light curves. The uncertainties in the photometric
measurements are shown as vertical bars with the following
colors representing the different bandpasses:
$B$ -- blue;
$V$ -- green;
$R$ -- dark grey;
$r'$ -- red;
$i'$ -- black;
$z'$ -- light grey.
Upper flux limits are indicated by downward facing arrows.
\label{fig13}}
\end{figure}

\begin{figure}
\includegraphics[width=0.40\textwidth]{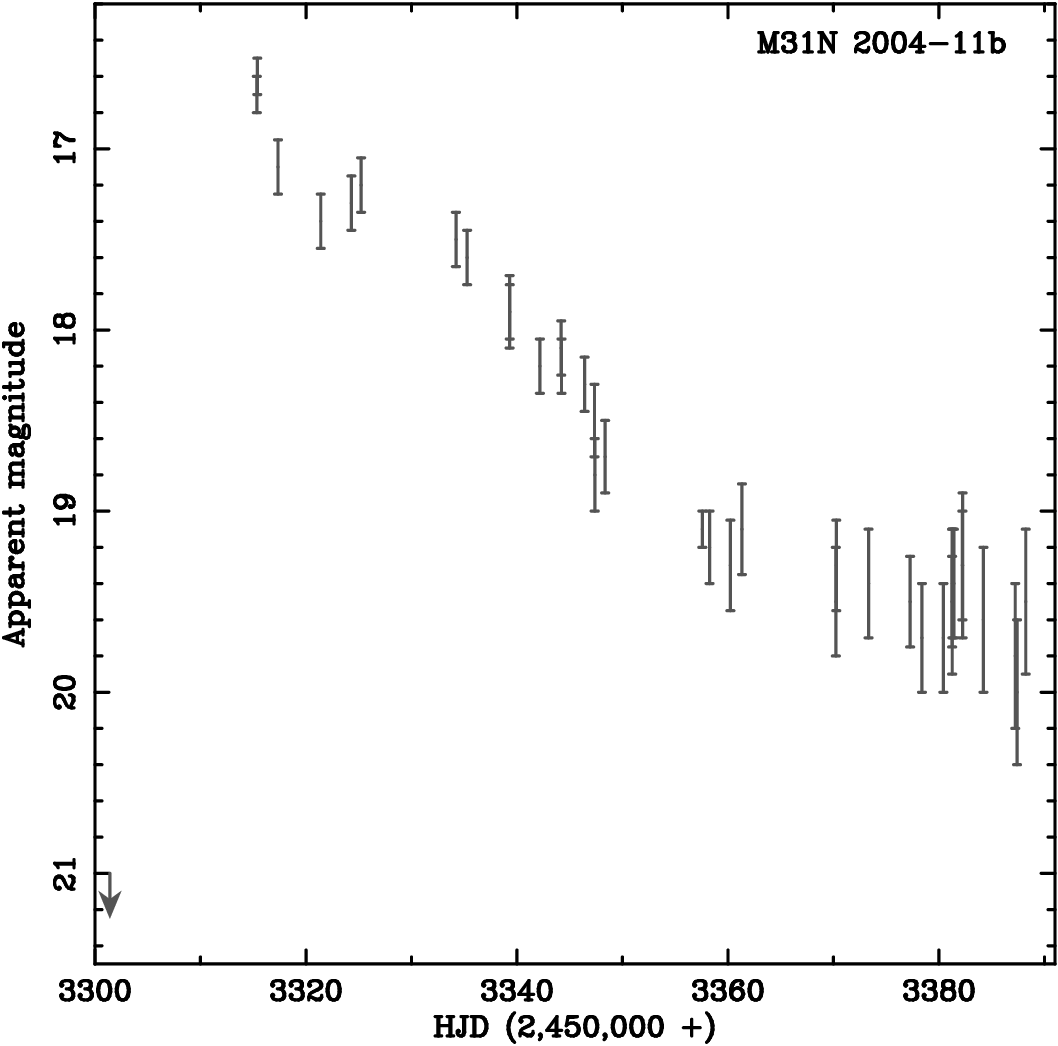}
\includegraphics[width=0.40\textwidth]{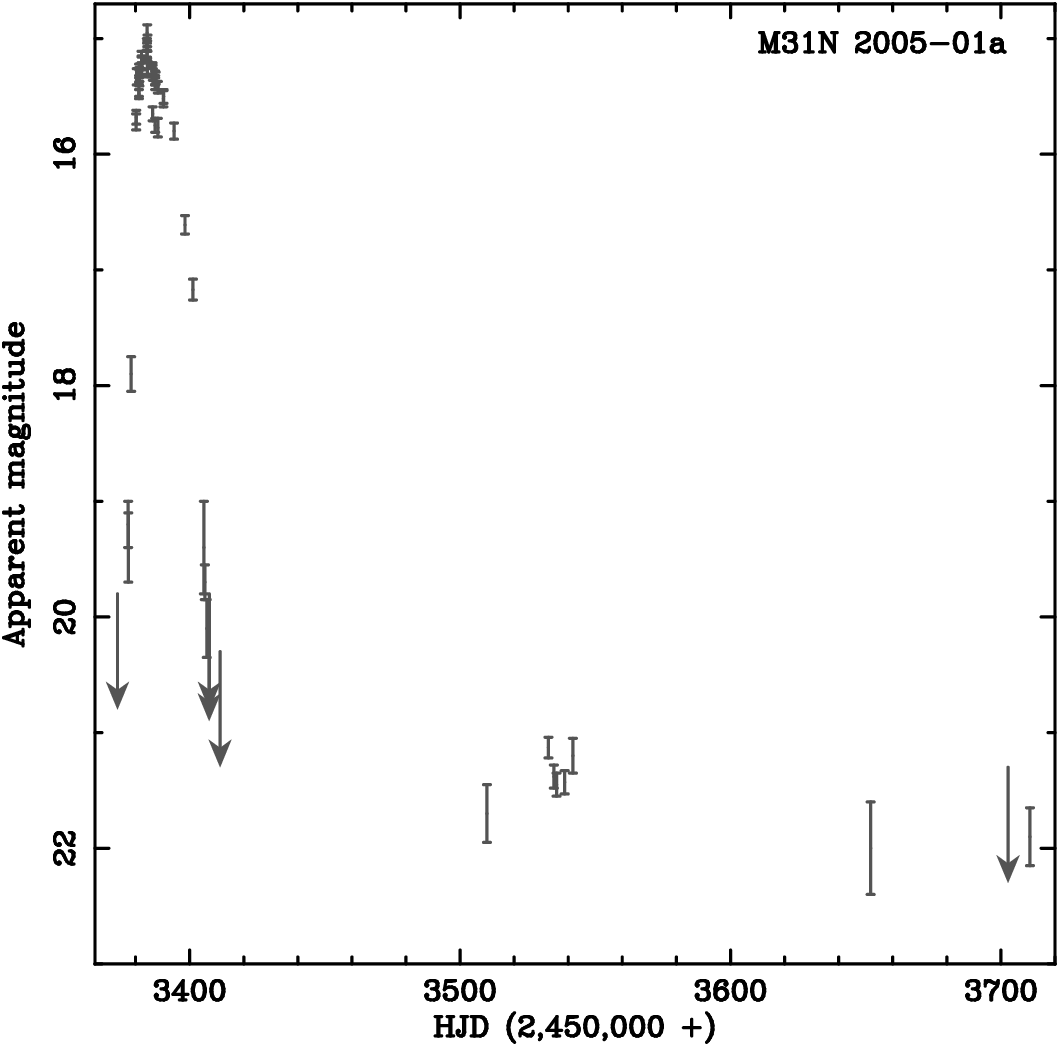}
\newline
\includegraphics[width=0.40\textwidth]{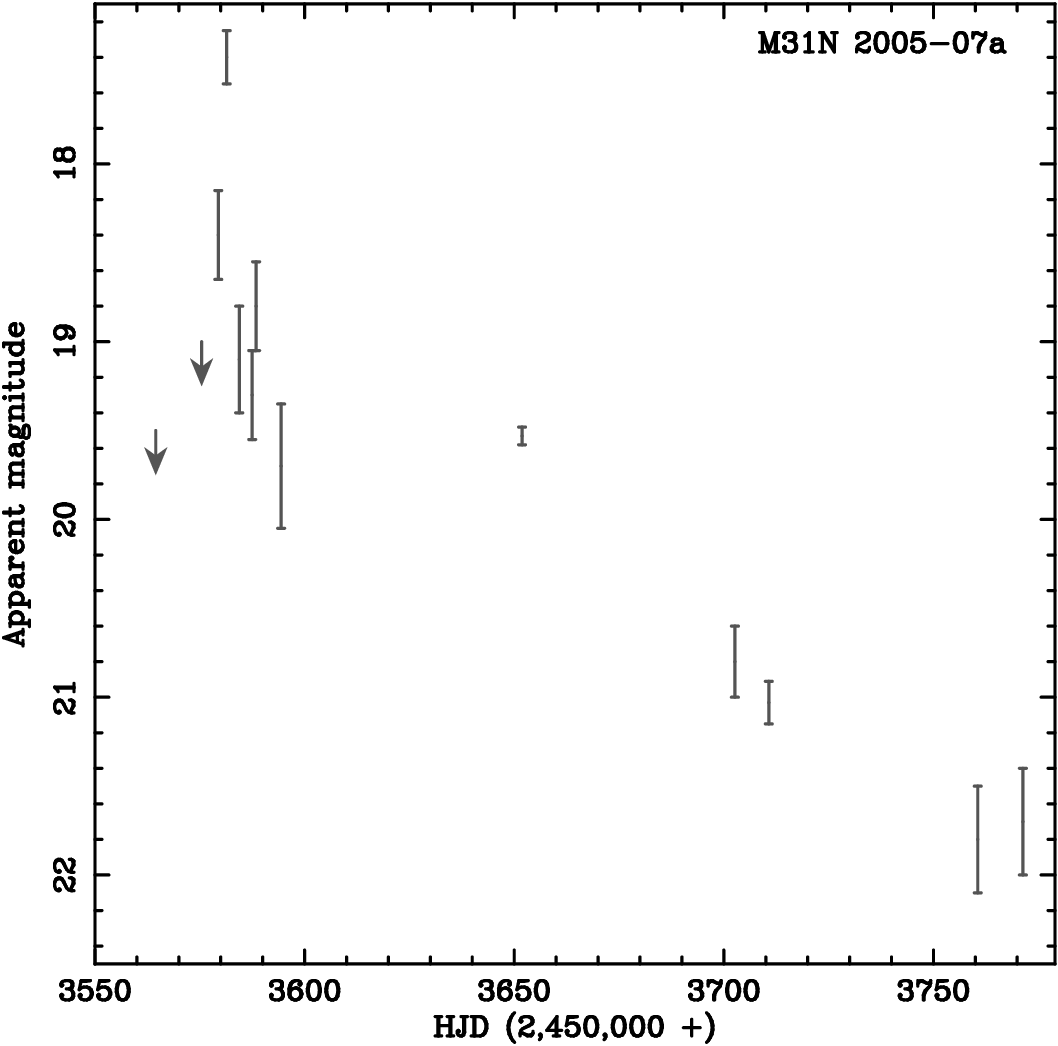}
\includegraphics[width=0.40\textwidth]{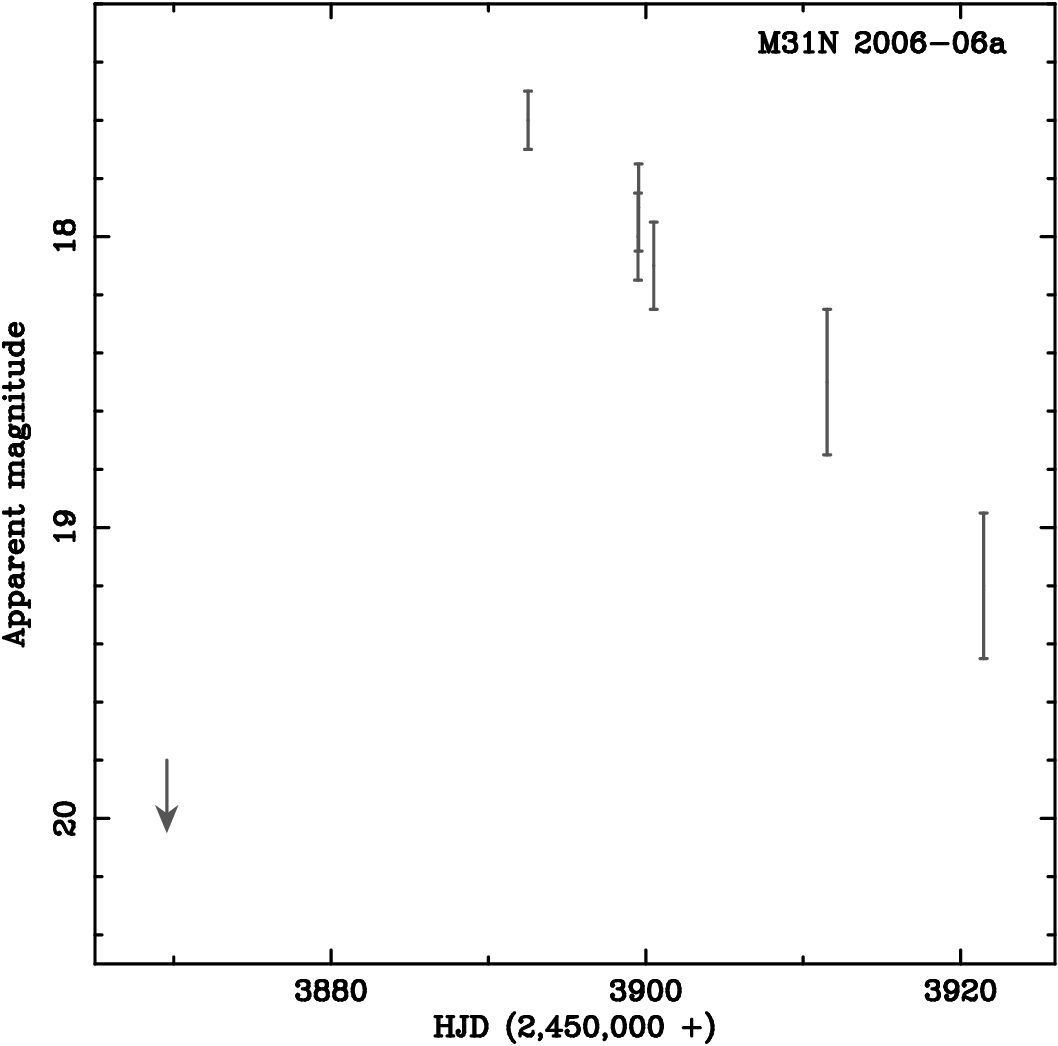}
\newline
\includegraphics[width=0.40\textwidth]{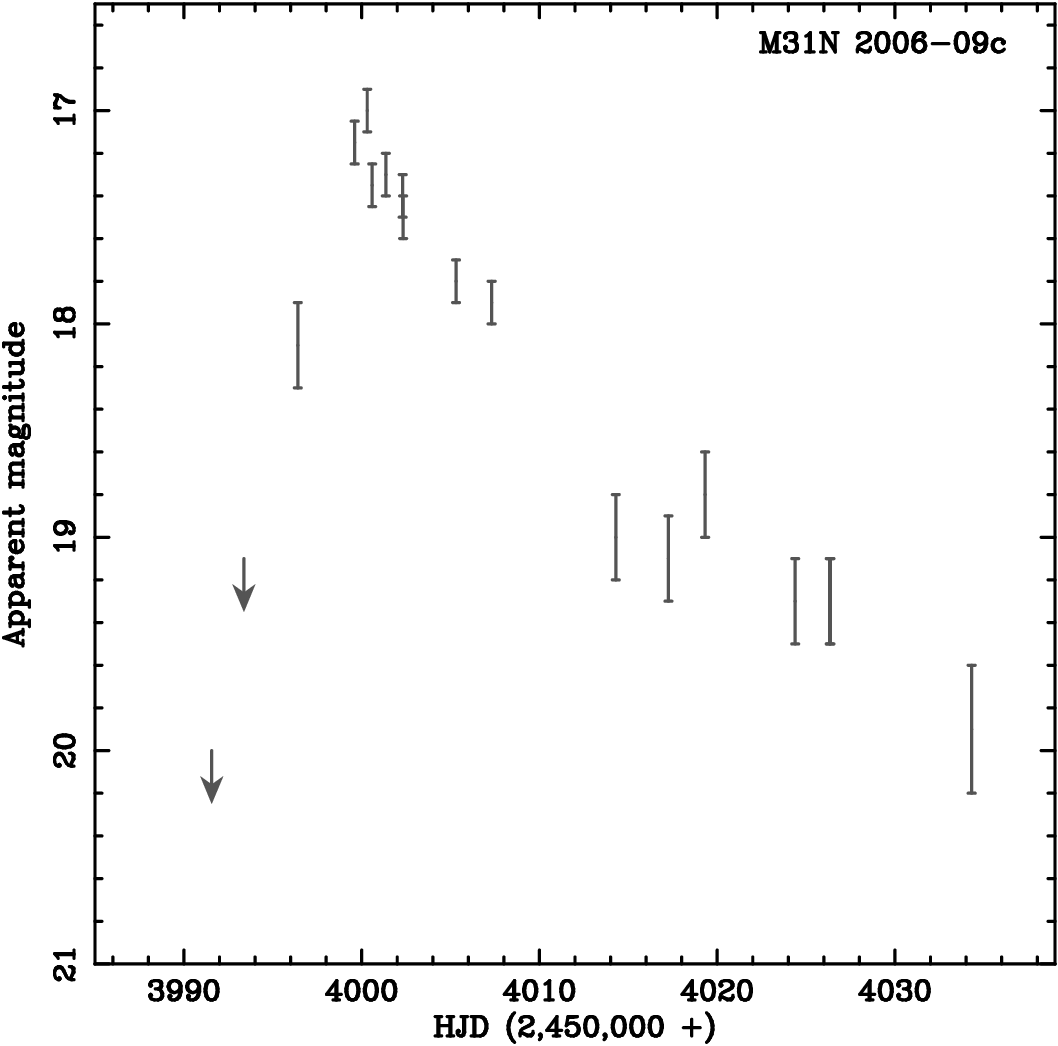}
\includegraphics[width=0.40\textwidth]{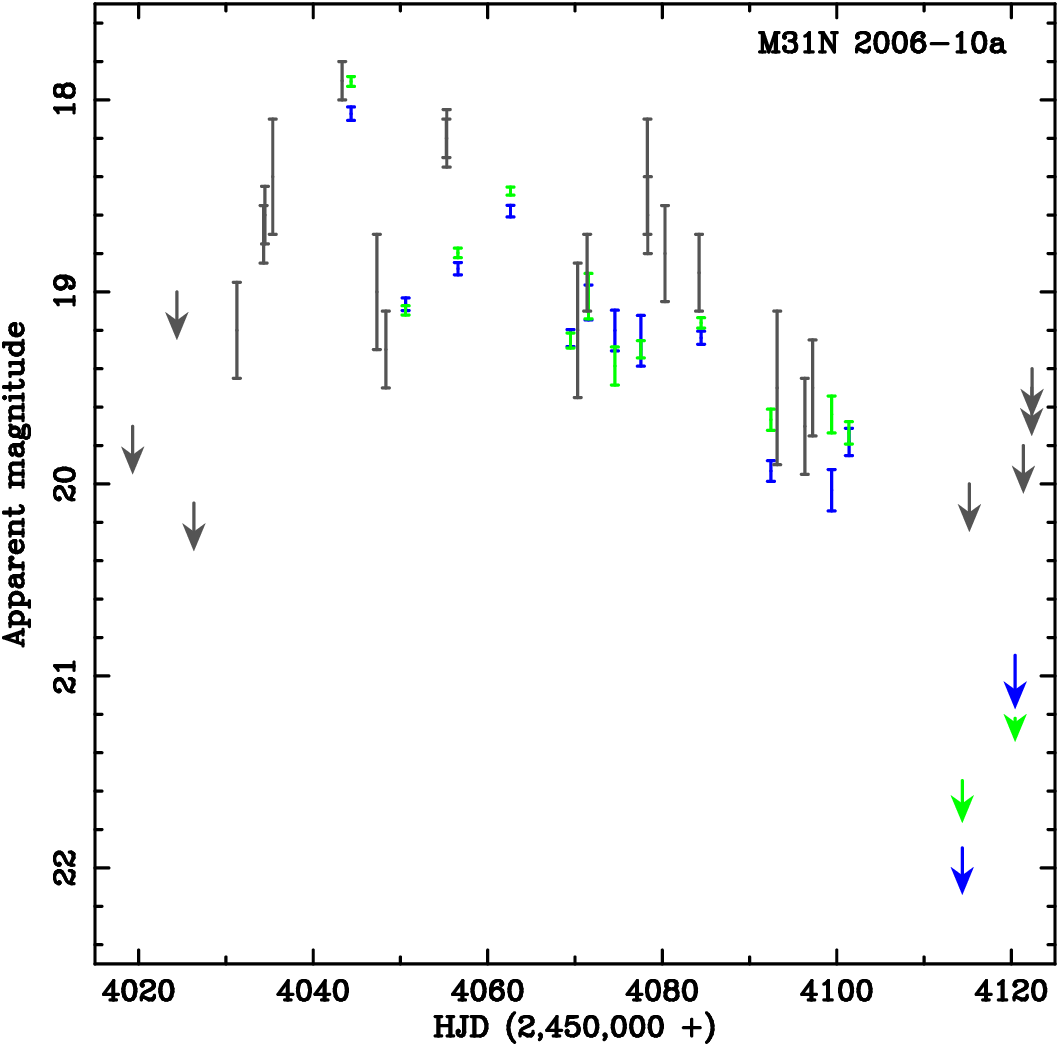}
\caption{Nova light curves (continued). See Fig.~\ref{fig13} for details.\label{fig14}}
\end{figure}

\begin{figure}
\includegraphics[width=0.40\textwidth]{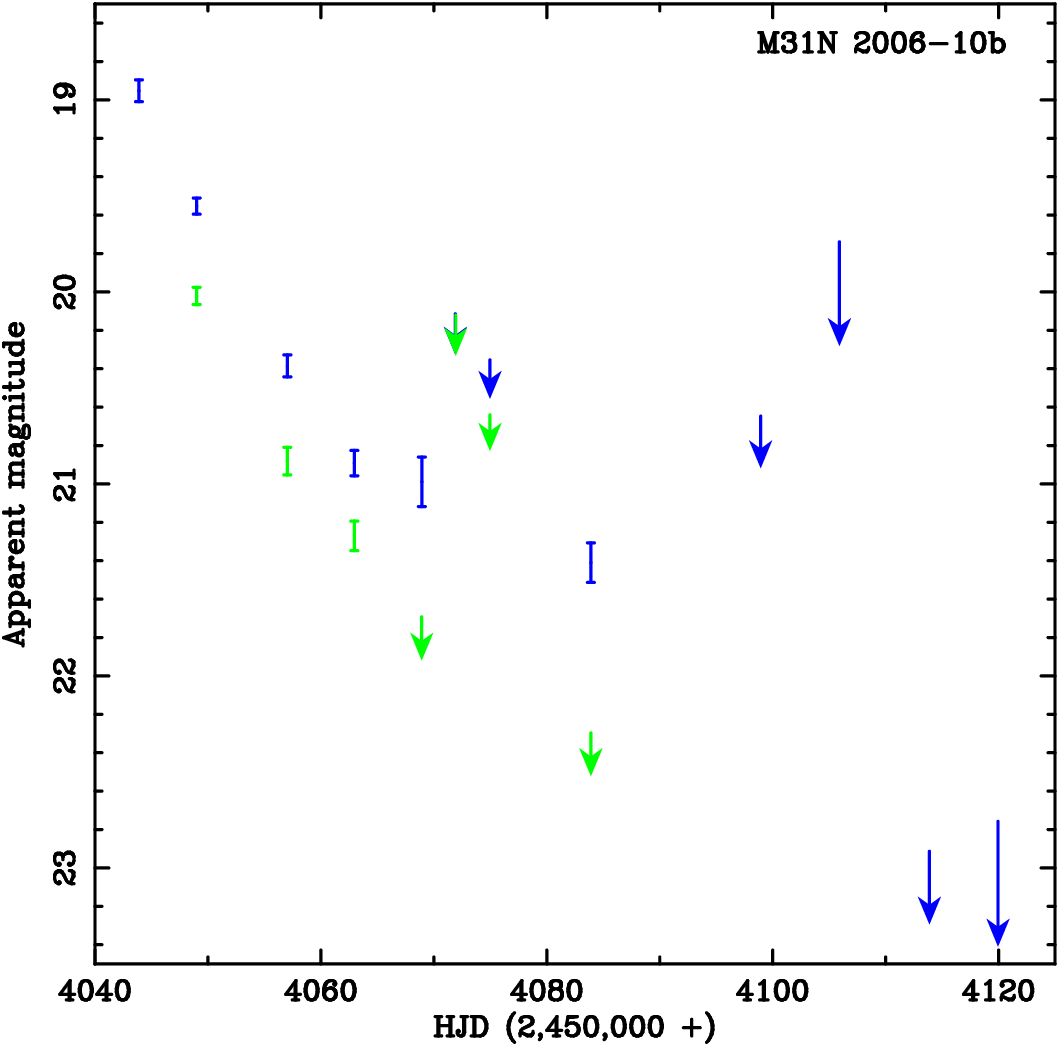}
\includegraphics[width=0.40\textwidth]{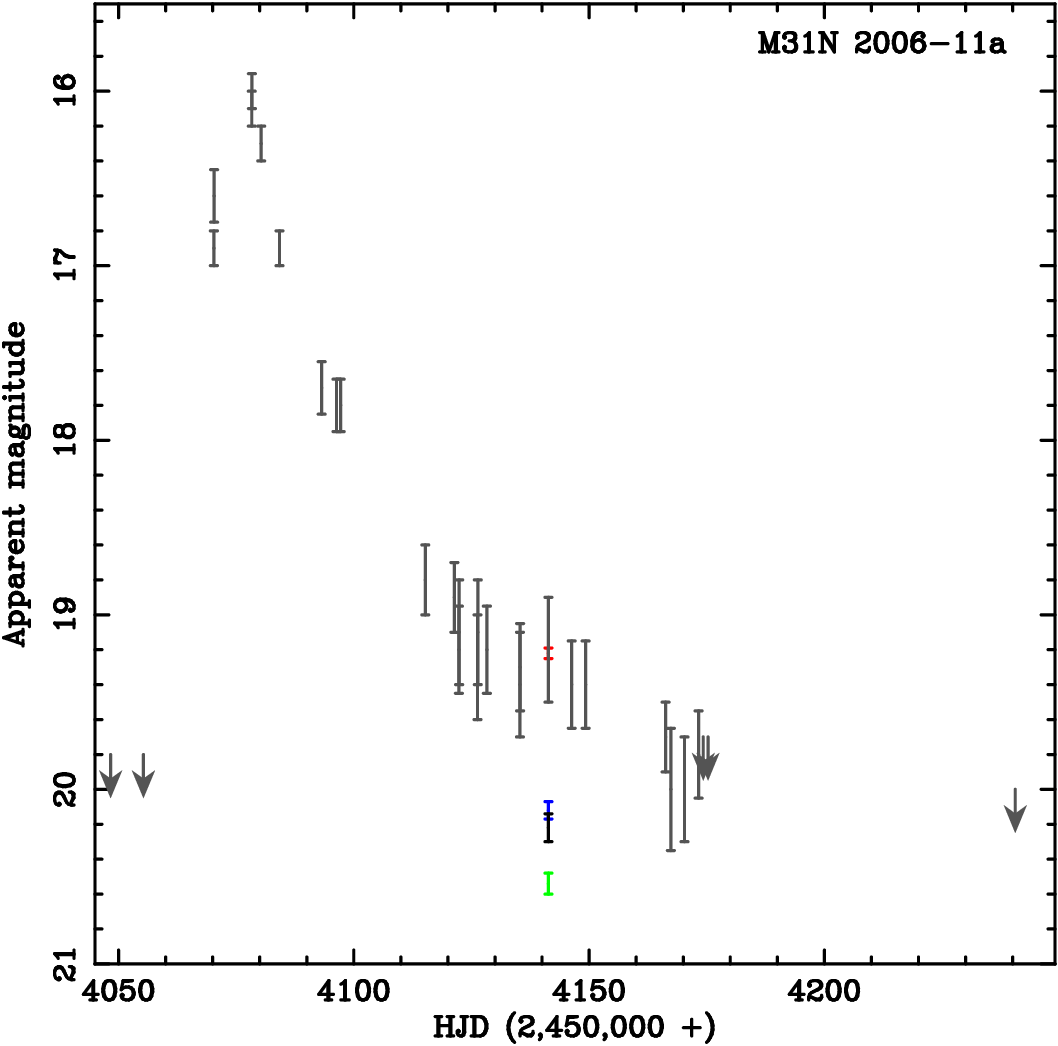}
\newline
\includegraphics[width=0.40\textwidth]{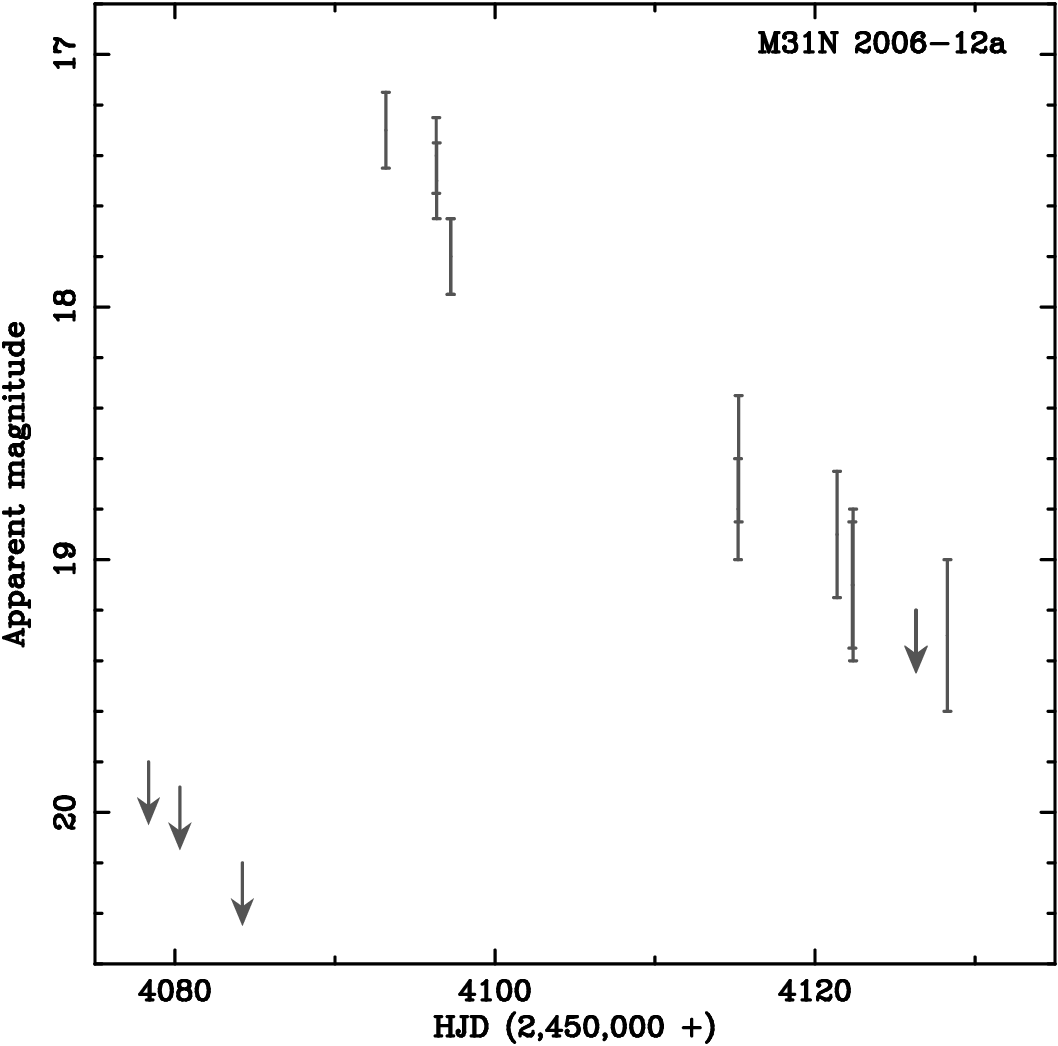}
\includegraphics[width=0.40\textwidth]{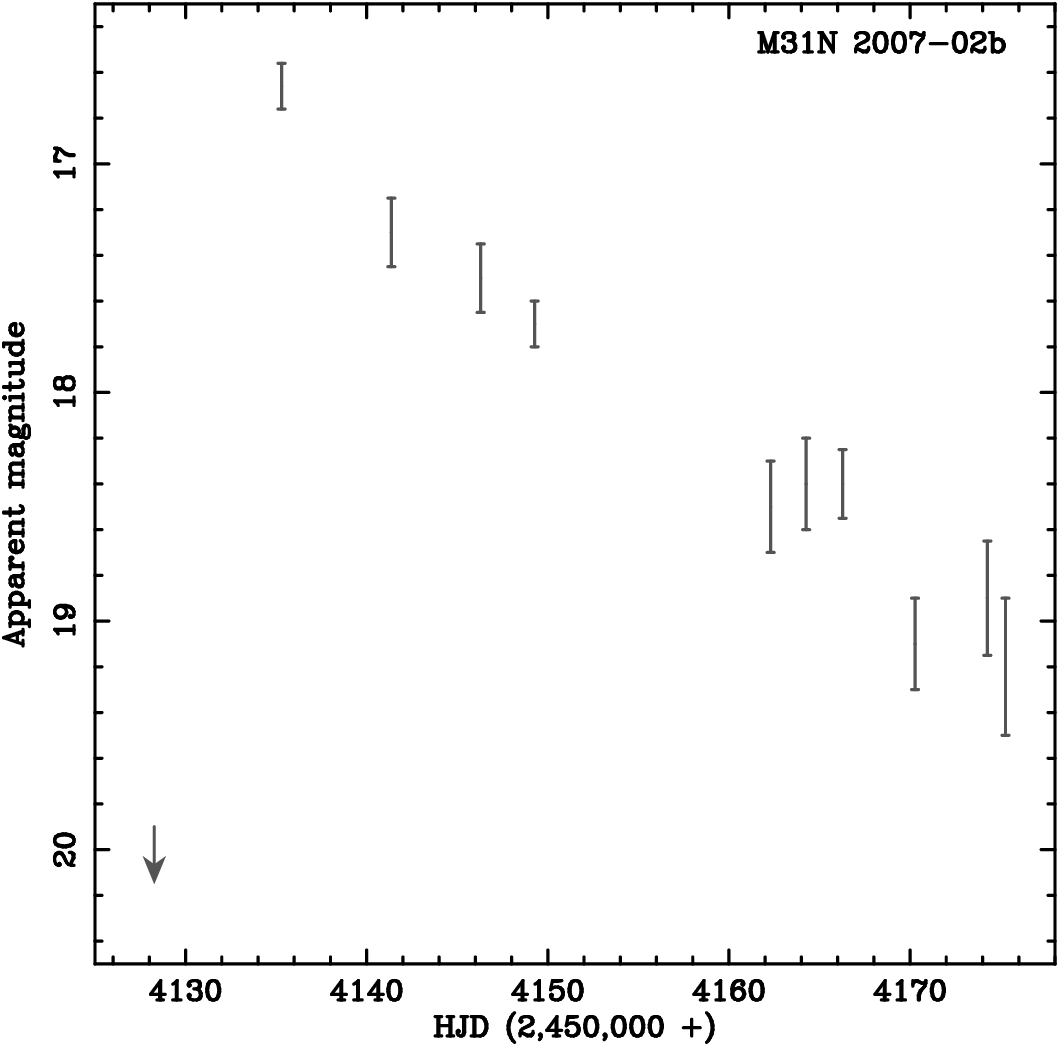}
\newline
\includegraphics[width=0.40\textwidth]{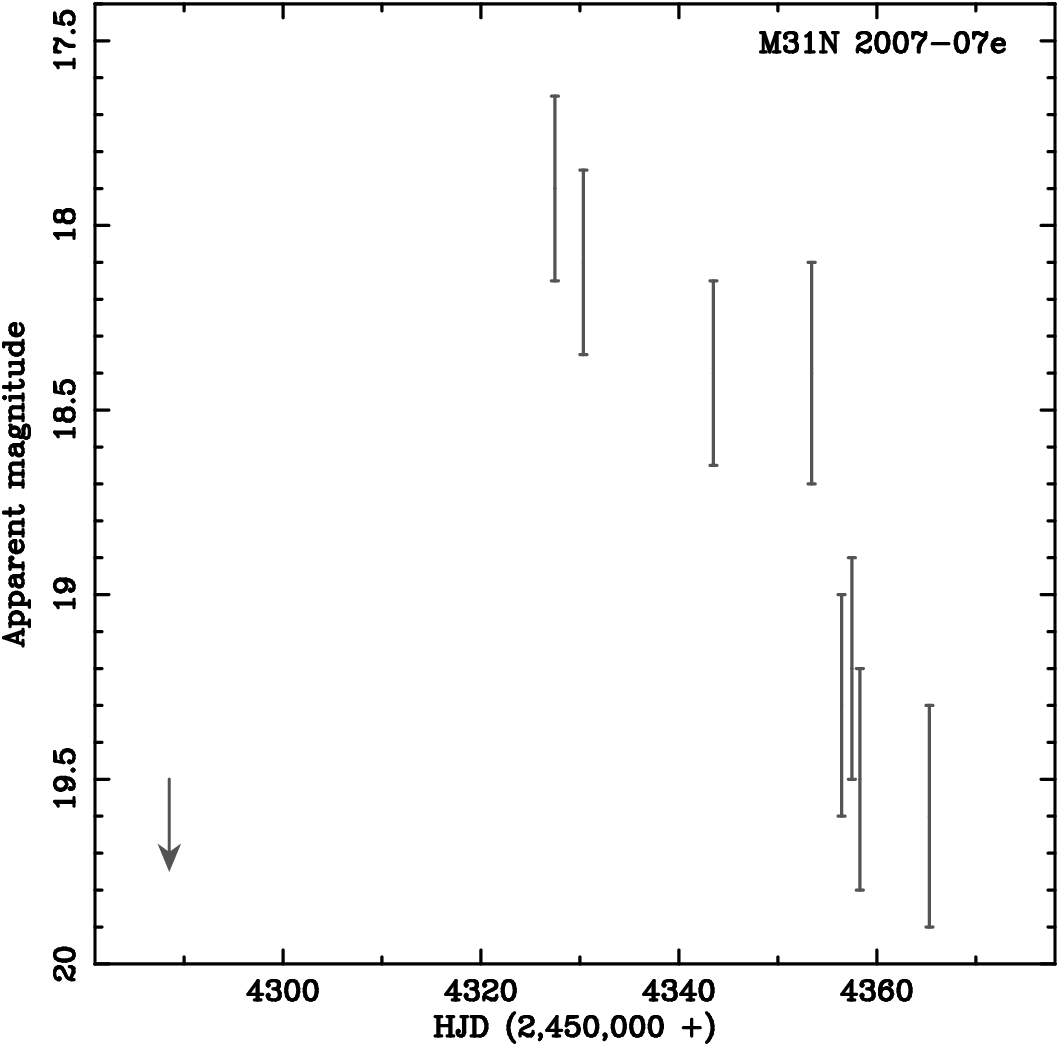}
\includegraphics[width=0.40\textwidth]{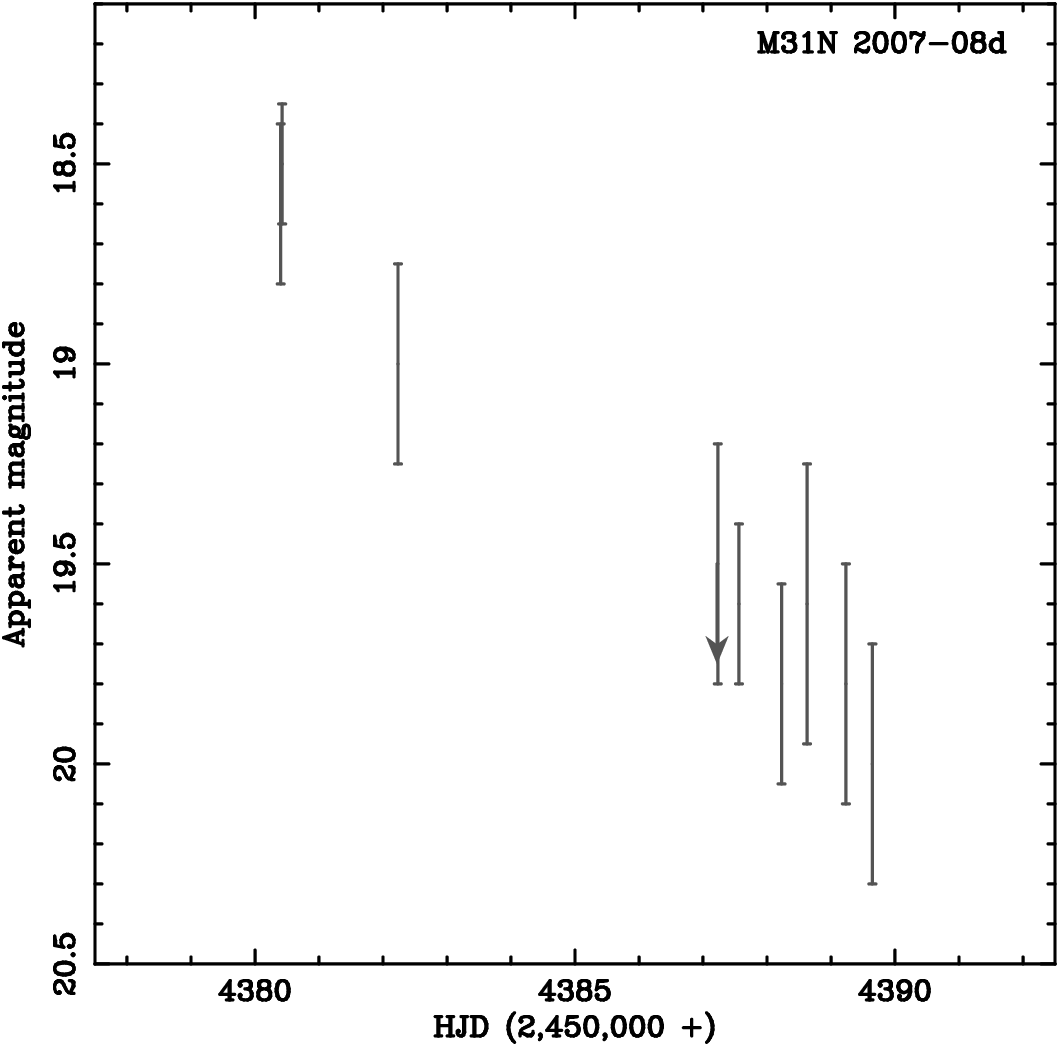}
\caption{Nova light curves (continued). See Fig.~\ref{fig13} for details.\label{fig15}}
\end{figure}

\begin{figure}
\includegraphics[width=0.40\textwidth]{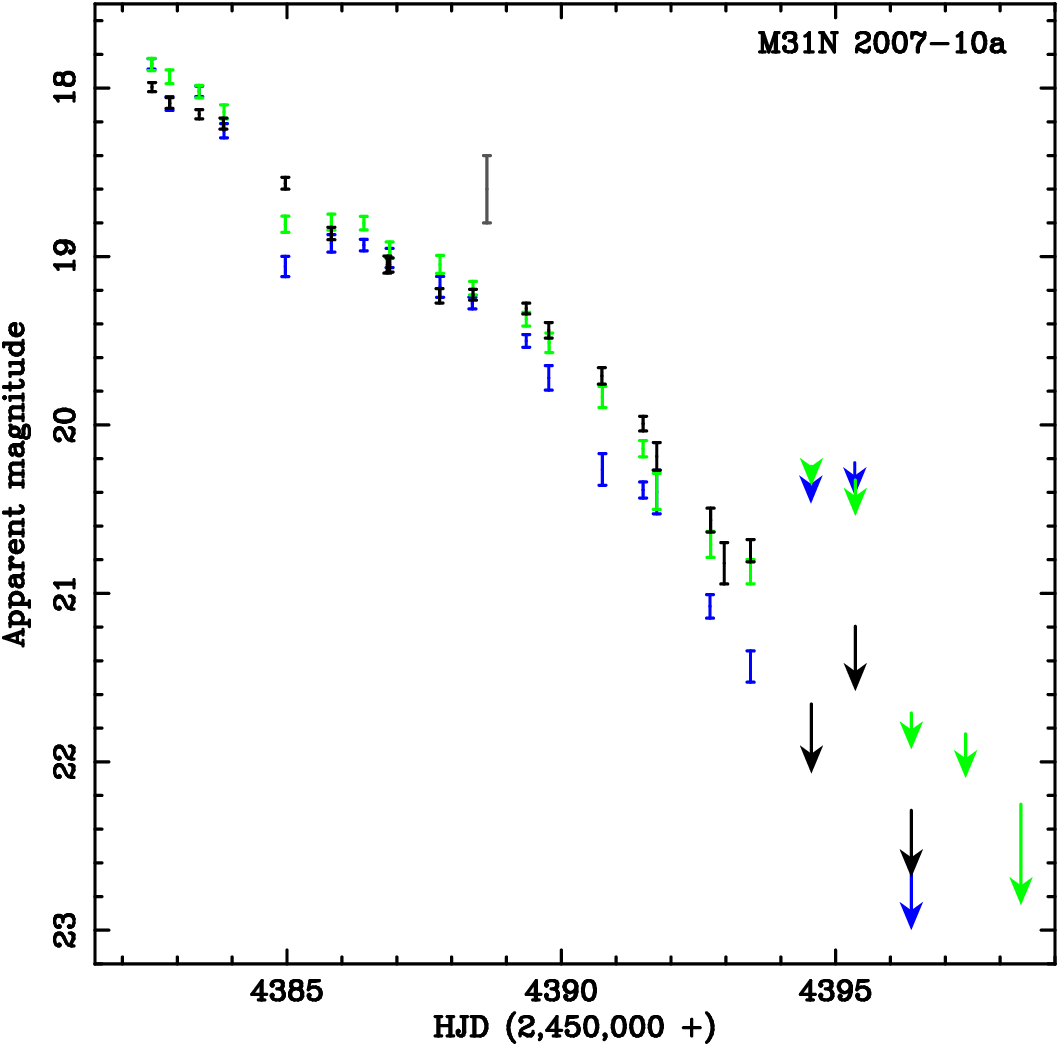}
\includegraphics[width=0.40\textwidth]{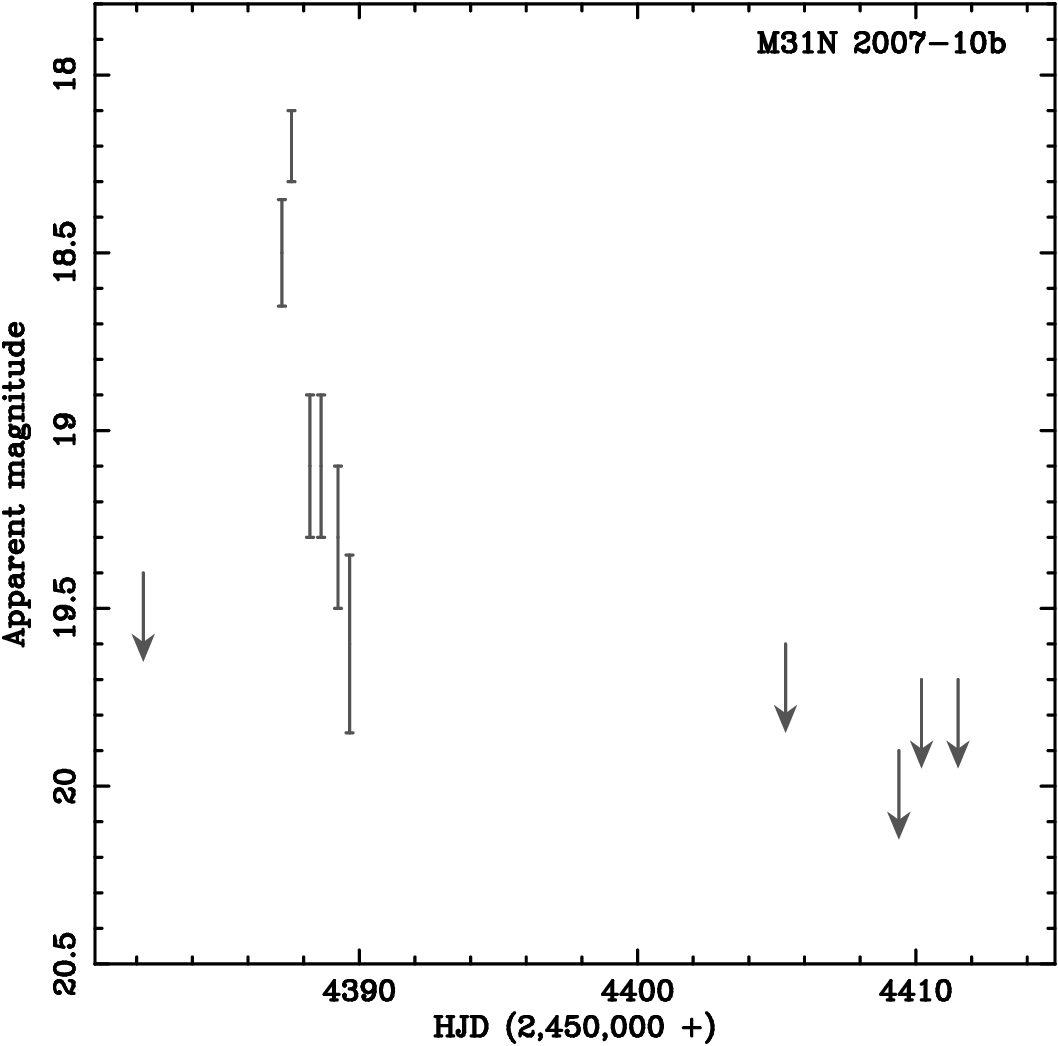}
\newline
\includegraphics[width=0.40\textwidth]{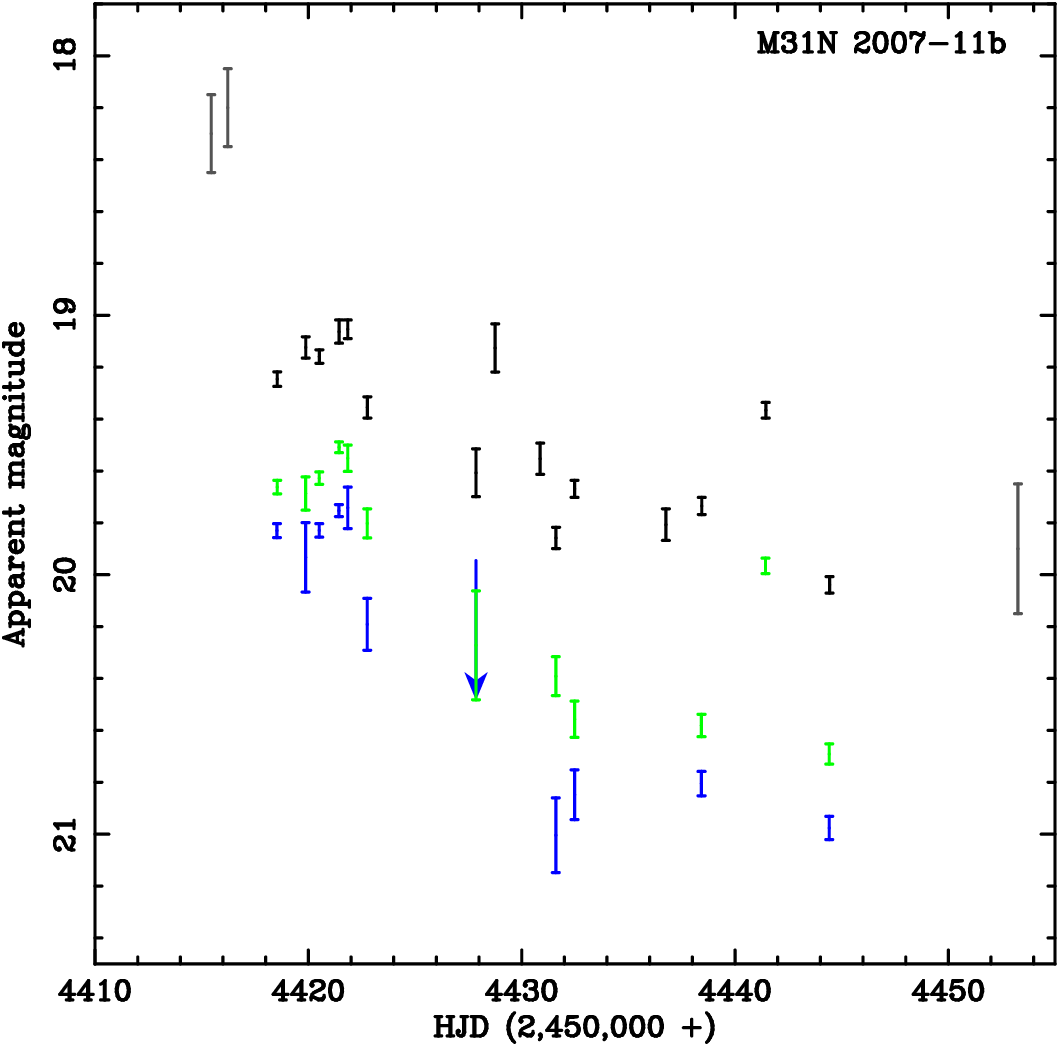}
\includegraphics[width=0.40\textwidth]{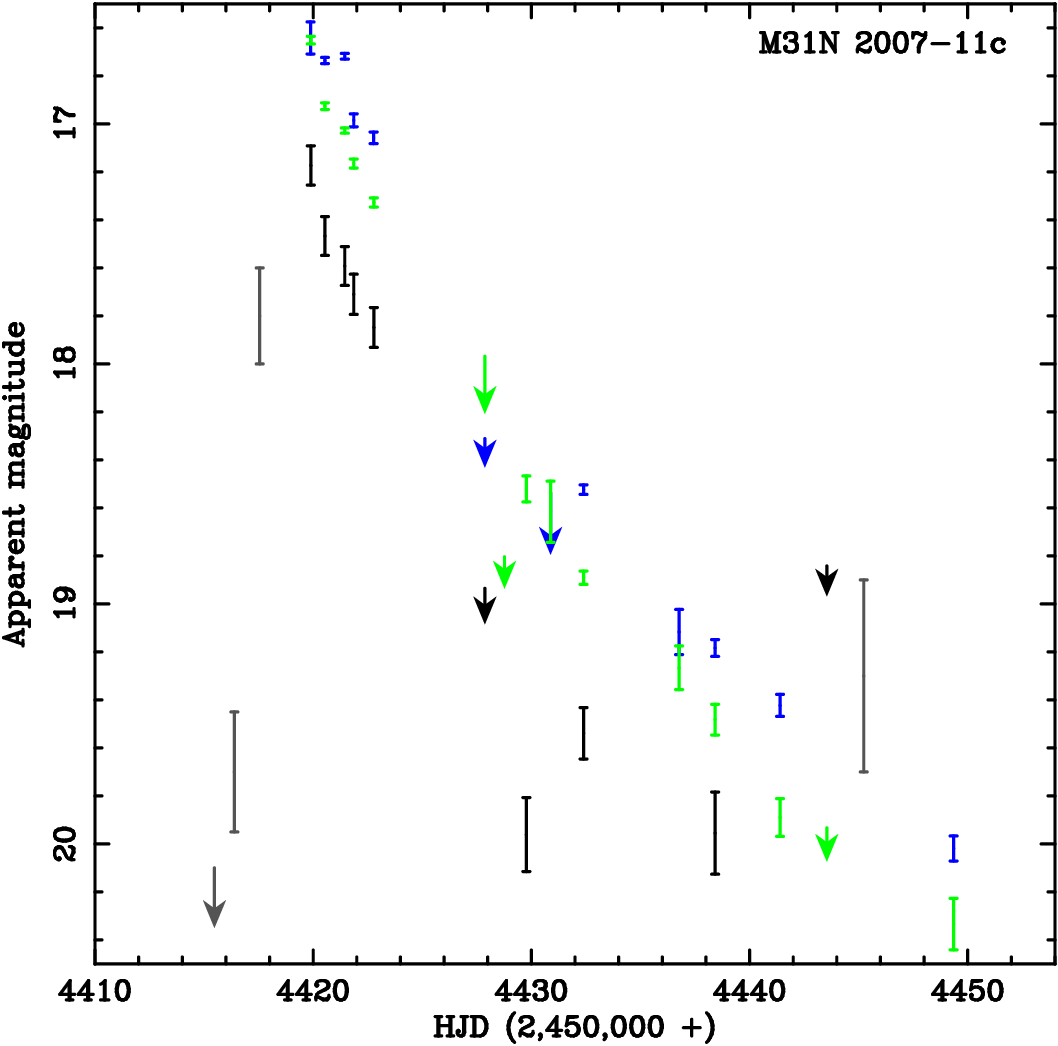}
\newline
\includegraphics[width=0.40\textwidth]{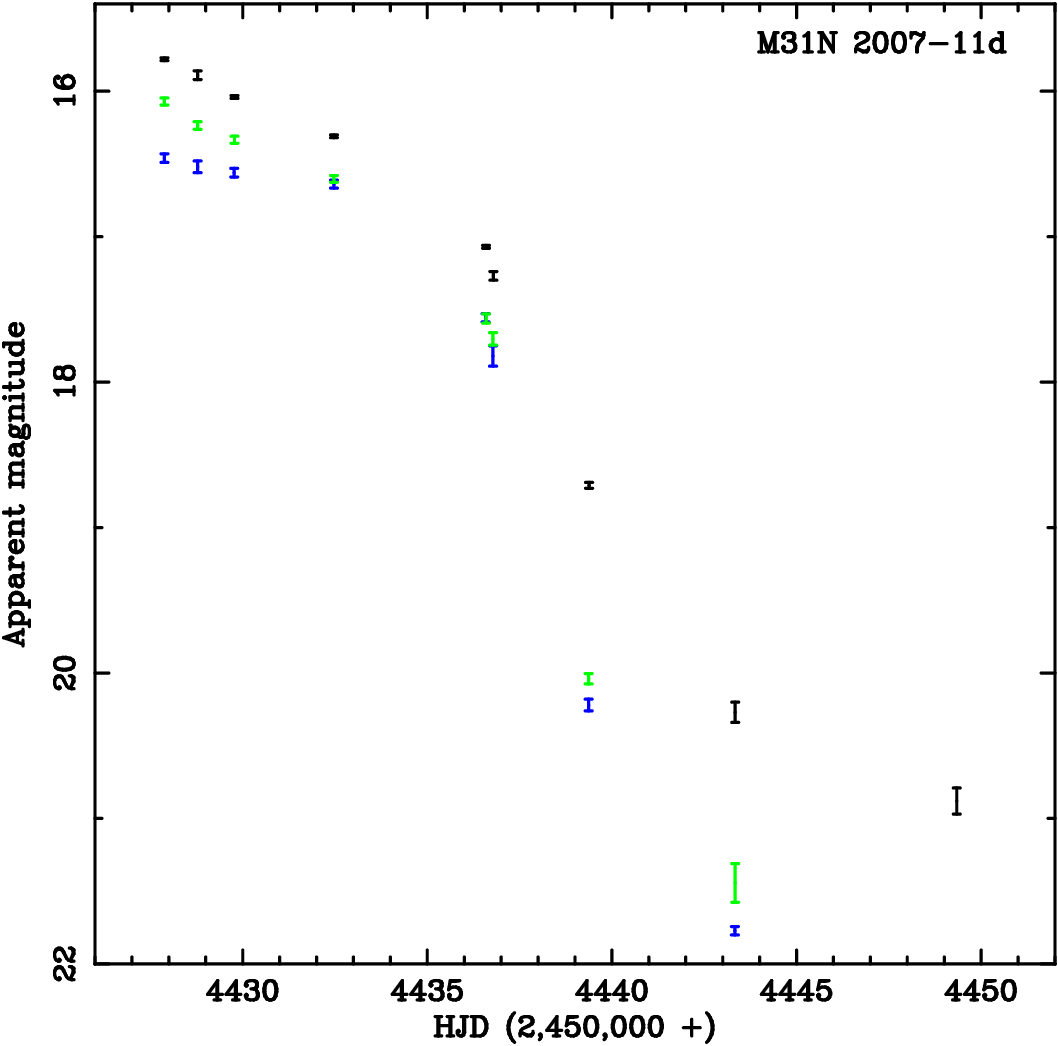}
\includegraphics[width=0.40\textwidth]{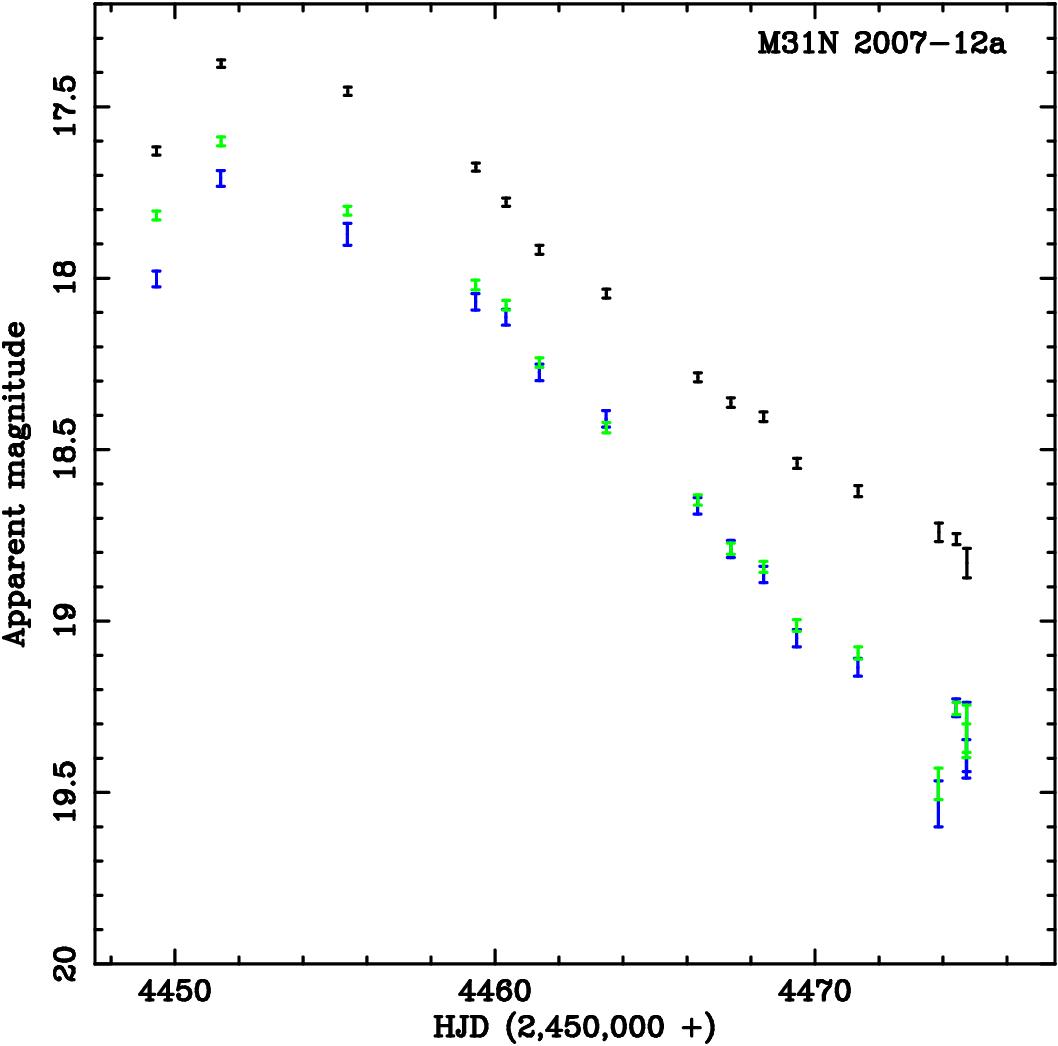}
\caption{Nova light curves (continued). See Fig.~\ref{fig13} for details.\label{fig16}}
\end{figure}

\begin{figure}
\includegraphics[width=0.40\textwidth]{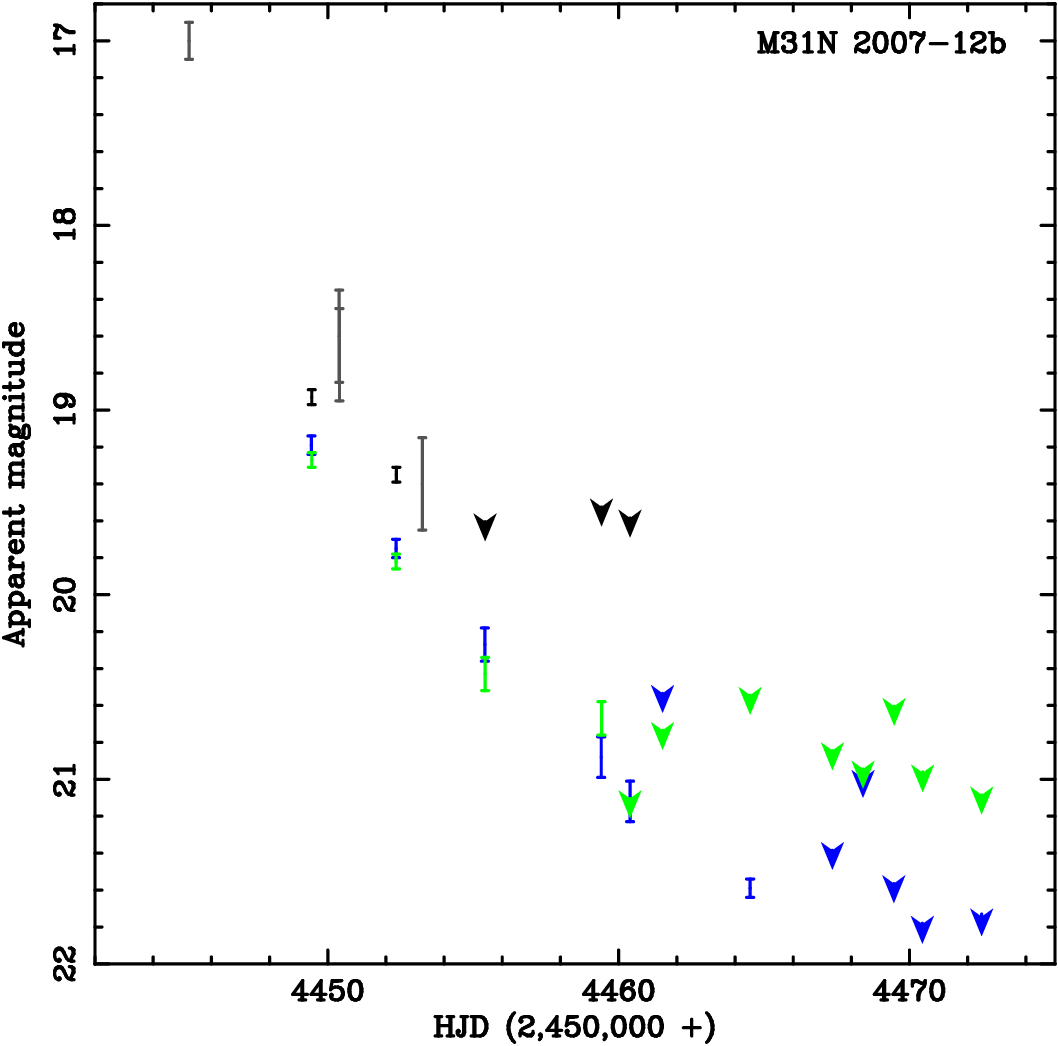}
\includegraphics[width=0.40\textwidth]{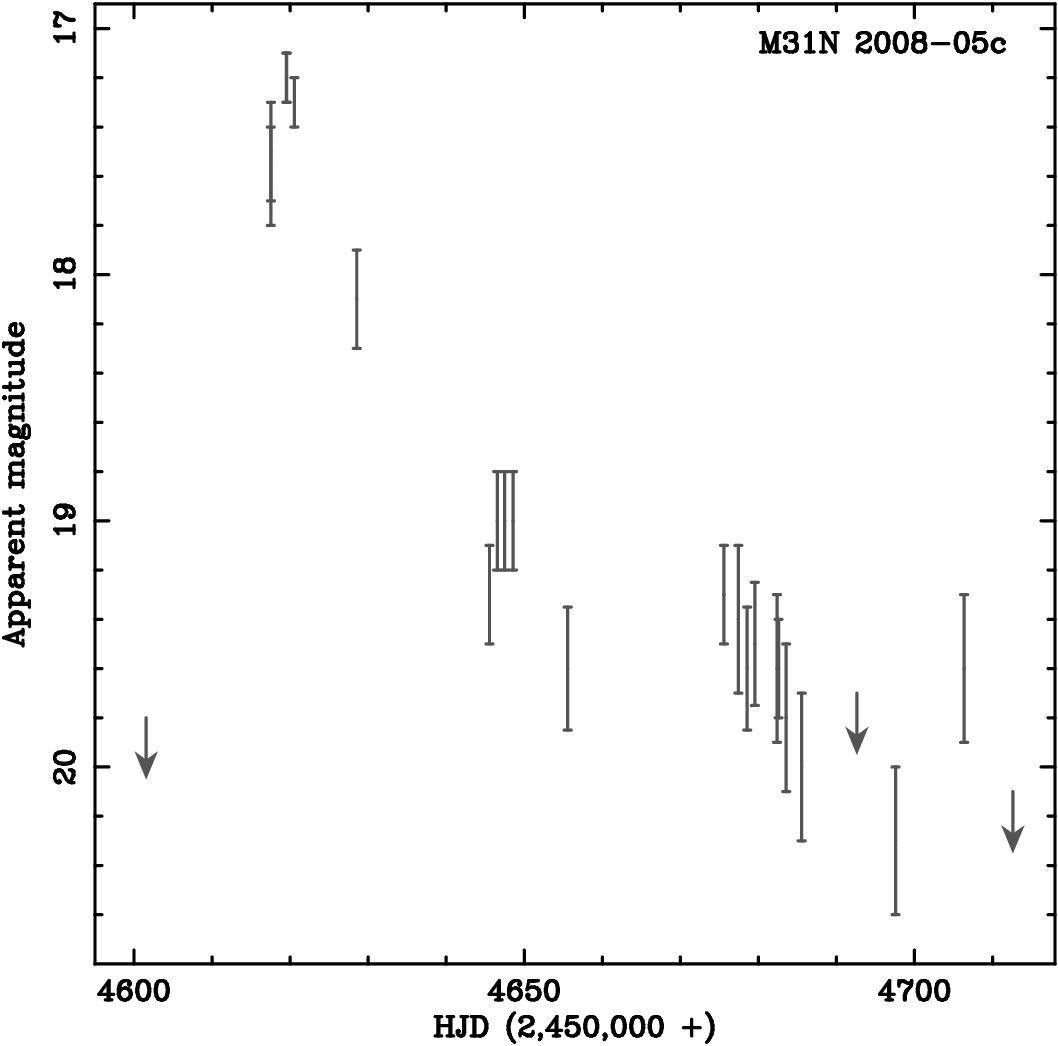}
\newline
\includegraphics[width=0.40\textwidth]{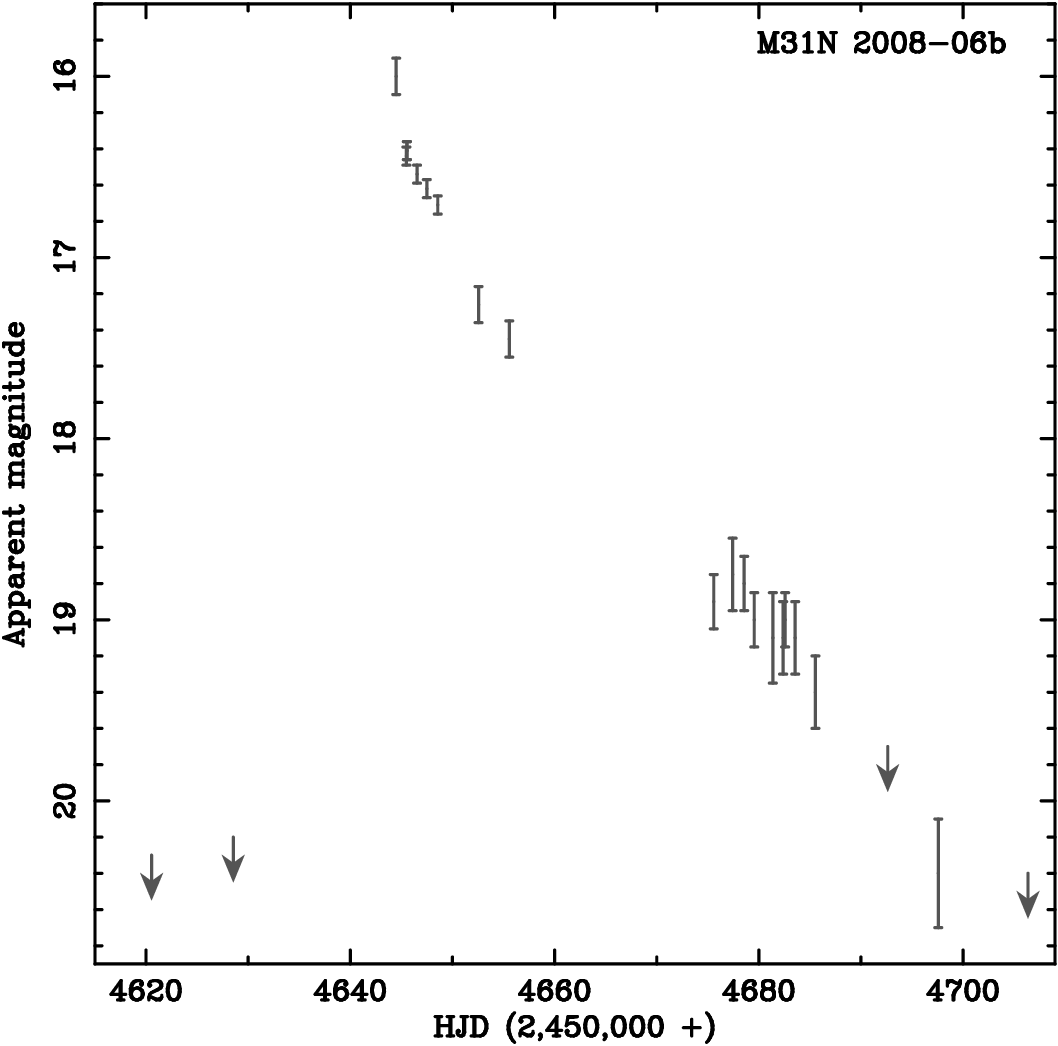}
\includegraphics[width=0.40\textwidth]{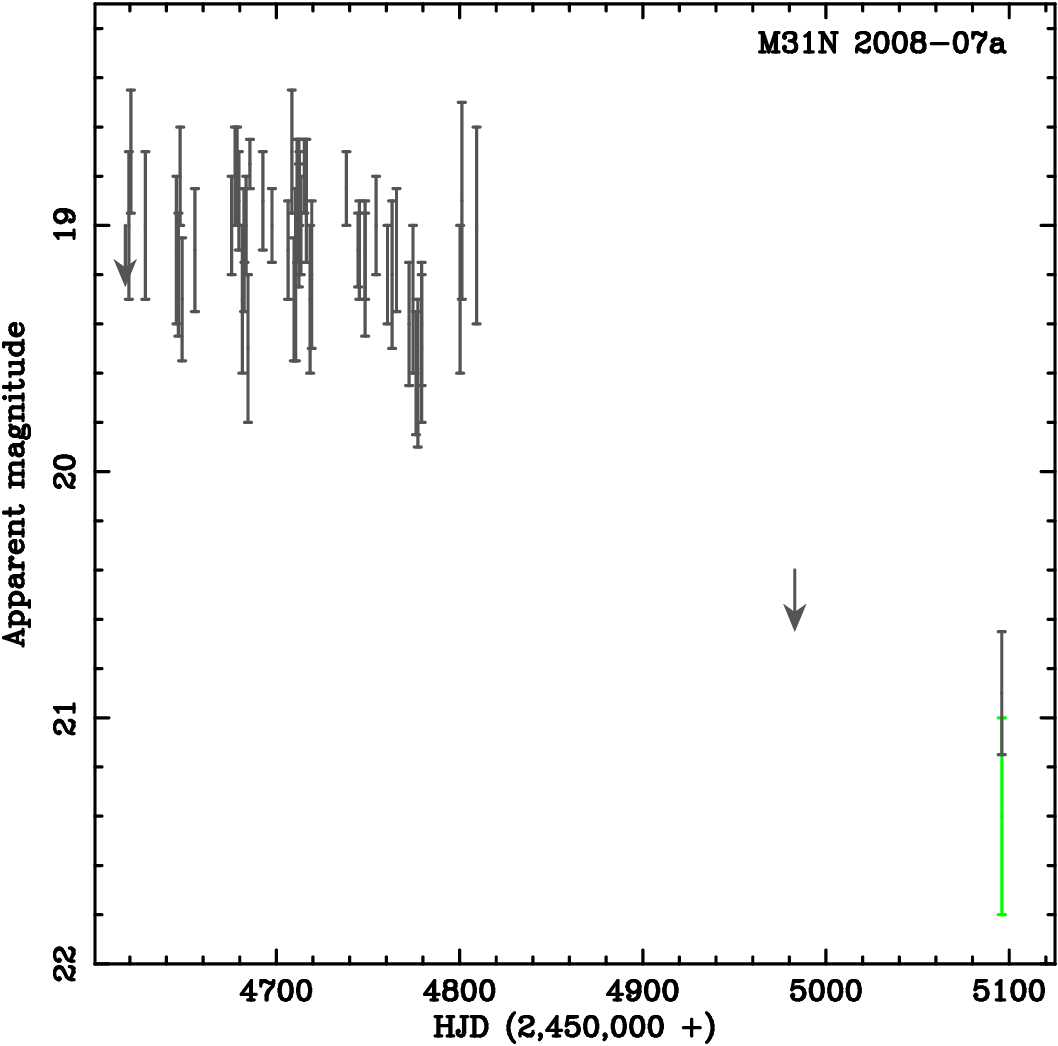}
\newline
\includegraphics[width=0.40\textwidth]{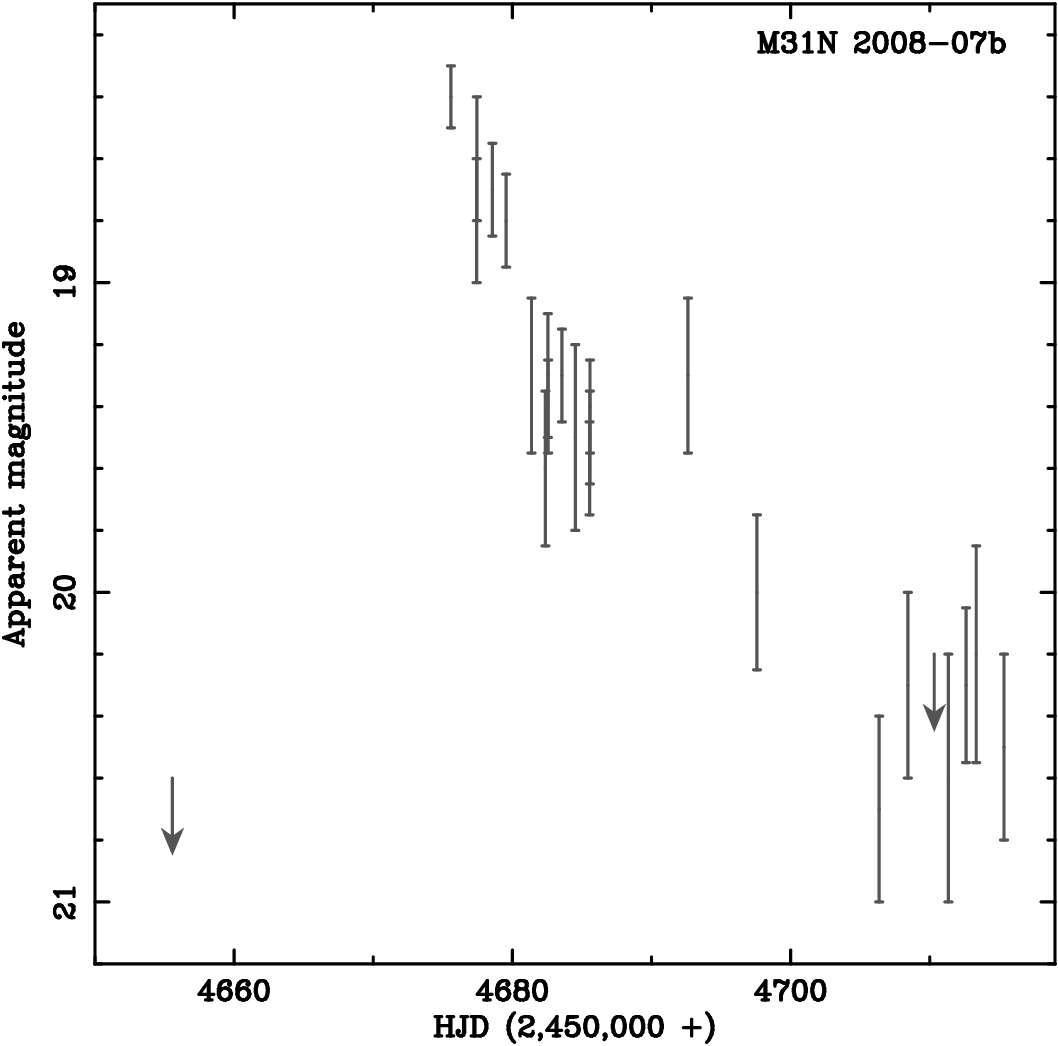}
\includegraphics[width=0.40\textwidth]{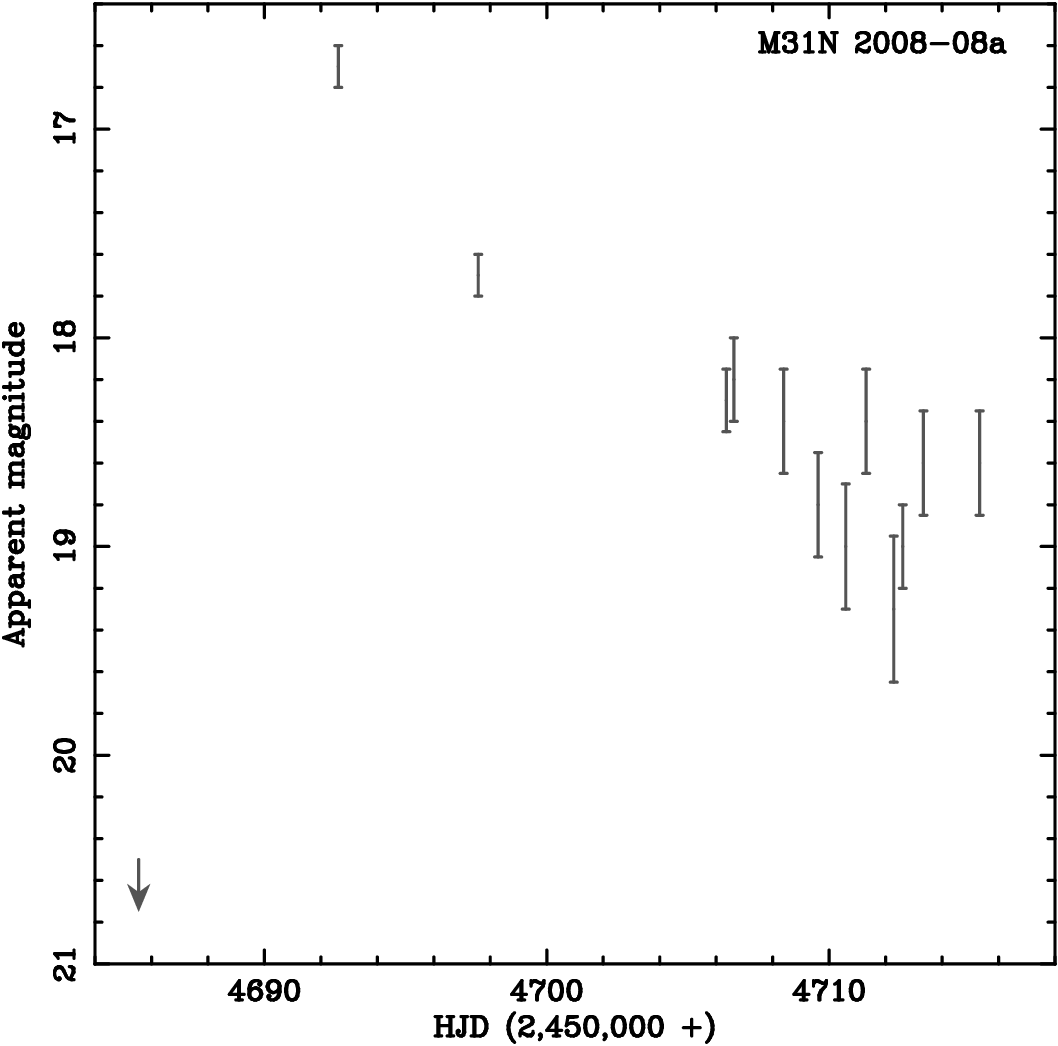}
\caption{Nova light curves (continued). See Fig.~\ref{fig13} for details.\label{fig17}}
\end{figure}

\begin{figure}
\includegraphics[width=0.40\textwidth]{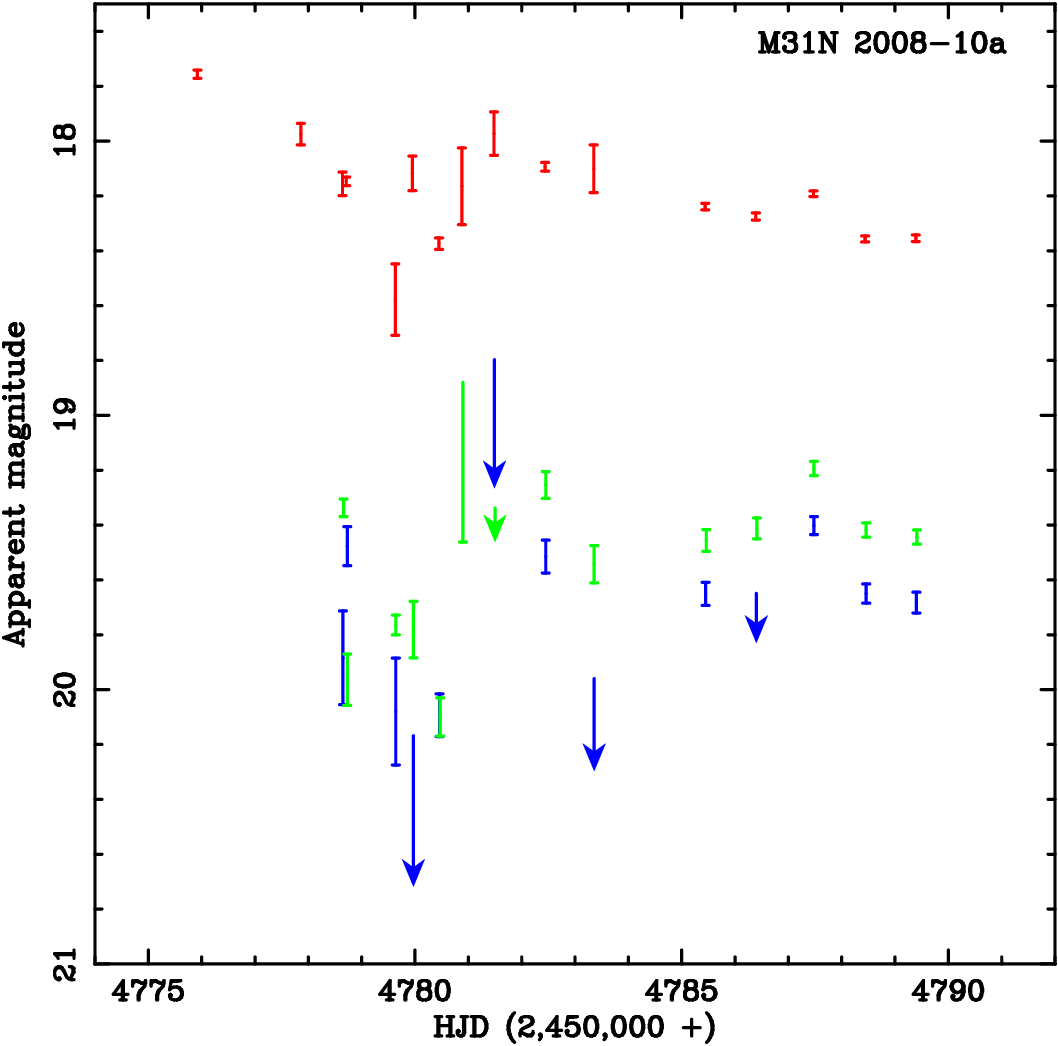}
\includegraphics[width=0.40\textwidth]{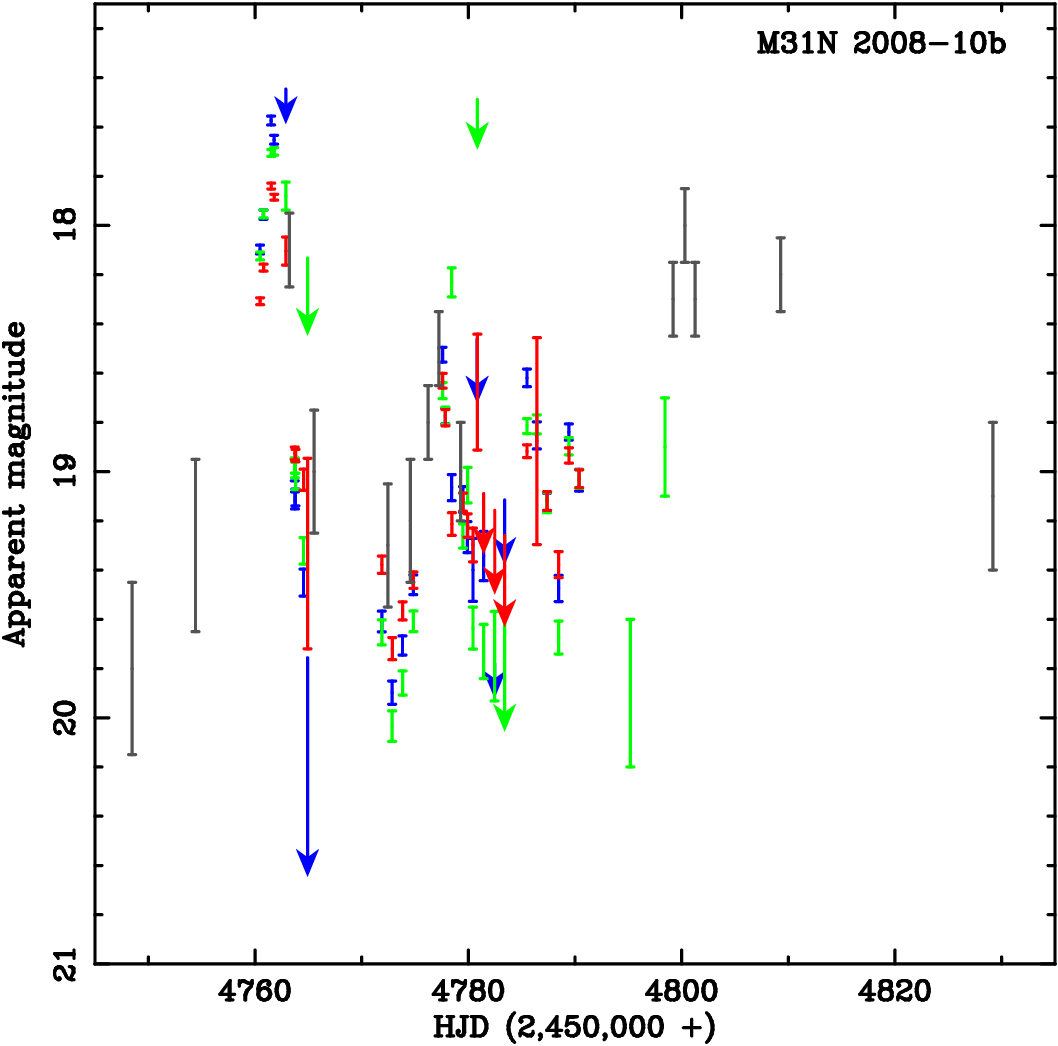}
\newline
\includegraphics[width=0.40\textwidth]{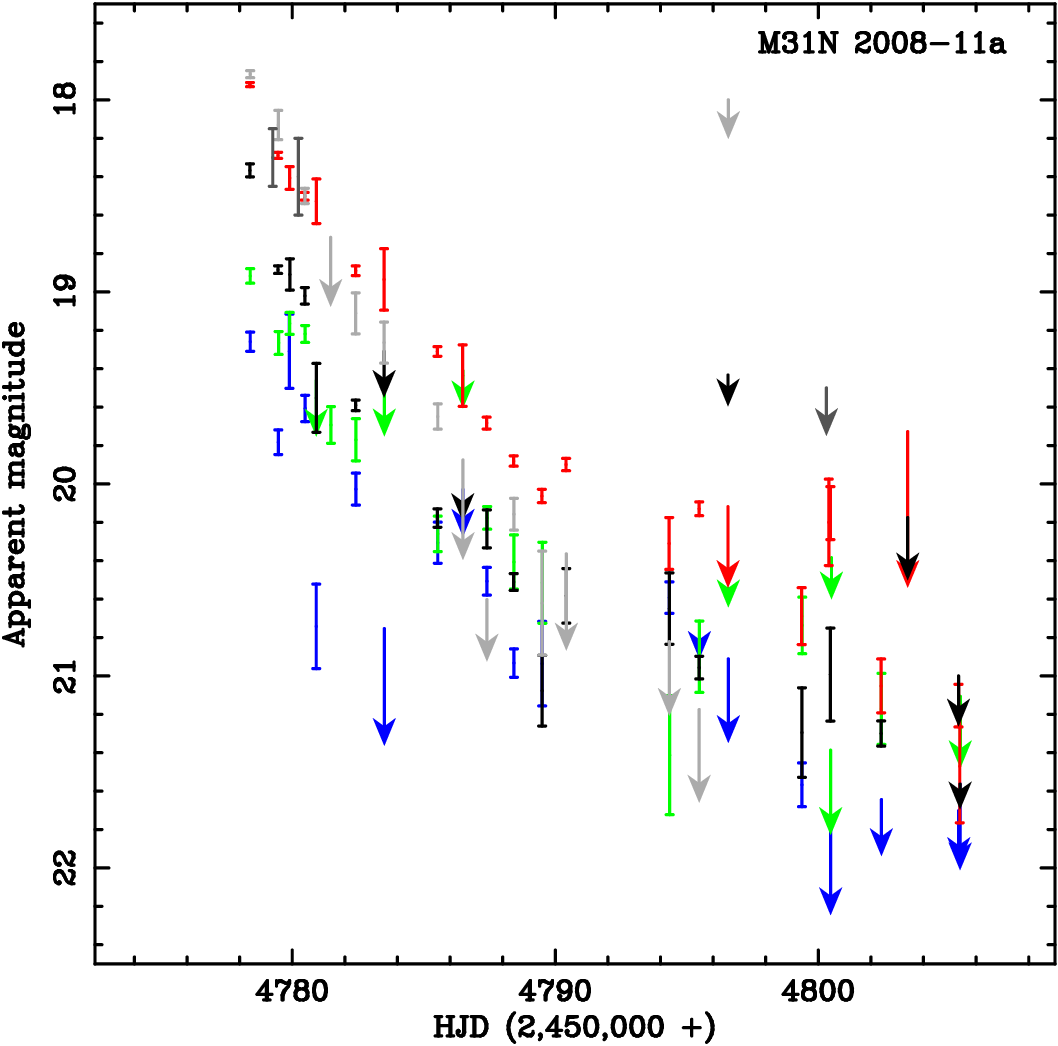}
\includegraphics[width=0.40\textwidth]{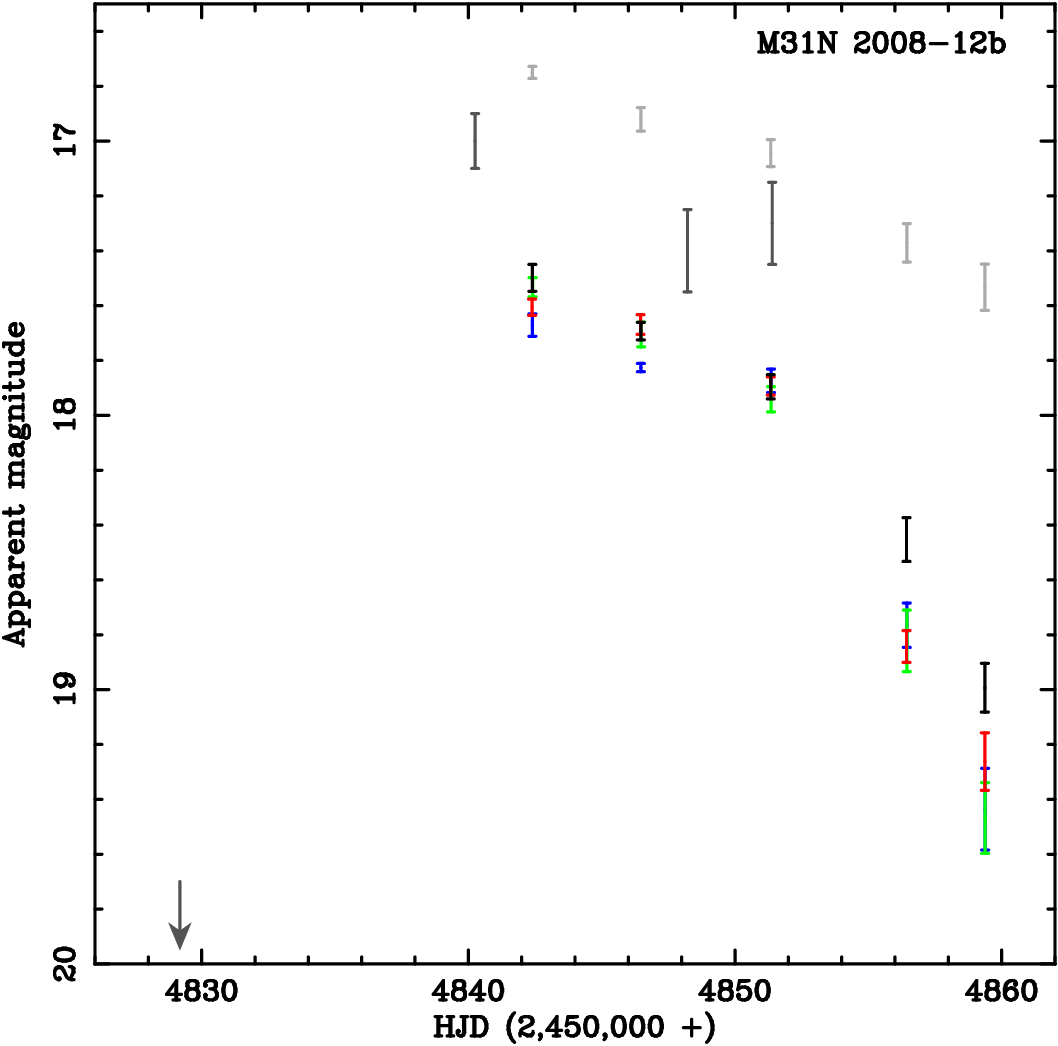}
\newline
\includegraphics[width=0.40\textwidth]{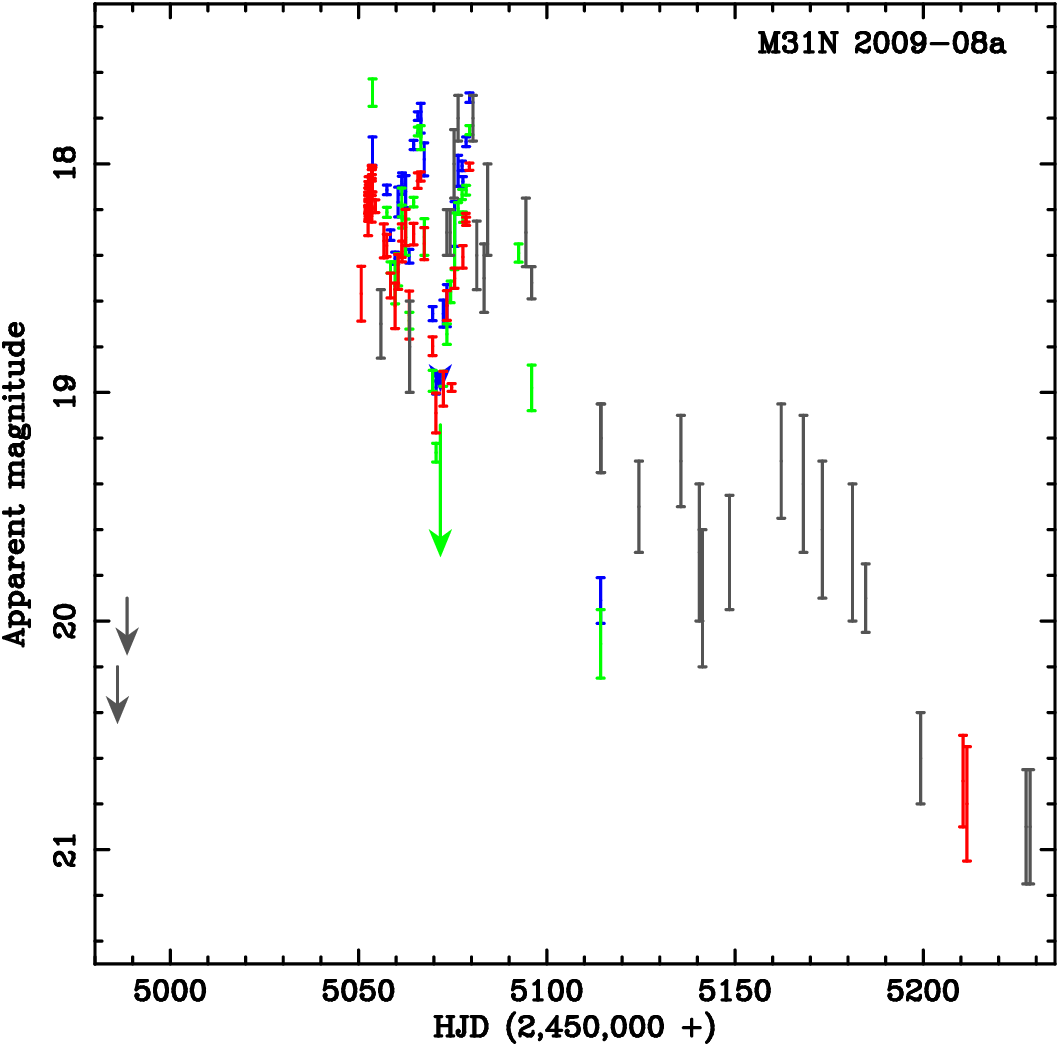}
\includegraphics[width=0.40\textwidth]{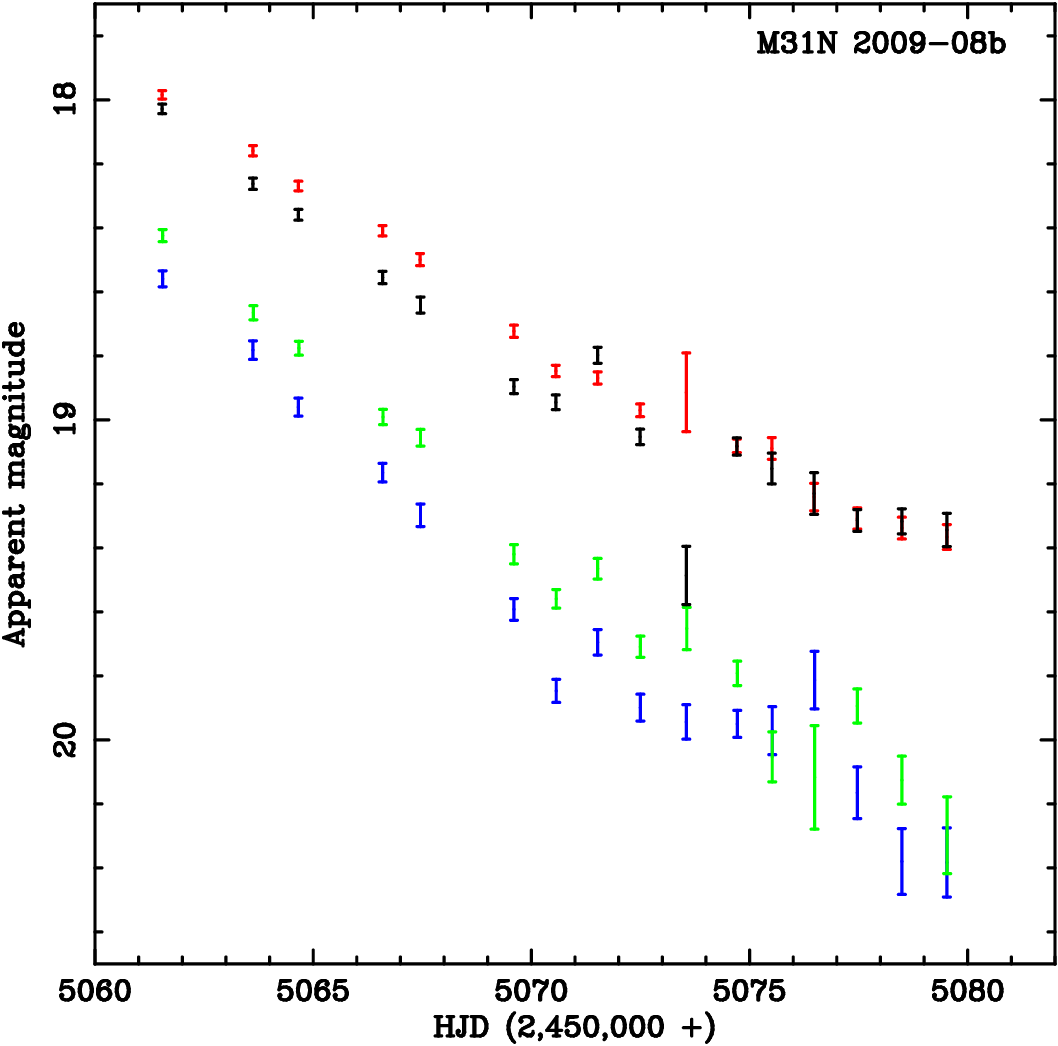}
\caption{Nova light curves (continued). See Fig.~\ref{fig13} for details.\label{fig18}}
\end{figure}

\begin{figure}
\includegraphics[width=0.40\textwidth]{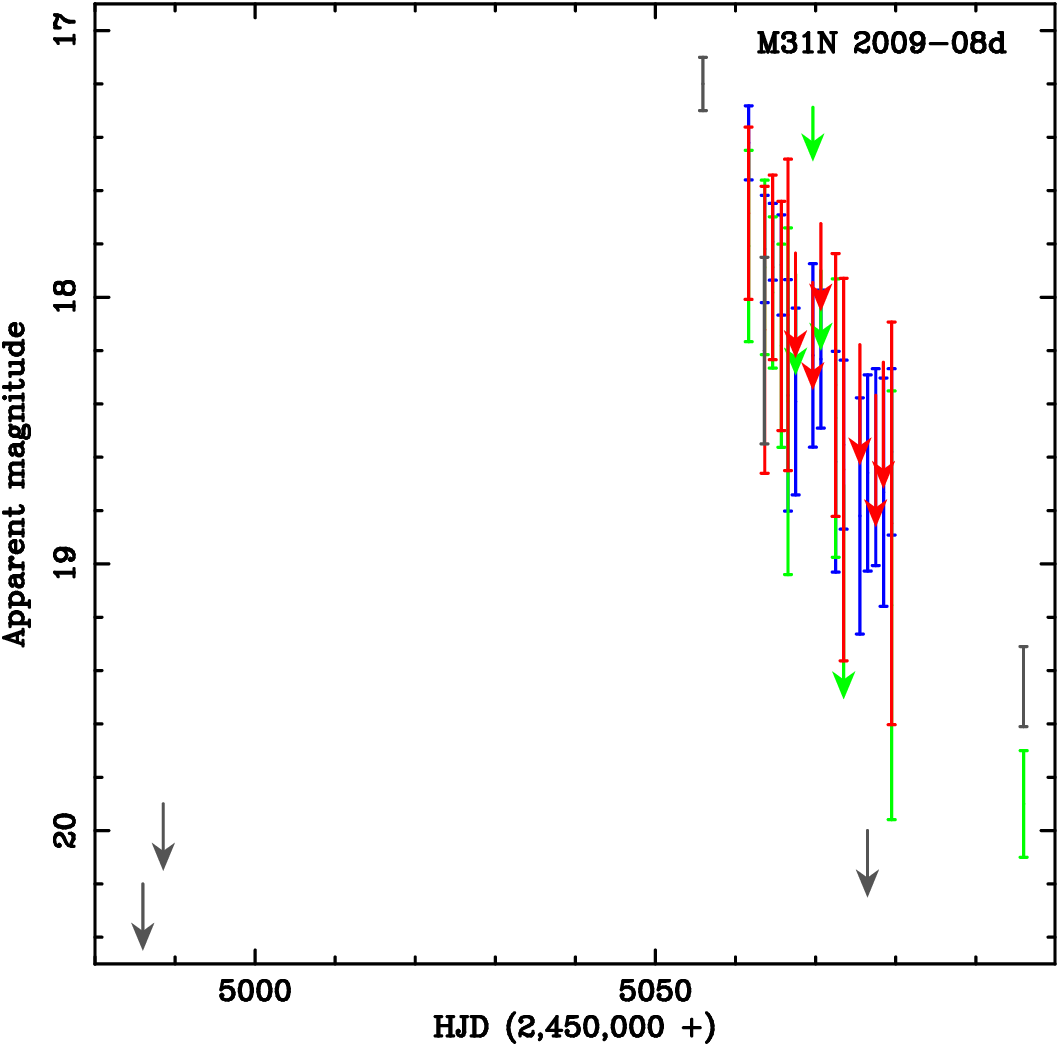}
\includegraphics[width=0.40\textwidth]{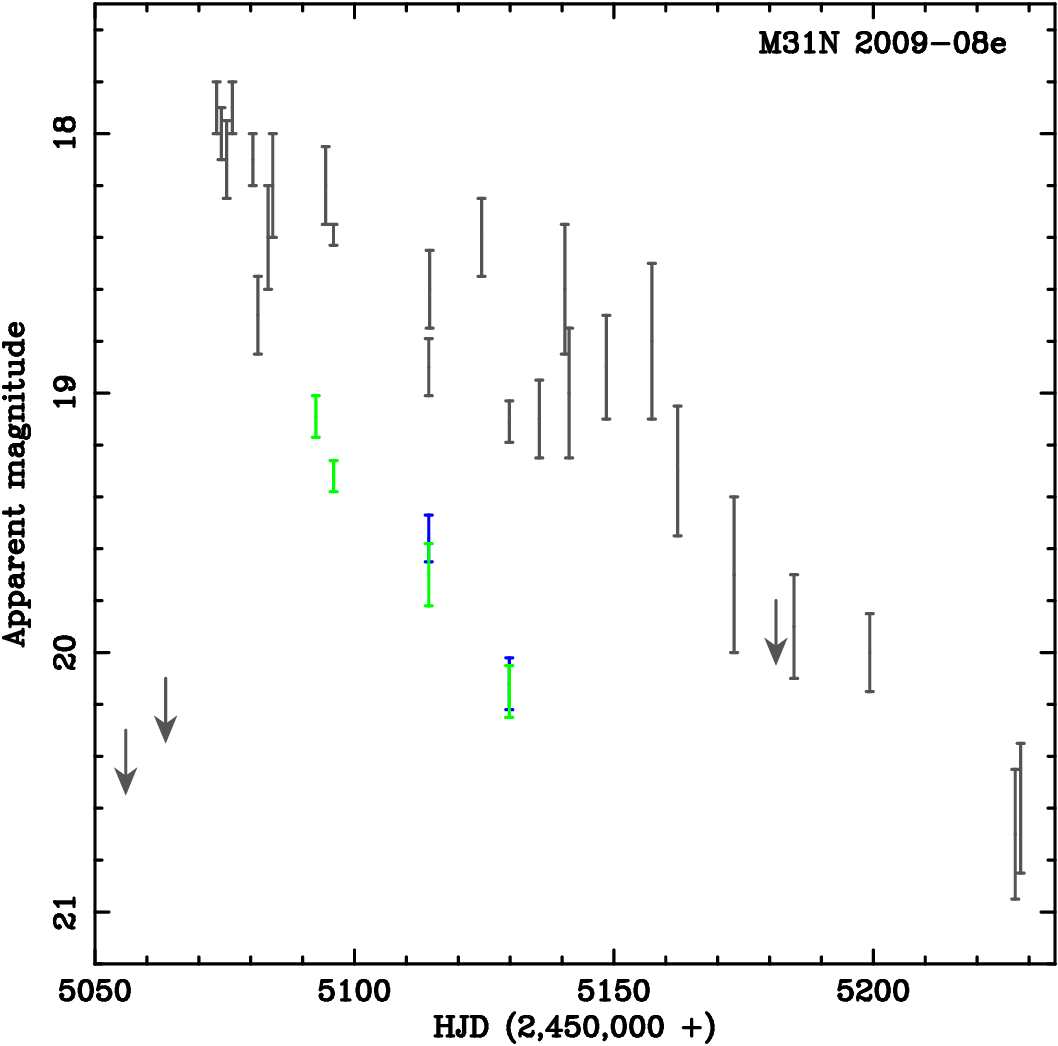}
\newline
\includegraphics[width=0.40\textwidth]{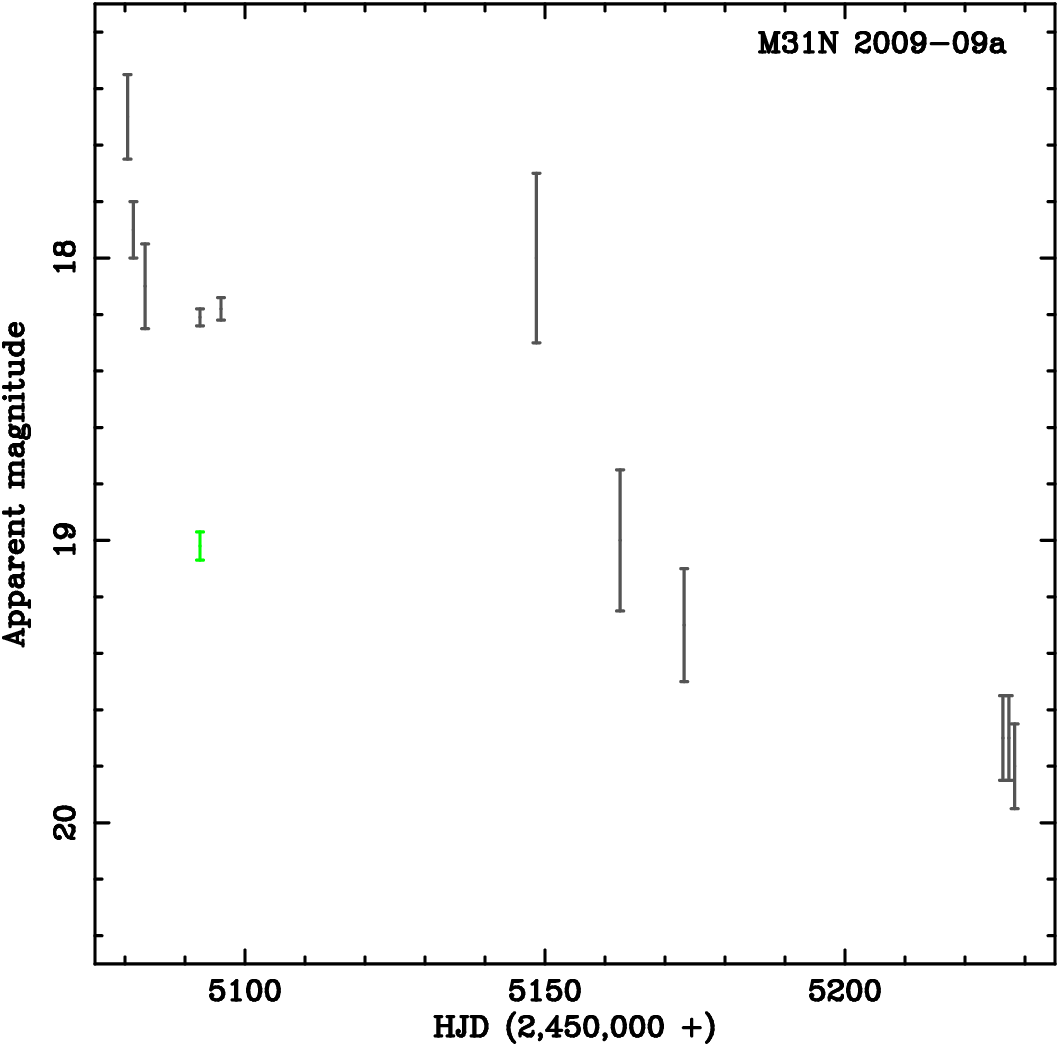}
\includegraphics[width=0.40\textwidth]{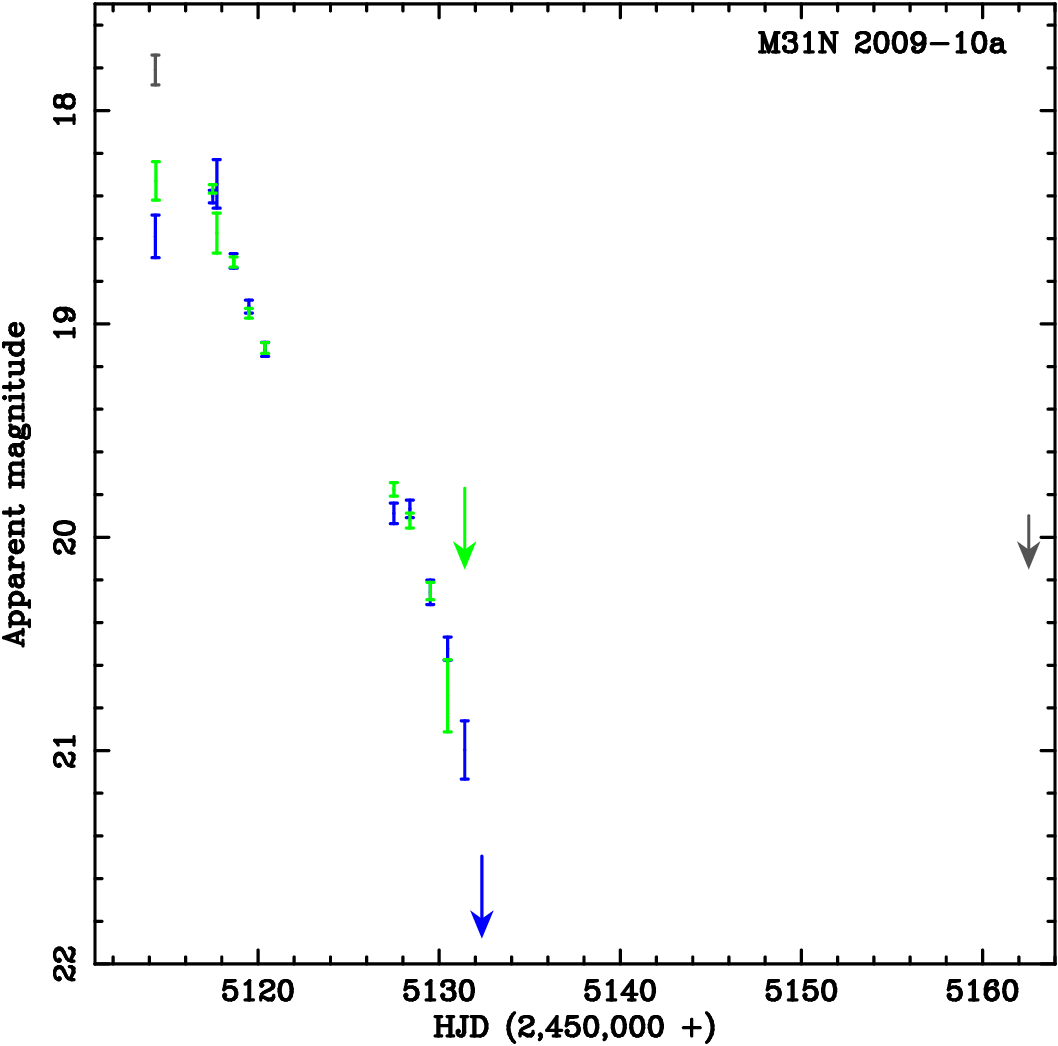}
\newline
\includegraphics[width=0.40\textwidth]{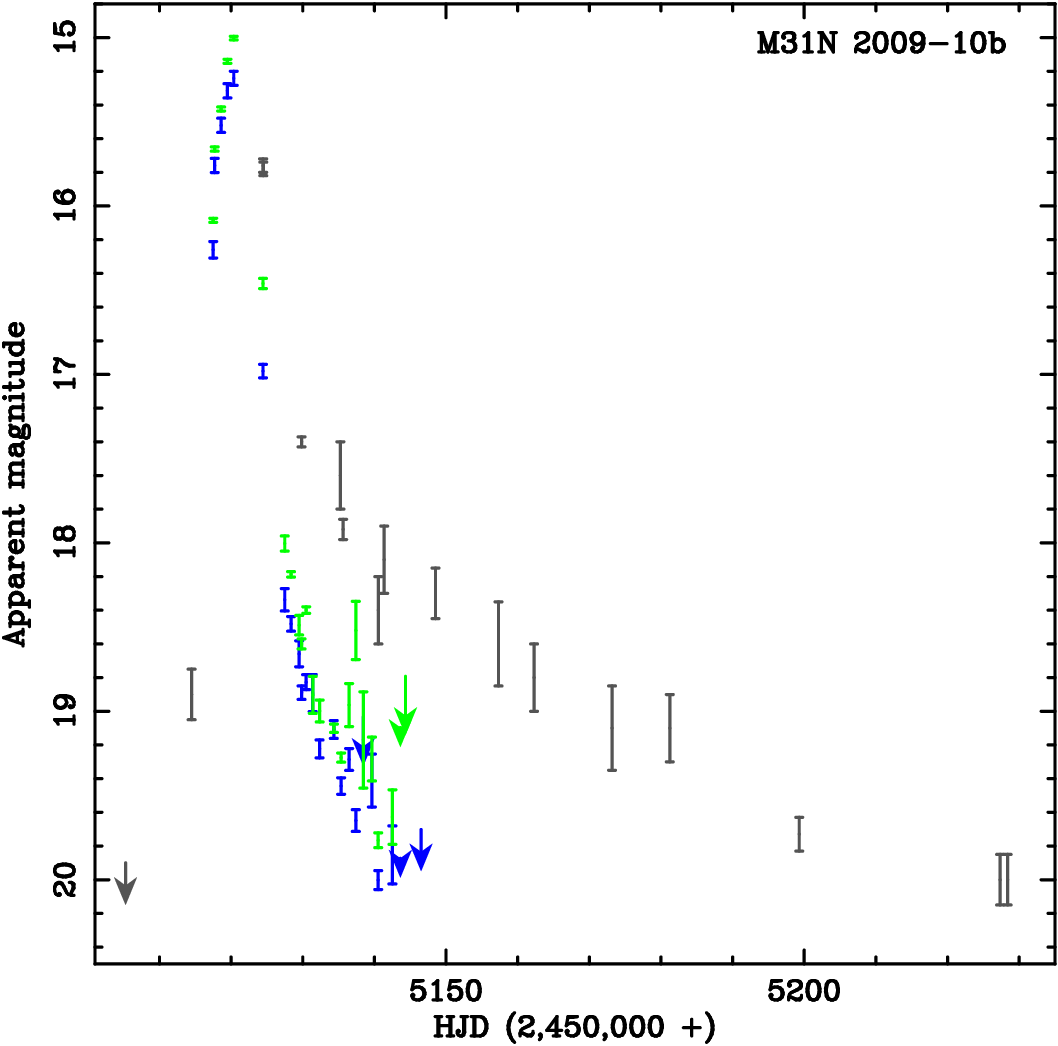}
\includegraphics[width=0.40\textwidth]{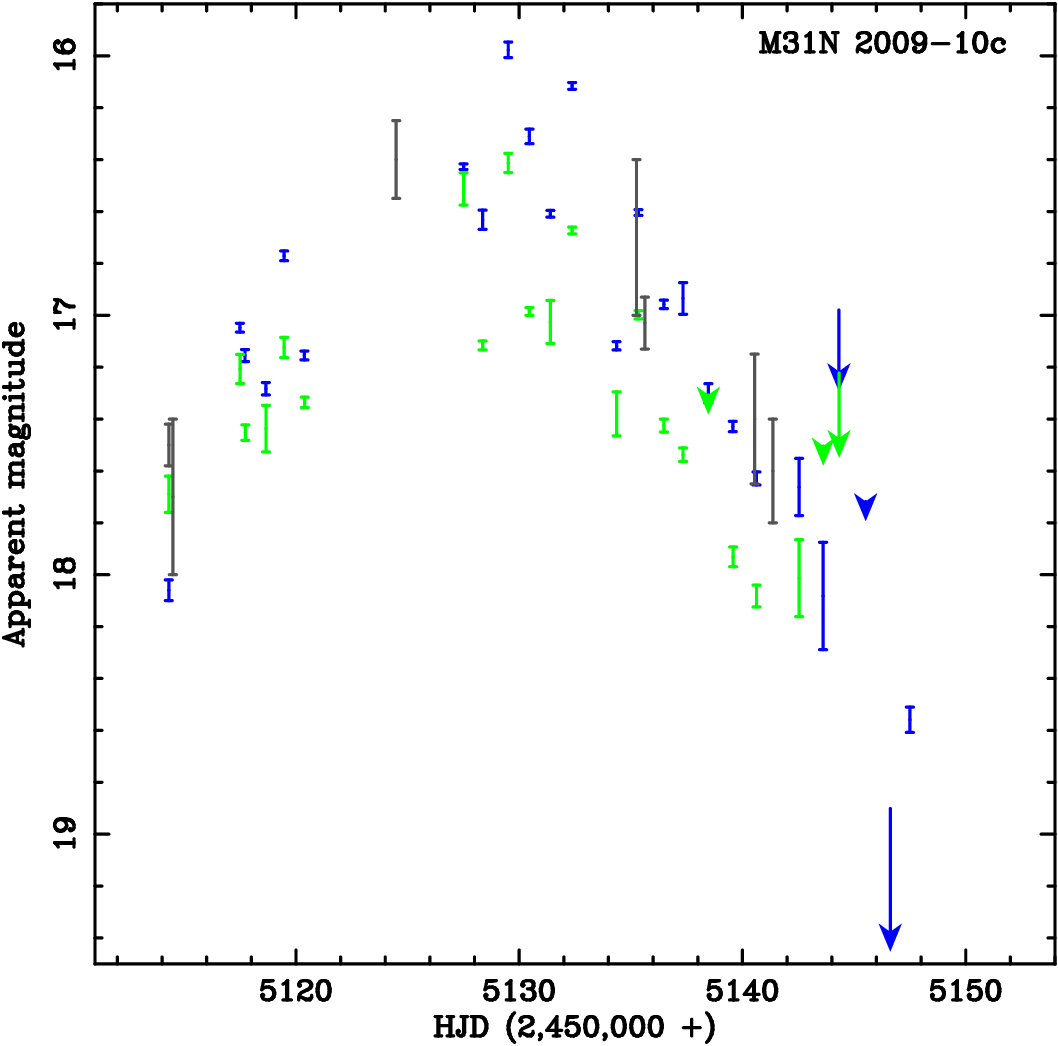}
\caption{Nova light curves (continued). See Fig.~\ref{fig13} for details.\label{fig19}}
\end{figure}

Between our targeted and archival observations, we were able to produce
light curves for a total of 47 of the 91 M31 novae with measured spectra
(46 spectra from the present survey and 45 from the literature).
The light curves are presented in Figures~\ref{fig13}--\ref{fig20}.
Before turning to a discussion of this wealth of data,
we briefly review our
present understanding of nova populations, both in the Galaxy and in M31.

\section{Nova Populations}

\subsection{Background}

It has been conjectured, based both on
Galactic and extragalactic observations, that there
may exist more than one population
of novae~\citep[e.g.,][and references therein]{del92,del98,sha08,kas11}.
Initially, Galactic observations suggested that novae
associated with the disk were on average more luminous and faded
more quickly
than novae thought to be associated with the bulge \citep[e.g.,][]
{due90, del92}.
However, the interpretation
of Galactic nova data is complicated by the need to correct for
interstellar extinction, which can be significant and varies widely with
the line of sight to a particular nova. Furthermore,
extinction hampers the discovery of a significant fraction of objects:
although Galactic novae occur at an estimated
rate of $\sim$30--35~yr$^{-1}$~\citep{sha97, sha02}, only about
one in five of these are discovered and subsequently studied in any detail.
Consequently, although much has been learned from the study of Galactic novae,
it is clear that these data are not ideal for establishing
the population characteristics of novae.

Nova eruptions are luminous enough to be detected as far away as the Virgo
cluster. However, despite the considerable data amassed in recent years, 
evidence for distinct nova populations in extragalactic systems is 
conflicting. \citet{cia90a} argued that a galaxy's nova rate
was independent of the galaxy's Hubble type, and therefore independent
of the galaxy's dominant stellar population. A few years later, based largely
on the same data, but with different assumptions regarding the
luminosity normalization,
\citet{del94} proposed that nova rates and
light-curve properties (e.g., rate of decline from maximum light) did in fact
vary between galaxies of differing Hubble types,
with late-type systems such as M33 and the Magellanic Clouds
having generally faster fading novae and
higher luminosity-specific nova rates.
However, subsequent studies
\citep[e.g.,][]{sha00,fer03,sha04}
have questioned these results,
arguing that (given the considerable uncertainties that plague
the determination of luminosity-specific nova rates)
the available data are insufficient to establish any significant 
correlation. Further, the broader implications of the 
argument by~\citet{del93}
that novae in the Large Magellanic Cloud (LMC) 
generally were brighter and faded faster
than those seen in the older
stellar populations of the Galaxy or M31's bulge,
has been
called into question by \citet{fer03} who showed that
novae in M49, the first-ranked Virgo elliptical galaxy,
generally faded at least as fast as novae in the LMC.

\subsection{The Spatial Distribution of Novae in M31}

The nearby Sb spiral, M31, is by far the most thoroughly studied
extragalactic system, with observations of novae going back to~\citet{hub29}.
Spanning $\sim4$~deg on the sky, M31 is well
resolved spatially, and
offers a convenient target for the study of nova populations.
In a classic nova survey from the mid 1950s,
\citet{arp56} reported the discovery of 30 M31 novae with the
60-in reflector at the Mount Wilson Observatory. Three principal
conclusions of Arp's study [augmented by data from~\citet{hub29}]
were that (1) the frequency
of novae dropped off more sharply than the galaxy's light,
(2) the nova density decreased within $\sim4'$ of the
nucleus, and (3) the frequency distribution of nova maximum magnitudes
is bimodal, with peaks $m_{pg} \approx 16.0$ and $m_{pg} \approx 17.5$ mag
\citep[also see][]{cap89}.
A subsequent survey by~\citet{cia87} using H$\alpha$ imaging
confirmed that the nova spatial distribution is more centrally
concentrated than the background light; in fact the distribution
was consistent with a purely bulge population. The central ``hole" in the
nova distribution noted by Arp was not
seen in the H$\alpha$ data, and was assumed
to be an artifact of the poor contrast against the bright nuclear background
in Arp's photographic images. 

Several more recent studies of the spatial
distribution of M31 novae have confirmed the association with M31's bulge
population
\citep[e.g.,][]{cap89, sha01, dar04, dar06}, but
a major uncertainty in these studies is whether a
significant fraction of disk novae are being missed due to
internal extinction within the galaxy~\citep[see also][]{hat97}.
Taken at face value,
the association of novae with M31's bulge population came as a surprise
given that Galactic novae have long been recognized to have a significant
disk population~\cite[e.g.,][]{due84, due90}. To address this discrepancy,
\citet{cia87} proposed the idea that a significant fraction of the novae
in M31's bulge could have been formed in the dense cores of the
galaxy's globular clusters and subsequently been ejected into the bulge through
three-body interactions within the clusters, through tidal disruption of
some clusters, or a combination of both processes.

\subsection{The Spectroscopic Classification of Galactic Novae}

A promising new approach for studying nova populations
is to consider the character of a nova's spectrum within a few weeks
after maximum light.
When analyzing a large sample of Galactic nova spectra,~\citet{wil92}
realized that the novae could be naturally
segregated into two principal spectroscopic
classes (Fe~II and He/N) based on the emission lines in their spectra.
Novae displaying prominent Fe~II emission (the Fe~II novae)
usually show P Cygni absorption profiles, evolve more slowly, 
have lower expansion velocities, and show lower levels of ionization,
compared to novae with strong lines of He and N (the He/N novae).
Complicating this division
is the fact that
a small fraction of novae initially exhibit Fe~II emission
lines along with broad (full width at half-maximum intensity 
[FWHM] $\grtsim 2500$~km~s$^{-1}$) Balmer emission
before going on to develop spectra typical of the He/N novae. Such
systems are referred to as either hybrid or Fe~IIb novae.
Both Fe~II and He/N novae (although more often the He/N systems)
occasionally go on to develop strong Ne emission in their post-outburst
spectra, which
suggests that the higher-mass ONe white dwarfs may be found
in both classes of novae.

\citet{del98} analyzed the spatial
distribution of a sample of 22
Galactic novae with data suitable for determining their spectroscopic class.
By restricting their sample to novae with well-determined
distances (i.e., from expansion parallax), they were able
to use the observed positions
to estimate the distance of each nova from the Galactic plane.
They discovered that the fastest and brightest novae
were primarily
associated with the He/N spectroscopic class, and that the
progenitors were preferentially located close to the
Galactic plane (i.e., $z\lessim100$~pc). Fainter and dimmer
novae, on the other hand, were more typically members
of the Fe~II class, and were found at much greater heights
(up to $z\grtsim1000$~pc). Thus, they concluded
that the progenitors of the He/N novae were associated with a younger
stellar population that are thought to contain a higher proportion of
massive white dwarfs.

The precise mechanism that leads to the formation of two distinct
post-eruption spectra is not well understood, but is thought to depend
most sensitively on the mass of the white dwarf. Systems with relatively
massive white dwarfs reach the critical density and temperature for a TNR
after accreting a relatively small amount of hydrogen-rich material
from the companion star~\citep{sta08,tow05}.
Once the TNR takes place, some fraction of this
small accreted mass is ejected at high velocity in the form of a discrete
shell. In systems with lower mass white dwarfs, more material is accreted
prior to the TNR, and this larger amount of gas is ejected in a combination
of a low-mass shell (early in the outburst),
and an optically thick wind (shortly thereafter).
It is this optically thick wind which is thought to be responsible
for the formation of the P~Cyg profiles seen in the lower excitation
Fe~II nova spectra.
Thus, the fact that spectroscopic observations appear to offer a powerful
discriminant between nova systems of varying white dwarf mass
from differing stellar
populations provided the principal motivation for our spectroscopic
survey of novae in M31.

\section{The M31 Nova Survey}

\subsection{Spectroscopic Class}

Through the end of 2009, a total of 837 nova candidates
have been discovered in M31 since the observations of~\citet{hub29} began
nearly a century ago.\citep{pie07e}
\footnote{\tt see also http://www.mpe.mpg.de/$\sim$m31novae/opt/m31/index.php}
Of these, spectra are now available for a total of
91 M31 novae, including the 46 from our present survey.
As described above,
novae spectra are, in principle, divisible into one of three primary classes:
Fe~II, He/N, and hybrid (or Fe~IIb) novae. In practice, however, it is often
difficult to make an unambiguous classification.
Spectra are taken at different times after eruption, and the signal-to-noise
ratio can vary widely from spectrum to spectrum. In addition,
a nova can, on occasion, show characteristics of more than one class.
For example,
during the course of our spectroscopic survey, we have identified three
novae (M31N 2007-10a, 2007-10b, and 2007-11b)
that might have been traditionally classified as He/N or hybrid,
but which do not share all of the characteristics of those classes.
These spectra are dominated by prominent but narrow (FWHM $<2000$~km~s$^{-1}$)
lines of H, He~I, and He~II,
with weaker N~III and Fe~II emission features occasionally seen.
In the basic scheme of \citet{wil92}, such novae would
be difficult to classify.
Henceforth, we will refer to these objects as narrow-line He/N, or He/Nn, 
novae.

The spectra of all novae included in our survey (see Figs.~\ref{fig1}--\ref{fig12})
were examined and subsequently classified into one
of six groups: Fe~II, likely Fe~II (Fe~II:), He/N, likely He/N (He/N:),
He/Nn, and hybrid (also known as broad-lined Fe~II or Fe~IIb novae). We found
a total of 30 Fe~II novae ($\sim$65\%), 6 likely Fe~II novae ($\sim$13\%),
3 He/N novae ($\sim$6.5\%), 3 likely He/N novae ($\sim$6.5\%), 3 He/Nn novae ($\sim$6.5\%),
and 1 hybrid/Fe~IIb nova ($\sim$2\%).\footnote{
Most, perhaps all, novae that are classified as He/N appear to
display some weak Fe~II emission near maximum light, and are therefore
technically members of the hybrid class. Rather than referring to
all of these novae as hybrid objects, we reserve the hybrid classification
for those novae with prominent Fe~II emission early on (e.g., M31N~2006-10b).}
When all 91 novae with measured spectra
are considered (see Table~\ref{spectsample}), the relative percentages remain similar
with $\sim$74\% (67 novae), $\sim$8\% (7 novae), $\sim$11\% (9 novae), $\sim$7\% (6 novae), and
$\sim$1\% (one nova) representing the Fe~II, Fe~II:, He/N, He/N: (including the
He/Nn systems), and hybrid classes, respectively.
Thus, when all the data are considered, approximately 4 out of 5 ($\sim82$\%)
of the total are likely Fe~II novae, with the remaining
systems falling in the He/N and related (He/Nn and hybrid/Fe~IIb) classes.

There is some exiguous evidence that the relative percentage
of He/N and hybrid/Fe~IIb
novae may be somewhat higher in the Milky Way. In his original paper,
\citet{wil92} indicated that $\sim40$\% of novae in his
Galactic sample belonged to
the He/N class. Similarly,~\citet{del98} found that as many as
10 out of 27 (37\%) in their sample of novae with well-determined
distances (from expansion parallax) were He/N or Fe~IIb systems.
More recently, however,~\citet{sha07a} has reviewed all available spectroscopic
data for Galactic novae, finding that only 20
out of the 94 systems ($\sim21$\%) with sufficient spectroscopic
data available for classification appeared to be He/N or hybrid systems.
This is consistent with the fraction found in our M31 survey.
The relatively large fraction of He/N and Fe~IIb systems included in the
\citet{del98} study, in particular,
may be the result of their sample selection,
which was restricted to novae with distances determined from
expansion parallax. Such novae are more likely to be nearby and thus located
in the Galactic disk.

\subsubsection{The Spatial Distribution of Spectroscopic Class}

The apparent concentration of He/N and hybrid/Fe~IIb novae
toward the disk of the Galaxy
\citep{del98}
is an intriguing finding, and it would be of considerable
interest if it could be confirmed
in M31. While it is not possible to determine
the height of a given nova above M31's galactic plane, we can explore differences
in the spatial distributions between the different spectroscopic classes. Based
on the Galactic results of \citet{del98}, one might expect that the
He/N and related systems would be preferentially associated
with the disk of M31, while the Fe~II systems would perhaps display
a more centrally concentrated, bulge-like distribution.

In Figure~\ref{fig21} we have plotted the projected
positions of the 91 M31 novae
with known spectroscopic class. Despite the expectation that the
He/N nova distribution might be more extended compared to the Fe~II systems,
there appears to be
no obvious dependence of spectroscopic type with spatial position
in the galaxy.
This impression can be misleading, however, since
the high inclination of M31 to the plane
of the sky ($i \approx 77^{\circ}$) makes
it difficult to assign an unambiguous position within M31 to a given nova.
This is particularly true for novae near the center of M31 where the foreground
disk is superimposed on the galactic bulge. On the other hand, novae
observed at a large galactocentric radius ($\grtsim15'$) are likely to
be associated with the disk of the galaxy. In order to 
approximate the true position of a nova within M31,
we have assigned each nova an isophotal radius, defined as
the length of the semimajor axis of an elliptical
isophote [computed from the $R$-band surface photometry of \citet{ken87}]
that passes through the observed position of the nova.

In Figure~\ref{fig22} we show the
cumulative distributions of the Fe~II and He/N (and hybrid) novae plotted
as a function of their isophotal radius.
Although it appears that the He/N-hybrid distribution may be slightly more extended
than the Fe~II distribution, a Kolmogorov-Smirnov (KS)
test reveals that the distributions
would be expected to differ by more than that
observed 81\% of the time if they were drawn
from the same parent population.
When only a subset of the novae with well-established spectroscopic types are
considered, this probability decreases slightly to 73\%.

The interpretation of these results is complicated
by the fact that our spectroscopic data are drawn from a
sample of M31 novae that may not be spatially complete.
As mentioned earlier, although the CCD surveys conducted in H$\alpha$ are
essentially complete in the innermost regions of the galaxy,
they did not typically
cover the full disk of M31. In recent years the situation has improved with
the availability of wide-field surveys, such as the ROTSE-IIIb program, which
have provided good coverage over most of the galaxy.
Given the nature of the M31 surveys, we suspect that
our spectroscopic sample may be biased somewhat
toward novae at smaller galactocentric radii (where, historically,
the galaxy has been more frequently monitored),
and thus may favor one spectroscopic class over the other.
Nevertheless, such a bias should not affect the distributions of
Figure~\ref{fig22} in a differential sense:
both Fe~II and He/N novae are detectable throughout the galaxy.
Another potential source of bias involves the fact that the He/N novae
are on average brighter and faster fading
than the Fe~II systems (see \S 4.2.1 below).
One could argue that the brighter He/N
novae might be easier to detect against the bright
background of the bulge. However, this advantage would be offset to some degree
by the fact that these novae generally fade more quickly,
making them more likely to be missed in synoptic surveys.
Consequently, we conclude,
as did \citet{dim10} based on a smaller sample of novae,
that there is no compelling evidence from the observed spatial
positions of the novae in our sample that the Fe~II and He/N novae
arise from different stellar populations in M31.

\subsubsection{Nova Expansion Velocity}

One of the defining properties of the He/N spectroscopic class is that the
emission-line widths are considerably broader than those seen in the Fe~II novae.
Specifically, \citet{wil92} found that the emission lines of Galactic
novae in the He/N class are characterized by a half-width at zero
intensity (HWZI) $>2500$ km~s$^{-1}$. Empirically,
we have found that for most nova line profiles, the HWZI roughly
equals the FWHM;
since the latter is the more
easily measured quantity, we have adopted it to characterize
the spectra in our survey.
The values of the FWHM and the equivalent widths of H$\alpha$
and H$\beta$ in our nova spectra are given in Table~\ref{balmerline}.

Although the emission-line width is expected to be correlated with the expansion velocity
of the nova ejecta, the FWHM does not necessarily yield the expansion velocity directly.
In an idealized nova, the broad
emission features typically seen in an He/N system are believed to be formed mainly in a
discrete, optically thin shell ejected at relatively high velocity
from near the white dwarf's surface.
In this case, the line profiles are expected to be flat-topped and nearly
rectangular in appearance, with the FWHM closely approximating the ejection velocity
of the shell.
In the Fe~II systems, however,
the lines are mainly produced in a wind, which originates
at a distance above the surface of the white dwarf that varies
as the outburst evolves. Thus, the escape velocity for this wind
is smaller than that
at the white dwarf's surface. As a result, the expansion velocity (and hence
line width) may depend on the time elapsed
since eruption.

When comparing the emission-line widths of the  novae in our sample, it must be
kept in mind that our spectroscopic data were obtained at varying times after eruption,
and thus do not necessarily reflect the relative expansion velocities accurately.
Nevertheless, as Figure~\ref{fig23} illustrates,
a clear difference between the ejection velocities
of the two principal classes of novae (Fe~II and He/N)
is apparent.
Without exception, the novae belonging to the He/N class are characterized by
H$\alpha$
FWHM~$>2500$~km~s$^{-1}$, while the Fe~II systems all have
H$\alpha$ FWHM less than this value.
Interestingly, although He/Nn novae
display prominent lines of helium as do the standard He/N novae, they are
narrow-line objects that resemble a typical Fe~II
nova at times (e.g., see M31N~2007-11b in Fig.~\ref{fig8}).
In addition, they often
do not display prominent lines of nitrogen as do the typical He/N systems.

\subsection{Photometric Properties}

To further explore the properties of the novae in our survey,
whenever possible we have augmented our spectroscopic data with
available photometric observations. Few
light curves are available for the novae in our spectroscopic sample
that erupted prior to the start of our HET survey in 2006. Nevertheless,
when the entire set of spectroscopic novae is considered,
we have sufficient photometric data to estimate decline rates
for half of the sample.

A convenient
and widely used parameterization of the decline rate
is $t_2$, which represents the time (in days) for a nova
to decline 2~mag from maximum light. According to the criteria
of \citet{war08}, novae
with $t_2\lessim25$ days are considered ``fast" or ``very fast,"
with the slowest novae characterized by $t_2$ values of
several months or longer.
Rates of decline, and corresponding values of $t_2$, have been measured
(for all novae with sufficient photometric coverage)
by performing weighted linear least-squares fits to the declining portion
of the light curves that extend up to 3 mag below peak.
In an attempt to account for systematic errors in the individual
photometric measurements, the weights used in the fits were composed
of the sum of the formal errors on the individual photometric
measurements plus a constant systematic error estimate of 0.1 mag.
The net effect of including the systematic error component was a
reduction of the relative weighting of points with small formal errors and a
corresponding increase in the formal errors of the best-fit parameters
and in the uncertainties in $t_2$ derived from them.

Because our photometric observations do not always
begin immediately after discovery, and the date of discovery does not
always represent the date of eruption,
we have made two modifications to our photometric data in order to
better estimate the light-curve parameters.
First, when available,
we have augmented our light-curve data with
the discovery dates and magnitudes given in the catalog
of \citet{pie07e}.\footnote{\tt see also http://www.mpe.mpg.de/$\sim$m31novae/opt/m31/index.php}
Second, for some novae we have
modified (brightened) the peak magnitude slightly
through an extrapolation of the declining portion
of the light curve up to 2.5 days pre-discovery in cases where
upper flux limits (within 5 days of discovery) are available.
The light-curve parameters resulting from our analysis
are given in Table~\ref{lcparam}.

\subsubsection{MMRD Relations}

If we adopt a distance modulus for M31 of $\mu_0=24.38$~mag \citep{fre01}
and a foreground reddening, $E(B-V)=0.062$~mag \citep{sch98}, we can
compute the absolute magnitude at maximum light, and thereby
produce calibrated maximum-magnitude versus rate-of-decline (MMRD) relations.
The MMRD relations (the peak absolute magnitude vs. $\rm{log}~t_2$)
for the $B$, $V$, and $R\/$ light curves are shown in
Figures~\ref{fig24}, \ref{fig25}, and \ref{fig26}, respectively.
These plots illustrate not only that the peak nova luminosity
is correlated with the rate of decline (i.e., the brightest novae generally
fade the fastest), as was
first studied extensively by~\citet{mcl45} for Galactic novae,
but that the He/N systems are typically
among the brightest and fastest novae.
Weighted, linear least-squares fits to our $B$, $V$, and $R\/$-band\footnote{The $R\/$-band MMRD relation includes $r'\/$-band data for novae where
$R\/$-band observations are not available. When both $R\/$ and $r'\/$
observations are available for a given nova, only the more extensive
data set is used.}
data yield

\begin{equation}
M_B = -9.75\pm0.11 + (1.69\pm0.085)~\mathrm{log}~t_2,
\end{equation}

\begin{equation}
M_V = -9.78\pm0.10 + (1.70\pm0.080)~\mathrm{log}~t_2,
\end{equation}

\noindent
and

\begin{equation}
M_R = -10.89\pm0.12 + (2.08\pm0.077)~\mathrm{log}~t_2,
\end{equation}

\noindent
respectively.
The peculiar He/Nn object M31N 2007-10b, which has
particularly scanty light-curve coverage, has a relatively large
uncertainty in the peak magnitude.
For comparison, in Figure~\ref{fig24}
we show the theoretical $M_B$ vs. $\rm{log}~t_2$ relation of \cite{liv92},
while in Figure~\ref{fig25} we include
the Galactic $M_V$ relation from \citet{dow00}:

\begin{equation}
M_V = -11.32\pm0.44 + (2.55\pm0.32)~\mathrm{log}~t_2.
\end{equation}

The slope of our M31 $V$-band MMRD relation is shallower than that
for the Galactic data,
and the M31 data are also systematically fainter
than expected
from the best-fitting Galactic relations. The latter discrepancy, in particular,
as well as some of the scatter generally seen in the MMRD relations, is likely due
to the fact that we have only corrected the M31 data for Galactic foreground
extinction, not for extinction internal to M31. Based on these comparisons,
it appears that the M31 nova sample perhaps suffers as much as 0.5~mag of
extinction from within M31 itself, especially in the disk of the galaxy.
This value is consistent with an estimate of $A(r')=0.5$~mag adopted by \citet{dar06}
based on the mean for Sb galaxies \citep{hol05}.
In addition, our estimates of maximum light are based upon the magnitude
at discovery, which will underestimate the peak luminosity in some cases.
If we divide the novae in our sample
into two groups, those with isophotal radii $r\leq10'$ and those
with $r>10'$, we find that the latter sample is slightly fainter at peak
by an average of $\sim0.4$ mag ($16.6\pm1.1$ vs. $17.2\pm0.9$ mag),
although the difference is not statistically significant.

In agreement with the results of the Galactic study by~\citet{del98},
it appears that the He/N novae are on average ``faster" than 
their Fe~II counterparts. Indeed,
although our sample is dominated by Fe~II systems,
three of the four fastest novae
are He/N or related (He/Nn) systems.
On the other hand,
with the exception of the He/Nn nova M31N 2007-10b,
we do not find strong evidence
for a significant population of fast, but relatively faint,
novae that apparently do not follow the classic MMRD relation.
As discussed by \citet{kas11}, it is possible that
these novae arise from progenitors containing
high $\dot M$ (hot) and relatively massive white dwarfs, similar
to what is expected for recurrent novae. Perhaps such novae are related
to the class of He/Nn novae described earlier.

Given that the He/N novae generally fade more quickly than the Fe~II
systems, and that He/N novae have significantly higher expansion 
velocities based on their emission-line widths, the 
expansion velocity should be inversely correlated
with the light-curve parameter, $t_2$. In Figure~\ref{fig27} we
have plotted the measured $t_2$ value (based on the $V$ band when possible)
versus the measured H$\alpha$ FWHM for the 25 novae in our survey where
it is possible to measure both parameters. As expected, there is a clear
trend of faster novae exhibiting higher expansion velocities. The one
exception is the He/Nn nova, M31N~2007-10a, which apparently evolved
quite quickly despite its relatively slow expansion velocity.
Based on these data (excluding M31N~2007-10a), a weighted linear least-squares
fit yields the following:

\begin{equation}
{\rm log}~t_2~\rm{(d)} = 6.84\pm0.10 - (1.68\pm0.02)~{\rm log}~[H\alpha~{\rm FWHM~(km/s)}].
\end{equation}

\noindent
This relation can be compared with a similar one for Galactic novae
found by \citet{mcl60}. A major factor in the discrepancy between the two
may arise because
in McLaughlin's relation the expansion velocities
are derived from the absorption-line minima (P~Cyg profiles) measured near
maximum light. Typically, such velocities are only 20\% to 50\% of those
inferred from the emission-line FWHM.
The scatter in our data, particularly for the slower novae,
is probably due in part to the time dependence of the derived velocities, as
referred to in \S 4.1.2 above.

\subsubsection{Spatial Distribution of Nova Speed Class}

The question of whether the photometric properties of novae in M31
(e.g., peak brightness, fade rate) vary with spatial position in the galaxy
(and possibly with stellar population) has yet to be thoroughly studied.
Most recent surveys, which have concentrated primarily
on the {\it discovery} of novae either
for the purpose of measuring the spatial distribution, the overall rate,
or both, lack the high cadence required to produce useful nova light curves.
Light-curve data, when available, often only cover H$\alpha$ or
the $R$ band where the relationship between fade rate and peak luminosity
is weak or absent~\citep[e.g.,][]{cia90b}. As mentioned
earlier, available broad-band nova data come largely from the
photographic surveys of~\citet{arp56}, \citet{ros64,ros73}, and~\citet{sha92}.
As noted above, the observation by~\citet{arp56} that the
apparent magnitude distribution for M31 appeared to be
bimodal, with peaks corresponding to $M_B \approx -8.5$ and 
$M_B \approx -7.0$ mag at the distance of M31, was later taken as
evidence for the existence of two nova populations \citep[e.g.,][]{cap89}.

The available light-curve data from previous surveys can be used
to augment the extensive photometry obtained as part of our survey
to study the variation of nova speed class with spatial position in M31.
In particular, \citet{cap89} has summarized the light-curve properties (peak magnitude
and rate of decline) for novae in the \citet{hub29}, \citet{arp56}, and
\cite{ros64,ros73} surveys. From this compilation, we have selected
the ``high quality" data from their Table~VI to supplement the light
curves given in Table~\ref{lcparam}. Using this combined data set, we have
assembled values of $t_2$ for a total of 74 novae. Of these, 35 
(approximately half the total) are characterized by
$t_2\le25$~days, and were categorized as either ``fast" or ``very fast"
according to the definition in \citet{war08}. For our purposes
we simply refer to this group as the ``fast" novae. We refer
to the remaining 35 novae with $t_2>25$~days as ``slow" novae.
Given the uncertainties involved in accurately measuring $t_2$,
we did not restrict our sample of fade rates to any particular color or bandpass;
however, when data were available in multiple colors, we chose the $t_2$
values based on $B\/$-band observations to be as consistent
as possible with data from earlier surveys.

In Figure~\ref{fig28} we have plotted separately the spatial distributions of the
``fast" and ``slow" novae.
It appears that the slower
novae (red circles) are perhaps more centrally concentrated than the fast sample
(blue squares). This impression is confirmed when we consider the cumulative
distributions shown in Figure~\ref{fig29}. A KS test reveals that the two
distributions would be expected to differ by more than they do 23\% of the
time if they were drawn from the same parent distribution. Thus, our data
are consistent with the notion that ``faster" novae,
both in the Galaxy \citep{due90,del92} and in M31, are
more associated with the disk population than are the
slower novae. We caution, however, that selection effects could potentially
complicate the interpretation of this result.
As was pointed out earlier in our discussion of spectroscopic class,
our nova sample is not likely to be spatially complete. It is possible
that we may be preferentially missing faster novae in the outer regions of
the galaxy where the temporal sampling of the surveys has perhaps
been less frequent. If so, our conclusion that the
faster novae appear to be more spatially extended
would actually be strengthened.
Taken at face value, our results suggest
that the photometric characteristics of novae are likely
affected by changes in the underlying stellar population.

\subsection{Discussion of Selected M31 Novae}

Below we highlight several individual novae of particular significance.
These objects have
either have been detected as a super-soft X-ray source (SSS), or
have been observed extensively, both photometrically and
spectroscopically, or have some peculiarity that warrants further discussion.

\subsubsection{M31N 1993-10g and 1993-11c}

As part of a program of follow-up spectroscopy of Local Group transients,
one of us (A.V.F.) obtained spectra of two novae in the bulge
of M31 on 1993 Nov. 8 and 17 (UT dates are used throughout this paper). 
The positions of the two objects are near that
of two novae discovered in the survey by \citet{sha01}, M31N 1993-10g and
1993-11c, which are separated by $\sim49''$. Unfortunately, the original
observing logs are no longer available, so we are unable to make an unambiguous
connection between the two spectra and the two novae. Based on approximate
coordinate information in the FITS header for the spectrum taken on 1993 Nov. 17,
we have made the tentative assignments indicated in Figures~\ref{fig1} and~\ref{fig2}.
The spectrum we have associated with M31N~1993-11h is clearly that of an Fe~II
nova, while that of 1993-11c is less certain but consistent with an
Fe~II classification.

Both novae have been identified as possible recurrent nova candidates by \citet{sha01}.
The position of M31N 1993-10g is coincident with 1964-01a to within $9.8''$,
while that of 1993-11c is within $3.6''$ and $6.5''$ of 1967-12a and 1923-02a,
respectively. Given that the coordinates for novae discovered on photographic plates
are not known precisely in many cases, these both appear to be plausible recurrent nova
candidates. However, both novae are located only $\sim2'$ from the nucleus of M31 where
the nova density is high, increasing the probability of a chance positional coincidence.
For a given observed separation $s$, we can
compute the probability of a chance coincidence, $P_C$, by
considering the nova density in an annulus of area $A$,
centered at the position of each nova. Specifically,

\begin{equation}
P_C = 1 - \rm{exp} \Big[ \sum_{i=1}^{n-1} \rm{ln} (1 - ix)\Big],
\end{equation}

\noindent
where $n$ is the number of novae in the annulus, and $x=\pi s^2/A$.
In both cases of interest here, we find that $P_C\grtsim0.95$, making it
highly likely that the two outbursts were a chance coincidence from
separate objects, and therefore not from a recurrent nova.

\subsubsection{M31N 1995-11e}

M31N 1995-11e was identified
during the nova survey of \citet{sha01}, who first recorded the object
on 1995 Nov. 28 at $m_{H\alpha}=18.1$ mag.
The object evolved quite slowly, reaching $m_{H\alpha}=17.9$ mag
after approximately a month before slowly fading.
Then, on 2008 July 6.8,
K. Nishiyama and F. Kabashima (Miyaki-Argenteus Observatory, Japan) found that
the object had appeared again at $m \approx 18.6$ mag (unfiltered) 
before reaching $m \approx 18$ mag on Sep. 09. 
Our spectrum presented here (see Fig.~\ref{fig12})
and originally reported by \cite{sha08b} clearly shows that the object
is a long-period red variable star (i.e., a Mira variable), and not a nova.

\subsubsection{M31N 2001-10a}

M31N 2001-10a was discovered as part of the POINT-AGAPE~\citep{dar04} and
the Naini Tal microlensing surveys \citep{jos04}.
Our spectroscopic data and $r'$-band photometry show that
M31N 2001-10a was a relatively
slowly evolving Fe~II nova characterized by $t_2=73$~days.
X-ray observations reported by \citet{hen10,hen11} show the nova
to be a long-lived SSS that was still detectable more than 7~yr
post outburst. The long duration of the SSS phase is indicative of
prolonged burning on the surface of the white dwarf. This is expected
for the relatively large accreted mass associated with slowly evolving
outbursts on a low-mass white dwarf.

\subsubsection{M31N 2005-01a}

M31N 2005-01a, discovered by \citet{hor05} on
2005~Jan.~07.89, was a particularly luminous nova that was well covered
photometrically near maximum light, reaching $R=15.04$ mag 
(see Fig.~\ref{fig14}). Our spectrum (see Fig.~\ref{fig4}), taken 8 days 
post discovery when the nova was near $R=15.3$ mag, 
shows that the object was clearly
an Fe~II system. It appears to be one of a small number of luminous
Fe~II novae ($M \lessim -9.0$ mag) similar to M31N 2007-11d and 2009-10b
(see Figs.~\ref{fig24}--\ref{fig26}).

\subsubsection{M31N 2005-07a}

M31N 2005-07a, discovered by K.H. on Jul 2005 27.909 at $R=18.4$ mag, 
reached $R=17.4$ mag on Jul 29.919.
Our spectrum (see Fig.~\ref{fig4}), taken $\sim$2 days post discovery
when the nova was near maximum, is characterized
by narrow H$\alpha$ emission and a number of extremely weak features
that may include He and N emission. We tentatively classify the
object as an Fe~II: system, but it is possibly related to the He/Nn novae.

\subsubsection{M31N 2005-09b}

M31N 2005-09b was discovered in the outskirts of M31
by \citet{qui05} on Sep. 1.23
using the 0.45-m ROTSE-IIIb telescope at the McDonald Observatory.
The nova reached $m=16.5$ mag (unfiltered)
a day later on Sep. 2.23. Optical spectra by Leonard (2005) and
Pietsch et al. (2006) reveal moderately broad Balmer
(H$\alpha$: FWHM $\approx 2000$ km~s$^{-1}$, HWZI $\approx 2200$ km~s$^{-1}$),
Fe~II, Na~D, and He~I emission features. The nova is consistent with membership
in the Fe~II spectroscopic class; however, the emission-line width is at the
high end of what is normally seen in Fe~II novae,
and the nova could be plausibly included in the Fe~IIb or hybrid class.

\subsubsection{M31N 2006-09c}

M31N 2006-09c was discovered independently by \citet{qui06},
K. Itagaki, P. Ku\v{s}nir\'ak, and K.H. on 2006 Sep. 18.\footnote{\tt
see http://www.cbat.eps.harvard.edu/CBAT\_M31.html\#2006-09c, and\\
http://www.mpe.mpg.de/$\sim$m31novae/opt/m31/M31\_table.html}
It was detected $\sim150$ days post discovery by both the 
IRAC and IRS instruments
on the {\it Spitzer Space Telescope\/} as part of an
infrared survey of selected M31 novae recently conducted by
\citet{sha11}. No evidence of an infrared excess characteristic of dust
formation was apparent at the time of these observations.
The nova was also detected as a weak SSS by
\citet{hen11} and originally classified as a Fe~II nova by \cite{sha06}.
Our $R$-band photometry shows that $t_2=26$~days, indicating a moderate
rate of decline typical of the Fe~II class.

\subsubsection{M31N 2006-10a}

M31N 2006-10a was a relatively faint nova
discovered at $R=18.7$ mag on 2006 Oct. 25.8 by K.H.\footnote{\tt
see http://www.cbat.eps.harvard.edu/CBAT\_M31.html\#2006-09c}
Our observations (see Figs.~\ref{fig5} and~\ref{fig14}) reveal the
object to be a slowly evolving Fe~II nova. It was
also observed by \citet{sha11} $\sim110$ days post discovery
as part of their {\it Spitzer\/} survey. M31N~2006-10a
showed the clearest evidence of the 10 novae observed by
{\it Spitzer\/} for a near-infrared excess
(in this case peaking at $\lambda \approx 4$~$\mu$m),
suggestive of dust formation. They went on to estimate the
total mass of dust formed to be 
$\sim2\times10^{-6}~{\rm M}_{\odot}$ under the assumption that the dust was
carbon based. \citet{hen11} found no evidence of
X-ray emission from this nova during the time of their observations.

\subsubsection{M31N 2006-10b}

M31N~2006-10b was discovered independently by 
K. Itagaki on 2006~Oct.~31.583 and by R. Quimby and F. Castro at
$m \approx 16.4$ mag on
unfiltered CCD images taken 2006~Oct.~31.09.\footnote{\tt 
see http://www.cbat.eps.harvard.edu/CBAT\_M31.html\#2006-10b}
The time of maximum light is well
constrained by Itagaki's image from Oct. 30.530 (limiting mag 20.0),
which shows no evidence of the nova. Although the light curve is not
complete near maximum light, the available evidence suggests that the nova
faded quite rapidly, with estimates of $t_2(B)=21$~days and $t_2(V)=11$~days.

We obtained two spectra of the nova, the first $\sim1$ day post discovery on
2006~Nov.~01, and the other $\sim 3$ weeks later on
2006~Nov.~23. M31N~2006-10b can be technically considered a ``hybrid" nova
given that Fe~II emission was seen the day after discovery, but
by the time of the second observation the spectrum
had clearly evolved into that of a classic He/N nova.
The object was observed 102 and 110 days post discovery by \citet{sha11} 
with the {\it Spitzer\/} IRS and IRAC, respectively, but not 
detected with either instrument.

\subsubsection{M31N 2006-11a}

M31N 2006-11a is a typical Fe~II nova (see Fig.~\ref{fig6}) that was discovered
by K. Itagaki at $m=17.4$ mag (unfiltered) on 2006 Nov. 25.494.\footnote{\tt
see http://www.cbat.eps.harvard.edu/CBAT\_M31.html\#2006-11a}
The object was observed by \citet{sha11} as part of their 
{\it Spitzer\/} survey.
It was marginally detected by the IRAC, but not with the IRS,
86 and 77 days after discovery, respectively.

\subsubsection{M31N 2007-02b}

This nova was discovered by one of us (K.H.) on 2007 Feb. 03.\footnote{\tt
see http://www.cbat.eps.harvard.edu/CBAT\_M31.html\#2007-02b} It was classified
as a likely hybrid nova by \citet{pie07a} based on possible He 
and N emission in the spectrum. Our spectrum, taken on 2007 Feb. 10.06,
suggests the object is an Fe~II system (see Fig.~\ref{fig7}),
although there appears to be 
a broader component in the
H$\alpha$ and (possibly) the H$\beta$ emission lines that
is often seen in the hybrid systems. The light curve measured in the $R$ band
yields $t_2(R)=35$~days, which is typical of an Fe~II nova but would be
somewhat slow for a hybrid system.
The object was also detected as a SSS by \cite{hen11},
and was still detectable 2~yr after eruption. 
Such a long active SSS phase is characteristic of a relatively large
accreted mass. We conclude that the object was likely an Fe~II nova.

\subsubsection{M31N 2007-06b}

M31N 2007-06b was discovered by \citet{qui07} as part of the ROTSE IIIb program
at McDonald Observatory on 2007 Jun 19.4 at $m=16.8$ mag (unfiltered)
and found to be spatially coincident with the M31 globular
cluster Bol~111. The nova faded by at least 1~mag in $\sim9$ days,
suggesting $t_2\lessim18$~days. Spectroscopic observations by \citet{sha07b}
revealed the nova to be a member of the He/N spectroscopic class.
The object was subsequently detected as a SSS by \citet{pie07b}.

\subsubsection{M31N 2007-07f}

M31N 2007-07f was a slowly evolving nova
discovered in the outskirts of M31
as part of the ROTSE-IIIb program by \citet{yua07}, and
is apparently a member of the Fe~II spectroscopic class (Quimby 2007,
private communication). The object exhibited a slow rise to
maximum light, reaching $m=17.7\pm0.3$ mag on 2007 Jul. 24.02. Approximately
203 days later the object was observed in the {\it Spitzer\/}
survey of \citet{sha11} and detected by the IRAC.
As discussed by \citet{sha11}, 2007-07f showed
evidence (although more marginal than for M31N 2006-10a)
for a weak infrared excess, consistent with that expected
from dust grains formed in the ejecta.

\subsubsection{M31N 2007-08d}

This object was discovered by \citet{pie07c} at $R=18.7$ mag 
on 2007 Aug. 24.081. Our spectrum (see Fig.~\ref{fig7}) and light curve
(Fig.~\ref{fig15}) reveal the object to be a member of the Fe~II
class with $t_2 \approx 63$~days. The nova was included in the \citet{sha11}
{\it Spitzer\/} survey and marginally detected with the IRAC, but not the IRS,
about 158 and 183 days post discovery, respectively.

\subsubsection{M31N 2007-10a}

M31N~2007-10a was discovered K. Itagaki on Oct. 5.606 and independently by
Pietsch et al.\footnote{\tt 
see http://www.cbat.eps.harvard.edu/CBAT\_M31.html\#2007-10a}
The spectrum was
classified by \citet{gal07} as an Fe~II nova, but our spectrum
(see Fig.~\ref{fig7}) reveals narrow Balmer,
He~I $\lambda\lambda$4921, 5016, 5876, and He~II $\lambda$4686 emission,
with only a trace of Fe~II, possibly blended
with He~I at 5169~\AA.
M31N 2007-10a is the prototype of our proposed He/Nn spectroscopic class.
The nova faded rapidly with $t_2(B) \approx t_2(V) \approx 7$~days, 
and was not subsequently detected as an SSS by \cite{hen11}, nor as an 
infrared source in the {\it Spitzer\/} survey of \citet{sha11}. 

\subsubsection{M31N 2007-10b}

M31N~2007-10b is the best example from our survey of a faint but fast nova
similar to the class of novae discovered by \citet{kas11}.
The nova was discovered by \citet{bur07} at $R=17.8$ mag, 
who were able to constrain
the time of maximum to within a day of 2007 Oct. 13.26, and independently
by K.H. and P. Ku\v{s}nir\'ak.\footnote{\tt
see http://www.cbat.eps.harvard.edu/CBAT\_M31.html\#2007-10b} 
The nova faded unusually quickly: our $B$, $V$, and $R$ light curves 
suggest $t_2 \approx 3$--4 days.
Rau et al. (2007) classified
the object as an He/N nova based on a spectrum obtained $\sim 3$ 
days post discovery,
and noted that the emission lines were unusually narrow
(FWHM H$\alpha = 1450\pm100$~km~s$^{-1}$) for this class. 
Based on the available
photometric and spectroscopic data, we suggest that the object
is a member of our proposed He/Nn class. Consistent with its rapid evolution,
the object was detected by \citet{hen11} as a SSS with an
X-ray duration of less than 100 days.

\subsubsection{M31N 2007-11b}

M31N 2007-11b, our third example of an He/Nn nova, was discovered by
\citet{pie07d} and independently by E. Ovcharov and A. Valcheva.\footnote{\tt
see http://www.cbat.eps.harvard.edu/CBAT\_M31.html\#2007-11b}
Although our spectrum
(see Fig.~\ref{fig8}) appears to be quite similar to
that of an Fe~II nova (albeit with weak
Fe~II emission), subsequent spectra by \citet{rau07}
and \citet{bar07b}
showed that the object quickly developed prominent,
narrow (FWHM H$\alpha = 1430\pm100$~km~s$^{-1}$) Balmer, He~I, 
and He~II emission lines.
Unlike the other He/Nn systems (M31N 2007-10a and 2007-10b),
the light-curve evolution of 2007-11b was not particularly fast, with
$t_2(B)=25$~days and $t_2(V)=45$~days, respectively.

\subsubsection{M31N 2007-11d}

M31N~2007-11d was an unusually bright and slowly rising
nova discovered by K. Nishiyama and F. Kabashima on Nov. 17.57, and
subsequently studied extensively by \citet{sha09}. The early spectrum
was that of a classic Fe~II system: narrow Balmer and Fe~II emission lines
flanked to the blue by pronounced P-Cyg absorption features. Another spectrum
obtained $\sim 2$ weeks later revealed moderately broad
Balmer and Fe~II emission lines (FWHM H$\alpha = 2260$~km~s$^{-1}$)
with weak He~I and O~I emission (see Fig.~\ref{fig8}).
The nova faded moderately rapidly ($t_2[V]=9.5$~days) for an Fe~II nova,
and was apparently
not detected as an SSS source despite what must have been a relatively large
accreted mass. It was also observed, but not detected, with the {\it Spitzer\/}
IRS \citep{sha11}.

\subsubsection{M31N 2007-11g}

M31N 2007-11g was discovered by \citet{ovc07} on 2007 Nov 28.716 at $R 
\approx 18.7$ mag. Our spectrum, presented in Fig.~\ref{fig12} and originally
reported by \cite{sha07c}, clearly shows that the object is not a nova,
but rather a long-period Mira variable in M31.

\subsubsection{M31N 2007-12b}

M31N~2007-12b was a relatively bright and rapidly evolving He/N nova. It was
discovered on 2007 Dec. 9.53 by K. Nishiyama
and F. Kabashima1 at $m=16.1$--16.2 mag (unfiltered), and independently
by K.H. on Dec. 10.73 at $R = 17.0$ mag.\footnote{\tt
see http://www.cbat.eps.harvard.edu/CBAT\_M31.html\#2007-12b}
Subsequent spectroscopic observations revealed the object to be a 
rapidly declining ($t_2 = 8.7[B], 7.2[V]$~days) 
He/N system (see Fig.~\ref{fig9} and Table~\ref{lcparam}).
Initially, the object was thought to be a recurrent nova given its close
proximity to the position of M31N~1969-08a; however, subsequent astrometry
established that the two novae were, in fact, distinct objects.

Archival {\it Hubble Space Telescope\/} observations of the pre-outburst
location of M31N 2007-12b \citep{bod09} revealed
the presence of a coincident stellar source
with magnitude and color very similar to those of the Galactic
recurrent nova RS Ophiuchi,
where the red giant secondary star dominates the light at quiescence.
This discovery, coupled with the rapid photometric evolution and
the object's detection by the {\it Swift\/} satellite \citep{bur05}
as an SSS \citep{kon08}, were interpreted by \citet{bod09}
as strong evidence that M31N 2007-12b is, nevertheless, a recurrent nova
system.

\subsubsection{M31N 2007-12d}

M31N~2007-12d was discovered independently by \citet{hen07} and by
K. Nishiyama \& F. Kabashima\footnote{\tt 
see http://www.cbat.eps.harvard.edu/CBAT\_M31.html\#2007-12d}
on 2007-Dec. 17.57. Although no detailed light-curve data exist, \citet{hen11}
estimated a rapid decline with $t_2 \approx 4$~days. Our spectrum
(see Fig.~\ref{fig9}) reveals the nova to be a classic He/N system with
broad emission lines of H, He, and N (FWHM H$\alpha \approx 5000$~km~s$^{-1}$)
indicating a high expansion velocity. The nova was detected briefly 
($<20$~days) as an SSS by \cite{hen11}.

\subsubsection{M31N 2009-10b}

M31N~2009-10b was discovered by K.H. and P. Ku\v{s}nir\'ak on 2009 Oct. 9.986
\citep{hor09}, and
independently by K. Itagaki on Oct. 11.414.\footnote{\tt 
see http://www.cbat.eps.harvard.edu/CBAT\_M31.html\#2009-10b}.
It was an unusually bright nova that reached $R=14.7$ mag 
(Hornoch et al. 2009b) before fading relatively rapidly 
($t_2 \approx 10$~days in $B$ and $V$).
Spectroscopic observations by \citet{dim09b} and \citet{bar09b}
show conclusively that the nova belongs to the Fe~II spectroscopic 
class, making it quite similar to M31N~2007-11d.

\section{Conclusions}

Whether there exist two distinct populations
of classical novae is an important but unanswered question.
In an attempt to gain further insight, we have conducted a major spectroscopic
survey of novae in the nearby galaxy M31. These data have allowed us to
determine spectroscopic classes for a total of 46 novae, more than doubling
the number previously available. Specifically, after combining our
data with published spectra, we have now been able to
compile a list of spectroscopic classes for
a total of 91 novae that erupted prior to 2010.
In addition, we have undertaken photometric observations of
many of the recent novae in this group
in order to measure their light curves (i.e.,
their peak brightness and rate of decline). Whenever possible, 
we have augmented our photometric observations with light-curve 
data from the literature.

Our combined spectroscopic and photometric survey has allowed us to explore
the spatial distribution of novae in M31 to a greater extent than
has been possible previously. An analysis of these data has enabled us 
to arrive at the following conclusions.

\noindent
$\bullet$ As part of this survey we have found that $\sim 80$\% of the M31
novae with available spectra belong to the Fe~II class. The remaining 
$\sim 20$\% are composed of novae whose spectra are characterized by
H, He, and N emission lines, with Fe~II emission features either weak 
or absent. Usually these latter systems display relatively broad 
(FWHM $\grtsim 2500$~km~s$^{-1}$) lines typical of the He/N 
spectroscopic class; however, a small fraction of these systems 
(e.g., M31N~2007-10a, 2007-10b, and 2007-11b) are characterized
by relatively narrow line widths (FWHM $\lessim$ 2000~km~s$^{-1}$). We refer
to the latter systems as narrow-line He/N, or He/Nn, novae.
The relative percentages of Fe~II and He/N (and related) novae are
similar to those found for Galactic novae \citep{sha07a,del98}.

\noindent
$\bullet$ We have presented photometric observations with sufficient 
coverage to determine light-curve parameters 
(peak brightness and rate of decline)
for most of the novae in our spectroscopic survey, and for
approximately half of all novae with known spectroscopic classifications.
These data have allowed us to confirm that novae in the He/N and He/Nn classes
have generally faster light-curve evolution than the more
common Fe~II objects. When the light-curve parameters for the
entire sample are considered, we find that the brighter novae generally
fade the fastest; they are consistent with an MMRD relation.
Similarly, we find that the ejection velocities inferred from line
widths are higher for the faster novae, as is the case for their
Galactic counterparts.

\noindent
$\bullet$ Under the assumption that there exist two separate populations of
novae (bulge and disk populations), it is believed that the disk population,
with their generally more massive white dwarfs,
should produce novae that are on average brighter and faster than
their counterparts in the bulge population \citep{due90,del92}. To
test this prediction further,
we have explored the photometric behavior (specifically $t_2$) of
novae as a function of spatial position in M31. After supplementing
photometric data from our survey with
decline rates from the ``high quality" light-curve sample given
by \citet{cap89}, we were able to generate a sample of 74 M31 novae
with well-determined fade rates. This sample was subsequently divided into two
groups: those with $t_2\le25$~days (the ``fast" sample) 
and those with $t_2>25$~days
(the ``slow" sample). A comparison of the spatial distributions for the two
samples shows that the fast novae are in fact more spatially extended
from the core of M31
than the slow novae, as expected in the two-population scenario.

\noindent
$\bullet$ We have also explored the possibility that the spectroscopic
class of M31 novae varies with spatial position in the galaxy, as would be
expected if the He/N and related novae contain more massive white dwarfs.
Surprisingly, as shown in Figures~\ref{fig21} and~\ref{fig22},
the spatial distribution shows only a hint that Fe~II objects may be more
centrally concentrated (i.e., associated with the bulge of M31), and there is
no compelling evidence for a
dependence on spectroscopic class.
Specifically, a KS test shows
a 81\% probability that the Fe~II and He/N
distributions would differ more than what is observed
if they were drawn from the same overall distribution.
This result suggests that the average white dwarf mass 
in nova systems may not be as strongly dependent on spatial
position (and hence stellar population) in M31 as suggested by the 
photometric data.

Taken together, our spectroscopic and photometric data
do not provide compelling evidence
in support of the hypothesis
that there exist two populations of novae in M31. Nevertheless, our
light-curve data can be interpreted as mildly suggestive of a weak
dependence of nova-speed 
class on spatial position (stellar population) within the galaxy.
Furthermore, our spectra are not inconsistent with the
possibility that spectroscopic type may be sensitive to stellar population.
Whatever sensitivity there may be, however, appears to be
weak, if it exists at all.

A major step forward in the understanding of nova populations generally,
and the spectroscopic classification specifically, will likely
require additional spectra and light-curve data for novae erupting
in galaxies spanning a range of morphological types.
For example, a sample of light curves and 
spectra from novae arising in the extreme Population II environment
of an elliptical galaxy will be particularly instructive.
The Large Synoptic Survey Telescope, when it becomes operational,
will generate a large sample of Virgo cluster nova light curves 
that will undoubtedly shed new light on the question of 
nova populations. Spectra of novae in Virgo cluster galaxies
can then be obtained with low-resolution
spectrographs currently available on 10-m class telescopes.

\acknowledgments

The work presented here was made possible through
observations obtained from
facilities based throughout the world. Spectra
were obtained with the Lick Observatory Shane 3-m
telescope operated by the University of California and with the Marcario
Low-Resolution Spectrograph on the Hobby-Eberly Telescope,
which is operated by McDonald Observatory on behalf of the University
of Texas at Austin, the Pennsylvania State University,
Stanford University, the Ludwig-Maximillians-Universit\"{a}t,
Munich, and the George-August-Universit\'{a}t, G\"{o}ttingen.
Photometric observations were made using the Liverpool 
Telescope, which is operated on the island of La Palma by Liverpool 
John Moores University (LJMU) in the Spanish Observatorio del
Roque de los Muchachos of the Instituto de Astrofisica de Canarias with
financial support from the UK Science and Technology Facilities Council. 
Faulkes Telescope North (FTN)
is operated by the Las Cumbres Observatory Global Telescope
network. Data from FTN were obtained as part of a joint programme
between Las Cumbres Observatory and the LJMU Astrophysics Research
Institute.
Photometric observations were also obtained at the Centro Astron\'omico
Hispano Alem\'an (CAHA) Observatory
at Calar Alto, operated jointly by the Max-Planck
Institut f\"{u}r Astronomie and the Instituto de Astrof\'isica de Andaluc\'ia
(CSIC), with the 6-m telescope
of the Special Astrophysical Observatory (SAO) of the Russian Academy
of Sciences (RAS), operated under the financial support of the Science
Department of Russia (registration number 01-43),
with the Vatican Advanced Technology Telescope (the Alice P. Lennon Telescope
and the Thomas J. Bannan Astrophysics Facility),
with the 1.3-m McGraw-Hill
and the 2.4-m Hiltner telescopes at the MDM Observatory,
with the 2.5-m Isaac Newton Telescope
operated on the island of La Palma by the Isaac Newton Group in the Spanish
Observatorio del Roque de los Muchachos of the Instituto de Astrofísica
de Canarias, with the
2.1-m telescope of the Kitt Peak National Observatory, National
Optical Astronomy Observatory, which is operated by the Association of Universities
for Research in Astronomy (AURA), Inc., under cooperative agreement with the National
Science Foundation,
and with the 0.84-m telescope of the Observatorio Astron\'omico
Nacional, San Pedro M\'artir.
We wish to thank the staff of all these facilities for their
assistance in obtaining the observations reported here.

Finally, we would like to thank the following individuals for
contributing images of M31: V. L. Afanasiev, Z. Bardon, M. Burleigh,
P. Caga\v{s}, S. Casewell, S. N. Dodonov, T. Farnham,
A. Gal\'ad, J. Gallagher, P. Garnavich, J. Gorosabel, T. Henych,
M. Jel\'inek, A. Karska, C. Kennedy, R. Khan, P. Kub\'anek, P. Ku\v{s}nir\'ak,
D. Mackey, K. Morhig, B. Mueller, O. Pejcha, J. Prieto, N. Samarasinha,
L. \v{S}arounov\'a, P. \v{S}edinov\'a, O. N. Sholukhova, K. Thorne, B. Tucker, 
M. Tukinsk\'a, A. Valeev, M. Wolf, and P. Zasche.
We are also grateful to the following
for assistance with the Lick spectroscopic observations and reductions:
A. Coil,
R. J. Foley,
M. Ganeshalingam,
S. Jha,
L. C. Ho,
J. Hoffman,
D. C. Leonard,
W. Li,
M. Papenkova,
F. J. D. Serduke,
J. C. Shields,
and J. M. Silverman.
Photometric reduction software was kindly provided by P. Caga\v{s} (SIMS),
F. Hroch (Munipack), and M.~Velen and P.~Pravec (Aphot).
This research has made use of the SIMBAD database, operated at CDS,
Strasbourg, France, and of NASA's Astrophysics Data System Bibliographic
Services.
A.V.F.'s group at UC Berkeley is grateful
for the financial support of the National Science Foundation
(most recently through grant AST-0908886) and the TABASGO Foundation.
A.W.S. is grateful to the University of Victoria for hospitality during
a recent sabbatical leave while this work was being completed,
and to the NSF for financial support through grants AST-0607682 
and AST-1009566.

\clearpage




\clearpage

\clearpage

\clearpage

\clearpage

\clearpage

\clearpage

\clearpage

\clearpage

\begin{figure}
\includegraphics[width=0.40\textwidth]{f20a.eps}
\includegraphics[width=0.40\textwidth]{f20b.eps}
\newline
\includegraphics[width=0.40\textwidth]{f20c.eps}
\includegraphics[width=0.40\textwidth]{f20d.eps}
\newline
\includegraphics[width=0.40\textwidth]{f20e.eps}
\caption{Nova light curves (continued). See Fig.~\ref{fig13} for details.\label{fig20}}
\end{figure}

\clearpage

\begin{figure}
\includegraphics[angle=-90,scale=.85]{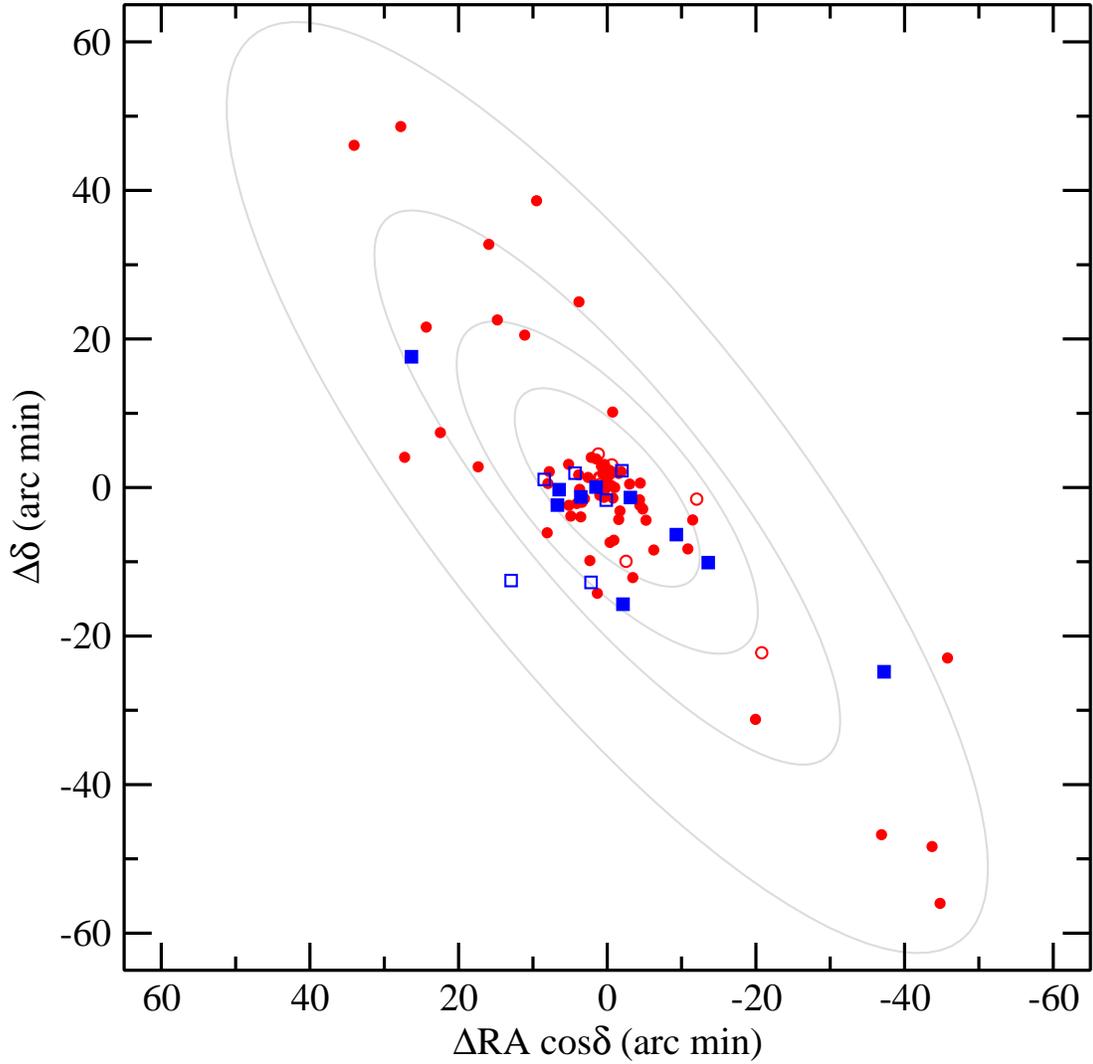}
\caption{The spatial distribution of the 91 M31
novae with known spectroscopic class (see Table~\ref{spectsample}).
The Fe~II and Fe~II: novae are
indicated by filled and open red circles, respectively. The
He/N and He/N: novae are represented by filled and open
blue squares, respectively. The gray ellipses
represent elliptical isophotes from the surface photometry of
\citet{ken87}.
\label{fig21}}
\end{figure}

\begin{figure}
\includegraphics[angle=-90,width=.7\textwidth]{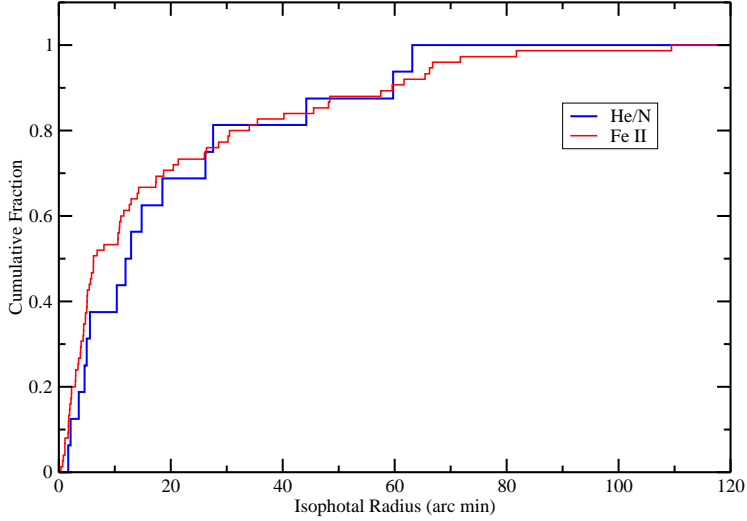}
\newline
\includegraphics[angle=-90,width=.7\textwidth]{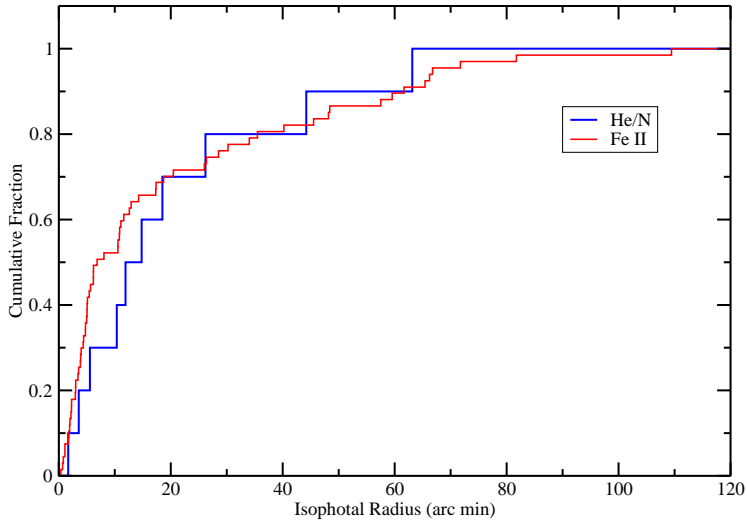}
\caption{The cumulative distributions of Fe~II novae compared with
that for He/N and related novae.
The top panel shows the Fe~II and Fe~II: systems
(red) compared with the He/N + hybrid and He/N: systems (blue). The bottom panel
compares only the well-established Fe~II and He/N + hybrid novae.
A KS test indicates a 81\% (73\% for bottom panel) probability that the distributions
would differ by more than they do if both
distributions come from the same parent population.
\label{fig22}}
\end{figure}
  
\begin{figure}
\includegraphics[angle=-90,scale=.60]{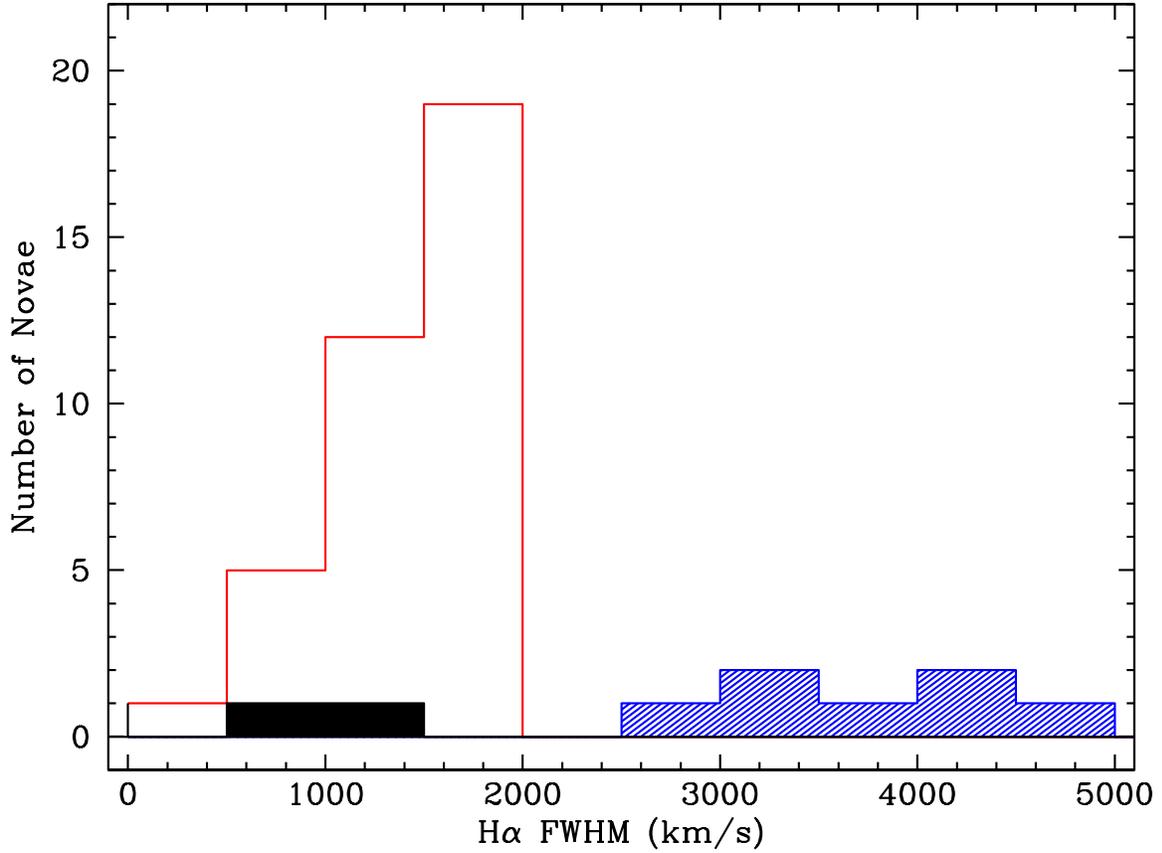}
\caption{The distribution of H$\alpha$ emission-line FWHM values from
the novae in our sample. The novae classified as He/N (cross-hatched blue histogram) are clearly
segregated from their Fe~II counterparts (red open histogram),
with the latter systems having FWHM~$\lessim2500$~km~s$^{-1}$.
Notable exceptions are two peculiar novae classified as He/Nn for which we
have FWHM measurements that are represented by the filled region.
\label{fig23}}
\end{figure}
  
\begin{figure}
\includegraphics[angle=-90,scale=0.65]{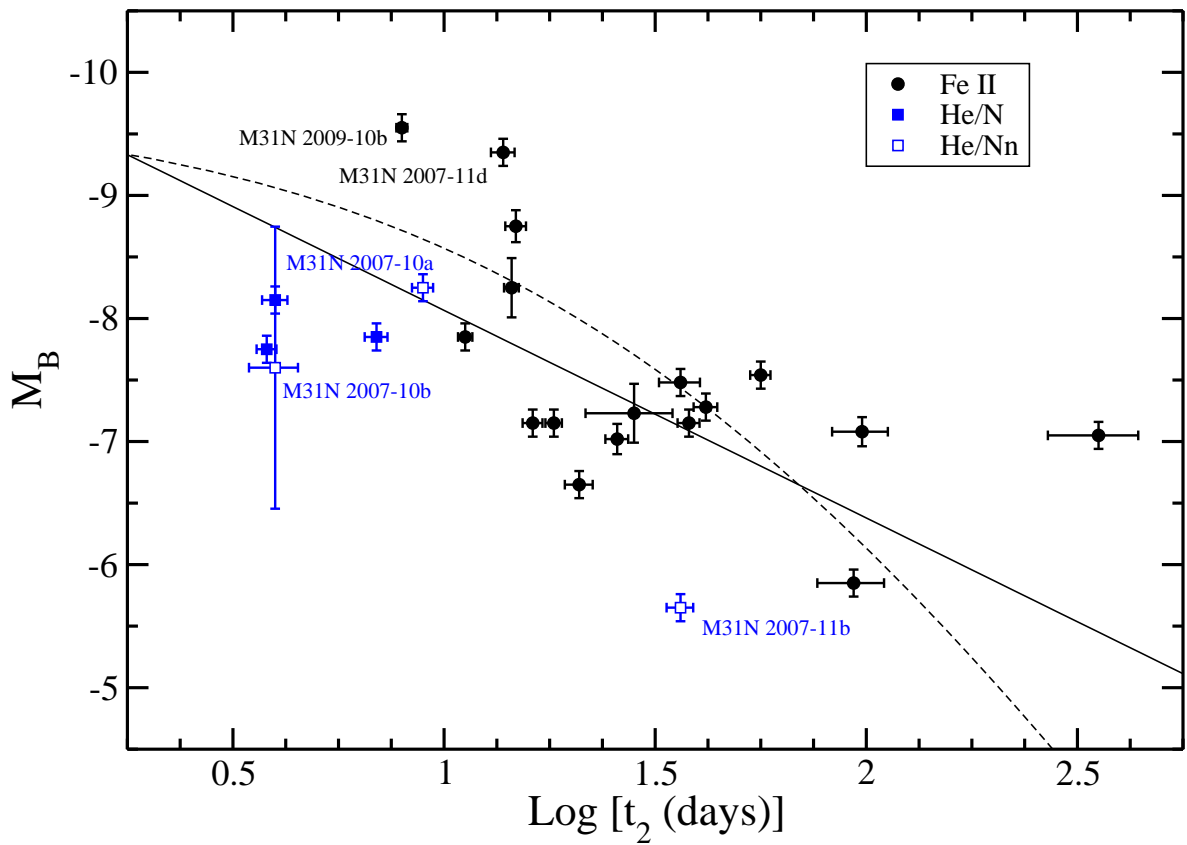}
\caption{The $B\/$-band maximum-magnitude vs. rate-of-decline relation
(MMRD) from our photometric survey. The Fe~II, He/N, and He/Nn novae are
represented by filled red circles, filled blue squares, and open blue
squares, respectively. The solid line represents the best-fit
relation determined from a weighted linear least-squares analysis
(Equation~1), while the dashed line represents the theoretical
relation from \citet{liv92}. Despite the considerable scatter, the data
follow the expected trend with the brightest novae generally
fading the fastest.\label{fig24}}
\end{figure}

\begin{figure}
\includegraphics[angle=-90,scale=0.65]{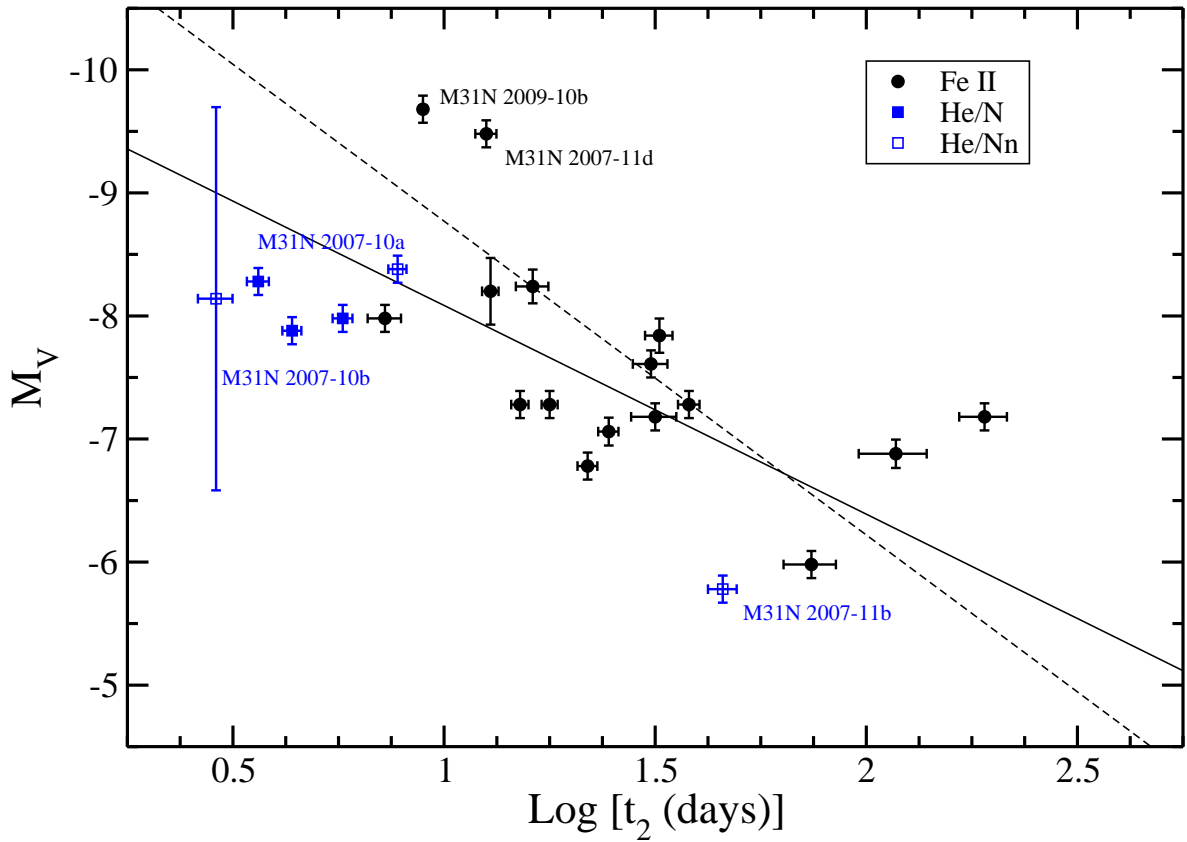}
\caption{The $V\/$-band MMRD relation
from our photometric survey. The symbols have the same meaning as in
Fig.~\ref{fig24}. The solid line is the best-fit relation
given by Equation~2, while
the dashed line represents the
Galactic $V\/$-band relation from \citet{dow00}.\label{fig25}}
\end{figure}

\begin{figure}
\includegraphics[angle=-90,scale=0.65]{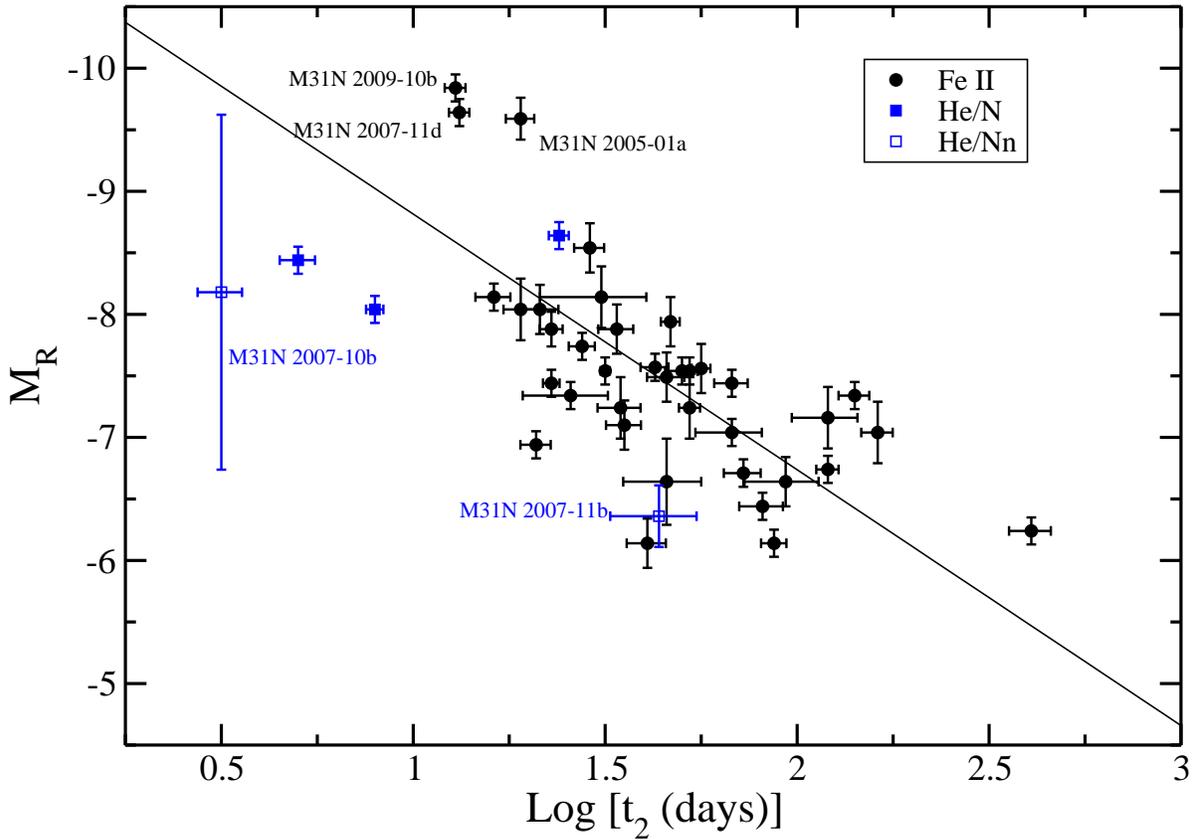}
\caption{The $R\/$-band MMRD relation.
The symbols have the same meaning as in Fig.~\ref{fig24}.
The best-fit relation is given by Equation~3. Note the tight group
of luminous Fe~II novae (M31N~2005-01a, 2007-11d, and 2009-10b)
with $M_V \lessim -9$ mag.\label{fig26}}
\end{figure}

\begin{figure}
\includegraphics[angle=-90,scale=.65]{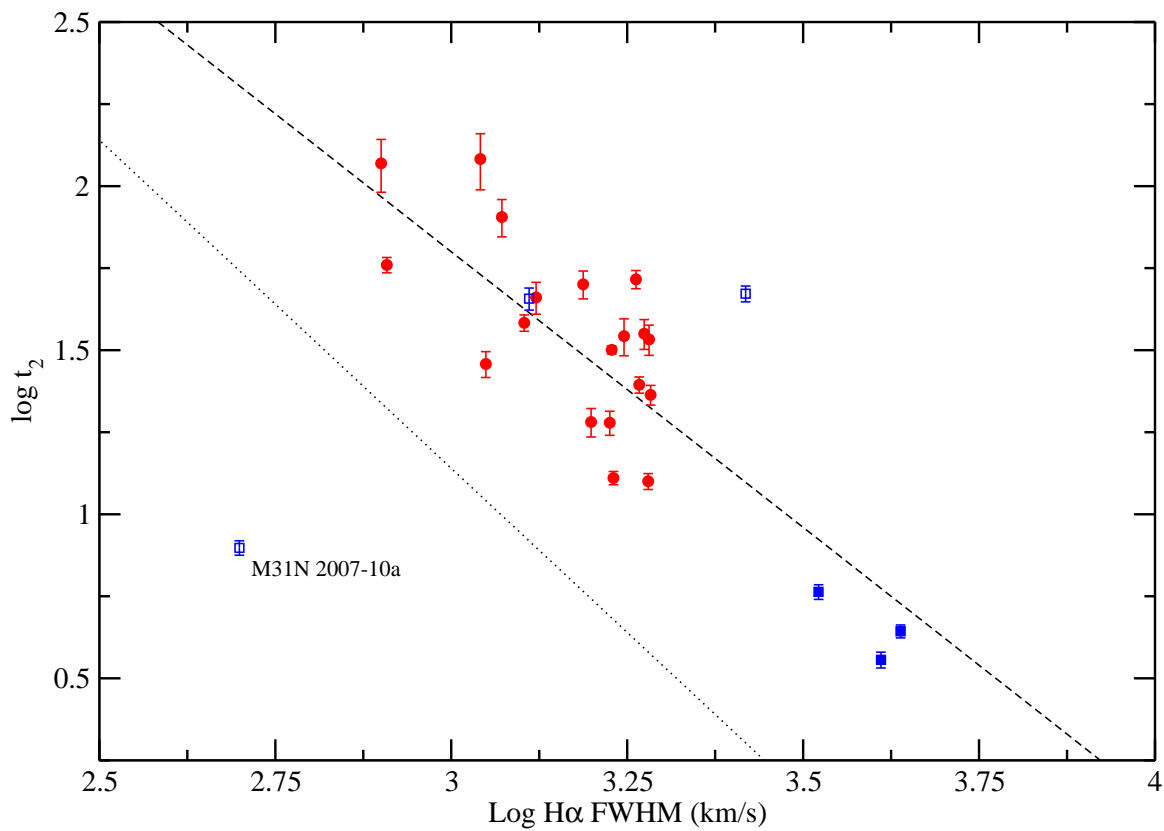}
\caption{The dependence of the
light-curve parameter $t_2$ on nova expansion velocity (as reflected by
the FWHM of H$\alpha$). With the exception of M31N 2007-10a, there is
a clear trend of decreasing $t_2$ with increasing H$\alpha$ emission-line
width. The red filled circles represent Fe~II novae, while the filled (open)
blue squares represent He/N (He/Nn and He/N:) novae, respectively.
The dashed line reflects the best-fit relation given in Equation~5,
while the dotted line gives the Galactic relation of \citet{mcl60}.
\label{fig27}}
\end{figure}

\begin{figure}
\includegraphics[angle=-90,scale=.85]{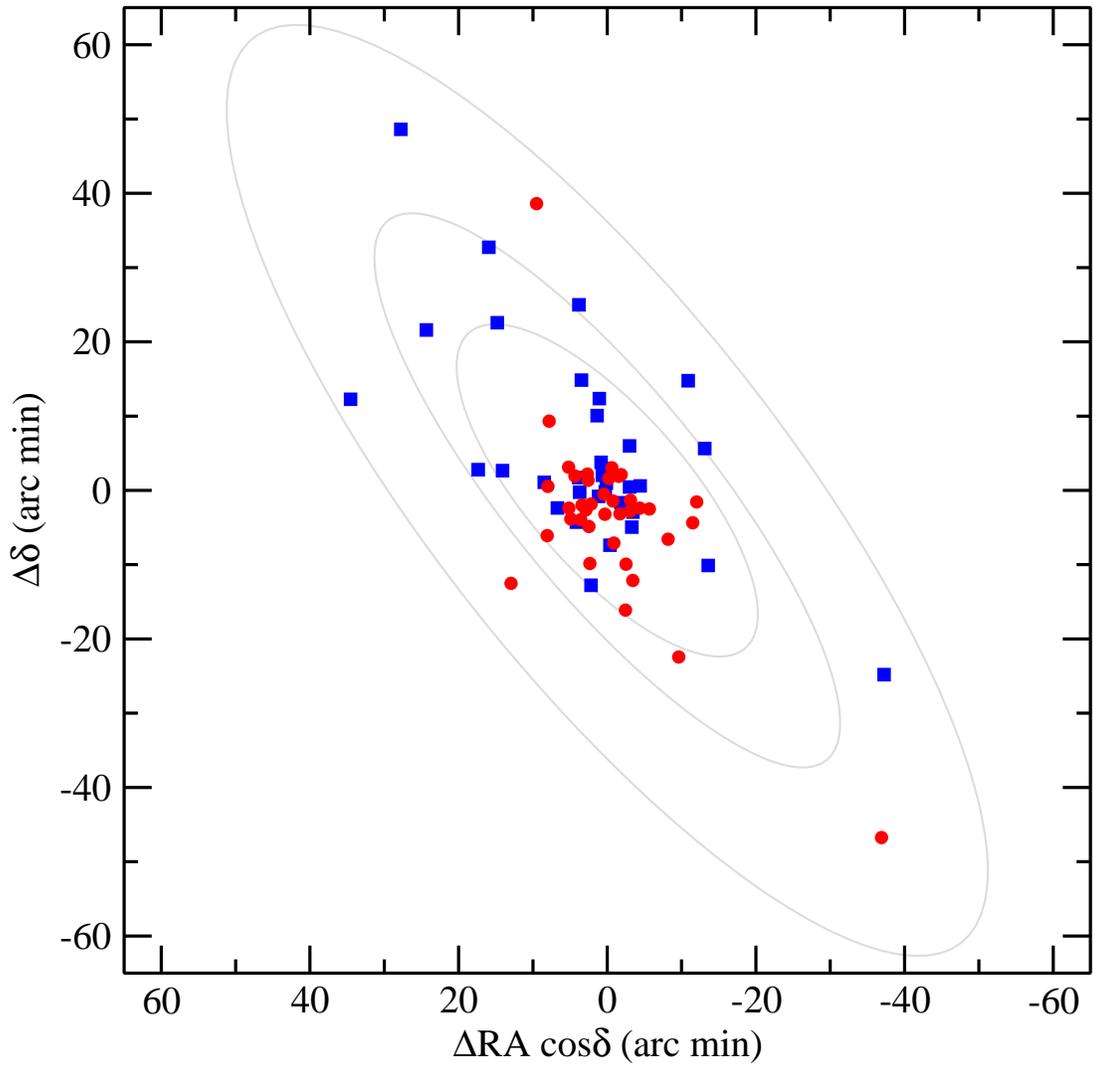}
\caption{The spatial distribution of the 47 M31
novae with measured decline rates from our survey supplemented by 27 decline rates
from the ``high quality" light-curve sample from the Hubble, Arp, and Rosino
surveys \cite{cap89}. The ``very fast" and ``fast" novae
($t_2\leq25$~days) are indicated by blue squares, with the slower
novae ($t_2>25$~days) are indicated by red circles. The gray ellipses
represent elliptical isophotes from the surface photometry of
\citet{ken87}.
\label{fig28}}
\end{figure}

\begin{figure}
\includegraphics[angle=-90,scale=.65]{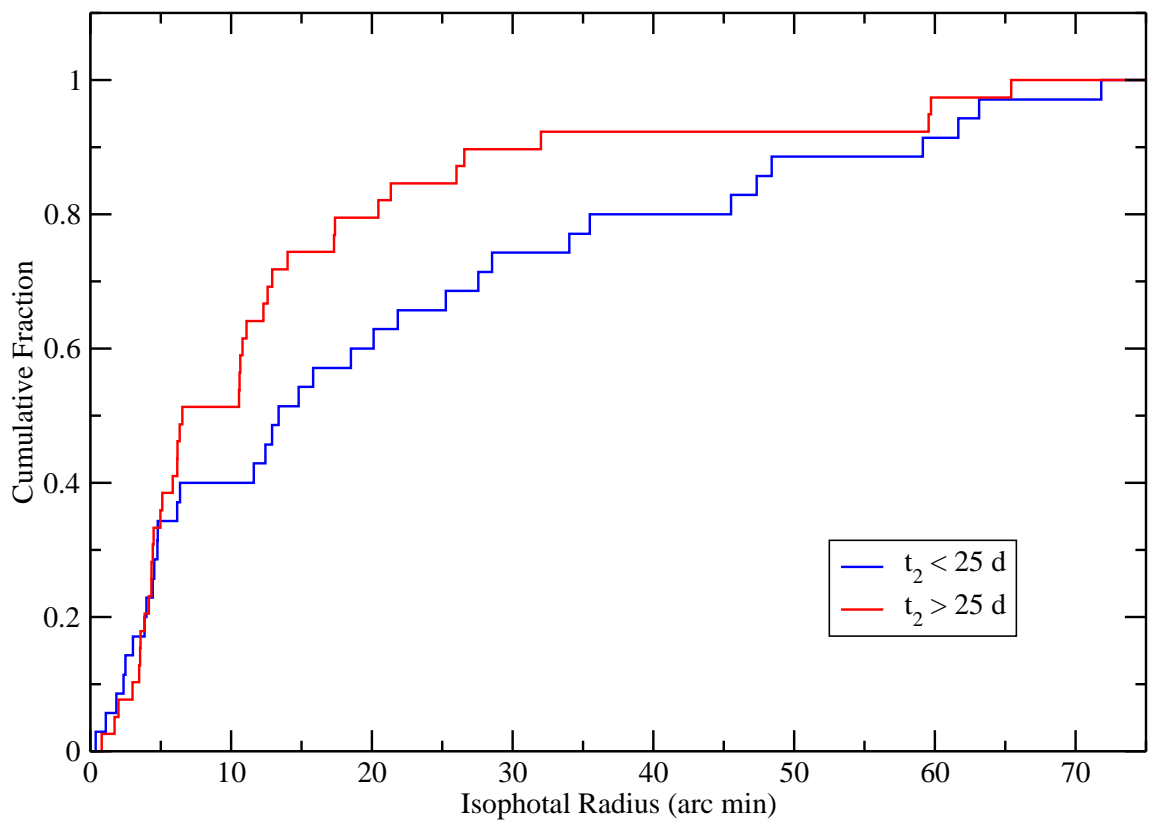}
\caption{The cumulative distributions of the two nova samples
from Figure~\ref{fig28}. The blue distribution represents ``very fast" 
and ``fast" novae ($t_2\leq25$~days), with the red distribution (broken 
lines) representing slower novae ($t_2>25$~days).
A KS test indicates a 23\% probability that the distributions
would differ by more than they do if both
distributions come from the same parent population.
Thus, it appears possible that the faster novae are more extended compared
with the slower declining systems.
\label{fig29}}
\end{figure}







\clearpage

\begin{deluxetable}{lcccr}
\tabletypesize{\scriptsize}
\tablenum{1}
\tablewidth{0pt}
\tablecolumns{5}
\tablecaption{Summary of Lick Spectroscopic Observations\label{lickdata}\tablenotemark{a}}
\tablehead{\colhead{} & \colhead{R.A.} & \colhead{Decl.} & \colhead{} & \colhead{Coverage} \\
\colhead{Nova} & \colhead{(J2000.0)} & \colhead{(J2000.0)} & \colhead{UT Date} & \colhead{(\AA)}}
\startdata
M31N 1990-10b  &  00$^{\rm h}42^{\rm m}36.2^{\rm s}$  & $41^\circ 11' 54.0''$ & 11 Nov. 1990 & 3900--7100 \\ 
M31N 1992-11b  &  00 42 36.2  & 41 11 54.0 & 18 Nov. 1992 & 3500--9500 \\ 
M31N 1993-06a  &  00 42 49.2  & 41 17 27.5 & 28 Jun. 1993 & 4200--7100 \\ 
M31N 1993-08a  &  00 42 45.1  & 41 14 27.0 & 12 Sep. 1993 & 3500--9500 \\ 
M31N 1993-10g\tablenotemark{b} &  00 42 47.7  & 41 18 01.0 & 08 Nov. 1993 & 3500--9500 \\ 
M31N 1993-11c\tablenotemark{b} &  00 42 50.1  & 41 17 28.0 & 17 Nov. 1993 & 4300--7100 \\ 
M31N 1998-09d  &  00 42 46.6  & 41 14 49.2 & 20 Sep. 1998 & 4300--7000 \\ 
M31N 1999-06a  &  00 42 49.7  & 41 15 05.6 & 10 Sep. 1999 & 4300--7000 \\ 
M31N 1999-08f  &  00 42 41.1  & 41 19 12.2 & 17 Sep. 1999 & 4300--7000 \\ 
M31N 1999-10a  &  00 42 49.7  & 41 16 32.0 & 08 Oct. 1999 & 4300--7000 \\ 
M31N 2001-10a  &  00 43 03.3  & 41 12 11.5 & 20 Oct. 2001 & 3500--9500 \\ 
M31N 2001-12a  &  00 42 41.4  & 41 16 24.5 & 11 Feb. 2002 & 3300--7900 \\ 
M31N 2002-01b  &  00 42 33.9  & 41 18 23.9 & 11 Feb. 2002 & 3300--7900 \\ 
M31N 2002-08a  &  00 42 30.92 & 41 06 13.1 & 13 Sep. 2002 & 3500--9500 \\
M31N 2004-08b  &  00 43 26.84 & 41 16 40.8 & 10 Sep. 2004 & 3500--9500 \\
M31N 2004-09a  &  00 42 40.27 & 41 14 42.5 & 10 Sep. 2004 & 3500--9500 \\
M31N 2004-11a  &  00 42 42.81 & 41 18 27.9 & 19 Nov. 2004 & 3500--9500 \\
M31N 2004-11b  &  00 43 07.45 & 41 18 04.7 & 19 Nov. 2004 & 3500--9500 \\
M31N 2005-01a  &  00 42 28.39 & 41 16 36.2 & 16 Jan. 2005 & 3500--9500 \\
M31N 2005-07a  &  00 42 50.79 & 41 20 39.8 & 01 Aug. 2005 & 3500--9500 \\
M31N 2008-08d  &  00 45 48.25 & 43 02 22.2 & 08 Sep. 2008 & 3500--9500 \\
\enddata
\tablenotetext{a}{All observations obtained with the Shane 3-m reflector.}
\tablenotetext{b}{Due to ambiguity in the data logs from November 1993, it is possible
that the dates of observation (and thus the spectra) for these two novae are reversed.}
\end{deluxetable}

\begin{deluxetable}{lcccccc}
\tabletypesize{\scriptsize}
\tablenum{2}
\tablewidth{0pt}
\tablecolumns{7}
\tablecaption{Summary of HET Spectroscopic Observations\label{hetdata}}
\tablehead{\colhead{} & \colhead{R.A.} & \colhead{Decl.} & \colhead{}&\colhead{Exp.}
 & \colhead{Coverage} & \colhead{}  \\
\colhead{Nova} & \colhead{(J2000.0)} & \colhead{(J2000.0)} & \colhead{UT Date} & \colhead{(sec)} & \colhead{(\AA)} & \colhead{Weather}}
\startdata
M31N 1995-11e  &  00$^{\rm h}$45$^{\rm m}$09.9$^{\rm s}$  & $41^\circ 52' 03.0''$ & 03.30 Nov. 2008 &1200 &4275--7250 &
Spec \\
M31N 2006-09c  &  00 42 42.38 & 41 08 45.5 & 24.18 Sep. 2006 &1800 &4275--7250 & Spec \\
M31N 2006-10a  &  00 41 43.23 & 41 11 45.9 & 30.31 Oct. 2006 &1500 &4275--7250 & Spec \\
M31N 2006-10b  &  00 39 27.38 & 40 51 09.8 & 01.09 Nov. 2006 &1800 &4275--7250 & Spec\\
               &              &            & 23.24 Nov. 2006 &1200 &4275--7250 & Phot\\
M31N 2006-11a  &  00 42 56.81 & 41 06 18.4 & 28.23 Nov. 2006 &1200 &4275--7250 & Spec\\
M31N 2006-12a  &  00 42 21.09 & 41 13 45.3 & 08.13 Jan. 2007 &1200 &4275--7250 & Spec\\
M31N 2006-12b  &  00 42 11.14 & 41 07 43.8 & 10.11 Jan. 2007 &3200 &4275--7250 & Cldy\\
M31N 2007-02a  &  00 40 59.02 & 40 44 52.7 & 09.07 Feb. 2007 &900 &4275--7250 & Spec\\
M31N 2007-02b  &  00 41 40.32 & 41 14 33.5 & 10.06 Feb. 2007 &900 &4275--7250 & Phot\\
M31N 2007-06b  &  00 42 33.14 & 41 00 25.9 & 25.34 Jul. 2007 &3600 &4275--7250 & Spec\\
M31N 2007-08d  &  00 39 30.27 & 40 29 14.2 & 13.44 Sep. 2007 &1200 &4275--7250 & Spec\\
M31N 2007-10a  &  00 42 55.95 & 41 03 22.0 & 19.13 Oct. 2007 &1400 &4275--7250 & Phot\\
M31N 2007-11b  &  00 43 52.99 & 41 03 36.2 & 11.30 Nov. 2007 &1200 &4100--8900 & Spec\\
M31N 2007-11c  &  00 43 04.14 & 41 15 54.3 & 16.28 Nov. 2007 &1200 &4100--8900 & Spec\\
M31N 2007-11d  &  00 44 54.60 & 41 37 40.0 & 21.27 Nov. 2007 &600 &4100--8900 & Phot\\
               &              &            & 04.22 Dec. 2007 &1200 &4100--8900 & Phot\\
M31N 2007-11e  &  00 45 47.74 & 42 02 03.5 & 05.23 Dec. 2007 &1200 &4100--8900 & Phot\\
M31N 2007-11g  &  00 44 15.80 & 41 13 50.3 & 18.28 Nov. 2008 &1200 &4275--7250 & Spec\\
M31N 2007-12a  &  00 44 03.51 & 41 38 41.1 & 15.20 Dec. 2007 &1200 &4100--8900 & Phot\\
M31N 2007-12b  &  00 43 19.94 & 41 13 46.6 & 15.18 Dec. 2007 &1000 &4100--8900 & Phot\\
M31N 2007-12d  &  00 41 54.96 & 41 09 47.3 & 21.17 Dec. 2007 &1800 &4100--8900 & Spec\\
M31N 2008-08d  &  00 45 48.25 & 43 02 22.2 & 20.13 Oct. 2008 &1200 &4100--8900 & Phot\\
M31N 2008-09a  &  00 41 46.72 & 41 07 52.1 & 22.41 Sep. 2008 &1200 &4100--8900 & Spec\\
M31N 2008-09c  &  00 42 51.42 & 41 01 54.0 & 20.18 Sep. 2008 &1200 &4100--8900 & Spec\\
M31N 2008-10a  &  00 43 35.46 & 41 54 44.2 & 18.34 Oct. 2008 &1200 &4100--8900 & Phot\\
M31N 2008-10b  &  00 43 02.42 & 41 14 09.9 & 20.11 Oct. 2008 &1200 &4100--8900 & Phot\\
               &              &            & 25.33 Oct. 2008 &1200 &4100--8900 & Spec\\
M31N 2008-11a  &  00 41 32.26 & 41 06 01.0 & 07.29 Nov. 2008 &1200 &4275--7250 & Phot\\
               &              &            & 25.25 Nov. 2008 &1800 &4275--7250 & Spec\\
M31N 2009-01a  &  00 44 44.03 & 41 23 28.3 & 02.07 Feb. 2009 &1350 &4100--8900 & Phot\\
M31N 2009-02a  &  00 43 43.81 & 41 36 38.8 & 07.07 Feb. 2009 &900 &4100--8900 & Spec\\
\enddata
\end{deluxetable}

\clearpage

\begin{planotable}{llcr}
\tabletypesize{\scriptsize}
\tablenum{3}
\tablewidth{0pt}
\tablecolumns{4}
\tablecaption{Photometric Observations\label{photobs}}
\tablehead{\colhead{$\mathrm{JD}$} & \colhead{} & \colhead{} & \colhead{}\\ \colhead{($2,450,000+$)} & \colhead{Mag} & \colhead{Filter} & \colhead{Notes\tablenotemark{a}}}
\startdata

\cutinhead{M31N 1999-08f}

1427.686 & $ 17.700 \pm 0.024 $ & $r'$ & (29)\\
1427.690 & $ 17.697 \pm 0.025 $ & $r'$ & (29)\\
1428.444 & $ 17.853 \pm 0.016 $ & $r'$ & (29)\\
1428.452 & $ 17.845 \pm 0.015 $ & $r'$ & (29)\\
1429.452 & $ 17.895 \pm 0.021 $ & $r'$ & (29)\\
1429.456 & $ 17.908 \pm 0.021 $ & $r'$ & (29)\\
1430.467 & $ 18.090 \pm 0.016 $ & $r'$ & (29)\\
1430.471 & $ 18.109 \pm 0.019 $ & $r'$ & (29)\\
1432.686 & $ 18.384 \pm 0.029 $ & $r'$ & (29)\\
1432.690 & $ 18.434 \pm 0.032 $ & $r'$ & (29)\\
1433.678 & $ 18.534 \pm 0.034 $ & $r'$ & (29)\\
1433.686 & $ 18.507 \pm 0.037 $ & $r'$ & (29)\\
1449.467 & $ 19.258 \pm 0.065 $ & $r'$ & (29)\\
1449.471 & $ 19.253 \pm 0.057 $ & $r'$ & (29)\\
1450.463 & $ 19.302 \pm 0.054 $ & $r'$ & (29)\\
1450.467 & $ 19.247 \pm 0.049 $ & $r'$ & (29)\\
1451.475 & $ 19.358 \pm 0.046 $ & $r'$ & (29)\\
1452.491 & $ 19.481 \pm 0.044 $ & $r'$ & (29)\\
1454.374 & $ 19.878 \pm 0.057 $ & $r'$ & (29)\\
1454.389 & $ 19.738 \pm 0.049 $ & $r'$ & (29)\\
1455.538 & $ 19.794 \pm 0.076 $ & $r'$ & (29)\\
1456.495 & $ 19.523 \pm 0.063 $ & $r'$ & (29)\\
1456.499 & $ 19.525 \pm 0.065 $ & $r'$ & (29)\\
1457.495 & $ 19.664 \pm 0.053 $ & $r'$ & (29)\\
1457.499 & $ 19.656 \pm 0.052 $ & $r'$ & (29)\\
1458.506 & $ 19.704 \pm 0.059 $ & $r'$ & (29)\\
1461.510 & $ 19.877 \pm 0.076 $ & $r'$ & (29)\\
1461.514 & $ 19.867 \pm 0.081 $ & $r'$ & (29)\\
1462.530 & $ 19.864 \pm 0.063 $ & $r'$ & (29)\\
1462.534 & $ 19.842 \pm 0.065 $ & $r'$ & (29)\\
1462.561 & $ 19.895 \pm 0.060 $ & $r'$ & (29)\\
1463.549 & $ 20.066 \pm 0.060 $ & $r'$ & (29)\\
1463.557 & $ 19.996 \pm 0.061 $ & $r'$ & (29)\\
1484.499 & $ 20.424 \pm 0.117 $ & $r'$ & (29)\\
1484.502 & $ 20.418 \pm 0.120 $ & $r'$ & (29)\\
1486.456 & $ 20.545 \pm 0.126 $ & $r'$ & (29)\\
\\
1426.694 & $ 19.544 \pm 0.146 $ & $i'$ & (29)\\
1426.702 & $ 19.507 \pm 0.132 $ & $i'$ & (29)\\
1427.690 & $ 19.124 \pm 0.089 $ & $i'$ & (29)\\
1427.698 & $ 19.148 \pm 0.092 $ & $i'$ & (29)\\
1428.702 & $ 19.220 \pm 0.106 $ & $i'$ & (29)\\

\cutinhead{M31N 2001-10a}

2191.616 & $ 17.512 \pm 0.008 $ & $r'$ & (29)\\
2191.620 & $ 17.481 \pm 0.011 $ & $r'$ & (29)\\
2194.620 & $ 17.378 \pm 0.014 $ & $r'$ & (29)\\
2194.624 & $ 17.402 \pm 0.013 $ & $r'$ & (29)\\
2195.640 & $ 17.968 \pm 0.021 $ & $r'$ & (29)\\
2195.643 & $ 17.971 \pm 0.026 $ & $r'$ & (29)\\
2196.382 & $ 18.227 \pm 0.017 $ & $r'$ & (29)\\
2196.386 & $ 18.207 \pm 0.018 $ & $r'$ & (29)\\
2197.628 & $ 18.043 \pm 0.020 $ & $r'$ & (29)\\
2197.632 & $ 18.047 \pm 0.020 $ & $r'$ & (29)\\
2199.452 & $ 17.430 \pm 0.013 $ & $r'$ & (29)\\
2199.456 & $ 17.433 \pm 0.013 $ & $r'$ & (29)\\
2227.339 & $ 18.845 \pm 0.034 $ & $r'$ & (29)\\
2271.405 & $ 20.056 \pm 0.078 $ & $r'$ & (29)\\
2271.409 & $ 20.065 \pm 0.096 $ & $r'$ & (29)\\
2292.429 & $ 20.273 \pm 0.105 $ & $r'$ & (29)\\
2295.343 & $ 20.171 \pm 0.068 $ & $r'$ & (29)\\
2295.350 & $ 20.240 \pm 0.076 $ & $r'$ & (29)\\
\\
2194.612 & $ 16.903 \pm 0.014 $ & $i'$ & (29)\\
2194.620 & $ 16.883 \pm 0.015 $ & $i'$ & (29)\\
2197.620 & $ 17.018 \pm 0.026 $ & $i'$ & (29)\\
2197.624 & $ 17.053 \pm 0.028 $ & $i'$ & (29)\\
2198.636 & $ 17.588 \pm 0.043 $ & $i'$ & (29)\\
2198.640 & $ 17.595 \pm 0.048 $ & $i'$ & (29)\\
2199.382 & $ 17.868 \pm 0.030 $ & $i'$ & (29)\\
2199.386 & $ 17.824 \pm 0.026 $ & $i'$ & (29)\\
2200.628 & $ 17.654 \pm 0.037 $ & $i'$ & (29)\\
2200.632 & $ 17.673 \pm 0.033 $ & $i'$ & (29)\\
2202.448 & $ 17.194 \pm 0.024 $ & $i'$ & (29)\\
2202.456 & $ 17.201 \pm 0.022 $ & $i'$ & (29)\\
2230.331 & $ 18.751 \pm 0.048 $ & $i'$ & (29)\\

\cutinhead{M31N 2002-08a}

2490.523 & $17.05\pm  0.1$ &$R$ &  (2)\\
2504.452 & $17.2\pm   0.1$ &$R$ &  (2)\\
2512.360 & $18.2\pm   0.2$ &$R$ &  (2)\\
2513.344 & $17.83\pm  0.09$&$R$ &  (2)\\
2516.506 & $18.1\pm   0.1$ &$R$ & (32)\\
2517.368 & $18.6\pm   0.2$ &$R$ &  (2)\\
2517.560 & $18.4\pm   0.15$&$R$ & (32)\\
2520.452 & $18.4\pm   0.2$ &$R$ &  (2)\\
2521.433 & $18.5\pm   0.2$ &$R$ &  (2)\\
2522.414 & $18.7\pm   0.2$ &$R$ &  (2)\\
2524.358 & $18.5\pm   0.15$&$R$ & (32)\\
2525.417 & $19.0\pm   0.25$&$R$ &  (2)\\
2529.435 & $18.6\pm   0.2$ &$R$ &  (2)\\
2530.438 & $18.7\pm   0.2$ &$R$ &  (2)\\
2530.637 & $18.7\pm   0.2$ &$R$ & (32)\\
2547.365 & $19.3\pm   0.3$ &$R$ &  (2)\\
2548.505 & $19.0\pm   0.3$ &$R$ &  (2)\\

\cutinhead{M31N 2004-08b}

3241.563 & $17.6\pm   0.15$&$V$&  (32)\\
\\
3220.474 &$>20.5$      &$R$ & (32)\\
3221.460 & $19.2\pm   0.4$ &$R$&   (2)\\
3222.401 & $18.8\pm   0.3$ &$R$&   (2)\\
3224.497 & $17.8\pm   0.15$&$R$&  (32)\\
3225.405 & $17.4\pm   0.2$ &$R$&   (2)\\
3225.482 & $17.3\pm   0.15$&$R$&  (32)\\
3226.570 & $17.8\pm   0.15$&$R$&  (32)\\
3227.505 & $18.3\pm   0.2$ &$R$&   (2)\\
3228.393 & $18.0\pm   0.2$ &$R$&   (2)\\
3228.456 & $17.7\pm   0.15$&$R$&   (2)\\
3229.557 & $18.3\pm   0.15$&$R$&  (32)\\
3233.443 & $17.6\pm   0.2$ &$R$ &  (2)\\
3235.370 & $17.6\pm   0.15$&$R$&   (2)\\
3236.410 & $17.7\pm   0.15$&$R$&   (2)\\
3236.586 & $17.7\pm   0.1$ &$R$&  (32)\\
3237.375 & $17.7\pm   0.15$&$R$&   (2)\\
3240.429 & $17.6\pm   0.2$ &$R$&   (2)\\
3241.402 & $17.5\pm   0.15$&$R$&   (2)\\
3241.560 & $17.6\pm   0.15$&$R$&  (32)\\
3246.342 & $18.3\pm   0.2$ &$R$&   (2)\\
3246.399 & $18.1\pm   0.15$&$R$&   (2)\\
3249.415 & $18.6\pm   0.2$ &$R$&   (2)\\
3249.464 & $18.5\pm   0.2$ &$R$&   (2)\\
3251.518 & $18.6\pm   0.2$ &$R$&  (32)\\
3252.312 & $18.7\pm   0.25$&$R$&   (2)\\
3252.368 & $18.7\pm   0.25$&$R$&   (2)\\
3253.319 & $19.0\pm   0.25$&$R$&   (2)\\
3253.558 & $19.0\pm   0.2$ &$R$&  (32)\\
3254.362 & $19.0\pm   0.3$ &$R$&   (2)\\
3255.430 & $19.0\pm   0.3$ &$R$&   (2)\\
3257.429 & $18.3\pm   0.2$ &$R$&   (2)\\
3257.453 & $18.4\pm   0.2$ &$R$&   (2)\\
3257.571 & $18.5\pm   0.15$&$R$&  (32)\\
3258.363 & $18.4\pm   0.15$&$R$&   (2)\\
3258.393 & $18.3\pm   0.2$ &$R$&   (2)\\
3259.384 & $18.6\pm   0.25$&$R$&   (2)\\
3259.413 & $18.4\pm   0.2$ &$R$&   (2)\\
3260.385 & $18.6\pm   0.2$ &$R$&  (32)\\
3262.453 & $18.6\pm   0.2$ &$R$&   (2)\\
3265.332 & $19.4\pm   0.3$ &$R$&   (2)\\
3265.352 & $19.5\pm   0.3$ &$R$&   (2)\\
3266.335 & $19.4\pm   0.3$ &$R$&   (2)\\
3270.330 & $19.6\pm   0.4$ &$R$&   (2)\\
3275.290 & $19.6\pm   0.3$ &$R$&   (2)\\
3279.356 & $19.6\pm   0.25$&$R$&  (32)\\
3282.281 & $19.4\pm   0.3$ &$R$&   (2)\\
3283.619 & $19.5\pm   0.25$&$R$&  (32)\\
3288.272 & $19.2\pm   0.25$&$R$&   (2)\\
3289.828 & $20.40\pm  0.1$ &$R$&  (33)\\
3301.408 & $20.2\pm   0.4$ &$R$&  (32)\\

\cutinhead{M31N 2004-09a}

3241.560 &$>19.8$     & $R$ & (32)\\
3246.399 &$>19.1$     & $R$ &  (2)\\
3249.415 & $18.3\pm   0.2$ &$R$ &  (2)\\
3249.438 & $18.0\pm   0.15$&$R$ &  (2)\\
3249.452 & $17.8\pm   0.15$&$R$ &  (2)\\
3251.518 & $17.5\pm   0.1$ &$R$ & (32)\\
3252.312 & $17.7\pm   0.15$&$R$ &  (2)\\
3252.368 & $17.6\pm   0.1$ &$R$ &  (2)\\
3253.319 & $18.3\pm   0.15$&$R$ &  (2)\\
3253.341 & $18.2\pm   0.15$&$R$ &  (2)\\
3253.441 & $18.0\pm   0.15$&$R$ &  (2)\\
3253.558 & $18.1\pm   0.1$ &$R$ & (32)\\
3254.362 & $18.1\pm   0.15$&$R$ &  (2)\\
3254.400 & $18.2\pm   0.15$&$R$ &  (2)\\
3255.430 & $18.0\pm   0.15$&$R$ &  (2)\\
3257.429 & $18.2\pm   0.15$&$R$ &  (2)\\
3257.453 & $18.1\pm   0.15$&$R$ &  (2)\\
3257.571 & $18.2\pm   0.1$ &$R$ & (32)\\
3258.363 & $18.3\pm   0.2$ &$R$ &  (2)\\
3258.393 & $18.4\pm   0.2$ &$R$ &  (2)\\
3259.384 & $18.6\pm   0.25$&$R$ &  (2)\\
3259.413 & $18.4\pm   0.25$&$R$ &  (2)\\
3260.385 & $18.6\pm   0.15$&$R$ & (32)\\
3262.453 & $18.7\pm   0.25$&$R$ &  (2)\\
3262.472 & $18.6\pm   0.2$ &$R$ &  (2)\\
3265.332 & $18.8\pm   0.25$&$R$ &  (2)\\
3265.352 & $18.5\pm   0.4$ &$R$ &  (2)\\
3266.335 & $19.2\pm   0.4$ &$R$ &  (2)\\
3270.330 & $18.6\pm   0.4$ &$R$ &  (2)\\
3270.351 & $19.1\pm   0.4$ &$R$ &  (2)\\
3275.290 & $19.5\pm   0.4$ &$R$ &  (2)\\
3279.356 & $19.3\pm   0.25$&$R$ & (32)\\
3279.398 & $19.4\pm   0.3$ &$R$ &  (2)\\
3282.257 & $19.2\pm   0.25$&$R$ &  (2)\\
3283.619 & $19.8\pm   0.4$ &$R$ & (32)\\
3288.272 & $19.6\pm   0.4$ &$R$ &  (2)\\
3288.292 & $20.1\pm   0.4$ &$R$ &  (2)\\
3289.802 & $20.8\pm   0.4$ &$R$ & (33)\\

\cutinhead{M31N 2004-11a}

3301.408 &$>21.0$         &$R$ & (2)\\
3315.347 &$16.6\pm   0.15$&$R$ & (2)\\
3315.390 &$16.5\pm   0.15$&$R$ & (2)\\
3317.352 &$17.0\pm   0.15$&$R$ & (2)\\
3321.404 &$18.1\pm   0.2 $&$R$ & (2)\\
3324.305 &$18.2\pm   0.2 $&$R$ & (2)\\
3325.218 &$18.4\pm   0.25$&$R$ & (2)\\
3334.218 &$19.0\pm   0.3 $&$R$ & (2)\\
3335.273 &$19.2\pm   0.3 $&$R$ & (2)\\
3339.296 &$19.4\pm   0.4 $&$R$ & (2)\\
3339.318 &$19.6\pm   0.4 $&$R$ & (2)\\
3342.192 &$19.3\pm   0.4 $&$R$ &(32)\\
3344.192 &$19.9\pm   0.4 $&$R$ & (2)\\
3344.214 &$19.3\pm   0.4 $&$R$ & (2)\\
3357.568 &$19.8\pm   0.15$&$R$ &(37)\\
3358.260 &$20.5\pm   0.4 $&$R$ &(32)\\

\cutinhead{M31N 2004-11b}

3381.253 &$ 19.5\pm   0.4 $&$V$& (32)\\
\\
3301.408 &$ >21.0       $&$R$&  (2)\\
3315.347 &$ 16.7\pm   0.1 $&$R$&  (2)\\
3315.390 &$ 16.6\pm   0.1 $&$R$&  (2)\\
3317.352 &$ 17.1\pm   0.15$&$R$&  (2)\\
3321.404 &$ 17.4\pm   0.15$&$R$&  (2)\\
3324.305 &$ 17.3\pm   0.15$&$R$&  (2)\\
3325.218 &$ 17.2\pm   0.15$&$R$&  (2)\\
3334.218 &$ 17.5\pm   0.15$&$R$&  (2)\\
3335.273 &$ 17.6\pm   0.15$&$R$&  (2)\\
3339.296 &$ 17.9\pm   0.2 $&$R$&  (2)\\
3339.318 &$ 17.9\pm   0.15$&$R$&  (2)\\
3342.172 &$ 18.2\pm   0.15$&$R$& (32)\\
3344.192 &$ 18.1\pm   0.15$&$R$&  (2)\\
3344.214 &$ 18.2\pm   0.15$&$R$&  (2)\\
3346.410 &$ 18.3\pm   0.15$&$R$&  (2)\\
3347.344 &$ 18.5\pm   0.2 $&$R$&  (2)\\
3347.370 &$ 18.8\pm   0.2 $&$R$&  (2)\\
3348.359 &$ 18.7\pm   0.2 $&$R$&  (2)\\
3357.568 &$ 19.1\pm   0.1 $&$R$& (37)\\
3358.260 &$ 19.2\pm   0.2 $&$R$& (32)\\
3360.236 &$ 19.3\pm   0.25$&$R$&  (2)\\
3361.324 &$ 19.1\pm   0.25$&$R$&  (2)\\
3370.230 &$ 19.5\pm   0.3 $&$R$&  (2)\\
3370.267 &$ 19.3\pm   0.25$&$R$&  (2)\\
3373.321 &$ 19.4\pm   0.3 $&$R$& (32)\\
3377.266 &$ 19.5\pm   0.25$&$R$& (32)\\
3378.391 &$ 19.7\pm   0.3 $&$R$&  (2)\\
3380.426 &$ 19.7\pm   0.3 $&$R$&  (2)\\
3381.249 &$ 19.5\pm   0.25$&$R$& (32)\\
3381.276 &$ 19.4\pm   0.3 $&$R$&  (2)\\
3381.417 &$ 19.4\pm   0.3 $&$R$&  (2)\\
3382.219 &$ 19.3\pm   0.3 $&$R$&  (2)\\
3382.246 &$ 19.3\pm   0.4 $&$R$& (32)\\
3384.212 &$ 19.6\pm   0.4 $&$R$&  (2)\\
3387.231 &$ 19.8\pm   0.4 $&$R$& (32)\\
3387.400 &$ 20.0\pm   0.4 $&$R$&  (2)\\
3388.224 &$ 19.5\pm   0.4 $&$R$& (32)\\

\cutinhead{M31N 2005-01a}

3381.253 & $15.45\pm  0.05$&$V$ & (32)\\
3382.251 & $15.26\pm  0.05$&$V$ & (32)\\
3384.333 & $15.26\pm  0.07$&$V$ & (32)\\
3386.305 & $15.65\pm  0.06$&$V$ & (32)\\
3387.237 & $15.75\pm  0.06$&$V$ & (32)\\
3388.233 & $15.77\pm  0.08$&$V$ & (32)\\
\\
3373.321 &$>19.8$    &  $R$ & (32)\\
3377.266 & $19.2\pm   0.2$ &$R$ & (32)\\
3377.293 & $19.4\pm   0.3$ &$R$ &  (2)\\
3378.391 & $17.9\pm   0.15$&$R$ &  (2)\\
3380.224 & $15.72\pm  0.07$&$R$ &  (2)\\
3380.253 & $15.68\pm  0.06$&$R$ &  (2)\\
3380.437 & $15.33\pm  0.07$&$R$ &  (2)\\
3381.249 & $15.48\pm  0.04$&$R$ & (32)\\
3381.266 & $15.32\pm  0.06$&$R$ &  (2)\\
3381.286 & $15.27\pm  0.05$&$R$ &  (2)\\
3381.306 & $15.30\pm  0.04$&$R$ &  (2)\\
3381.408 & $15.34\pm  0.07$&$R$ &  (2)\\
3381.426 & $15.31\pm  0.06$&$R$ &  (2)\\
3382.219 & $15.16\pm  0.05$&$R$ &  (2)\\
3382.246 & $15.21\pm  0.05$&$R$ & (32)\\
3382.257 & $15.21\pm  0.06$&$R$ &  (2)\\
3384.212 & $15.05\pm  0.05$&$R$ &  (2)\\
3384.252 & $15.10\pm  0.06$&$R$ &  (2)\\
3384.326 & $15.12\pm  0.05$&$R$ & (32)\\
3384.405 & $15.04\pm  0.07$&$R$ &  (2)\\
3386.202 & $15.27\pm  0.06$&$R$ &  (2)\\
3386.228 & $15.26\pm  0.05$&$R$ &  (2)\\
3386.300 & $15.27\pm  0.04$&$R$ & (32)\\
3386.424 & $15.31\pm  0.05$&$R$ &  (2)\\
3387.212 & $15.36\pm  0.04$&$R$ &  (2)\\
3387.231 & $15.39\pm  0.05$&$R$ & (32)\\
3387.249 & $15.34\pm  0.06$&$R$ &  (2)\\
3387.400 & $15.34\pm  0.05$&$R$ &  (2)\\
3388.224 & $15.42\pm  0.05$&$R$ & (32)\\
3390.287 & $15.52\pm  0.07$&$R$ &  (2)\\
3390.327 & $15.50\pm  0.06$&$R$ &  (2)\\
3394.234 & $15.80\pm  0.07$&$R$ &  (2)\\
3398.274 & $16.61\pm  0.08$&$R$ &  (2)\\
3401.212 & $17.17\pm  0.09$&$R$ &  (2)\\
3405.273 & $19.4\pm   0.4$ &$R$ & (32)\\
3405.633 & $19.7\pm   0.15$&$R$ & (20)\\
3406.326 & $20.1\pm   0.25$&$R$ & (32)\\
3407.231 &$>19.9$   &  $ R  $&  (2)\\
3407.283 &$>19.8$   &  $ R  $&  (2)\\
3509.952 & $21.7\pm   0.25$&$R$&  (33)\\
3411.257 &$>20.3$   &   $R$ & (32)\\
3532.699 & $21.13\pm  0.09$&$R$&  (38)\\
3534.701 & $21.38\pm  0.1$ &$R$&  (38)\\
3535.681 & $21.45\pm  0.1$ &$R$&  (38)\\
3538.697 & $21.43\pm  0.1$ &$R$&  (38)\\
3541.718 & $21.2\pm   0.15$&$R$&  (38)\\
3651.821 & $22.0\pm   0.4$ &$R$&  (35)\\
3702.638 &$>21.3$    & $ R$ & (33)\\
3710.730 & $21.9\pm   0.25$&$R$&  (36)\\
3996.839 &$>23.0$    &  $R$ & (37)\\
\\
3384.315 & $14.95\pm  0.07$&$I$ &  (2)\\

\cutinhead{M31N 2005-07a}

3564.493& $>19.5        $&$R$& (32)\\
3575.429& $>19.0        $&$R$&  (2)\\
3579.409& $ 18.4\pm0.25 $&$R$&  (2)\\
3581.419& $ 17.4\pm0.15 $&$R$&  (2)\\
3584.404& $ 19.1\pm0.3  $&$R$&  (2)\\
3587.509& $ 19.3\pm0.25 $&$R$& (32)\\
3588.415& $ 18.8\pm0.25 $&$R$&  (2)\\
3594.390& $ 19.7\pm0.35 $&$R$& (32)\\
3651.862& $ 19.53\pm0.05 $&$R$& (32)\\
3702.638& $ 20.8\pm0.2  $&$R$& (33)\\
3710.730& $ 21.03\pm0.12 $&$R$& (36)\\
3760.586& $ 21.8\pm0.3  $&$R$& (36)\\
3771.346& $ 21.7\pm0.3  $&$R$& (21)\\

\cutinhead{M31N 2006-06a}

3771.264 & $>19.2         $  & $R$ &   (2)\\
3771.346 & $>22           $  & $R$ &  (21)\\
3869.565 & $>19.8         $  & $R$ &   (1)\\
3892.518 & $ 17.6 \pm0.1  $ & $R$ &   (3)\\
3899.502 & $ 18.0 \pm0.15 $ & $R$ &   (1)\\
3899.540 & $ 17.9 \pm0.15 $ & $R$ &   (1)\\
3900.502 & $ 18.1 \pm0.15 $ & $R$ &   (1)\\
3911.522 & $ 18.5 \pm0.25 $ & $R$ &   (2)\\
3921.464 & $ 19.2 \pm0.25 $ & $R$ &   (2)\\

\cutinhead{M31N 2006-09c}

4256.677 & $>21.2$ & $B$ & (30)\\
4260.700 & $>21.7$ & $B$ & (30)\\
\\
4260.705 & $>21.0$ & $V$ & (30)\\
\\
3892.967 & $>21.1         $  & $R$ &  (14)\\
3991.566 & $>20.0         $  & $R$ &   (4)\\
3993.376 & $>19.1         $  & $R$ &   (2)\\
3996.404 & $ 18.1 \pm0.2  $ & $R$ &   (3)\\
3999.609 & $ 17.15\pm0.1  $ & $R$ &   (3)\\
4000.317 & $ 17.0 \pm0.1  $ & $R$ &   (2)\\
4000.590 & $ 17.35\pm0.1  $ & $R$ &   (4)\\
4001.360 & $ 17.3 \pm0.1  $ & $R$ &   (2)\\
4002.302 & $ 17.4 \pm0.1  $ & $R$ &   (2)\\
4002.329 & $ 17.5 \pm0.1  $ & $R$ &   (2)\\
4005.312 & $ 17.8 \pm0.1  $ & $R$ &   (2)\\
4007.312 & $ 17.9 \pm0.1  $ & $R$ &   (2)\\
4014.295 & $ 19.0 \pm0.2  $ & $R$ &   (2)\\
4017.258 & $ 19.1 \pm0.2  $ & $R$ &   (2)\\
4019.319 & $ 18.8 \pm0.2  $ & $R$ &   (2)\\
4024.383 & $ 19.3 \pm0.2  $ & $R$ &   (2)\\
4026.330 & $ 19.3 \pm0.2  $ & $R$ &   (2)\\
4026.364 & $ 19.3 \pm0.2  $ & $R$ &   (2)\\
4034.312 & $ 19.9 \pm0.3  $ & $R$ &   (2)\\
\\
4260.690 & $>21.4$ & $r'$ & (30)\\
\\
4260.695 & $ 21.800 \pm 0.500 $ & $i'$ & (30)\\

\cutinhead{M31N 2006-10a}

4044.337 & $ 18.072 \pm 0.035 $ & $B$ & (30)\\
4050.589 & $ 19.064 \pm 0.033 $ & $B$ & (30)\\
4056.585 & $ 18.879 \pm 0.032 $ & $B$ & (30)\\
4062.606 & $ 18.579 \pm 0.030 $ & $B$ & (30)\\
4069.484 & $ 19.240 \pm 0.044 $ & $B$ & (30)\\
4071.549 & $ 19.055 \pm 0.091 $ & $B$ & (30)\\
4074.575 & $ 19.201 \pm 0.106 $ & $B$ & (30)\\
4077.542 & $ 19.255 \pm 0.132 $ & $B$ & (30)\\
4084.460 & $ 19.239 \pm 0.034 $ & $B$ & (30)\\
4092.454 & $ 19.933 \pm 0.054 $ & $B$ & (30)\\
4099.407 & $ 20.033 \pm 0.108 $ & $B$ & (30)\\
4101.393 & $ 19.781 \pm 0.071 $ & $B$ & (30)\\
4114.372 & $>21.896$ & $B$ & (30)\\
4120.432 & $>20.893$ & $B$ & (30)\\
4254.685 & $>17.6$ & $B$ & (30)\\
\\
4044.334 & $ 17.904 \pm 0.026 $ & $V$ & (30)\\
4050.586 & $ 19.096 \pm 0.024 $ & $V$ & (30)\\
4056.582 & $ 18.797 \pm 0.025 $ & $V$ & (30)\\
4062.604 & $ 18.475 \pm 0.021 $ & $V$ & (30)\\
4069.481 & $ 19.253 \pm 0.039 $ & $V$ & (30)\\
4071.546 & $ 19.022 \pm 0.118 $ & $V$ & (30)\\
4074.572 & $ 19.387 \pm 0.099 $ & $V$ & (30)\\
4077.539 & $ 19.299 \pm 0.045 $ & $V$ & (30)\\
4084.457 & $ 19.161 \pm 0.027 $ & $V$ & (30)\\
4092.451 & $ 19.666 \pm 0.055 $ & $V$ & (30)\\
4099.404 & $ 19.638 \pm 0.096 $ & $V$ & (30)\\
4101.391 & $ 19.734 \pm 0.058 $ & $V$ & (30)\\
4114.370 & $>21.545$ & $V$ & (30)\\
4120.429 & $>21.221$ & $V$ & (30)\\
4254.692 & $>18.4$ & $V$ & (30)\\
\\
3771.346 & $>22           $  & $R$ &  (21)\\
4019.319 & $>19.7         $  & $R$ &   (2)\\
4024.383 & $>19.0         $  & $R$ &   (2)\\
4026.330 & $>20.1         $  & $R$ &   (2)\\
4031.251 & $ 19.2 \pm0.25 $ & $R$ &   (2)\\
4034.312 & $ 18.7 \pm0.15 $ & $R$ &   (2)\\
4034.470 & $ 18.6 \pm0.15 $ & $R$ &   (3)\\
4035.360 & $ 18.4 \pm0.3  $ & $R$ &   (2)\\
4043.331 & $ 17.9 \pm0.1  $ & $R$ &   (2)\\
4047.288 & $ 19.0 \pm0.3  $ & $R$ &   (2)\\
4048.324 & $ 19.3 \pm0.2  $ & $R$ &   (2)\\
4055.296 & $ 18.2 \pm0.15 $ & $R$ &   (2)\\
4055.262 & $ 18.2 \pm0.1  $ & $R$ &  (23)\\
4070.308 & $ 19.2 \pm0.35 $ & $R$ &   (6)\\
4071.385 & $ 18.9 \pm0.2  $ & $R$ &   (3)\\
4078.308 & $ 18.4 \pm0.3  $ & $R$ &   (2)\\
4078.343 & $ 18.6 \pm0.2  $ & $R$ &   (2)\\
4080.306 & $ 18.8 \pm0.25 $ & $R$ &   (2)\\
4084.212 & $ 18.9 \pm0.2  $ & $R$ &   (2)\\
4093.174 & $ 19.5:\pm0.4  $ & $R$ &   (2)\\
4096.325 & $ 19.7 \pm0.25 $ & $R$ &   (2)\\
4097.222 & $ 19.5 \pm0.25 $ & $R$ &   (2)\\
4115.194 & $>20.0         $  & $R$ &   (2)\\
4121.381 & $>19.8         $  & $R$ &   (2)\\
4122.331 & $>19.5         $  & $R$ &   (2)\\
4122.377 & $>19.4         $  & $R$ &   (2)\\
\\
4254.671 & $>19.0$ & $r'$ & (30)\\

\cutinhead{M31N 2006-10b}

4044.388 & $ 18.952 \pm 0.057 $ & $B$ & (30)\\
4049.501 & $ 19.553 \pm 0.042 $ & $B$ & (30)\\
4057.534 & $ 20.385 \pm 0.057 $ & $B$ & (30)\\
4063.463 & $ 20.892 \pm 0.066 $ & $B$ & (30)\\
4069.445 & $ 20.989 \pm 0.129 $ & $B$ & (30)\\
4072.440 & $>20.144$ & $B$ & (30)\\
4075.473 & $>20.354$ & $B$ & (30)\\
4084.411 & $ 21.410 \pm 0.103 $ & $B$ & (30)\\
4099.438 & $>20.647$ & $B$ & (30)\\
4106.415 & $>19.739$ & $B$ & (30)\\
4114.383 & $>22.913$ & $B$ & (30)\\
4120.458 & $>21.830$ & $B$ & (30)\\
4248.704 & $>22.422$ & $B$ & (30)\\
\\
4049.498 & $ 20.021 \pm 0.045 $ & $V$ & (30)\\
4054.317 & $>17.332$ & $V$ & (30)\\
4057.531 & $ 20.881 \pm 0.072 $ & $V$ & (30)\\
4063.460 & $ 21.271 \pm 0.077 $ & $V$ & (30)\\
4069.442 & $>21.692$ & $V$ & (30)\\
4072.437 & $>20.121$ & $V$ & (30)\\
4075.470 & $>20.641$ & $V$ & (30)\\
4084.408 & $>22.297$ & $V$ & (30)\\
4099.436 & $>20.506$ & $V$ & (30)\\
4102.385 & $>20.134$ & $V$ & (30)\\
4106.412 & $>17.613$ & $V$ & (30)\\
4108.486 & $>18.404$ & $V$ & (30)\\
4114.380 & $>21.955$ & $V$ & (30)\\
4120.455 & $>22.758$ & $V$ & (30)\\
4248.709 & $>22.304$ & $V$ & (30)\\
\\
4248.693 & $>22.9 $& $r'$ & (30)\\
\\
4248.699 & $>21.5 $& $i'$ & (30)\\

\cutinhead{M31N 2006-11a}

4141.368 & $ 20.120 \pm 0.050 $ & $B$ & (30)\\
\\
4141.372 & $ 20.540 \pm 0.060 $ & $V$ & (30)\\
\\
3771.346 & $>22           $  & $R$ &  (21)\\
4048.324 & $>19.8         $  & $R$ &   (2)\\
4055.296 & $>19.8         $  & $R$ &   (2)\\
4070.263 & $ 16.9 \pm0.1  $ & $R$ &   (6)\\
4070.308 & $ 16.6 \pm0.15 $ & $R$ &   (6)\\
4078.308 & $ 16.1 \pm0.1  $ & $R$ &   (2)\\
4078.343 & $ 16.0 \pm0.1  $ & $R$ &   (2)\\
4080.306 & $ 16.3 \pm0.1  $ & $R$ &   (2)\\
4084.212 & $ 16.9 \pm0.1  $ & $R$ &   (2)\\
4093.174 & $ 17.7 \pm0.15 $ & $R$ &   (2)\\
4096.325 & $ 17.8 \pm0.15 $ & $R$ &   (2)\\
4097.222 & $ 17.8 \pm0.15 $ & $R$ &   (2)\\
4115.194 & $ 18.8 \pm0.2  $ & $R$ &   (2)\\
4121.381 & $ 18.9 \pm0.2  $ & $R$ &   (2)\\
4122.331 & $ 19.2 \pm0.25 $ & $R$ &   (2)\\
4122.377 & $ 19.1 \pm0.3  $ & $R$ &   (2)\\
4126.289 & $ 19.3 \pm0.3  $ & $R$ &   (2)\\
4126.339 & $ 19.1 \pm0.3  $ & $R$ &   (2)\\
4128.275 & $ 19.2 \pm0.25 $ & $R$ &   (2)\\
4135.298 & $ 19.4 \pm0.3  $ & $R$ &   (2)\\
4135.334 & $ 19.3 \pm0.25 $ & $R$ &   (2)\\
4141.362 & $ 19.2 \pm0.3  $ & $R$ &   (2)\\
4146.295 & $ 19.4 \pm0.25 $ & $R$ &   (2)\\
4149.273 & $ 19.4 \pm0.25 $ & $R$ &   (2)\\
4166.247 & $ 19.7 \pm0.2  $ & $R$ &   (1)\\
4167.366 & $ 20.0 \pm0.35 $ & $R$ &   (1)\\
4170.264 & $ 20.0 \pm0.3  $ & $R$ &   (1)\\
4173.269 & $ 19.8 \pm0.25 $ & $R$ &  (25)\\
4174.268 & $>19.7         $  & $R$ &   (2)\\
4175.267 & $>19.7         $  & $R$ &   (4)\\
4240.561 & $>20.0         $  & $R$ &   (1)\\
\\
4141.356 & $ 19.220 \pm 0.030 $ & $r'$ & (30)\\
\\
4141.362 & $ 20.220 \pm 0.080 $ & $i'$ & (30)\\

\cutinhead{M31N 2006-12a}

3771.346 & $>22           $  & $R$ &  (21)\\
4078.343 & $>19.8         $  & $R$ &   (2)\\
4080.306 & $>19.9         $  & $R$ &   (2)\\
4084.212 & $>20.2         $  & $R$ &   (2)\\
4093.174 & $ 17.3 \pm0.15 $ & $R$ &   (2)\\
4096.325 & $ 17.4 \pm0.15 $ & $R$ &   (2)\\
4096.357 & $ 17.5 \pm0.15 $ & $R$ &   (2)\\
4097.222 & $ 17.8 \pm0.15 $ & $R$ &   (2)\\
4097.241 & $ 17.8 \pm0.15 $ & $R$ &   (2)\\
4115.194 & $ 18.8 \pm0.2  $ & $R$ &   (2)\\
4115.226 & $ 18.6 \pm0.25 $ & $R$ &   (2)\\
4121.381 & $ 18.9 \pm0.25 $ & $R$ &   (2)\\
4122.331 & $ 19.1 \pm0.25 $ & $R$ &   (2)\\
4122.377 & $ 19.1 \pm0.3  $ & $R$ &   (2)\\
4126.289 & $>19.2         $  & $R$ &   (2)\\
4126.339 & $>19.2         $  & $R$ &   (2)\\
4128.275 & $ 19.3 \pm0.3  $ & $R$ &   (2)\\

\cutinhead{M31N 2007-02b}

3771.346 & $>22           $  & $R$ &  (21)\\
4128.275 & $>19.9         $  & $R$ &   (2)\\
4135.298 & $ 16.66\pm0.1  $ & $R$ &   (2)\\
4141.362 & $ 17.3 \pm0.15 $ & $R$ &   (2)\\
4146.295 & $ 17.5 \pm0.15 $ & $R$ &   (2)\\
4149.273 & $ 17.7 \pm0.1  $ & $R$ &   (2)\\
4162.295 & $ 18.5 \pm0.2  $ & $R$ &   (1)\\
4164.263 & $ 18.4 \pm0.2  $ & $R$ &   (1)\\
4166.274 & $ 18.4 \pm0.15 $ & $R$ &   (1)\\
4170.274 & $ 19.1 \pm0.2  $ & $R$ &   (1)\\
4174.268 & $ 18.9 \pm0.25 $ & $R$ &   (2)\\
4175.267 & $ 19.2 \pm0.3  $ & $R$ &   (4)\\
4238.560 & $>19.7         $  & $R$ &   (1)\\

\cutinhead{M31N 2007-07c}
 
3771.346 & $>22           $  & $R$ &  (21)\\
4288.506 & $>20.0         $  & $R$ &   (1)\\
4327.469 & $ 19.2 \pm0.25 $ & $R$ &   (1)\\
4330.344 & $ 19.2 \pm0.25 $ & $R$ &   (1)\\

\cutinhead{M31N 2007-07e}

3771.346 & $>22           $  & $R$ &  (21)\\
4288.506 & $>19.5         $  & $R$ &   (1)\\
4327.469 & $ 17.9 \pm0.25 $ & $R$ &   (1)\\
4330.344 & $ 18.1 \pm0.25 $ & $R$ &   (1)\\
4343.487 & $ 18.4 \pm0.25 $ & $R$ &   (4)\\
4353.409 & $ 18.4 \pm0.3  $ & $R$ &   (1)\\
4356.430 & $ 19.3 \pm0.3  $ & $R$ &   (1)\\
4357.473 & $ 19.2 \pm0.3  $ & $R$ &   (1)\\
4358.284 & $ 19.5 \pm0.3  $ & $R$ &   (1)\\
4365.319 & $ 19.6 \pm0.3  $ & $R$ &   (3)\\

\cutinhead{M31N 2007-08d}

3771.346 & $>22           $  & $R$ &  (21)\\
4380.401 & $ 18.6 \pm0.2  $ & $R$ &  (24)\\
4380.424 & $ 18.5 \pm0.15 $ & $R$ &  (24)\\
4382.235 & $ 19.0 \pm0.25 $ & $R$ &  (24)\\
4387.219 & $>19.5         $  & $R$ &  (24)\\
4387.231 & $ 19.5 \pm0.3  $ & $R$ &  (24)\\
4387.561 & $ 19.6 \pm0.2  $ & $R$ &   (1)\\
4388.227 & $ 19.8 \pm0.25 $ & $R$ &  (24)\\
4388.626 & $ 19.6 \pm0.35 $ & $R$ &   (1)\\
4389.233 & $ 19.8 \pm0.3  $ & $R$ &  (24)\\
4389.646 & $ 20.0 \pm0.3  $ & $R$ &   (1)\\

\cutinhead{M31N 2007-10a}

4383.529 & $ 17.856 \pm 0.031 $ & $B$ & (30)\\
4384.398 & $ 18.019 \pm 0.031 $ & $B$ & (30)\\
4387.398 & $ 18.932 \pm 0.034 $ & $B$ & (30)\\
4389.384 & $ 19.276 \pm 0.035 $ & $B$ & (30)\\
4390.355 & $ 19.501 \pm 0.038 $ & $B$ & (30)\\
4392.486 & $ 20.386 \pm 0.048 $ & $B$ & (30)\\
4393.714 & $ 21.077 \pm 0.070 $ & $B$ & (30)\\
4394.448 & $ 21.434 \pm 0.093 $ & $B$ & (30)\\
4395.553 & $>20.359$ & $B$ & (30)\\
4396.352 & $>20.224$ & $B$ & (30)\\
4397.375 & $>22.588$ & $B$ & (30)\\
4398.367 & $>22.575$ & $B$ & (30)\\
4399.380 & $>21.985$ & $B$ & (30)\\
4400.693 & $>20.718$ & $B$ & (30)\\
4402.606 & $>22.359$ & $B$ & (30)\\
4405.649 & $>22.927$ & $B$ & (30)\\
4406.679 & $>21.795$ & $B$ & (30)\\
4407.573 & $>22.505$ & $B$ & (30)\\
4410.416 & $>22.726$ & $B$ & (30)\\
\\
4383.860 & $ 18.091 \pm 0.041 $ & $B$ & (31)\\
4384.846 & $ 18.253 \pm 0.042 $ & $B$ & (31)\\
4385.970 & $ 19.059 \pm 0.060 $ & $B$ & (31)\\
4386.811 & $ 18.922 \pm 0.052 $ & $B$ & (31)\\
4387.869 & $ 19.009 \pm 0.057 $ & $B$ & (31)\\
4388.786 & $ 19.180 \pm 0.062 $ & $B$ & (31)\\
4389.991 & $ 17.651 \pm 0.037 $ & $B$ & (31)\\
4390.773 & $ 19.720 \pm 0.073 $ & $B$ & (31)\\
4391.745 & $ 20.264 \pm 0.094 $ & $B$ & (31)\\
4392.739 & $ 20.398 \pm 0.129 $ & $B$ & (31)\\
4393.971 & $>21.016$ & $B$ & (3)\\
4394.793 & $>21.294$ & $B$ & (31)\\
\\
4383.532 & $ 17.859 \pm 0.036 $ & $V$ & (30)\\
4384.401 & $ 18.021 \pm 0.037 $ & $V$ & (30)\\
4387.403 & $ 18.802 \pm 0.040 $ & $V$ & (30)\\
4389.386 & $ 19.188 \pm 0.040 $ & $V$ & (30)\\
4390.358 & $ 19.372 \pm 0.041 $ & $V$ & (30)\\
4392.489 & $ 20.141 \pm 0.048 $ & $V$ & (30)\\
4393.717 & $ 20.709 \pm 0.077 $ & $V$ & (30)\\
4394.451 & $ 20.871 \pm 0.072 $ & $V$ & (30)\\
4395.555 & $>20.255$ & $V$ & (30)\\
4396.355 & $>20.328$ & $V$ & (30)\\
4397.378 & $>21.711$ & $V$ & (30)\\
4398.370 & $>21.835$ & $V$ & (30)\\
4399.383 & $>22.252$ & $V$ & (30)\\
4400.696 & $>21.472$ & $V$ & (30)\\
4402.609 & $>21.735$ & $V$ & (30)\\
4405.652 & $>22.490$ & $V$ & (30)\\
4406.682 & $>22.278$ & $V$ & (30)\\
4407.576 & $>21.828$ & $V$ & (30)\\
4410.419 & $>21.934$ & $V$ & (30)\\
\\
4383.862 & $ 17.933 \pm 0.041 $ & $V$ & (31)\\
4384.849 & $ 18.141 \pm 0.042 $ & $V$ & (31)\\
4385.973 & $ 18.808 \pm 0.048 $ & $V$ & (31)\\
4386.813 & $ 18.796 \pm 0.047 $ & $V$ & (31)\\
4387.872 & $ 18.962 \pm 0.049 $ & $V$ & (31)\\
4388.789 & $ 19.047 \pm 0.053 $ & $V$ & (31)\\
4389.994 & $ 18.163 \pm 0.042 $ & $V$ & (31)\\
4390.777 & $ 19.512 \pm 0.058 $ & $V$ & (31)\\
4391.748 & $ 19.834 \pm 0.063 $ & $V$ & (31)\\
4392.742 & $ 20.394 \pm 0.107 $ & $V$ & (31)\\
4393.975 & $>21.105$ & $V$ & (31)\\
4394.796 & $>20.904$ & $V$ & (31)\\
\\
3771.346 & $>22           $  & $R$ &  (21)\\
4388.644 & $ 18.6 \pm0.2  $ & $R$ &   (1)\\
\\
4383.535 & $ 17.994 \pm 0.027 $ & $i'$ & (30)\\
4384.404 & $ 18.155 \pm 0.027 $ & $i'$ & (30)\\
4389.389 & $ 19.227 \pm 0.032 $ & $i'$ & (30)\\
4390.361 & $ 19.309 \pm 0.032 $ & $i'$ & (30)\\
4392.492 & $ 19.992 \pm 0.044 $ & $i'$ & (30)\\
4393.720 & $ 20.565 \pm 0.071 $ & $i'$ & (30)\\
4394.453 & $ 20.746 \pm 0.066 $ & $i'$ & (30)\\
4395.558 & $>21.657$ & $i'$ & (30)\\
4396.358 & $>21.196$ & $i'$ & (30)\\
4397.381 & $>22.289$ & $i'$ & (30)\\
4398.373 & $>21.485$ & $i'$ & (30)\\
4399.386 & $>21.672$ & $i'$ & (30)\\
4400.698 & $>21.071$ & $i'$ & (30)\\
4402.612 & $>21.385$ & $i'$ & (30)\\
4405.655 & $>21.920$ & $i'$ & (30)\\
4406.685 & $>21.647$ & $i'$ & (30)\\
4407.578 & $>21.377$ & $i'$ & (30)\\
4410.422 & $>21.955$ & $i'$ & (30)\\
\\
4383.857 & $ 18.089 \pm 0.032 $ & $i'$ & (31)\\
4384.844 & $ 18.211 \pm 0.033 $ & $i'$ & (31)\\
4385.967 & $ 18.565 \pm 0.035 $ & $i'$ & (31)\\
4386.808 & $ 18.863 \pm 0.037 $ & $i'$ & (31)\\
4387.835 & $ 19.048 \pm 0.051 $ & $i'$ & (31)\\
4387.866 & $ 19.050 \pm 0.042 $ & $i'$ & (31)\\
4388.783 & $ 19.233 \pm 0.043 $ & $i'$ & (31)\\
4389.988 & $ 18.186 \pm 0.031 $ & $i'$ & (31)\\
4390.771 & $ 19.438 \pm 0.046 $ & $i'$ & (31)\\
4391.743 & $ 19.709 \pm 0.049 $ & $i'$ & (31)\\
4392.736 & $ 20.186 \pm 0.082 $ & $i'$ & (31)\\
4393.968 & $ 20.821 \pm 0.123 $ & $i'$ & (31)\\
4394.790 & $>20.323$ & $i'$ & (31)\\
4395.810 & $>20.778$ & $i'$ & (31)\\

\cutinhead{M31N 2007-10b}

4389.507 & $ 19.603 \pm 0.049 $ & $B$ & (30)\\
4392.397 & $>22.280$ & $B$ & (30)\\
4393.519 & $>21.828$ & $B$ & (30)\\
4394.436 & $>22.276$ & $B$ & (30)\\
4395.483 & $>21.807$ & $B$ & (30)\\
\\
4389.504 & $ 19.927 \pm 0.038 $ & $V$ & (30)\\
4392.394 & $>21.704$ & $V$ & (30)\\
4393.516 & $>21.935$ & $V$ & (30)\\
4394.433 & $>22.293$ & $V$ & (30)\\
4395.480 & $>22.240$ & $V$ & (30)\\
\\
3771.346 & $>22           $  & $R$ &  (21)\\
4382.235 & $>19.4         $  & $R$ &   (1)\\
4387.219 & $ 18.5 \pm0.15 $ & $R$ &  (24)\\
4387.561 & $ 18.2 \pm0.1  $ & $R$ &   (1)\\
4388.227 & $ 19.1 \pm0.2  $ & $R$ &  (24)\\
4388.626 & $ 19.1 \pm0.2  $ & $R$ &   (1)\\
4389.233 & $ 19.3 \pm0.2  $ & $R$ &  (24)\\
4389.646 & $ 19.6 \pm0.25 $ & $R$ &   (1)\\
4405.318 & $>19.6         $  & $R$ &  (28)\\
4409.388 & $>19.9         $  & $R$ &   (1)\\
4410.204 & $>19.7 \pm0.25 $ & $R$ &  (27)\\
4411.517 & $>19.7 \pm0.25 $ & $R$ &  (27)\\
\\ 
4389.501 & $ 20.148 \pm 0.062 $ & $i'$ & (30)\\
4390.406 & $ 20.649 \pm 0.096 $ & $i'$ & (30)\\
4392.391 & $>20.840$ & $i'$ & (30)\\
4393.513 & $>20.811$ & $i'$ & (30)\\
4394.430 & $>21.046$ & $i'$ & (30)\\
4395.477 & $>20.406$ & $i'$ & (30)\\

\cutinhead{M31N 2007-11b}

4418.534 & $ 19.830 \pm 0.027 $ & $B$ & (30)\\
4420.510 & $ 19.829 \pm 0.026 $ & $B$ & (30)\\
4421.434 & $ 19.753 \pm 0.023 $ & $B$ & (30)\\
4431.597 & $ 21.004 \pm 0.144 $ & $B$ & (30)\\
4432.478 & $ 20.848 \pm 0.096 $ & $B$ & (30)\\
4438.431 & $ 20.805 \pm 0.047 $ & $B$ & (30)\\
4444.422 & $ 20.976 \pm 0.045 $ & $B$ & (30)\\
\\
4419.872 & $ 19.933 \pm 0.134 $ & $B$ & (31)\\
4421.846 & $ 19.742 \pm 0.080 $ & $B$ & (31)\\
4422.762 & $ 20.191 \pm 0.100 $ & $B$ & (31)\\
4427.856 & $>19.946$ & $B$ & (31)\\
\\
4418.537 & $ 19.662 \pm 0.026 $ & $V$ & (30)\\
4420.513 & $ 19.628 \pm 0.024 $ & $V$ & (30)\\
4421.437 & $ 19.509 \pm 0.021 $ & $V$ & (30)\\
4431.602 & $ 20.391 \pm 0.075 $ & $V$ & (30)\\
4432.481 & $ 20.557 \pm 0.070 $ & $V$ & (30)\\
4438.434 & $ 20.581 \pm 0.043 $ & $V$ & (30)\\
4441.433 & $ 19.966 \pm 0.030 $ & $V$ & (30)\\
4444.424 & $ 20.691 \pm 0.039 $ & $V$ & (30)\\
\\
4419.874 & $ 19.687 \pm 0.064 $ & $V$ & (31)\\
4421.849 & $ 19.551 \pm 0.051 $ & $V$ & (31)\\
4422.765 & $ 19.802 \pm 0.056 $ & $V$ & (31)\\
4427.859 & $ 20.272 \pm 0.210 $ & $V$ & (31)\\
\\
3771.346 & $>21.0         $  & $R$ &  (21)\\
4415.446 & $ 18.3 \pm0.15 $ & $R$ &   (1)\\
4416.223 & $ 18.2 \pm0.15 $ & $R$ &   (1)\\
4453.263 & $ 19.9 \pm0.25 $ & $R$ &   (1)\\
\\
4418.540 & $ 19.246 \pm 0.028 $ & $i'$ & (30)\\
4420.516 & $ 19.159 \pm 0.026 $ & $i'$ & (30)\\
4421.440 & $ 19.063 \pm 0.045 $ & $i'$ & (30)\\
4431.607 & $ 19.858 \pm 0.041 $ & $i'$ & (30)\\
4432.484 & $ 19.669 \pm 0.033 $ & $i'$ & (30)\\
4438.437 & $ 19.735 \pm 0.033 $ & $i'$ & (30)\\
4441.436 & $ 19.366 \pm 0.030 $ & $i'$ & (30)\\
4444.427 & $ 20.039 \pm 0.032 $ & $i'$ & (30)\\
\\
4419.877 & $ 19.124 \pm 0.041 $ & $i'$ & (31)\\
4421.852 & $ 19.054 \pm 0.036 $ & $i'$ & (31)\\
4422.768 & $ 19.355 \pm 0.041 $ & $i'$ & (31)\\
4427.862 & $ 19.607 \pm 0.092 $ & $i'$ & (31)\\
4428.758 & $ 19.126 \pm 0.093 $ & $i'$ & (31)\\
4430.869 & $ 19.553 \pm 0.060 $ & $i'$ & (31)\\
4436.765 & $ 19.807 \pm 0.061 $ & $i'$ & (31)\\

\cutinhead{M31N 2007-11c}

4419.883 & $ 16.642 \pm 0.067 $ & $B$ & (31)\\
4420.536 & $ 16.736 \pm 0.013 $ & $B$ & (30)\\
4421.447 & $ 16.718 \pm 0.012 $ & $B$ & (30)\\
4421.858 & $ 16.985 \pm 0.027 $ & $B$ & (31)\\
4422.774 & $ 17.058 \pm 0.024 $ & $B$ & (31)\\
4427.868 & $>18.311$ & $B$ & (31)\\
4430.875 & $>18.536$ & $B$ & (31)\\
4432.385 & $ 18.524 \pm 0.020 $ & $B$ & (30)\\
4436.771 & $ 19.117 \pm 0.094 $ & $B$ & (30)\\
4438.418 & $ 19.184 \pm 0.035 $ & $B$ & (30)\\
4441.397 & $ 19.423 \pm 0.046 $ & $B$ & (30)\\
4443.534 & $>20.583$ & $B$ & (30)\\
4449.347 & $ 20.019 \pm 0.052 $ & $B$ & (30)\\
\\
4419.886 & $ 16.651 \pm 0.016 $ & $V$ & (31)\\
4420.539 & $ 16.926 \pm 0.014 $ & $V$ & (30)\\
4421.450 & $ 17.028 \pm 0.011 $ & $V$ & (30)\\
4421.860 & $ 17.165 \pm 0.019 $ & $V$ & (31)\\
4422.776 & $ 17.327 \pm 0.019 $ & $V$ & (31)\\
4427.870 & $>17.968$ & $V$ & (31)\\
4428.767 & $>18.804$ & $V$ & (31)\\
4429.766 & $ 18.521 \pm 0.054 $ & $V$ & (31)\\
4430.878 & $ 18.616 \pm 0.127 $ & $V$ & (31)\\
4432.388 & $ 18.891 \pm 0.028 $ & $V$ & (30)\\
4436.773 & $ 19.266 \pm 0.091 $ & $V$ & (31)\\
4438.421 & $ 19.482 \pm 0.064 $ & $V$ & (30)\\
4441.400 & $ 19.890 \pm 0.079 $ & $V$ & (30)\\
4443.537 & $>19.933$ & $V$ & (30)\\
4449.350 & $ 20.334 \pm 0.107 $ & $V$ & (30)\\
\\
3771.346 & $>21.5         $  & $R$ &  (21)\\
4415.475 & $>20.1         $  & $R$ &   (1)\\
4416.387 & $ 19.7 \pm0.25 $ & $R$ &   (1)\\
4417.547 & $ 17.8 \pm0.2  $ & $R$ &   (1)\\
4445.234 & $ 19.3:\pm0.4  $ & $R$ &   (1)\\
\\
4419.888 & $ 17.173 \pm 0.082 $ & $i'$ & (31)\\
4420.542 & $ 17.467 \pm 0.081 $ & $i'$ & (30)\\
4421.453 & $ 17.592 \pm 0.081 $ & $i'$ & (30)\\
4421.863 & $ 17.710 \pm 0.084 $ & $i'$ & (31)\\
4422.779 & $ 17.848 \pm 0.083 $ & $i'$ & (31)\\
4427.873 & $>18.935$ & $i'$ & (31)\\
4429.769 & $ 19.961 \pm 0.154 $ & $i'$ & (31)\\
4432.391 & $ 19.539 \pm 0.107 $ & $i'$ & (30)\\
4438.424 & $ 19.955 \pm 0.171 $ & $i'$ & (30)\\
4443.540 & $>18.842$ & $i'$ & (30)\\
4449.352 & $>20.690$ & $i'$ & (30)\\

\cutinhead{M31N 2007-11d}

4427.880 & $ 16.460 \pm 0.030 $ & $B$ & (30)\\
4428.780 & $ 16.520 \pm 0.040 $ & $B$ & (30)\\
4429.770 & $ 16.560 \pm 0.030 $ & $B$ & (30)\\
4432.470 & $ 16.639 \pm 0.027 $ & $B$ & (31)\\
4436.580 & $ 17.558 \pm 0.028 $ & $B$ & (31)\\
4436.780 & $ 17.820 \pm 0.070 $ & $B$ & (30)\\
4439.370 & $ 20.220 \pm 0.040 $ & $B$ & (31)\\
4443.330 & $ 21.772 \pm 0.029 $ & $B$ & (31)\\
\\
4427.880 & $ 16.071 \pm 0.024 $ & $V$ & (30)\\
4428.780 & $ 16.236 \pm 0.026 $ & $V$ & (30)\\
4429.780 & $ 16.334 \pm 0.024 $ & $V$ & (30)\\
4432.470 & $ 16.603 \pm 0.023 $ & $V$ & (31)\\
4436.590 & $ 17.563 \pm 0.031 $ & $V$ & (31)\\
4436.780 & $ 17.703 \pm 0.043 $ & $V$ & (30)\\
4439.370 & $ 20.039 \pm 0.036 $ & $V$ & (31)\\
4443.340 & $ 21.443 \pm 0.133 $ & $V$ & (31)\\
\\
4427.880 & $ 15.780 \pm 0.010 $ & $i'$ & (30)\\
4428.780 & $ 15.890 \pm 0.030 $ & $i'$ & (30)\\
4429.780 & $ 16.040 \pm 0.010 $ & $i'$ & (30)\\
4432.470 & $ 16.310 \pm 0.010 $ & $i'$ & (31)\\
4436.590 & $ 17.070 \pm 0.010 $ & $i'$ & (31)\\
4436.790 & $ 17.270 \pm 0.030 $ & $i'$ & (30)\\
4439.380 & $ 18.710 \pm 0.020 $ & $i'$ & (31)\\
4443.340 & $ 20.270 \pm 0.070 $ & $i'$ & (31)\\
4449.340 & $ 20.880 \pm 0.090 $ & $i'$ & (31)\\

\cutinhead{M31N 2007-12a}

4449.415 & $ 18.002 \pm 0.023 $ & $B$ & (30)\\
4451.434 & $ 17.709 \pm 0.023 $ & $B$ & (30)\\
4455.392 & $ 17.872 \pm 0.032 $ & $B$ & (30)\\
4459.390 & $ 18.069 \pm 0.024 $ & $B$ & (30)\\
4460.344 & $ 18.114 \pm 0.023 $ & $B$ & (30)\\
4461.386 & $ 18.275 \pm 0.024 $ & $B$ & (30)\\
4463.473 & $ 18.410 \pm 0.024 $ & $B$ & (30)\\
4466.335 & $ 18.664 \pm 0.024 $ & $B$ & (30)\\
4467.367 & $ 18.790 \pm 0.025 $ & $B$ & (30)\\
4468.388 & $ 18.864 \pm 0.024 $ & $B$ & (30)\\
4469.432 & $ 19.050 \pm 0.025 $ & $B$ & (30)\\
4471.343 & $ 19.135 \pm 0.026 $ & $B$ & (30)\\
4473.862 & $ 19.533 \pm 0.067 $ & $B$ & (31)\\
4474.413 & $ 19.253 \pm 0.026 $ & $B$ & (30)\\
4474.732 & $ 19.402 \pm 0.056 $ & $B$ & (31)\\
4474.741 & $ 19.338 \pm 0.101 $ & $B$ & (31)\\
\\
4449.418 & $ 17.817 \pm 0.013 $ & $V$ & (30)\\
4451.437 & $ 17.601 \pm 0.013 $ & $V$ & (30)\\
4455.395 & $ 17.803 \pm 0.013 $ & $V$ & (30)\\
4459.393 & $ 18.020 \pm 0.014 $ & $V$ & (30)\\
4460.347 & $ 18.079 \pm 0.014 $ & $V$ & (30)\\
4461.389 & $ 18.246 \pm 0.014 $ & $V$ & (30)\\
4463.476 & $ 18.436 \pm 0.015 $ & $V$ & (30)\\
4466.338 & $ 18.647 \pm 0.015 $ & $V$ & (30)\\
4467.370 & $ 18.789 \pm 0.016 $ & $V$ & (30)\\
4468.391 & $ 18.842 \pm 0.016 $ & $V$ & (30)\\
4469.435 & $ 19.013 \pm 0.017 $ & $V$ & (30)\\
4471.346 & $ 19.093 \pm 0.018 $ & $V$ & (30)\\
4473.865 & $ 19.475 \pm 0.046 $ & $V$ & (31)\\
4474.416 & $ 19.255 \pm 0.018 $ & $V$ & (30)\\
4474.734 & $ 19.349 \pm 0.049 $ & $V$ & (31)\\
4474.744 & $ 19.314 \pm 0.069 $ & $V$ & (31)\\
\\
4449.421 & $ 17.629 \pm 0.012 $ & $i'$ & (30)\\
4451.439 & $ 17.374 \pm 0.011 $ & $i'$ & (30)\\
4455.398 & $ 17.455 \pm 0.012 $ & $i'$ & (30)\\
4459.396 & $ 17.676 \pm 0.012 $ & $i'$ & (30)\\
4460.350 & $ 17.778 \pm 0.012 $ & $i'$ & (30)\\
4461.392 & $ 17.917 \pm 0.013 $ & $i'$ & (30)\\
4463.479 & $ 18.045 \pm 0.013 $ & $i'$ & (30)\\
4466.341 & $ 18.289 \pm 0.013 $ & $i'$ & (30)\\
4467.373 & $ 18.363 \pm 0.014 $ & $i'$ & (30)\\
4468.394 & $ 18.404 \pm 0.014 $ & $i'$ & (30)\\
4469.437 & $ 18.540 \pm 0.015 $ & $i'$ & (30)\\
4471.349 & $ 18.621 \pm 0.016 $ & $i'$ & (30)\\
4473.868 & $ 18.741 \pm 0.027 $ & $i'$ & (31)\\
4474.419 & $ 18.761 \pm 0.016 $ & $i'$ & (30)\\
4474.747 & $ 18.831 \pm 0.043 $ & $i'$ & (31)\\

\cutinhead{M31N 2007-12b}

4449.440 & $ 19.190 \pm 0.050 $ & $B$ & (30)\\
4452.350 & $ 19.750 \pm 0.050 $ & $B$ & (30)\\
4455.400 & $ 20.270 \pm 0.090 $ & $B$ & (30)\\
4459.400 & $ 20.880 \pm 0.110 $ & $B$ & (30)\\
4460.390 & $ 21.120 \pm 0.110 $ & $B$ & (30)\\
4461.510 & $>20.54$ & $B$ & (30)\\
4464.520 & $ 21.590 \pm 0.050 $ & $B$ & (30)\\
4467.350 & $>21.4$ & $B$ & (30)\\
4468.400 & $>21.01$ & $B$ & (30)\\
4469.460 & $>21.63$ & $B$ & (30)\\
4470.440 & $>21.78$ & $B$ & (30)\\
4472.480 & $>21.73$ & $B$ & (30)\\
\\
4449.450 & $ 19.270 \pm 0.040 $ & $V$ & (30)\\
4452.350 & $ 19.820 \pm 0.040 $ & $V$ & (30)\\
4455.410 & $ 20.430 \pm 0.090 $ & $V$ & (30)\\
4459.410 & $ 20.670 \pm 0.090 $ & $V$ & (30)\\
4460.390 & $>21.11$ & $V$ & (30)\\
4461.510 & $>20.74$ & $V$ & (30)\\
4464.520 & $>20.55$ & $V$ & (30)\\
4467.350 & $>20.88$ & $V$ & (30)\\
4468.400 & $>20.97$ & $V$ & (30)\\
4469.470 & $>20.62$ & $V$ & (30)\\
4470.450 & $>20.96$ & $V$ & (30)\\
4472.480 & $>21.09$ & $V$ & (30)\\
\\
3771.346 & $>21.5         $  & $R$ &  (21)\\
4417.547 & $>19.1         $  & $R$ &   (1)\\
4445.234 & $ 17.0 \pm0.1  $ & $R$ &   (1)\\
4450.392 & $ 18.6 \pm0.25 $ & $R$ &   (1)\\
4450.402 & $ 18.7 \pm0.25 $ & $R$ &   (1)\\
4453.250 & $ 19.4 \pm0.25 $ & $R$ &   (1)\\
\\
4449.450 & $ 18.930 \pm 0.040 $ & $i'$ & (30)\\
4452.360 & $ 19.350 \pm 0.040 $ & $i'$ & (30)\\
4455.410 & $>19.65$ & $i'$ & (30)\\
4459.410 & $>19.58$ & $i'$ & (30)\\
4460.390 & $>19.64$ & $i'$ & (30)\\

\cutinhead{M31N 2007-12d}

3771.346 & $>21.5         $  & $R$ &  (21)\\
4453.238 & $ 17.8 \pm0.2  $ & $R$ &   (1)\\

\cutinhead{M31N 2008-05c}

4528.246 & $>19.6         $  & $R$ &   (1)\\
4601.558 & $>19.8         $  & $R$ &  (28)\\
4617.523 & $ 17.6 \pm0.2  $ & $R$ &   (1)\\
4617.549 & $ 17.5 \pm0.2  $ & $R$ &   (1)\\
4619.529 & $ 17.2 \pm0.1  $ & $R$ &   (1)\\
4619.554 & $ 17.2 \pm0.1  $ & $R$ &   (1)\\
4620.543 & $ 17.3 \pm0.1  $ & $R$ &   (1)\\
4628.541 & $ 18.1 \pm0.2  $ & $R$ &   (3)\\
4645.550 & $ 19.3 \pm0.2  $ & $R$ &   (1)\\
4646.535 & $ 19.0 \pm0.2  $ & $R$ &   (1)\\
4647.492 & $ 19.0 \pm0.2  $ & $R$ &   (1)\\
4648.557 & $ 19.0 \pm0.2  $ & $R$ &   (1)\\
4655.567 & $ 19.6 \pm0.25 $ & $R$ &   (3)\\
4675.581 & $ 19.3 \pm0.2  $ & $R$ &   (3)\\
4677.432 & $ 19.4 \pm0.3  $ & $R$ &   (1)\\
4678.562 & $ 19.6 \pm0.25 $ & $R$ &   (1)\\
4679.551 & $ 19.5 \pm0.25 $ & $R$ &   (1)\\
4682.379 & $ 19.6 \pm0.3  $ & $R$ &   (1)\\
4682.574 & $ 19.6 \pm0.2  $ & $R$ &   (1)\\
4683.562 & $ 19.8 \pm0.3  $ & $R$ &   (1)\\
4685.548 & $ 20.0 \pm0.3  $ & $R$ &   (1)\\
4692.616 & $>19.7         $  & $R$ &  (28)\\
4697.573 & $ 20.3 \pm0.3  $ & $R$ &   (3)\\
4706.362 & $ 19.6 \pm0.3  $ & $R$ &   (1)\\
4712.608 & $>20.1         $  & $R$ &   (1)\\

\cutinhead{M31N 2008-06b}

4505.226 & $>20.0         $  & $R$ &   (1)\\
4620.542 & $>20.3         $  & $R$ &   (1)\\
4628.541 & $>20.2         $  & $R$ &   (3)\\
4644.498 & $ 16.0 \pm0.1  $ & $R$ &   (5)\\
4645.489 & $ 16.44\pm0.05 $ & $R$ &   (1)\\
4645.529 & $ 16.41\pm0.05 $ & $R$ &   (1)\\
4645.550 & $ 16.41\pm0.05 $ & $R$ &   (1)\\
4646.535 & $ 16.54\pm0.05 $ & $R$ &   (1)\\
4647.492 & $ 16.62\pm0.05 $ & $R$ &   (1)\\
4648.557 & $ 16.71\pm0.05 $ & $R$ &   (1)\\
4652.560 & $ 17.26\pm0.1  $ & $R$ &   (3)\\
4655.567 & $ 17.45\pm0.1  $ & $R$ &   (3)\\
4675.581 & $ 18.9 \pm0.15 $ & $R$ &   (3)\\
4677.432 & $ 18.75\pm0.2  $ & $R$ &   (1)\\
4678.562 & $ 18.8 \pm0.15 $ & $R$ &   (1)\\
4679.551 & $ 19.0 \pm0.15 $ & $R$ &   (1)\\
4681.370 & $ 19.1 \pm0.25 $ & $R$ &   (1)\\
4682.379 & $ 19.1 \pm0.2  $ & $R$ &   (1)\\
4682.574 & $ 19.0 \pm0.15 $ & $R$ &   (1)\\
4683.562 & $ 19.1 \pm0.2  $ & $R$ &   (1)\\
4685.548 & $ 19.4 \pm0.2  $ & $R$ &   (1)\\
4692.616 & $>19.7         $  & $R$ &  (28)\\
4697.573 & $ 20.4 \pm0.3  $ & $R$ &   (3)\\
4706.362 & $>20.4         $  & $R$ &   (1)\\

\cutinhead{M31N 2008-07a}

5095.981 & $ 21.4 \pm0.4  $ V  (13)\\
\\
4505.226 & $>19.7         $  & $R$ &   (1)\\
4617.523 & $>19.0         $  & $R$ &   (1)\\
4619.529 & $ 19.0 \pm0.3  $ & $R$ &   (1)\\
4620.542 & $ 18.7 \pm0.25 $ & $R$ &   (1)\\
4628.541 & $ 19.0 \pm0.3  $ & $R$ &   (3)\\
4645.488 & $ 19.1 \pm0.3  $ & $R$ &   (1)\\
4646.535 & $ 19.2 \pm0.25 $ & $R$ &   (1)\\
4647.492 & $ 18.8 \pm0.2  $ & $R$ &   (1)\\
4648.557 & $ 19.3 \pm0.25 $ & $R$ &   (1)\\
4655.567 & $ 19.1 \pm0.25 $ & $R$ &   (3)\\
4675.581 & $ 19.0 \pm0.2  $ & $R$ &   (3)\\
4677.432 & $ 18.8 \pm0.2  $ & $R$ &   (1)\\
4678.562 & $ 18.8 \pm0.2  $ & $R$ &   (1)\\
4679.551 & $ 18.9 \pm0.2  $ & $R$ &   (1)\\
4681.370 & $ 19.3 \pm0.3  $ & $R$ &   (1)\\
4682.379 & $ 19.1 \pm0.25 $ & $R$ &   (1)\\
4682.574 & $ 19.0 \pm0.15 $ & $R$ &   (1)\\
4683.562 & $ 19.0 \pm0.2  $ & $R$ &   (1)\\
4684.525 & $ 19.5 \pm0.3  $ & $R$ &   (1)\\
4685.548 & $ 18.75\pm0.1  $ & $R$ &   (1)\\
4692.616 & $ 18.9 \pm0.2  $ & $R$ &  (28)\\
4697.573 & $ 19.0 \pm0.15 $ & $R$ &   (3)\\
4706.362 & $ 19.1 \pm0.2  $ & $R$ &   (1)\\
4708.387 & $ 18.7 \pm0.25 $ & $R$ &   (1)\\
4709.613 & $ 19.3 \pm0.25 $ & $R$ &   (1)\\
4710.589 & $ 19.2 \pm0.35 $ & $R$ &   (1)\\
4711.310 & $ 18.9 \pm0.25 $ & $R$ &   (1)\\
4712.291 & $ 19.0 \pm0.25 $ & $R$ &   (1)\\
4712.337 & $ 18.8 \pm0.15 $ & $R$ &   (1)\\
4712.608 & $ 18.85\pm0.15 $ & $R$ &   (1)\\
4713.343 & $ 19.0 \pm0.2  $ & $R$ &   (1)\\
4715.330 & $ 18.8 \pm0.15 $ & $R$ &   (1)\\
4716.334 & $ 18.9 \pm0.25 $ & $R$ &   (1)\\
4718.378 & $ 19.3 \pm0.3  $ & $R$ &   (3)\\
4719.299 & $ 19.2 \pm0.3  $ & $R$ &   (3)\\
4738.221 & $ 18.85\pm0.15 $ & $R$ &   (3)\\
4744.604 & $ 19.1 \pm0.15 $ & $R$ &   (1)\\
4745.257 & $ 19.1 \pm0.2  $ & $R$ &   (1)\\
4748.472 & $ 19.1 \pm0.2  $ & $R$ &   (7)\\
4748.485 & $ 19.2 \pm0.25 $ & $R$ &   (7)\\
4754.400 & $ 19.0 \pm0.2  $ & $R$ &   (3)\\
4760.628 & $ 19.2 \pm0.2  $ & $R$ &   (3)\\
4763.219 & $ 19.2 \pm0.3  $ & $R$ &   (1)\\
4765.545 & $ 19.1 \pm0.25 $ & $R$ &   (1)\\
4772.448 & $ 19.4 \pm0.25 $ & $R$ &   (1)\\
4774.545 & $ 19.3 \pm0.3  $ & $R$ &   (1)\\
4776.228 & $ 19.6 \pm0.25 $ & $R$ &   (1)\\
4777.215 & $ 19.6 \pm0.3  $ & $R$ &   (1)\\
4779.283 & $ 19.4 \pm0.25 $ & $R$ &   (1)\\
4779.318 & $ 19.5 \pm0.3  $ & $R$ &   (1)\\
4800.301 & $ 19.3 \pm0.3  $ & $R$ &   (1)\\
4801.226 & $ 18.9:\pm0.4  $ & $R$ &   (1)\\
4809.285 & $ 19.0:\pm0.4  $ & $R$ &   (1)\\
4982.972 & $>20.4         $  & $R$ &   (9)\\
5095.968 & $ 20.9 \pm0.25 $ & $R$ &  (13)\\

\cutinhead{M31N 2008-07b}
 
4505.226 & $>19.7         $  & $R$ &   (1)\\
4655.567 & $>20.6         $  & $R$ &   (3)\\
4675.581 & $ 18.4 \pm0.1  $ & $R$ &   (3)\\
4677.432 & $ 18.8 \pm0.2  $ & $R$ &   (1)\\
4677.448 & $ 18.6 \pm0.2  $ & $R$ &   (1)\\
4678.562 & $ 18.7 \pm0.15 $ & $R$ &   (1)\\
4679.551 & $ 18.8 \pm0.15 $ & $R$ &   (1)\\
4681.370 & $ 19.3 \pm0.25 $ & $R$ &   (1)\\
4682.379 & $ 19.6 \pm0.25 $ & $R$ &   (1)\\
4682.550 & $ 19.3 \pm0.2  $ & $R$ &   (1)\\
4682.574 & $ 19.4 \pm0.15 $ & $R$ &   (1)\\
4683.562 & $ 19.3 \pm0.15 $ & $R$ &   (1)\\
4684.525 & $ 19.5 \pm0.3  $ & $R$ &   (1)\\
4685.548 & $ 19.6 \pm0.15 $ & $R$ &   (1)\\
4685.564 & $ 19.5 \pm0.15 $ & $R$ &   (1)\\
4685.577 & $ 19.4 \pm0.15 $ & $R$ &   (1)\\
4692.616 & $ 19.3 \pm0.25 $ & $R$ &  (28)\\
4697.573 & $ 20.0 \pm0.25 $ & $R$ &   (3)\\
4706.362 & $ 20.7 \pm0.3  $ & $R$ &   (1)\\
4708.424 & $ 20.3 \pm0.3  $ & $R$ &   (1)\\
4710.323 & $>20.2         $  & $R$ &   (1)\\
4711.335 & $ 20.6 \pm0.4  $ & $R$ &   (1)\\
4712.608 & $ 20.3 \pm0.25 $ & $R$ &   (1)\\
4713.343 & $ 20.2 \pm0.35 $ & $R$ &   (1)\\
4715.330 & $ 20.5 \pm0.3  $ & $R$ &   (1)\\

\cutinhead{M31N 2008-08a}

4685.548 & $>20.5         $  & $R$ &   (1)\\
4692.616 & $ 16.7 \pm0.1  $ & $R$ &  (28)\\
4697.573 & $ 17.7 \pm0.1  $ & $R$ &   (3)\\
4706.362 & $ 18.3 \pm0.15 $ & $R$ &   (1)\\
4706.627 & $ 18.2 \pm0.2  $ & $R$ &   (1)\\
4708.387 & $ 18.4 \pm0.25 $ & $R$ &   (1)\\
4709.613 & $ 18.8 \pm0.25 $ & $R$ &   (1)\\
4710.589 & $ 19.0 \pm0.3  $ & $R$ &   (1)\\
4711.310 & $ 18.4 \pm0.25 $ & $R$ &   (1)\\
4712.291 & $ 19.3 \pm0.35 $ & $R$ &   (1)\\
4712.608 & $ 19.0 \pm0.2  $ & $R$ &   (1)\\
4713.343 & $ 18.6 \pm0.25 $ & $R$ &   (1)\\
4715.330 & $ 18.6 \pm0.25 $ & $R$ &   (1)\\

\cutinhead{M31N 2008-08b}

4685.548 & $>20.5         $  & $R$ &   (1)\\
4692.616 & $ 17.8 \pm0.1  $ & $R$ &  (28)\\
4697.573 & $ 19.3 \pm0.2  $ & $R$ &   (3)\\

\cutinhead{M31N 2008-09a}

4744.340 & $ 18.4 \pm0.2  $ & $R$ &   (1)\\

\cutinhead{M31N 2008-10a}

4778.655 & $ 19.884 \pm 0.171 $ & $B$ & (30)\\
4778.727 & $ 19.477 \pm 0.071 $ & $B$ & (31)\\
4779.967 & $>20.168$ & $B$ & (31)\\
4779.637 & $ 20.080 \pm 0.195 $ & $B$ & (30)\\
4780.463 & $ 20.093 \pm 0.078 $ & $B$ & (30)\\
4781.491 & $>18.797$ & $B$ & (30)\\
4782.449 & $ 19.515 \pm 0.060 $ & $B$ & (30)\\
4783.360 & $>19.960$ & $B$ & (30)\\
4785.451 & $ 19.651 \pm 0.042 $ & $B$ & (30)\\
4786.404 & $>19.649$ & $B$ & (30)\\
4787.479 & $ 19.402 \pm 0.033 $ & $B$ & (30)\\
4788.455 & $ 19.650 \pm 0.035 $ & $B$ & (30)\\
4789.403 & $ 19.683 \pm 0.038 $ & $B$ & (30)\\
\\
4778.658 & $ 19.337 \pm 0.032 $ & $V$ & (30)\\
4778.732 & $ 19.964 \pm 0.094 $ & $V$ & (31)\\
4779.641 & $ 19.764 \pm 0.036 $ & $V$ & (30)\\
4779.972 & $ 19.781 \pm 0.103 $ & $V$ & (31)\\
4780.466 & $ 20.100 \pm 0.070 $ & $V$ & (30)\\
4780.900 & $ 18.880 \pm 0.582 $ & $V$ & (31)\\
4781.495 & $>19.338$ & $V$ & (30)\\
4782.453 & $ 19.254 \pm 0.049 $ & $V$ & (30)\\
4783.364 & $ 19.543 \pm 0.068 $ & $V$ & (30)\\
4785.456 & $ 19.456 \pm 0.040 $ & $V$ & (30)\\
4786.407 & $ 19.412 \pm 0.038 $ & $V$ & (30)\\
4787.483 & $ 19.194 \pm 0.026 $ & $V$ & (30)\\
4788.458 & $ 19.418 \pm 0.026 $ & $V$ & (30)\\
4789.406 & $ 19.444 \pm 0.026 $ & $V$ & (30)\\
\\
4775.921 & $ 17.756 \pm 0.015 $ & $r'$ & (31)\\
4777.861 & $ 17.975 \pm 0.039 $ & $r'$ & (31)\\
4778.644 & $ 18.156 \pm 0.043 $ & $r'$ & (30)\\
4778.712 & $ 18.147 \pm 0.016 $ & $r'$ & (31)\\
4779.626 & $ 18.578 \pm 0.130 $ & $r'$ & (30)\\
4779.952 & $ 18.118 \pm 0.063 $ & $r'$ & (31)\\
4780.452 & $ 18.374 \pm 0.021 $ & $r'$ & (30)\\
4780.877 & $ 18.165 \pm 0.140 $ & $r'$ & (31)\\
4781.480 & $ 17.973 \pm 0.079 $ & $r'$ & (30)\\
4782.438 & $ 18.094 \pm 0.016 $ & $r'$ & (30)\\
4783.349 & $ 18.101 \pm 0.087 $ & $r'$ & (30)\\
4785.440 & $ 18.239 \pm 0.012 $ & $r'$ & (30)\\
4786.393 & $ 18.275 \pm 0.013 $ & $r'$ & (30)\\
4787.468 & $ 18.192 \pm 0.010 $ & $r'$ & (30)\\
4788.444 & $ 18.357 \pm 0.011 $ & $r'$ & (30)\\
4789.392 & $ 18.354 \pm 0.012 $ & $r'$ & (30)\\

\cutinhead{M31N 2008-10b}

4716.407 & $>20.1         $  & $R$ &   (1)\\
4748.479 & $ 19.8 \pm0.35 $ & $R$ &   (4)\\
4754.419 & $ 19.3 \pm0.35 $ & $R$ &   (3)\\
4763.218 & $ 18.1 \pm0.15 $ & $R$ &   (1)\\
4765.546 & $ 19.0 \pm0.25 $ & $R$ &   (1)\\
4772.448 & $ 19.3 \pm0.25 $ & $R$ &   (1)\\
4774.571 & $ 19.2 \pm0.25 $ & $R$ &   (1)\\
4776.228 & $ 18.8 \pm0.15 $ & $R$ &   (1)\\
4777.229 & $ 18.5 \pm0.15 $ & $R$ &   (1)\\
4779.283 & $ 19.0 \pm0.2  $ & $R$ &   (1)\\
4795.192 & $ 19.9 \pm0.3  $ & $V$ &   (1)\\
4798.431 & $ 18.9 \pm0.2  $ & $V$ &   (5)\\
4799.199 & $ 18.3 \pm0.15 $ & $R$ &   (1)\\
4800.301 & $ 18.0 \pm0.15 $ & $R$ &   (1)\\
4801.267 & $ 18.3 \pm0.15 $ & $R$ &   (1)\\
4809.285 & $ 18.2 \pm0.15 $ & $R$ &   (1)\\
4829.183 & $ 19.1 \pm0.3  $ & $R$ &   (1)\\
\\
4760.472 & $ 18.098 \pm 0.018 $ & $B$ & (30)\\
4761.514 & $ 17.574 \pm 0.018 $ & $B$ & (30)\\
4764.549 & $ 19.451 \pm 0.055 $ & $B$ & (30)\\
4777.584 & $ 18.525 \pm 0.030 $ & $B$ & (30)\\
4778.441 & $ 19.065 \pm 0.053 $ & $B$ & (30)\\
4779.485 & $ 19.113 \pm 0.052 $ & $B$ & (30)\\
4780.429 & $ 19.399 \pm 0.128 $ & $B$ & (30)\\
4781.435 & $ 19.343 \pm 0.100 $ & $B$ & (30)\\
4782.461 & $>19.780$ & $B$ & (30)\\
4783.401 & $>19.115$ & $B$ & (30)\\
4785.495 & $ 18.619 \pm 0.036 $ & $B$ & (30)\\
4786.422 & $ 18.853 \pm 0.055 $ & $B$ & (30)\\
4787.370 & $ 19.123 \pm 0.036 $ & $B$ & (30)\\
4788.467 & $ 19.475 \pm 0.053 $ & $B$ & (30)\\
4789.418 & $ 18.839 \pm 0.033 $ & $B$ & (30)\\
4790.383 & $ 19.035 \pm 0.044 $ & $B$ & (30)\\
\\
4760.793 & $ 17.956 \pm 0.019 $ & $B$ & (31)\\
4761.793 & $ 17.652 \pm 0.018 $ & $B$ & (31)\\
4762.894 & $>17.446$ & $B$ & (31)\\
4763.727 & $ 18.993 \pm 0.044 $ & $B$ & (31)\\
4763.740 & $ 19.117 \pm 0.035 $ & $B$ & (31)\\
4763.793 & $ 19.105 \pm 0.034 $ & $B$ & (31)\\
4764.919 & $>19.756$ & $B$ & (31)\\
4771.889 & $ 19.609 \pm 0.042 $ & $B$ & (31)\\
4772.862 & $ 19.898 \pm 0.047 $ & $B$ & (31)\\
4773.835 & $ 19.706 \pm 0.039 $ & $B$ & (31)\\
4774.842 & $ 19.460 \pm 0.040 $ & $B$ & (31)\\
4777.843 & $ 18.774 \pm 0.032 $ & $B$ & (31)\\
4779.922 & $ 19.266 \pm 0.063 $ & $B$ & (31)\\
4780.835 & $>18.461$ & $B$ & (31)\\
\\
4760.476 & $ 18.124 \pm 0.016 $ & $V$ & (30)\\
4761.517 & $ 17.706 \pm 0.014 $ & $V$ & (30)\\
4764.553 & $ 19.322 \pm 0.054 $ & $V$ & (30)\\
4777.588 & $ 18.671 \pm 0.033 $ & $V$ & (30)\\
4778.444 & $ 18.231 \pm 0.059 $ & $V$ & (30)\\
4779.487 & $ 19.262 \pm 0.049 $ & $V$ & (30)\\
4780.433 & $ 19.636 \pm 0.085 $ & $V$ & (30)\\
4781.438 & $ 19.731 \pm 0.110 $ & $V$ & (30)\\
4782.465 & $ 19.750 \pm 0.182 $ & $V$ & (30)\\
4783.405 & $>19.558$ & $V$ & (30)\\
4785.499 & $ 18.815 \pm 0.030 $ & $V$ & (30)\\
4786.426 & $ 18.808 \pm 0.039 $ & $V$ & (30)\\
4787.374 & $ 19.126 \pm 0.041 $ & $V$ & (30)\\
4788.471 & $ 19.674 \pm 0.067 $ & $V$ & (30)\\
4789.421 & $ 18.897 \pm 0.035 $ & $V$ & (30)\\
4790.387 & $ 19.030 \pm 0.038 $ & $V$ & (30)\\
\\
4760.796 & $ 17.953 \pm 0.016 $ & $V$ & (31)\\
4761.796 & $ 17.699 \pm 0.015 $ & $V$ & (31)\\
4762.895 & $ 17.881 \pm 0.057 $ & $V$ & (31)\\
4763.730 & $ 18.975 \pm 0.031 $ & $V$ & (31)\\
4763.743 & $ 18.993 \pm 0.032 $ & $V$ & (31)\\
4763.796 & $ 19.040 \pm 0.033 $ & $V$ & (31)\\
4764.923 & $>18.132$ & $V$ & (31)\\
4771.892 & $ 19.653 \pm 0.051 $ & $V$ & (31)\\
4772.865 & $ 20.034 \pm 0.062 $ & $V$ & (31)\\
4773.837 & $ 19.859 \pm 0.049 $ & $V$ & (31)\\
4774.845 & $ 19.608 \pm 0.042 $ & $V$ & (31)\\
4777.846 & $ 18.773 \pm 0.035 $ & $V$ & (31)\\
4779.925 & $ 19.055 \pm 0.072 $ & $V$ & (31)\\
4780.838 & $>17.488$ & $V$ & (31)\\
\\
4760.478 & $ 18.308 \pm 0.014 $ & $r'$ & (30)\\
4761.520 & $ 17.840 \pm 0.012 $ & $r'$ & (30)\\
4764.556 & $ 19.033 \pm 0.043 $ & $r'$ & (30)\\
4777.591 & $ 18.631 \pm 0.030 $ & $r'$ & (30)\\
4778.448 & $ 19.213 \pm 0.046 $ & $r'$ & (30)\\
4779.492 & $ 19.124 \pm 0.037 $ & $r'$ & (30)\\
4780.437 & $ 19.298 \pm 0.068 $ & $r'$ & (30)\\
4781.442 & $>19.089$ & $r'$ & (30)\\
4782.469 & $>19.158$ & $r'$ & (30)\\
4783.409 & $>19.260$ & $r'$ & (30)\\
4785.503 & $ 18.917 \pm 0.026 $ & $r'$ & (30)\\
4786.430 & $ 18.876 \pm 0.420 $ & $r'$ & (30)\\
4787.378 & $ 19.119 \pm 0.038 $ & $r'$ & (30)\\
4788.474 & $ 19.378 \pm 0.053 $ & $r'$ & (30)\\
4789.425 & $ 18.934 \pm 0.031 $ & $r'$ & (30)\\
4790.391 & $ 19.029 \pm 0.036 $ & $r'$ & (30)\\
\\
4760.799 & $ 18.171 \pm 0.014 $ & $r'$ & (31)\\
4761.799 & $ 17.885 \pm 0.012 $ & $r'$ & (31)\\
4762.898 & $ 18.104 \pm 0.057 $ & $r'$ & (31)\\
4763.733 & $ 18.926 \pm 0.026 $ & $r'$ & (31)\\
4763.799 & $ 18.935 \pm 0.025 $ & $r'$ & (31)\\
4764.925 & $ 19.333 \pm 0.387 $ & $r'$ & (31)\\
4771.895 & $ 19.378 \pm 0.035 $ & $r'$ & (31)\\
4772.868 & $ 19.719 \pm 0.045 $ & $r'$ & (31)\\
4773.841 & $ 19.566 \pm 0.037 $ & $r'$ & (31)\\
4774.848 & $ 19.441 \pm 0.033 $ & $r'$ & (31)\\
4777.850 & $ 18.781 \pm 0.033 $ & $r'$ & (31)\\
4779.928 & $ 19.219 \pm 0.048 $ & $r'$ & (31)\\
4780.841 & $ 18.677 \pm 0.235 $ & $r'$ & (31)\\

\cutinhead{M31N 2008-11a}

4778.400 & $ 19.259 \pm 0.050 $ & $B$ & (30)\\
4779.472 & $ 19.783 \pm 0.064 $ & $B$ & (30)\\
4779.891 & $ 19.309 \pm 0.193 $ & $B$ & (31)\\
4780.486 & $ 19.607 \pm 0.069 $ & $B$ & (30)\\
4780.907 & $ 20.742 \pm 0.220 $ & $B$ & (31)\\
4782.418 & $ 20.027 \pm 0.083 $ & $B$ & (30)\\
4783.496 & $>20.753$ & $B$ & (30)\\
4785.529 & $ 20.306 \pm 0.107 $ & $B$ & (30)\\
4786.490 & $>20.029$ & $B$ & (30)\\
4787.404 & $ 20.507 \pm 0.072 $ & $B$ & (30)\\
4788.428 & $ 20.933 \pm 0.074 $ & $B$ & (30)\\
4789.503 & $ 20.936 \pm 0.220 $ & $B$ & (30)\\
4794.344 & $ 20.592 \pm 0.082 $ & $B$ & (30)\\
4795.478 & $>20.830$ & $B$ & (30)\\
4796.581 & $>20.911$ & $B$ & (30)\\
4799.384 & $ 21.567 \pm 0.114 $ & $B$ & (30)\\
4800.470 & $>21.775$ & $B$ & (30)\\
4802.396 & $>21.644$ & $B$ & (30)\\
4805.339 & $>21.701$ & $B$ & (30)\\
4805.395 & $>21.562$ & $B$ & (30)\\
\\
4778.404 & $ 18.917 \pm 0.038 $ & $V$ & (30)\\
4779.476 & $ 19.266 \pm 0.060 $ & $V$ & (30)\\
4779.896 & $ 19.164 \pm 0.057 $ & $V$ & (31)\\
4780.489 & $ 19.219 \pm 0.044 $ & $V$ & (30)\\
4780.912 & $>19.462$ & $V$ & (31)\\
4781.472 & $ 19.693 \pm 0.095 $ & $V$ & (30)\\
4782.422 & $ 19.770 \pm 0.110 $ & $V$ & (30)\\
4783.499 & $>19.457$ & $V$ & (30)\\
4785.532 & $ 20.260 \pm 0.092 $ & $V$ & (30)\\
4786.494 & $>19.410$ & $V$ & (30)\\
4787.407 & $ 20.177 \pm 0.059 $ & $V$ & (30)\\
4788.432 & $ 20.407 \pm 0.141 $ & $V$ & (30)\\
4789.508 & $ 20.515 \pm 0.211 $ & $V$ & (30)\\
4794.348 & $ 21.413 \pm 0.310 $ & $V$ & (30)\\
4795.482 & $ 20.900 \pm 0.186 $ & $V$ & (30)\\
4796.585 & $>20.561$ & $V$ & (30)\\
4799.387 & $ 20.737 \pm 0.147 $ & $V$ & (30)\\
4800.474 & $>21.387$ & $V$ & (30)\\
4800.498 & $>20.384$ & $V$ & (30)\\
4802.400 & $ 21.172 \pm 0.185 $ & $V$ & (30)\\
4805.399 & $>21.106$ & $V$ & (30)\\
\\
4744.340 & $>19.9         $  & $R$ &   (1)\\
4775.218 & $ 16.5 \pm0.1  $ & $R$ &   (1)\\
4776.217 & $ 17.4 \pm0.1  $ & $R$ &   (1)\\
4779.260 & $ 18.3 \pm0.15 $ & $R$ &   (1)\\
4780.234 & $ 18.4 \pm0.2  $ & $R$ &   (1)\\
4800.310 & $>19.5         $  & $R$ &   (1)\\
\\
4778.389 & $ 17.920 \pm 0.011 $ & $r'$ & (30)\\
4779.461 & $ 18.289 \pm 0.016 $ & $r'$ & (30)\\
4779.901 & $ 18.407 \pm 0.059 $ & $r'$ & (31)\\
4780.475 & $ 18.502 \pm 0.020 $ & $r'$ & (30)\\
4780.917 & $ 18.528 \pm 0.116 $ & $r'$ & (31)\\
4782.407 & $ 18.890 \pm 0.025 $ & $r'$ & (30)\\
4783.485 & $ 18.935 \pm 0.160 $ & $r'$ & (30)\\
4785.518 & $ 19.310 \pm 0.025 $ & $r'$ & (30)\\
4786.479 & $ 19.436 \pm 0.160 $ & $r'$ & (30)\\
4787.393 & $ 19.683 \pm 0.031 $ & $r'$ & (30)\\
4788.418 & $ 19.881 \pm 0.027 $ & $r'$ & (30)\\
4789.492 & $ 20.063 \pm 0.035 $ & $r'$ & (30)\\
4790.412 & $ 19.899 \pm 0.032 $ & $r'$ & (30)\\
4794.333 & $ 20.310 \pm 0.135 $ & $r'$ & (30)\\
4795.467 & $ 20.130 \pm 0.036 $ & $r'$ & (30)\\
4796.570 & $>20.119$ & $r'$ & (30)\\
4799.373 & $ 20.689 \pm 0.148 $ & $r'$ & (30)\\
4800.403 & $ 20.200 \pm 0.225 $ & $r'$ & (30)\\
4800.459 & $ 20.152 \pm 0.138 $ & $r'$ & (30)\\
4802.385 & $ 21.052 \pm 0.140 $ & $r'$ & (30)\\
4803.398 & $>19.727$ & $r'$ & (30)\\
4805.328 & $ 21.155 \pm 0.111 $ & $r'$ & (30)\\
4805.384 & $ 21.470 \pm 0.295 $ & $r'$ & (30)\\
\\
4778.393 & $ 18.367 \pm 0.034 $ & $i'$ & (30)\\
4779.465 & $ 18.884 \pm 0.020 $ & $i'$ & (30)\\
4779.906 & $ 18.909 \pm 0.082 $ & $i'$ & (31)\\
4780.478 & $ 19.020 \pm 0.043 $ & $i'$ & (30)\\
4780.922 & $ 19.552 \pm 0.179 $ & $i'$ & (31)\\
4782.411 & $ 19.591 \pm 0.028 $ & $i'$ & (30)\\
4783.489 & $>19.309$ & $i'$ & (30)\\
4785.521 & $ 20.178 \pm 0.048 $ & $i'$ & (30)\\
4786.483 & $>20.126$ & $i'$ & (30)\\
4787.396 & $ 20.234 \pm 0.099 $ & $i'$ & (30)\\
4788.421 & $ 20.511 \pm 0.044 $ & $i'$ & (30)\\
4789.497 & $ 21.077 \pm 0.184 $ & $i'$ & (30)\\
4790.415 & $ 20.583 \pm 0.142 $ & $i'$ & (30)\\
4794.337 & $ 20.649 \pm 0.186 $ & $i'$ & (30)\\
4795.471 & $ 20.957 \pm 0.059 $ & $i'$ & (30)\\
4796.574 & $>19.433$ & $i'$ & (30)\\
4799.376 & $ 21.295 \pm 0.233 $ & $i'$ & (30)\\
4800.464 & $ 20.993 \pm 0.242 $ & $i'$ & (30)\\
4802.389 & $ 21.300 \pm 0.066 $ & $i'$ & (30)\\
4803.401 & $>20.175$ & $i'$ & (30)\\
4805.331 & $>21.000$ & $i'$ & (30)\\
4805.388 & $>21.564$ & $i'$ & (30)\\
\\
4778.396 & $ 17.866 \pm 0.018 $ & $z'$ & (30)\\
4779.468 & $ 18.131 \pm 0.076 $ & $z'$ & (30)\\
4780.482 & $ 18.500 \pm 0.040 $ & $z'$ & (30)\\
4781.465 & $>18.716$ & $z'$ & (30)\\
4782.414 & $ 19.112 \pm 0.107 $ & $z'$ & (30)\\
4783.492 & $ 19.264 \pm 0.107 $ & $z'$ & (30)\\
4785.525 & $ 19.649 \pm 0.066 $ & $z'$ & (30)\\
4786.486 & $>19.875$ & $z'$ & (30)\\
4787.400 & $>20.602$ & $z'$ & (30)\\
4788.425 & $ 20.158 \pm 0.083 $ & $z'$ & (30)\\
4789.499 & $ 20.621 \pm 0.271 $ & $z'$ & (30)\\
4790.419 & $>20.364$ & $z'$ & (30)\\
4794.340 & $>20.821$ & $z'$ & (30)\\
4795.474 & $>21.175$ & $z'$ & (30)\\
4796.578 & $>17.999$ & $z'$ & (30)\\

\cutinhead{M31N 2008-12b}

4842.404 & $ 17.671 \pm 0.041 $ & $B$ & (30)\\
4846.474 & $ 17.826 \pm 0.015 $ & $B$ & (30)\\
4851.347 & $ 17.874 \pm 0.043 $ & $B$ & (30)\\
4856.441 & $ 18.765 \pm 0.081 $ & $B$ & (30)\\
4859.377 & $ 19.436 \pm 0.149 $ & $B$ & (30)\\
\\
4842.408 & $ 17.533 \pm 0.035 $ & $V$ & (30)\\
4846.478 & $ 17.705 \pm 0.045 $ & $V$ & (30)\\
4851.351 & $ 17.941 \pm 0.046 $ & $V$ & (30)\\
4856.444 & $ 18.822 \pm 0.112 $ & $V$ & (30)\\
4859.381 & $ 19.468 \pm 0.129 $ & $V$ & (30)\\
\\
4829.183 & $>19.7         $  & $R$ &   (1)\\
4840.253 & $ 17.0 \pm0.1  $ & $R$ &   (4)\\
4848.217 & $ 17.4 \pm0.15 $ & $R$ &  (11)\\
4851.394 & $ 17.3 \pm0.15 $ & $R$ &   (1)\\
\\ 
4842.394 & $ 17.606 \pm 0.030 $ & $r'$ & (30)\\
4846.464 & $ 17.669 \pm 0.036 $ & $r'$ & (30)\\
4851.337 & $ 17.893 \pm 0.033 $ & $r'$ & (30)\\
4856.430 & $ 18.843 \pm 0.058 $ & $r'$ & (30)\\
4859.366 & $ 19.262 \pm 0.105 $ & $r'$ & (30)\\
\\
4842.397 & $ 17.499 \pm 0.049 $ & $i'$ & (30)\\
4846.467 & $ 17.693 \pm 0.032 $ & $i'$ & (30)\\
4851.340 & $ 17.896 \pm 0.044 $ & $i'$ & (30)\\
4856.433 & $ 18.453 \pm 0.080 $ & $i'$ & (30)\\
4859.370 & $ 18.993 \pm 0.089 $ & $i'$ & (30)\\
\\
4842.400 & $ 16.750 \pm 0.022 $ & $z'$ & (30)\\
4846.471 & $ 16.921 \pm 0.043 $ & $z'$ & (30)\\
4851.344 & $ 17.044 \pm 0.049 $ & $z'$ & (30)\\
4856.437 & $ 17.371 \pm 0.070 $ & $z'$ & (30)\\
4859.374 & $ 17.533 \pm 0.084 $ & $z'$ & (30)\\

\cutinhead{M31N 2009-01a}

4872.241 & $>19.3         $  & $R$ &  (11)\\

\cutinhead{M31N 2009-02a}

4872.254 & $ 16.91\pm0.1  $ & $R$ &  (11)\\

\cutinhead{M31N 2009-08a}

5053.724 & $ 17.944 \pm 0.062 $ & $B$ & (30)\\
5057.516 & $ 18.114 \pm 0.021 $ & $B$ & (30)\\
5058.498 & $ 18.312 \pm 0.023 $ & $B$ & (30)\\
5059.719 & $ 18.413 \pm 0.027 $ & $B$ & (30)\\
5060.533 & $ 18.167 \pm 0.065 $ & $B$ & (30)\\
5061.484 & $ 18.076 \pm 0.022 $ & $B$ & (30)\\
5061.562 & $ 18.086 \pm 0.047 $ & $B$ & (30)\\
5062.507 & $ 18.121 \pm 0.070 $ & $B$ & (30)\\
5063.485 & $ 18.404 \pm 0.030 $ & $B$ & (30)\\
5064.630 & $ 17.917 \pm 0.020 $ & $B$ & (30)\\
5065.739 & $ 17.791 \pm 0.019 $ & $B$ & (30)\\
5066.525 & $ 17.800 \pm 0.065 $ & $B$ & (30)\\
5067.474 & $ 17.980 \pm 0.072 $ & $B$ & (30)\\
5069.678 & $ 18.655 \pm 0.031 $ & $B$ & (30)\\
5070.586 & $ 18.978 \pm 0.029 $ & $B$ & (30)\\
5071.746 & $>18.911$ & $B$ & (30)\\
5072.526 & $ 18.655 \pm 0.059 $ & $B$ & (30)\\
5073.467 & $ 18.620 \pm 0.092 $ & $B$ & (30)\\
5075.501 & $ 18.263 \pm 0.098 $ & $B$ & (30)\\
5076.459 & $ 18.030 \pm 0.068 $ & $B$ & (30)\\
5077.718 & $ 18.077 \pm 0.021 $ & $B$ & (30)\\
5077.487 & $ 18.009 \pm 0.021 $ & $B$ & (30)\\
5078.572 & $ 17.903 \pm 0.021 $ & $B$ & (30)\\
5079.469 & $ 17.710 \pm 0.021 $ & $B$ & (30)\\
5114.306 & $ 19.91\pm0.10 $ & $B$ &  (17)\\
\\
5053.729 & $ 17.688 \pm 0.060 $ & $V$ & (30)\\
5057.520 & $ 18.212 \pm 0.022 $ & $V$ & (30)\\
5058.501 & $ 18.453 \pm 0.025 $ & $V$ & (30)\\
5059.723 & $ 18.520 \pm 0.092 $ & $V$ & (30)\\
5060.536 & $ 18.463 \pm 0.071 $ & $V$ & (30)\\
5061.488 & $ 18.169 \pm 0.063 $ & $V$ & (30)\\
5061.566 & $ 18.231 \pm 0.050 $ & $V$ & (30)\\
5062.511 & $ 18.321 \pm 0.079 $ & $V$ & (30)\\
5063.489 & $ 18.686 \pm 0.037 $ & $V$ & (30)\\
5064.633 & $ 18.167 \pm 0.021 $ & $V$ & (30)\\
5065.742 & $ 17.858 \pm 0.019 $ & $V$ & (30)\\
5066.528 & $ 17.885 \pm 0.052 $ & $V$ & (30)\\
5067.478 & $ 18.320 \pm 0.080 $ & $V$ & (30)\\
5069.681 & $ 18.949 \pm 0.046 $ & $V$ & (30)\\
5070.590 & $ 19.263 \pm 0.041 $ & $V$ & (30)\\
5071.749 & $>19.142$ & $V$ & (30)\\
5072.529 & $ 18.940 \pm 0.033 $ & $V$ & (30)\\
5073.470 & $ 18.745 \pm 0.045 $ & $V$ & (30)\\
5074.483 & $ 18.560 \pm 0.047 $ & $V$ & (30)\\
5075.504 & $ 18.338 \pm 0.124 $ & $V$ & (30)\\
5076.463 & $ 18.195 \pm 0.026 $ & $V$ & (30)\\
5077.721 & $ 18.233 \pm 0.023 $ & $V$ & (30)\\
5077.490 & $ 18.134 \pm 0.022 $ & $V$ & (30)\\
5078.575 & $ 18.115 \pm 0.021 $ & $V$ & (30)\\
5079.472 & $ 17.852 \pm 0.020 $ & $V$ & (30)\\
5092.525 & $ 18.39\pm0.04 $ & $V$ &  (17)\\
5095.981 & $ 18.98\pm0.1  $ & $V$ &  (13)\\
5114.308 & $ 20.1 \pm0.15 $ & $V$ &  (17)\\
\\
4985.974 & $>20.2         $  & $R$ &   (9)\\
4988.527 & $>19.9         $  & $R$ &   (1)\\
5055.923 & $ 18.7 \pm0.15 $ & $R$ &   (8)\\
5063.605 & $ 18.8 \pm0.2  $ & $R$ &  (22)\\
5073.441 & $ 18.3 \pm0.1  $ & $R$ &   (1)\\
5074.366 & $ 18.3 \pm0.1  $ & $R$ &   (1)\\
5075.384 & $ 18.0 \pm0.15 $ & $R$ &   (1)\\
5076.478 & $ 17.8 \pm0.1  $ & $R$ &   (1)\\
5080.413 & $ 17.8 \pm0.1  $ & $R$ &   (1)\\
5081.393 & $ 18.4 \pm0.15 $ & $R$ &   (1)\\
5083.357 & $ 18.5 \pm0.15 $ & $R$ &   (1)\\
5084.267 & $ 18.2 \pm0.2  $ & $R$ &   (1)\\
5094.451 & $ 18.3 \pm0.15 $ & $R$ &   (3)\\
5095.968 & $ 18.52\pm0.07 $ & $R$ &  (13)\\
5114.304 & $ 19.2 \pm0.15 $ & $R$ &  (17)\\
5114.486 & $ 19.2 \pm0.15 $ & $R$ &   (3)\\
5124.491 & $ 19.5 \pm0.2  $ & $R$ &   (3)\\
5135.632 & $ 19.3 \pm0.2  $ & $R$ &  (12)\\
5140.546 & $ 19.7 \pm0.3  $ & $R$ &   (1)\\
5141.359 & $ 19.9 \pm0.3  $ & $R$ &   (1)\\
5148.536 & $ 19.7 \pm0.25 $ & $R$ &   (1)\\
5162.281 & $ 19.3 \pm0.25 $ & $R$ &   (5)\\
5168.173 & $ 19.4 \pm0.3  $ & $R$ &   (7)\\
5173.192 & $ 19.6 \pm0.3  $ & $R$ &   (1)\\
5181.226 & $ 19.7 \pm0.3  $ & $R$ &   (1)\\
5184.717 & $ 19.9 \pm0.15 $ & $R$ &  (19)\\
5199.299 & $ 20.6 \pm0.2  $ & $R$ &  (12)\\
5227.345 & $ 20.9 \pm0.25 $ & $R$ &  (12)\\
5228.383 & $ 20.9 \pm0.25 $ & $R$ &  (12)\\
\\
5050.721 & $ 18.568 \pm 0.120 $ & $r'$ & (30)\\
5052.551 & $ 18.194 \pm 0.057 $ & $r'$ & (30)\\
5052.556 & $ 18.248 \pm 0.066 $ & $r'$ & (30)\\
5052.570 & $ 18.170 \pm 0.047 $ & $r'$ & (30)\\
5052.585 & $ 18.119 \pm 0.042 $ & $r'$ & (30)\\
5052.591 & $ 18.198 \pm 0.044 $ & $r'$ & (30)\\
5052.605 & $ 18.147 \pm 0.042 $ & $r'$ & (30)\\
5052.620 & $ 18.202 \pm 0.019 $ & $r'$ & (30)\\
5052.626 & $ 18.186 \pm 0.020 $ & $r'$ & (30)\\
5052.640 & $ 18.171 \pm 0.041 $ & $r'$ & (30)\\
5052.652 & $ 18.184 \pm 0.046 $ & $r'$ & (30)\\
5052.660 & $ 18.098 \pm 0.042 $ & $r'$ & (30)\\
5052.682 & $ 18.147 \pm 0.043 $ & $r'$ & (30)\\
5052.711 & $ 18.152 \pm 0.061 $ & $r'$ & (30)\\
5052.721 & $ 18.154 \pm 0.045 $ & $r'$ & (30)\\
5052.728 & $ 18.152 \pm 0.044 $ & $r'$ & (30)\\
5053.716 & $ 18.069 \pm 0.054 $ & $r'$ & (30)\\
5053.505 & $ 18.160 \pm 0.095 $ & $r'$ & (30)\\
5053.509 & $ 18.109 \pm 0.084 $ & $r'$ & (30)\\
5053.514 & $ 18.037 \pm 0.026 $ & $r'$ & (30)\\
5053.528 & $ 18.143 \pm 0.068 $ & $r'$ & (30)\\
5053.532 & $ 18.064 \pm 0.058 $ & $r'$ & (30)\\
5053.537 & $ 18.081 \pm 0.057 $ & $r'$ & (30)\\
5053.541 & $ 18.101 \pm 0.054 $ & $r'$ & (30)\\
5054.478 & $ 18.185 \pm 0.028 $ & $r'$ & (30)\\
5056.718 & $ 18.337 \pm 0.074 $ & $r'$ & (30)\\
5057.510 & $ 18.357 \pm 0.049 $ & $r'$ & (30)\\
5058.491 & $ 18.532 \pm 0.054 $ & $r'$ & (30)\\
5059.713 & $ 18.621 \pm 0.099 $ & $r'$ & (30)\\
5060.526 & $ 18.472 \pm 0.077 $ & $r'$ & (30)\\
5061.478 & $ 18.336 \pm 0.073 $ & $r'$ & (30)\\
5061.555 & $ 18.383 \pm 0.045 $ & $r'$ & (30)\\
5062.501 & $ 18.278 \pm 0.079 $ & $r'$ & (30)\\
5063.479 & $ 18.661 \pm 0.105 $ & $r'$ & (30)\\
5064.623 & $ 18.307 \pm 0.047 $ & $r'$ & (30)\\
5065.733 & $ 18.073 \pm 0.034 $ & $r'$ & (30)\\
5066.518 & $ 18.055 \pm 0.020 $ & $r'$ & (30)\\
5067.468 & $ 18.349 \pm 0.070 $ & $r'$ & (30)\\
5069.672 & $ 18.797 \pm 0.041 $ & $r'$ & (30)\\
5070.580 & $ 19.090 \pm 0.087 $ & $r'$ & (30)\\
5072.519 & $ 18.984 \pm 0.076 $ & $r'$ & (30)\\
5073.460 & $ 18.620 \pm 0.065 $ & $r'$ & (30)\\
5074.733 & $ 18.978 \pm 0.017 $ & $r'$ & (30)\\
5075.495 & $ 18.500 \pm 0.044 $ & $r'$ & (30)\\
5077.712 & $ 18.407 \pm 0.049 $ & $r'$ & (30)\\
5078.566 & $ 18.251 \pm 0.018 $ & $r'$ & (30)\\
5078.515 & $ 18.235 \pm 0.019 $ & $r'$ & (30)\\
5079.463 & $ 18.012 \pm 0.016 $ & $r'$ & (30)\\
5210.587 & $ 20.7 \pm0.2  $ & $r'$ &  (16)\\
5211.600 & $ 20.8 \pm0.25 $ & $r'$ &  (16)\\

\cutinhead{M31N 2009-08b}

5061.546 & $ 18.559 \pm 0.025 $ & $B$ & (30)\\
5063.625 & $ 18.782 \pm 0.029 $ & $B$ & (30)\\
5064.662 & $ 18.960 \pm 0.028 $ & $B$ & (30)\\
5066.594 & $ 19.165 \pm 0.029 $ & $B$ & (30)\\
5067.459 & $ 19.298 \pm 0.035 $ & $B$ & (30)\\
5069.600 & $ 19.592 \pm 0.034 $ & $B$ & (30)\\
5070.567 & $ 19.847 \pm 0.036 $ & $B$ & (30)\\
5071.522 & $ 19.695 \pm 0.040 $ & $B$ & (30)\\
5072.496 & $ 19.899 \pm 0.042 $ & $B$ & (30)\\
5073.554 & $ 19.944 \pm 0.054 $ & $B$ & (30)\\
5074.717 & $ 19.950 \pm 0.042 $ & $B$ & (30)\\
5075.517 & $ 19.971 \pm 0.075 $ & $B$ & (30)\\
5076.485 & $ 19.813 \pm 0.090 $ & $B$ & (30)\\
5077.471 & $ 20.165 \pm 0.081 $ & $B$ & (30)\\
5078.492 & $ 20.380 \pm 0.103 $ & $B$ & (30)\\
5079.523 & $ 20.383 \pm 0.108 $ & $B$ & (30)\\
\\
5061.549 & $ 18.424 \pm 0.019 $ & $V$ & (30)\\
5063.628 & $ 18.665 \pm 0.022 $ & $V$ & (30)\\
5064.665 & $ 18.776 \pm 0.022 $ & $V$ & (30)\\
5066.596 & $ 18.991 \pm 0.024 $ & $V$ & (30)\\
5067.462 & $ 19.056 \pm 0.026 $ & $V$ & (30)\\
5069.603 & $ 19.420 \pm 0.030 $ & $V$ & (30)\\
5070.570 & $ 19.559 \pm 0.029 $ & $V$ & (30)\\
5071.523 & $ 19.465 \pm 0.032 $ & $V$ & (30)\\
5072.499 & $ 19.709 \pm 0.033 $ & $V$ & (30)\\
5073.556 & $ 19.652 \pm 0.066 $ & $V$ & (30)\\
5074.719 & $ 19.792 \pm 0.038 $ & $V$ & (30)\\
5075.520 & $ 20.053 \pm 0.078 $ & $V$ & (30)\\
5076.487 & $ 20.117 \pm 0.162 $ & $V$ & (30)\\
5077.474 & $ 19.894 \pm 0.053 $ & $V$ & (30)\\
5078.494 & $ 20.126 \pm 0.075 $ & $V$ & (30)\\
5079.526 & $ 20.298 \pm 0.120 $ & $V$ & (30)\\
\\
5061.541 & $ 17.984 \pm 0.013 $ & $r'$ & (30)\\
5063.620 & $ 18.159 \pm 0.016 $ & $r'$ & (30)\\
5064.657 & $ 18.269 \pm 0.015 $ & $r'$ & (30)\\
5066.588 & $ 18.409 \pm 0.016 $ & $r'$ & (30)\\
5067.454 & $ 18.499 \pm 0.019 $ & $r'$ & (30)\\
5069.595 & $ 18.723 \pm 0.019 $ & $r'$ & (30)\\
5070.562 & $ 18.847 \pm 0.018 $ & $r'$ & (30)\\
5071.516 & $ 18.869 \pm 0.019 $ & $r'$ & (30)\\
5072.491 & $ 18.970 \pm 0.020 $ & $r'$ & (30)\\
5073.548 & $ 18.914 \pm 0.123 $ & $r'$ & (30)\\
5074.711 & $ 19.081 \pm 0.021 $ & $r'$ & (30)\\
5075.512 & $ 19.089 \pm 0.034 $ & $r'$ & (30)\\
5076.479 & $ 19.241 \pm 0.043 $ & $r'$ & (30)\\
5077.466 & $ 19.308 \pm 0.033 $ & $r'$ & (30)\\
5078.487 & $ 19.338 \pm 0.034 $ & $r'$ & (30)\\
5079.518 & $ 19.366 \pm 0.039 $ & $r'$ & (30)\\
\\
5061.544 & $ 18.028 \pm 0.015 $ & $i'$ & (30)\\
5063.622 & $ 18.262 \pm 0.018 $ & $i'$ & (30)\\
5064.659 & $ 18.359 \pm 0.017 $ & $i'$ & (30)\\
5066.591 & $ 18.555 \pm 0.019 $ & $i'$ & (30)\\
5067.456 & $ 18.641 \pm 0.025 $ & $i'$ & (30)\\
5069.598 & $ 18.896 \pm 0.022 $ & $i'$ & (30)\\
5070.564 & $ 18.945 \pm 0.023 $ & $i'$ & (30)\\
5071.517 & $ 18.798 \pm 0.025 $ & $i'$ & (30)\\
5072.494 & $ 19.053 \pm 0.024 $ & $i'$ & (30)\\
5073.551 & $ 19.486 \pm 0.091 $ & $i'$ & (30)\\
5074.714 & $ 19.083 \pm 0.027 $ & $i'$ & (30)\\
5075.514 & $ 19.152 \pm 0.048 $ & $i'$ & (30)\\
5076.482 & $ 19.230 \pm 0.065 $ & $i'$ & (30)\\
5077.469 & $ 19.314 \pm 0.034 $ & $i'$ & (30)\\
5078.489 & $ 19.317 \pm 0.039 $ & $i'$ & (30)\\
5079.521 & $ 19.344 \pm 0.052 $ & $i'$ & (30)\\

\cutinhead{M31N 2009-08d}

5061.639 & $ 17.421 \pm 0.139 $ & $B$ & (30)\\
5063.650 & $ 17.819 \pm 0.201 $ & $B$ & (30)\\
5064.648 & $ 17.792 \pm 0.144 $ & $B$ & (30)\\
5065.724 & $ 17.879 \pm 0.188 $ & $B$ & (30)\\
5066.555 & $ 18.368 \pm 0.434 $ & $B$ & (30)\\
5067.499 & $ 18.391 \pm 0.350 $ & $B$ & (30)\\
5069.658 & $ 18.218 \pm 0.344 $ & $B$ & (30)\\
5070.656 & $ 18.232 \pm 0.259 $ & $B$ & (30)\\
5071.730 & $>15.710$ & $B$ & (30)\\
5072.510 & $ 18.617 \pm 0.414 $ & $B$ & (30)\\
5073.509 & $ 18.553 \pm 0.317 $ & $B$ & (30)\\
5075.531 & $ 18.820 \pm 0.443 $ & $B$ & (30)\\
5076.514 & $ 18.659 \pm 0.368 $ & $B$ & (30)\\
5077.512 & $ 18.637 \pm 0.369 $ & $B$ & (30)\\
5078.478 & $ 18.731 \pm 0.428 $ & $B$ & (30)\\
5079.494 & $ 18.580 \pm 0.312 $ & $B$ & (30)\\
\\
5061.641 & $ 17.808 \pm 0.359 $ & $V$ & (30)\\
5063.653 & $ 17.888 \pm 0.327 $ & $V$ & (30)\\
5064.651 & $ 17.982 \pm 0.284 $ & $V$ & (30)\\
5065.727 & $ 18.182 \pm 0.381 $ & $V$ & (30)\\
5066.558 & $ 18.390 \pm 0.650 $ & $V$ & (30)\\
5067.502 & $>17.915$ & $V$ & (30)\\
5069.661 & $>17.288$ & $V$ & (30)\\
5070.659 & $>17.900$ & $V$ & (30)\\
5072.513 & $ 18.453 \pm 0.522 $ & $V$ & (30)\\
5073.512 & $>18.705$ & $V$ & (30)\\
5079.497 & $ 19.155 \pm 0.804 $ & $V$ & (30)\\
5095.981 & $ 19.9 \pm0.2  $ & $V$ &  (13)\\
\\
4985.974 & $>20.2         $  & $R$ &   (9)\\
4988.527 & $>19.9         $  & $R$ &   (1)\\
5055.923 & $ 17.2 \pm0.1  $ & $R$ &   (8)\\
5063.605 & $ 18.2 \pm0.35 $ & $R$ &  (22)\\
5076.478 & $>20.0         $  & $R$ &   (1)\\
5095.968 & $ 19.46\pm0.15 $ & $R$ &  (13)\\
\\
5061.633 & $ 17.685 \pm 0.323 $ & $r'$ & (30)\\
5063.644 & $ 18.122 \pm 0.538 $ & $r'$ & (30)\\
5064.643 & $ 17.888 \pm 0.346 $ & $r'$ & (30)\\
5065.719 & $ 18.070 \pm 0.430 $ & $r'$ & (30)\\
5066.550 & $ 18.066 \pm 0.584 $ & $r'$ & (30)\\
5067.494 & $>17.835$ & $r'$ & (30)\\
5069.653 & $>17.945$ & $r'$ & (30)\\
5070.651 & $>17.724$ & $r'$ & (30)\\
5072.505 & $ 18.329 \pm 0.493 $ & $r'$ & (30)\\
5073.504 & $ 18.646 \pm 0.717 $ & $r'$ & (30)\\
5075.526 & $>18.178$ & $r'$ & (30)\\
5077.507 & $>18.368$ & $r'$ & (30)\\
5078.472 & $>18.244$ & $r'$ & (30)\\
5079.489 & $ 18.848 \pm 0.755 $ & $r'$ & (30)\\

\cutinhead{M31N 2009-08e}
 
5114.306 & $ 19.56\pm0.09 $ & $B$ &  (17)\\
5129.835 & $ 20.12\pm0.1  $ & $B$ &  (15)\\
\\
5092.543 & $ 19.09\pm0.08 $ & $V$ &  (17)\\
5095.981 & $ 19.32\pm0.06 $ & $V$ &  (13)\\
5114.308 & $ 19.7 \pm0.12 $ & $V$ &  (17)\\
5129.824 & $ 20.15\pm0.1  $ & $V$ &  (15)\\
\\
5055.923 & $>20.3         $  & $R$ &   (8)\\
5063.605 & $>20.1         $  & $R$ &  (22)\\
5073.441 & $ 17.9 \pm0.1  $ & $R$ &   (1)\\
5074.366 & $ 18.0 \pm0.1  $ & $R$ &   (1)\\
5075.384 & $ 18.1 \pm0.15 $ & $R$ &   (1)\\
5076.478 & $ 17.9 \pm0.1  $ & $R$ &   (1)\\
5080.413 & $ 18.1 \pm0.1  $ & $R$ &   (1)\\
5081.393 & $ 18.7 \pm0.15 $ & $R$ &   (1)\\
5083.357 & $ 18.4 \pm0.2  $ & $R$ &   (1)\\
5084.267 & $ 18.2 \pm0.2  $ & $R$ &   (1)\\
5094.451 & $ 18.2 \pm0.15 $ & $R$ &   (3)\\
5095.968 & $ 18.39\pm0.04 $ & $R$ &  (13)\\
5114.304 & $ 18.9 \pm0.11 $ & $R$ &  (17)\\
5114.486 & $ 18.6 \pm0.15 $ & $R$ &   (3)\\
5124.491 & $ 18.4 \pm0.15 $ & $R$ &   (3)\\
5129.829 & $ 19.11\pm0.08 $ & $R$ &  (15)\\
5135.632 & $ 19.1 \pm0.15 $ & $R$ &  (12)\\
5140.546 & $ 18.6 \pm0.25 $ & $R$ &   (1)\\
5141.359 & $ 19.0 \pm0.25 $ & $R$ &   (1)\\
5148.536 & $ 18.9 \pm0.2  $ & $R$ &   (1)\\
5157.316 & $ 18.8 \pm0.3  $ & $R$ &  (24)\\
5162.281 & $ 19.3 \pm0.25 $ & $R$ &   (5)\\
5173.177 & $ 19.7 \pm0.3  $ & $R$ &   (1)\\
5181.226 & $>19.8         $  & $R$ &   (1)\\
5184.717 & $ 19.9 \pm0.2  $ & $R$ &  (19)\\
5199.299 & $ 20.0 \pm0.15 $ & $R$ &  (12)\\
5227.345 & $ 20.7 \pm0.25 $ & $R$ &  (12)\\
5228.383 & $ 20.6 \pm0.25 $ & $R$ &  (12)\\

\cutinhead{M31N 2009-09a}

5092.506 & $ 19.02\pm0.05 $ & $V$ & (17)\\
\\
5080.413 & $ 17.5 \pm0.15 $ & $R$ &   (1)\\
5081.393 & $ 17.9 \pm0.1  $ & $R$ &   (1)\\
5083.357 & $ 18.1 \pm0.15 $ & $R$ &   (1)\\
5092.506 & $ 18.21\pm0.03 $ & $R$ &  (17)\\
5095.993 & $ 18.18\pm0.04 $ & $R$ &  (13)\\
5148.578 & $ 18.0 \pm0.3  $ & $R$ &   (1)\\
5162.513 & $ 19.0 \pm0.25 $ & $R$ &   (5)\\
5173.203 & $ 19.3 \pm0.2  $ & $R$ &   (1)\\
5226.321 & $ 19.7 \pm0.15 $ & $R$ &  (12)\\
5227.306 & $ 19.7 \pm0.15 $ & $R$ &  (12)\\
5228.284 & $ 19.8 \pm0.15 $ & $R$ &  (12)\\
 
\cutinhead{M31N 2009-10a}
 
5114.341 & $ 18.59\pm0.10 $ & $B$ &  (17)\\
5117.718 & $ 18.344 \pm 0.114 $ & $B$ & (30)\\
5117.501 & $ 18.403 \pm 0.029 $ & $B$ & (30)\\
5118.647 & $ 18.705 \pm 0.034 $ & $B$ & (30)\\
5119.495 & $ 18.919 \pm 0.030 $ & $B$ & (30)\\
5120.389 & $ 19.119 \pm 0.033 $ & $B$ & (30)\\
5127.496 & $ 19.888 \pm 0.048 $ & $B$ & (30)\\
5128.375 & $ 19.867 \pm 0.041 $ & $B$ & (30)\\
5129.515 & $ 20.258 \pm 0.057 $ & $B$ & (30)\\
5130.468 & $ 20.522 \pm 0.054 $ & $B$ & (30)\\
5131.408 & $ 20.997 \pm 0.137 $ & $B$ & (30)\\
5132.359 & $>21.495$ & $B$ & (30)\\
\\
5114.361 & $ 18.33\pm0.09 $ & $V$ &  (17)\\
5117.721 & $ 18.574 \pm 0.094 $ & $V$ & (30)\\
5117.505 & $ 18.367 \pm 0.020 $ & $V$ & (30)\\
5118.650 & $ 18.710 \pm 0.024 $ & $V$ & (30)\\
5119.498 & $ 18.950 \pm 0.023 $ & $V$ & (30)\\
5120.392 & $ 19.113 \pm 0.025 $ & $V$ & (30)\\
5127.499 & $ 19.776 \pm 0.032 $ & $V$ & (30)\\
5128.378 & $ 19.922 \pm 0.035 $ & $V$ & (30)\\
5129.518 & $ 20.252 \pm 0.041 $ & $V$ & (30)\\
5130.472 & $ 20.743 \pm 0.169 $ & $V$ & (30)\\
5131.412 & $>19.771$ & $V$ & (30)\\
\\
5114.340 & $ 17.81\pm0.07 $ & $R$ &  (17)\\
5162.549 & $>19.9         $  & $R$ &   (5)\\

\cutinhead{M31N 2009-10b}
 
5117.487 & $ 16.260 \pm 0.049 $ & $B$ & (30)\\
5117.711 & $ 15.759 \pm 0.042 $ & $B$ & (30)\\
5118.603 & $ 15.521 \pm 0.042 $ & $B$ & (30)\\
5119.488 & $ 15.316 \pm 0.042 $ & $B$ & (30)\\
5120.370 & $ 15.242 \pm 0.042 $ & $B$ & (30)\\
5124.457 & $ 16.98 \pm0.04 $ & $B$ &  (18)\\
5127.489 & $ 18.338 \pm 0.066 $ & $B$ & (30)\\
5128.368 & $ 18.481 \pm 0.044 $ & $B$ & (30)\\
5129.500 & $ 18.659 \pm 0.077 $ & $B$ & (30)\\
5130.475 & $ 18.827 \pm 0.044 $ & $B$ & (30)\\
5131.393 & $ 18.892 \pm 0.110 $ & $B$ & (30)\\
5132.352 & $ 19.223 \pm 0.053 $ & $B$ & (30)\\
5134.341 & $ 19.108 \pm 0.053 $ & $B$ & (30)\\
5135.354 & $ 19.444 \pm 0.049 $ & $B$ & (30)\\
5136.461 & $ 19.286 \pm 0.065 $ & $B$ & (30)\\
5137.400 & $ 19.649 \pm 0.065 $ & $B$ & (30)\\
5138.449 & $>19.033$ & $B$ & (30)\\
5139.652 & $ 19.411 \pm 0.157 $ & $B$ & (30)\\
5140.529 & $ 20.002 \pm 0.056 $ & $B$ & (30)\\
5142.515 & $ 19.853 \pm 0.172 $ & $B$ & (30)\\
5143.620 & $>19.917$ & $B$ & (30)\\
5146.507 & $>19.702$ & $B$ & (30)\\
\\
5117.714 & $ 15.662 \pm 0.013 $ & $V$ & (30)\\
5117.490 & $ 16.085 \pm 0.012 $ & $V$ & (30)\\
5118.606 & $ 15.424 \pm 0.012 $ & $V$ & (30)\\
5119.491 & $ 15.141 \pm 0.012 $ & $V$ & (30)\\
5120.374 & $ 15.003 \pm 0.011 $ & $V$ & (30)\\
5124.458 & $ 16.46\pm0.03 $ & $V$ &  (18)\\
5127.493 & $ 18.004 \pm 0.045 $ & $V$ & (30)\\
5128.371 & $ 18.187 \pm 0.016 $ & $V$ & (30)\\
5129.503 & $ 18.489 \pm 0.058 $ & $V$ & (30)\\
5129.824 & $ 18.60\pm0.03 $ & $V$ &  (15)\\
5130.479 & $ 18.399 \pm 0.020 $ & $V$ & (30)\\
5131.397 & $ 18.903 \pm 0.110 $ & $V$ & (30)\\
5132.355 & $ 18.998 \pm 0.065 $ & $V$ & (30)\\
5134.345 & $ 19.101 \pm 0.025 $ & $V$ & (30)\\
5135.357 & $ 19.275 \pm 0.027 $ & $V$ & (30)\\
5136.465 & $ 18.963 \pm 0.128 $ & $V$ & (30)\\
5137.403 & $ 18.520 \pm 0.173 $ & $V$ & (30)\\
5138.452 & $ 19.170 \pm 0.286 $ & $V$ & (30)\\
5139.656 & $ 19.283 \pm 0.130 $ & $V$ & (30)\\
5140.532 & $ 19.766 \pm 0.044 $ & $V$ & (30)\\
5142.519 & $ 19.628 \pm 0.162 $ & $V$ & (30)\\
5143.623 & $>19.170$ & $V$ & (30)\\
5144.343 & $>18.793$ & $V$ & (30)\\
\\
4985.974 & $>20.2         $  & $R$ &   (9)\\
4988.527 & $>20.0         $  & $R$ &   (1)\\
5105.262 & $>19.9         $  & $R$ &   (3)\\
5114.486 & $ 18.9 \pm0.15 $ & $R$ &   (3)\\
5124.460 & $ 15.76\pm0.04 $ & $R$ &  (18)\\
5124.491 & $ 15.78\pm0.04 $ & $R$ &  (26)\\
5129.829 & $ 17.40\pm0.03 $ & $R$ &  (15)\\
5129.835 & $ 18.89\pm0.04 $ & $B$ &  (15)\\
5135.257 & $ 17.6 \pm0.2  $ & $R$ &   (3)\\
5135.632 & $ 17.92\pm0.06 $ & $R$ &  (12)\\
5140.546 & $ 18.4 \pm0.2  $ & $R$ &   (1)\\
5141.359 & $ 18.1 \pm0.2  $ & $R$ &   (1)\\
5148.536 & $ 18.3 \pm0.15 $ & $R$ &   (1)\\
5157.316 & $ 18.6 \pm0.25 $ & $R$ &  (24)\\
5162.281 & $ 18.8 \pm0.2  $ & $R$ &   (5)\\
5173.177 & $ 19.1 \pm0.25 $ & $R$ &   (1)\\
5181.226 & $ 19.1 \pm0.2  $ & $R$ &   (1)\\
5199.299 & $ 19.73\pm0.1  $ & $R$ &  (12)\\
5227.345 & $ 20.0 \pm0.15 $ & $R$ &  (12)\\
5228.383 & $ 20.0 \pm0.15 $ & $R$ &  (12)\\

\cutinhead{M31N 2009-10c}
 
5114.306 & $ 18.06\pm0.04 $ & $B$ &  (17)\\
5117.725 & $ 17.155 \pm 0.023 $ & $B$ & (30)\\
5117.494 & $ 17.048 \pm 0.017 $ & $B$ & (30)\\
5118.654 & $ 17.283 \pm 0.024 $ & $B$ & (30)\\
5119.471 & $ 16.771 \pm 0.019 $ & $B$ & (30)\\
5120.382 & $ 17.155 \pm 0.017 $ & $B$ & (30)\\
5127.511 & $ 16.427 \pm 0.011 $ & $B$ & (30)\\
5128.361 & $ 16.632 \pm 0.037 $ & $B$ & (30)\\
5129.507 & $ 15.977 \pm 0.030 $ & $B$ & (30)\\
5130.461 & $ 16.311 \pm 0.028 $ & $B$ & (30)\\
5131.401 & $ 16.609 \pm 0.013 $ & $B$ & (30)\\
5132.366 & $ 16.116 \pm 0.013 $ & $B$ & (30)\\
5134.356 & $ 17.118 \pm 0.016 $ & $B$ & (30)\\
5135.340 & $ 16.604 \pm 0.011 $ & $B$ & (30)\\
5136.476 & $ 16.958 \pm 0.016 $ & $B$ & (30)\\
5137.340 & $ 16.935 \pm 0.061 $ & $B$ & (30)\\
5138.469 & $ 17.301 \pm 0.037 $ & $B$ & (30)\\
5139.574 & $ 17.429 \pm 0.020 $ & $B$ & (30)\\
5140.627 & $ 17.629 \pm 0.025 $ & $B$ & (30)\\
5142.540 & $ 17.662 \pm 0.110 $ & $B$ & (30)\\
5143.613 & $ 18.082 \pm 0.207 $ & $B$ & (30)\\
5144.325 & $>16.980$ & $B$ & (30)\\
5145.520 & $>17.737$ & $B$ & (30)\\
5146.619 & $>18.901$ & $B$ & (30)\\
5147.498 & $ 18.559 \pm 0.049 $ & $B$ & (30)\\
\\
5114.308 & $ 17.69\pm0.07 $ & $V$ &  (17)\\
5117.728 & $ 17.452 \pm 0.030 $ & $V$ & (30)\\
5117.497 & $ 17.207 \pm 0.056 $ & $V$ & (30)\\
5118.657 & $ 17.437 \pm 0.090 $ & $V$ & (30)\\
5119.475 & $ 17.124 \pm 0.039 $ & $V$ & (30)\\
5120.385 & $ 17.336 \pm 0.020 $ & $V$ & (30)\\
5127.515 & $ 16.513 \pm 0.062 $ & $V$ & (30)\\
5128.364 & $ 17.116 \pm 0.017 $ & $V$ & (30)\\
5129.511 & $ 16.413 \pm 0.037 $ & $V$ & (30)\\
5130.464 & $ 16.985 \pm 0.015 $ & $V$ & (30)\\
5131.404 & $ 17.026 \pm 0.083 $ & $V$ & (30)\\
5132.369 & $ 16.673 \pm 0.013 $ & $V$ & (30)\\
5134.359 & $ 17.380 \pm 0.085 $ & $V$ & (30)\\
5135.343 & $ 16.998 \pm 0.016 $ & $V$ & (30)\\
5136.479 & $ 17.425 \pm 0.025 $ & $V$ & (30)\\
5137.343 & $ 17.537 \pm 0.026 $ & $V$ & (30)\\
5138.472 & $>17.347$ & $V$ & (30)\\
5139.578 & $ 17.931 \pm 0.038 $ & $V$ & (30)\\
5140.630 & $ 18.082 \pm 0.042 $ & $V$ & (30)\\
5142.543 & $ 18.013 \pm 0.148 $ & $V$ & (30)\\
5143.616 & $>17.542$ & $V$ & (30)\\
5144.328 & $>17.223$ & $V$ & (30)\\
\\
5114.304 & $ 17.50\pm0.08 $ & $R$ &  (17)\\
5114.486 & $ 17.7 \pm0.3  $ & $R$ &   (3)\\
5124.491 & $ 16.4 \pm0.15 $ & $R$ &   (3)\\
5135.257 & $ 16.7 \pm0.3  $ & $R$ &   (3)\\
5135.632 & $ 17.03\pm0.1  $ & $R$ &  (12)\\
5140.546 & $ 17.4 \pm0.25 $ & $R$ &   (1)\\
5141.359 & $ 17.6 \pm0.2  $ & $R$ &   (1)\\

\cutinhead{M31N 2009-11a}
 
5142.606 & $ 18.504 \pm 0.078 $ & $B$ & (30)\\
5142.522 & $ 18.411 \pm 0.073 $ & $B$ & (30)\\
5143.606 & $ 18.858 \pm 0.132 $ & $B$ & (30)\\
5144.311 & $ 18.533 \pm 0.180 $ & $B$ & (30)\\
5145.513 & $ 19.017 \pm 0.038 $ & $B$ & (30)\\
5146.498 & $ 19.211 \pm 0.211 $ & $B$ & (30)\\
5147.477 & $ 19.119 \pm 0.057 $ & $B$ & (30)\\
5149.358 & $ 19.325 \pm 0.025 $ & $B$ & (30)\\
5150.410 & $ 19.610 \pm 0.086 $ & $B$ & (30)\\
5151.559 & $ 19.356 \pm 0.159 $ & $B$ & (30)\\
5154.592 & $ 19.705 \pm 0.043 $ & $B$ & (30)\\
5155.334 & $ 19.532 \pm 0.098 $ & $B$ & (30)\\
5156.309 & $ 19.550 \pm 0.184 $ & $B$ & (30)\\
5157.572 & $ 19.840 \pm 0.184 $ & $B$ & (30)\\
5159.438 & $ 19.810 \pm 0.110 $ & $B$ & (30)\\
5161.539 & $ 20.012 \pm 0.239 $ & $B$ & (30)\\
5164.398 & $ 21.234 \pm 0.317 $ & $B$ & (30)\\
5168.433 & $>20.525$ & $B$ & (30)\\
5169.522 & $>20.686$ & $B$ & (30)\\
\\
5142.609 & $ 18.146 \pm 0.049 $ & $V$ & (30)\\
5143.609 & $ 18.206 \pm 0.088 $ & $V$ & (30)\\
5144.315 & $ 18.801 \pm 0.165 $ & $V$ & (30)\\
5145.516 & $ 18.639 \pm 0.216 $ & $V$ & (30)\\
5146.501 & $ 18.686 \pm 0.169 $ & $V$ & (30)\\
5147.480 & $ 18.835 \pm 0.050 $ & $V$ & (30)\\
5149.361 & $ 19.187 \pm 0.023 $ & $V$ & (30)\\
5150.413 & $ 19.482 \pm 0.025 $ & $V$ & (30)\\
5151.563 & $ 19.392 \pm 0.119 $ & $V$ & (30)\\
5154.595 & $ 19.310 \pm 0.141 $ & $V$ & (30)\\
5155.338 & $ 19.563 \pm 0.030 $ & $V$ & (30)\\
5156.312 & $ 19.557 \pm 0.130 $ & $V$ & (30)\\
5157.575 & $ 19.885 \pm 0.055 $ & $V$ & (30)\\
5159.441 & $ 19.896 \pm 0.086 $ & $V$ & (30)\\
5161.542 & $ 20.160 \pm 0.063 $ & $V$ & (30)\\
5164.401 & $ 20.249 \pm 0.059 $ & $V$ & (30)\\
5168.436 & $>19.597$ & $V$ & (30)\\
5169.525 & $ 19.910 \pm 0.134 $ & $V$ & (30)\\
\\
5140.463 & $ 17.6 \pm0.1  $ & $R$ &   (1)\\
5141.221 & $ 17.74\pm0.05 $ & $R$ &   (5)\\
5148.564 & $ 18.4 \pm0.25 $ & $R$ &   (1)\\
5162.534 & $ 19.4 \pm0.3  $ & $R$ &   (5)\\
\\
5160.803 & $ 19.71\pm0.08 $ & $r'$ &  (10)\\
5209.656 & $ 21.5 \pm0.15 $ & $r'$ &  (16)\\
5160.806 & $ 20.35\pm0.07 $ & $g'$ &  (10)\\

\cutinhead{M31N 2009-11b (recurrent nova)}
 
5154.412 & $ 19.994 \pm 0.119 $ & $B$ & (30)\\
5156.323 & $ 19.511 \pm 0.120 $ & $B$ & (30)\\
5159.406 & $ 19.859 \pm 0.043 $ & $B$ & (30)\\
5162.503 & $ 20.347 \pm 0.123 $ & $B$ & (30)\\
5168.480 & $>19.123$ & $B$ & (30)\\
5169.340 & $>19.175$ & $B$ & (30)\\
5170.551 & $ 19.519 \pm 0.098 $ & $B$ & (30)\\
5171.365 & $ 19.439 \pm 0.037 $ & $B$ & (30)\\
5172.330 & $ 19.659 \pm 0.036 $ & $B$ & (30)\\
\\
5154.415 & $ 20.079 \pm 0.139 $ & $V$ & (30)\\
5156.326 & $ 19.299 \pm 0.033 $ & $V$ & (30)\\
5159.409 & $ 19.972 \pm 0.040 $ & $V$ & (30)\\
5162.507 & $ 20.339 \pm 0.055 $ & $V$ & (30)\\
5168.483 & $>18.741$ & $V$ & (30)\\
5169.343 & $ 19.372 \pm 0.145 $ & $V$ & (30)\\
5170.554 & $ 19.482 \pm 0.039 $ & $V$ & (30)\\
5171.368 & $ 19.490 \pm 0.086 $ & $V$ & (30)\\
5172.333 & $ 19.564 \pm 0.032 $ & $V$ & (30)\\
\\
5148.536 & $ 18.9 \pm0.15 $ & $R$ &   (1)\\
5135.632 & $ 18.6 \pm0.1  $ & $R$ &  (12)\\
5162.281 & $ 19.3 \pm0.25 $ & $R$ &   (5)\\
5162.513 & $ 19.6 \pm0.3  $ & $R$ &   (5)\\
5173.177 & $ 19.3 \pm0.2  $ & $R$ &   (1)\\
5173.203 & $ 19.2 \pm0.2  $ & $R$ &   (1)\\
5181.226 & $ 19.7 \pm0.25 $ & $R$ &   (1)\\
5199.299 & $ 20.05\pm0.1  $ & $R$ &  (12)\\
5227.306 & $ 20.0 \pm0.15 $ & $R$ &  (12)\\
5228.284 & $ 20.0 \pm0.15 $ & $R$ &  (12)\\
\\ 
5209.616 & $ 20.9 \pm0.15 $ & $g'$ &  (16)\\
5207.591 & $ 21.2 \pm0.1  $ & $g'$ &  (16)\\
\\
5207.599 & $ 20.66\pm0.1  $ & $r'$ &  (16)\\
5209.621 & $ 20.6 \pm0.1  $ & $r'$ &  (16)\\

\cutinhead{M31N 2009-11c}
 
5145.323 & $ 19.144 \pm 0.029 $ & $B$ & (30)\\
5146.516 & $ 18.614 \pm 0.023 $ & $B$ & (30)\\
5147.487 & $ 18.203 \pm 0.018 $ & $B$ & (30)\\
5148.383 & $ 17.453 \pm 0.016 $ & $B$ & (30)\\
5149.503 & $ 18.981 \pm 0.022 $ & $B$ & (30)\\
5150.403 & $ 18.864 \pm 0.022 $ & $B$ & (30)\\
5154.422 & $ 18.636 \pm 0.019 $ & $B$ & (30)\\
5156.337 & $ 18.650 \pm 0.021 $ & $B$ & (30)\\
5161.546 & $ 18.883 \pm 0.027 $ & $B$ & (30)\\
5162.535 & $ 18.775 \pm 0.026 $ & $B$ & (30)\\
5168.457 & $ 19.075 \pm 0.256 $ & $B$ & (30)\\
5169.333 & $>18.858$ & $B$ & (30)\\
5170.584 & $ 19.676 \pm 0.056 $ & $B$ & (30)\\
5171.373 & $ 19.393 \pm 0.027 $ & $B$ & (30)\\
5172.363 & $ 19.375 \pm 0.028 $ & $B$ & (30)\\
5173.509 & $ 19.635 \pm 0.042 $ & $B$ & (30)\\
\\
5145.326 & $ 18.409 \pm 0.043 $ & $V$ & (30)\\
5146.519 & $ 17.879 \pm 0.040 $ & $V$ & (30)\\
5147.490 & $ 17.497 \pm 0.039 $ & $V$ & (30)\\
5148.386 & $ 16.770 \pm 0.039 $ & $V$ & (30)\\
5149.506 & $ 18.122 \pm 0.040 $ & $V$ & (30)\\
5150.406 & $ 18.178 \pm 0.040 $ & $V$ & (30)\\
5154.425 & $ 17.959 \pm 0.040 $ & $V$ & (30)\\
5156.340 & $ 18.172 \pm 0.040 $ & $V$ & (30)\\
5161.549 & $ 18.710 \pm 0.044 $ & $V$ & (30)\\
5162.538 & $ 18.292 \pm 0.042 $ & $V$ & (30)\\
5168.461 & $ 18.196 \pm 0.138 $ & $V$ & (30)\\
5169.336 & $ 18.742 \pm 0.280 $ & $V$ & (30)\\
5170.587 & $ 19.065 \pm 0.053 $ & $V$ & (30)\\
5171.375 & $ 18.870 \pm 0.044 $ & $V$ & (30)\\
5172.366 & $ 19.070 \pm 0.044 $ & $V$ & (30)\\
5173.512 & $ 19.392 \pm 0.062 $ & $V$ & (30)\\
\\
5135.632 & $>21.4         $  & $R$ &  (12)\\
5140.463 & $>20.3         $  & $R$ &   (1)\\
5141.359 & $ 19.9 \pm0.3  $ & $R$ &   (1)\\
5148.536 & $ 17.03\pm0.1  $ & $R$ &   (1)\\
5154.424 & $ 17.8 \pm0.15 $ & $R$ &   (7)\\
5157.316 & $ 17.9 \pm0.1  $ & $R$ &  (24)\\
5158.542 & $ 17.9 \pm0.1  $ & $R$ &   (1)\\
5161.164 & $ 18.4 \pm0.15 $ & $R$ &   (7)\\
5162.281 & $ 18.1 \pm0.15 $ & $R$ &   (5)\\
5162.513 & $ 18.2 \pm0.15 $ & $R$ &   (5)\\
5168.173 & $ 18.3 \pm0.1  $ & $R$ &   (7)\\
5169.202 & $ 18.6 \pm0.2  $ & $R$ &   (1)\\
5173.177 & $ 18.6 \pm0.15 $ & $R$ &   (1)\\
5173.192 & $ 18.7 \pm0.15 $ & $R$ &   (1)\\
5173.203 & $ 18.5 \pm0.15 $ & $R$ &   (1)\\
5181.226 & $ 18.9 \pm0.15 $ & $R$ &   (1)\\
5184.717 & $ 19.22\pm0.1  $ & $R$ &  (19)\\
5227.306 & $>22.0         $  & $R$ &  (12)\\
\\
5156.756 & $ 18.86\pm0.08 $ & $g'$ &  (10)\\
\\
5156.759 & $ 18.32\pm0.05 $ & $r'$ &  (10)\\
5209.631 & $>21.1         $  & $r'$ &  (16)\\

\cutinhead{M31N 2009-11d}
 
5156.330 & $ 17.278 \pm 0.016 $ & $B$ & (30)\\
5157.579 & $ 18.659 \pm 0.102 $ & $B$ & (30)\\
5159.445 & $ 18.123 \pm 0.019 $ & $B$ & (30)\\
5161.532 & $ 18.554 \pm 0.023 $ & $B$ & (30)\\
5168.464 & $>19.753$ & $B$ & (30)\\
5169.515 & $ 19.699 \pm 0.040 $ & $B$ & (30)\\
5170.529 & $ 20.139 \pm 0.063 $ & $B$ & (30)\\
5171.358 & $ 20.038 \pm 0.035 $ & $B$ & (30)\\
5172.556 & $ 19.787 \pm 0.112 $ & $B$ & (30)\\
5173.365 & $ 19.997 \pm 0.032 $ & $B$ & (30)\\
5176.376 & $>19.722$ & $B$ & (30)\\
\\
5156.333 & $ 17.050 \pm 0.012 $ & $V$ & (30)\\
5157.582 & $ 18.568 \pm 0.101 $ & $V$ & (30)\\
5159.448 & $ 18.047 \pm 0.014 $ & $V$ & (30)\\
5161.535 & $ 18.288 \pm 0.075 $ & $V$ & (30)\\
5168.468 & $>19.415$ & $V$ & (30)\\
5169.518 & $ 20.034 \pm 0.139 $ & $V$ & (30)\\
5170.532 & $ 20.696 \pm 0.077 $ & $V$ & (30)\\
5171.361 & $ 19.912 \pm 0.098 $ & $V$ & (30)\\
5172.559 & $ 20.501 \pm 0.055 $ & $V$ & (30)\\
5173.368 & $ 20.375 \pm 0.035 $ & $V$ & (30)\\
5176.378 & $ 20.523 \pm 0.077 $ & $V$ & (30)\\
\\
5157.299 & $ 17.26\pm0.10 $ & $R$ &  (24)\\
5158.560 & $ 17.3 \pm0.2  $ & $R$ &   (1)\\
5162.525 & $ 17.6 \pm0.15 $ & $R$ &   (5)\\
5168.505 & $ 18.1 \pm0.2  $ & $R$ &   (7)\\
5173.192 & $ 18.7 \pm0.15 $ & $R$ &   (1)\\
5227.379 & $>21.5         $  & $R$ &  (12)\\
\\
5156.787 & $ 16.53\pm0.10 $ & $g'$ &  (10)\\
5207.619 & $ 21.3 \pm0.1  $ & $g'$ &  (16)\\
\\
5156.785 & $ 16.43\pm0.08 $ & $r'$ &  (10)\\
5207.626 & $ 21.3 \pm0.1  $ & $r'$ &  (16)\\

\cutinhead{M31N 2009-11e}
 
5161.575 & $ 18.122 \pm 0.074 $ & $B$ & (30)\\
5162.524 & $ 17.445 \pm 0.041 $ & $B$ & (30)\\
5168.445 & $ 17.496 \pm 0.018 $ & $B$ & (30)\\
5169.326 & $ 17.279 \pm 0.079 $ & $B$ & (30)\\
5170.572 & $ 17.110 \pm 0.014 $ & $B$ & (30)\\
5171.379 & $ 16.869 \pm 0.011 $ & $B$ & (30)\\
5172.355 & $ 16.982 \pm 0.011 $ & $B$ & (30)\\
5173.411 & $ 18.485 \pm 0.064 $ & $B$ & (30)\\
5176.408 & $ 18.423 \pm 0.033 $ & $B$ & (30)\\
\\
5161.578 & $ 18.254 \pm 0.066 $ & $V$ & (30)\\
5162.528 & $ 17.786 \pm 0.014 $ & $V$ & (30)\\
5168.448 & $ 17.810 \pm 0.061 $ & $V$ & (30)\\
5169.329 & $ 17.590 \pm 0.104 $ & $V$ & (30)\\
5170.575 & $ 17.418 \pm 0.038 $ & $V$ & (30)\\
5171.382 & $ 17.073 \pm 0.010 $ & $V$ & (30)\\
5173.415 & $ 19.005 \pm 0.094 $ & $V$ & (30)\\
5176.412 & $ 19.185 \pm 0.269 $ & $V$ & (30)\\
\\
5148.536 & $>20.6         $  & $R$ &   (1)\\
5157.316 & $ 17.8 \pm0.15 $ & $R$ &  (24)\\
5158.542 & $ 17.0 \pm0.1  $ & $R$ &   (1)\\
5161.164 & $ 17.7 \pm0.1  $ & $R$ &   (7)\\
5162.281 & $ 17.4 \pm0.1  $ & $R$ &   (5)\\
5162.513 & $ 17.5 \pm0.1  $ & $R$ &   (5)\\
5168.173 & $ 17.5 \pm0.1  $ & $R$ &   (7)\\
5169.202 & $ 17.2 \pm0.2  $ & $R$ &   (1)\\
5173.177 & $ 18.1 \pm0.1  $ & $R$ &   (1)\\
5173.203 & $ 18.1 \pm0.1  $ & $R$ &   (1)\\
5181.226 & $ 18.5 \pm0.15 $ & $R$ &   (1)\\
5184.717 & $ 18.96\pm0.09 $ & $R$ &  (19)\\
5186.730 & $ 18.8 \pm0.25 $ & $R$ &  (20)\\
5199.299 & $ 19.09\pm0.08 $ & $R$ &  (12)\\
5227.345 & $ 19.92\pm0.10 $ & $R$ &  (12)\\
5228.383 & $ 19.40\pm0.10 $ & $R$ &  (12)\\
\\
5157.691 & $ 17.80\pm0.09 $ & $g'$ &  (10)\\
5158.715 & $ 17.29\pm0.10 $ & $g'$ &  (10)\\
5207.591 & $ 19.10\pm0.09 $ & $g'$ &  (16)\\
5210.592 & $ 19.65\pm0.10 $ & $g'$ &  (16)\\
\\
5157.688 & $ 17.80\pm0.04 $ & $r'$ &  (10)\\
5158.713 & $ 17.33\pm0.03 $ & $r'$ &  (10)\\
5207.599 & $ 18.91\pm0.08 $ & $r'$ &  (16)\\
5210.598 & $ 19.45\pm0.09 $ & $r'$ &  (16)\\

\enddata
\tablenotetext{a}{Observers:
(1)  K. Hornoch, Ond\v{r}ejov 0.65-m;
(2)  K. Hornoch, Lelekovice 0.35-m;
(3)  P. Ku\v{s}nir\'ak, Ond\v{r}ejov 0.65-m;
(4)  M. Wolf,  Ond\v{r}ejov 0.65-m; 
(5)  K. Hornoch \& M. Wolf, Ond\v{r}ejov 0.65-m;
(6)  M. Wolf i\& P. Zasche, Ond\v{r}ejov 0.65-m;
(7)  K. Hornoch \& P. Zasche, Ond\v{r}ejov 0.65-m;
(8)  P. Zasche, San Pedro M\'artir 0.84-m;
(9)  O. Pejcha, MDM 1.3-m;
(10)  O. Pejcha, MDM 2.4-m;
(11)  K. Hornoch \& P. \v{S}edinov\'a, Ond\v{r}ejov 0.65-m;
(12)  P. Kub\'anek, J. Gorosabel i\& M. Jel\'inek, Calar Alto 1.23-m;
(13)  P. Kub\'anek, MDM 2.4-m;
(14)  P. Garnavich \& A. Karska, MGIO 1.83-m VATT;
(15)  P. Garnavich, K. Thorne \& K. Morhig, MGIO 1.83-m VATT;
(16)  J. Prieto \& R. Khan, MDM 2.4-m;
(17)  A. Valeev \& O. Sholukhova, SAO 6-m;
(18)  V. L. Afanasiev \& S. N. Dodonov, SAO 6-m;
(19)  T. Farnham \& B. Mueller, KPNO 2.1-m;
(20)  N. Samarasinha \& B. Mueller, KPNO 2.1-m;
(21)  M. Burleigh \& S. Casewell, La Palma 2.5-m INT;
(22)  A. Gal\'ad, AGO Modra 0.60-m;
(23)  P. Caga\v{s}, Zl\'{\i}n 0.26-m;
(24)  K. Hornoch \& P. Ku\v{s}nir\'ak, Ond\v{r}ejov 0.65-m;
(25)  P. Ku\v{s}nir\'ak \& T. Henych, Ond\v{r}ejov 0.65-m;
(26)  P. Ku\v{s}nir\'ak \& Z. Bardon, Ond\v{r}ejov 0.65-m;
(27)  K. Hornoch \& M. Tukinsk\'a, Ond\v{r}ejov 0.65-m;
(28)  P. Zasche \& Ond\v{r}ejov, 0.65-m;
(29)  Darnley et al. 2004; La Palma 2.5-m INT;
(30)  La Palma 2.0-m LT;
(31)  Mt. Haleakala 2.0-m FTN;
(32)  P. Ku\v{s}nir\'ak, L. \v{S}arounov\'a \& M. Wolf, Ond\v{r}ejov 0.65-m;
(33)  P. Garnavich, MGIO 1.83-m VATT;
(34)  P. Garnavich \& B. Tucker, MGIO 1.83-m VATT;
(35)  P. Garnavich \& C. Kennedy, MGIO 1.83-m VATT;
(36)  P. Garnavich \& J. Gallagher, MGIO 1.83-m VATT;
(37)  P. Garnavich, KPNO 3.5-m WIYN;
(38)  D. Mackey, La Palma 2.5-m INT.
}
\end{planotable}

\clearpage

\begin{planotable}{lrrrclr}
\tabletypesize{\scriptsize}
\tablenum{4}
\tablewidth{0pt}
\tablecolumns{7}
\tablecaption{Full Spectroscopic Sample of M31 Novae\label{spectsample}}
\tablehead{\colhead{} & \colhead{$\Delta\alpha~cos\delta$} & \colhead{$\Delta\delta$} & \colhead{$a$}&\colhead{Discovery} & \colhead{} & \colhead{}  \\
\colhead{Nova} & \colhead{($'$)} & \colhead{($'$)} & \colhead{($'$)} & \colhead{mag (Filter)} & \colhead{Type} & \colhead{References\tablenotemark{a}}}
\startdata
M31N 1981-09a  & $-4.76$& $-2.89$&   5.67  &  16.4(H) &  Fe II    &   1 \\
M31N 1981-09b  & $-0.65$&   2.10 &   2.96  &  14.9(H) &  Fe II    &   1 \\
M31N 1981-09c  &  0.78 &   2.87 &   3.43  &  15.0(H) &  Fe II    &   1 \\
M31N 1981-09d  &  4.11 & $-2.16$&   8.07  &  15.8(H) &  Fe II    &   1 \\
M31N 1986-08a  & $-4.35$& $-1.64$&   4.98  &  16.3(H) &  Fe II    &   2 \\
M31N 1987-09a  &  7.80 &   2.13 &  10.57  &  19.4(B) &  Fe II    &   2 \\
M31N 1987-10a  & $-0.33$&   1.75 &   2.28  &  18.4(B) &  Fe II    &   2 \\
M31N 1989-08a  &  6.47 & $-0.29$&  10.36  &  17.9(B) &  He/N     &   2 \\
M31N 1989-08b  & $-1.01$&   0.00 &   1.10  &  19.3(B) &  Fe II    &   2 \\
M31N 1989-08c  & $-5.22$& $-4.41$&   6.84  &  17.9(B) &  Fe II    &   2 \\
M31N 1989-09a  &  2.15 &   4.05 &   5.06  &  15.5(H) &  Fe II    &   2 \\
M31N 1989-10a  &  0.36 &   3.10 &   3.91  &  17.6(B) &  Fe II    &   2 \\
M31N 1990-10b  & $-1.53$& $-4.32$&   5.44  &  17.6(B) &  Fe II    &   3 \\
M31N 1992-11b  &  3.07 & $-1.49$&   5.04  &  17.2(V) &  Fe II    &   3 \\
M31N 1993-06a  &  0.91 & 1.31 &   1.64  &  15.8(R) &  Fe II:    &   3 \\
M31N 1993-08a  &  0.14 & $-1.70$ &   1.64  &  15.8(R) &  He/N:    &   3 \\
M31N 1993-10g  &  0.63 &   1.87  &  2.17   & 16.7(H) &  Fe II    &  3 \\
M31N 1993-11c  &  1.08 &   1.32  &  1.73   & 15.8(H) &  Fe II:   &  3 \\
M31N 1998-09d  &  0.43 & $-1.33$ &  1.78   & 16.5(B) &  Fe II    &   3 \\
M31N 1999-06a  &  0.97 & $-1.05$ &  1.89  &  17.8(R) &  Fe II    &   3  \\
M31N 1999-08f  & $-0.61$&   3.04 &   4.44  &  17.0(w) &  Fe II:   &  3 \\
M31N 1999-10a  &  1.00 &   0.39 &   1.10  &  17.5(w) &  Fe II:   &   3  \\
M31N 2001-10a  &  3.56 & $-3.96$&  10.81  &  17.0(R) &  Fe II    &   3  \\
M31N 2001-12a  & $-0.55$&   0.26 &   0.68  &  15.5(R) & Fe II    &   3  \\
M31N 2002-01b  & $-1.97$&   2.25 &   4.60  &  16.8(R) & He/N:    &   3  \\
M31N 2002-08a  & $-2.53$ &$-9.93$ & 14.01   & 17.1(R) & Fe II:   &  3 \\
M31N 2004-08b  &  7.98  &  0.54  & 12.60   & 17.3(R) & Fe II    &  3 \\
M31N 2004-09a  &$-0.77$ &$-1.44$ &  1.72   & 17.5(R) & Fe II    &  3 \\
M31N 2004-11a  &$-0.29$ & 2.32   &  3.03   & 16.5(R) & Fe II    &  3 \\
M31N 2004-11b  & 4.34   & 1.93   &  4.97   & 16.6(R) & He/N:    &  3 \\
M31N 2005-01a  &$-3.00$ &  0.46  &  3.98   & 15.0(R) & Fe II    &  3 \\
M31N 2005-06d  & 26.35 &  17.61 &  44.21  &  15.7(w) &  He/N     &  4 \\
M31N 2005-07a  &  1.21 &   4.52 &   5.85  &  17.4(R) &  FeII:    &  3 \\
M31N 2005-09a  &  1.48 &   3.84 &   4.76  &  16.7(R) &  Fe II    &  4 \\
M31N 2005-09b  &$-44.77$&$-56.00$&  71.73  &  16.5(w) &  Fe II\tablenotemark{b}   &  4 \\
M31N 2005-09c  &$-43.70$&$-48.34$&  66.76  &  16.0(w) &  Fe II    &  5 \\
M31N 2006-06a  &  5.16 & $-2.40$&  11.09  &  17.5(R) &  Fe II    &  4 \\
M31N 2006-09c  & $-0.37$& $-7.39$&  11.61  &  16.8(w) &  Fe II    & 3 \\
M31N 2006-10a  &$-11.50$& $-4.36$&  17.31  &  18.7(R) &  Fe II    &  3  \\
M31N 2006-10b  &$-37.25$&$-24.81$&  63.14  &  16.4(w) &  Hy       & 3 \\
M31N 2006-11a  &  2.35 & $-9.84$&  20.46  &  17.3(w) &  Fe II    &  3 \\
M31N 2006-12a  & $-4.38$& $-2.39$&   5.11  &  17.8(R) &  Fe II    & 3 \\
M31N 2006-12b  & $-6.26$& $-8.41$&  10.88  &  18.0(R) &  Fe II    & 3 \\
M31N 2007-02a  &$-19.95$&$-31.22$&  40.21  &  16.3(w) &  Fe II    & 3 \\
M31N 2007-02b  &$-12.04$& $-1.57$&  21.34  &  16.7(R) &  Fe II:   & 3,6  \\
M31N 2007-06b  & $-2.12$&$-15.71$&  26.20  &  16.8(w) &  He/N     & 3,7 \\
M31N 2007-07b  &  0.29 &   1.92 &   2.26  &  17.7(R) &  Fe II    & 8 \\
M31N 2007-07c  &  3.56 & $-1.26$&   5.55  &  15.8(R) &  He/N     & 8,9 \\
M31N 2007-07e  & $-0.20$&   1.59 &   2.00  &  18.1(R) &  Fe II    & 8 \\
M31N 2007-07f  & $-45.77$& $-22.95$& 81.74 &  17.4(w) & Fe II    & 10 \\
M31N 2007-08a  &$-20.78$&$-22.25$&  30.49  &  17.6(w) &  Fe II:  & 8 \\
M31N 2007-08d  &$-36.91$&$-46.74$&  59.56  &  18.1(R) &  Fe II    & 3 \\
M31N 2007-10a  &  2.19 &$-12.78$&  27.57  &  16.0(w) &  He/Nn    & 3,11 \\
M31N 2007-10b  &  8.48 &   1.09 &  12.91  &  17.8(w) &  He/Nn    & 12 \\
M31N 2007-11b  & 12.94 &$-12.52$&  59.72  &  18.6(R) &  He/Nn    & 3,13,14 \\
M31N 2007-11c  &  3.72 & $-0.24$&   4.77  &  17.4(R) &  Fe II    & 14,15\\
M31N 2007-11d  & 24.34 &  21.60 &  34.03  &  14.9(w) &  Fe II    & 3,16 \\
M31N 2007-11e  & 34.06 &  46.07 &  57.52  &  16.4(w) &  Fe II    & 3,17 \\
M31N 2007-12a  & 14.79 &  22.57 &  28.53  &  17.8(w) &  Fe II    & 3    \\
M31N 2007-12b  &  6.70 & $-2.37$&  14.79  &  16.1(w) &  He/N     & 3,18 \\
M31N 2007-12c  & 27.26 &   4.08 &  66.22  &  16.4(w) &  Fe II    & 19 \\
M31N 2007-12d  & $-9.30$& $-6.35$&  11.92  &  17.2(R) &  He/N     & 3 \\
M31N 2008-05c  &  5.20 &   3.12 &   6.18  &  17.0(R) &  Fe II    & 20 \\
M31N 2008-06b  & $-3.11$& $-1.34$&   3.56  &  15.9(R) &  He/N     & 21 \\
M31N 2008-07a  & $-1.87$&   2.12 &   4.34  &  18.3(R) &  Fe II    & 22 \\
M31N 2008-07b  &  8.08 & $-6.08$&  26.01  &  19.0(g) &  Fe II    & 23 \\
M31N 2008-08a  &  0.12 &   0.98 &   1.09  &  16.8(R) &  Fe II    & 24,25 \\
M31N 2008-08b  &  1.51 &   0.07 &   1.69  &  16.4(R) &  He/N     & 24,25 \\
M31N 2008-08c  & $-0.72$ & 10.15 &  18.72  &  16.8(R) &  Fe II    & 25 \\
M31N 2008-08d  & 33.62 & 106.41 & 109.45  &  18.1(w) &  Fe II    & 26 \\
M31N 2008-09a  &$-10.85$& $-8.26$&  14.26  &  18.1(g) &  Fe II    & 3 \\
M31N 2008-09c  &  1.33 &$-14.25$&  30.22  &  17.6(g) &  Fe II    & 3 \\
M31N 2008-10a  &  9.51 &  38.60 &  65.42  &  17.1(w) &  Fe II    & 3 \\
M31N 2008-10b  &  3.40 & $-1.98$&   6.19  &  18.3(R) &  Fe II    & 25,27 \\
M31N 2008-11a  &$-13.58$&$-10.11$&  18.51  &  16.5(R) &  He/N     & 3 \\
M31N 2008-12b  &  3.85 &   1.72 &   4.43  &  16.8(w) &  Fe II    & 28 \\
M31N 2009-01a  & 22.44 &   7.39 &  48.17  &  18.5(w) &  Fe II    & 3 \\
M31N 2009-02a  & 11.11 &  20.52 &  26.38  &  16.8(w) &  Fe II    & 3 \\
M31N 2009-08a  &  2.57 &   1.35 &   3.00  &  17.2(H) &  Fe II    & 29 \\
M31N 2009-08b  & 15.93 &  32.73 &  45.52  &  17.1(w) &  Fe II    & 30 \\
M31N 2009-08d  &  0.45 & $-0.52$&   0.81  &  17.2(R) &  Fe II    & 31 \\
M31N 2009-08e  & $-1.54$&   1.89 &   3.54  &  17.8(w) &  Fe II   & 32 \\
M31N 2009-09a  & $-3.45$&$-12.13$&  17.38  &  17.6(w) &  Fe II   & 33 \\
M31N 2009-10a  & 27.78 &  48.61 &  61.67  &  17.1(w) &  Fe II    & 34 \\
M31N 2009-10b  & $-4.43$&   0.60 &   6.17  &  14.7(R) &  Fe II   & 35 \\
M31N 2009-10c  &  0.26 & $-0.19$&   0.37  &  17.2(H) &  Fe II    & 36 \\
M31N 2009-11a  &  3.81 &  24.99 &  48.41  &  17.6(R) &  Fe II    & 37 \\
M31N 2009-11b  & $-0.89$&$-7.09$&  10.56  &  18.4(R) &  Fe II    & 23,38 \\
M31N 2009-11c  &  4.91 & $-3.83$&  12.92  &  17.0(R) &  Fe II    & 39 \\
M31N 2009-11d  & 17.37 &   2.79 &  35.48  &  16.4(r) &  Fe II    & 40 \\
M31N 2009-11e  & $-1.70$& $-3.16$&   3.86  &  17.4(R) &  Fe II   & 41 \\
\enddata
\tablenotetext{a}{REFERENCES: (1) \citet{cia83}; (2) \citet{tom92}; (3) this work; (4) \cite{pie05}; (5) \citet{hat07}; (6) \citet{pie07a}; (7) \citet{sha07b}; (8) \citet{bar07a}; (9) \citet{rau07a}; (10) Quimby (2007, private communication); (11) \citet{gal07}; (12) \citet{rau07b}; (13) \citet{rau07}; (14) \citet{bar07b}; (15) \citet{cir07}; (16) \citet{sha09}; (17) \citet{dim07}; (18) \citet{bod09}; (19) \citet{rau07c}; (20) \citet{rau08} (21) \citet{rei08}; (22) \citet{bar08}; (23) \citet{kas11}; (24) \citet{dim08a}; (25) \citet{dim10}; (26) \citet{cho08}; (27) \citet{dim08b}; (28) \citet{kas08}; (29) \citet{val09}; (30) \citet{rod09}; (31) \citet{dim09a}; (32) \citet{med09}; (33) \citet{bar09a}; (34) \citet{fab09a}; (35) \citet{bar09b}; (36) \citet{fab09b}; (37) \citet{hor09a}; (38) \citet{kas09}; (39) \citet{hor09b}; (40) \citet{hor09c}; (41) \citet{hor09d}}
\tablenotetext{b}{The line with of 4400~km~s$^{-1}$ reported by D. C. Leonard referred to an estimate of the full width at zero intensity, not the FWHM. Analysis of the original spectrum reveals the object to be a classic Fe~II nova}
\end{planotable}

\begin{deluxetable}{lcccc}
\tabletypesize{\scriptsize}
\tablenum{5}
\tablewidth{0pt}
\tablecolumns{5}
\tablecaption{Balmer Emission-Line Properties\label{balmerline}}
\tablehead{
\colhead{} & \multicolumn{2}{c}{EW (\AA)} & \multicolumn{2}{c}{FWHM (km~s$^{-1}$)} \\
\colhead{Nova} & \colhead{H$\beta$} & \colhead{H$\alpha$} & \colhead{H$\beta$} & \colhead{H$\alpha$}}
\startdata
M31N 1990-10b  & $-267$ & $-541$ & 1600 & 1550 \\ 
M31N 1992-11b  & $-250$ & $-1219$& 1670 & 1740 \\
M31N 1993-06a  & $-36$  & $-392$ & 1840 & 1770 \\
M31N 1993-08a  & \dots  & \dots & \dots & 4350 \\
M31N 1993-10g\tablenotemark{a} & $-87$  & $-324$ & 1630 & 1680 \\
M31N 1993-11c\tablenotemark{a} & $-118$ & $-828$ & 1450 & 1560 \\
M31N 1998-09d  & $-92$  & $-323$ & 1720 & 1840 \\
M31N 1999-06a  & $-322$ & $-1200$& 1110 & 920 \\
M31N 1999-08f  & $-135$ & $-704$ & 1640 & 1690 \\
M31N 1999-10a  & $-230$ & $-565$ & 1780 & 1930 \\
M31N 2001-10a  & $-279$ & $-945$ & 1540 & 1540 \\
M31N 2001-12a  & \dots  & $-1500$& \dots& 1310 \\
M31N 2002-01b  & \dots  & $-760$ & \dots& 3430 \\
M31N 2002-08a  & $-94$  & $-470$ & 1360 & 1320 \\
M31N 2004-08b  & $-90$  & $-410$ & 1970 & 1830 \\
M31N 2004-09a  & $-54$  & $-250$ & 2000 & 1880 \\
M31N 2004-11a  & $-175$  & $-830$ & 1230 & 1580 \\
M31N 2004-11b  & $-80$  & $-1020$ & 2500 & 2620 \\
M31N 2005-01a  & $-50$  & $-107$ & 1810 & 1680 \\
M31N 2005-07a  & \dots  & $-120$ & \dots& 1100 \\
M31N 2006-09c  & $-127$ & $-470$ & 1910 & 1920 \\
M31N 2006-10a  & $-44$  & $-90$  & 950  & 810 \\
M31N 2006-10b  & $-64$  & $-102$ & 3330 & 3090 \\
               & $-133$ & $-2130$& 3030 & 3562 \\
M31N 2006-11a  & $-35$  & $-57$  & 1420 & 1120 \\
M31N 2006-12a  & $-107$ & $-330$ & 1850 & 1760 \\
M31N 2006-12b  & $-86 $ & $-226$ & 1230 & 1020 \\
M31N 2007-02a  & $-35 $ & $-61$  & 1530 & 1310 \\
M31N 2007-02b  & $-39 $ & $-304$ & 1460 & 1910 \\
M31N 2007-06b  & $-51 $ & $-168$ & 2940 & 2870 \\
M31N 2007-08d  & $-68 $ & $-284$ & 1160 & 1180 \\
M31N 2007-10a  & $-37 $ & $-154$ &  470 & 500 \\
M31N 2007-11b  & $-147$ & $-535$ & 1480 & 1290 \\
M31N 2007-11c  & $-75 $ & $-137$ & 1950 & 1700 \\
M31N 2007-11d  & $-11 $ & $-16 $ & 1630 & 1550 \\
               & $-297$ & $-1223$& 2060 & 2260 \\
M31N 2007-11e  & $-137$ & $-306$ & 1750 & 1600 \\
M31N 2007-11g  & $-18 $ & $-13 $ & 340  & 300 \\
M31N 2007-12a  & $-159$ & $-309$ & 2050 & 1850 \\
M31N 2007-12b  & $-247$ & $-990$ & 4070 & 4080 \\
M31N 2007-12d  & $-271$ & $-1210$& 5220 & 4980 \\
M31N 2008-08d  & $-130 $ & $-440$ & 1150 & 1050 \\
               & $-63 $ & $-634$ & 1360 & 1480 \\
M31N 2008-09a  & $-113$ & $-262$ & 1540 & 1460 \\
M31N 2008-09c  & $-16 $ & $-23 $ & 1590 & 1010 \\
M31N 2008-10a  & $-97 $ & $-300$ & 1390 & 1270 \\
M31N 2008-10b  & $-2.8$ & $-19 $ & 660  & 640 \\
               & $-24 $ & $-95 $ & 1220 & 950 \\
M31N 2008-11a  & $-188$ & $-636$ & 4510 & 4350 \\
               & $-33 $ & $-370$ & 1460 & 3160 \\
M31N 2009-01a  & $-5  $ & $-25 $ & 560  & 550 \\
M31N 2009-02a  & $-4  $ & $-15 $ & 1280 & 1410 \\
\enddata
\tablenotetext{a}{Due to ambiguity in the data logs from November 1993, it is possible
that the data for these two novae are reversed.}
\end{deluxetable}

\clearpage

\begin{planotable}{lcccc}
\tabletypesize{\scriptsize}
\tablenum{6}
\tablewidth{0pt}
\tablecolumns{5}
\tablecaption{Light-Curve Parameters\label{lcparam}}
\tablehead{\colhead{Nova} & \colhead{Filter} & \colhead{$M_{\rm max}$} & \colhead{Fade Rate (mag day$^{-1}$)} & \colhead{$t_2$ (days)}}
\startdata
M31N 1999-08f   &$r'$ & $ -7.54\pm  0.11$& $ 0.063\pm 0.002$& $  31.7\pm  0.9$ \\
M31N 2001-10a   &$r'$ & $ -7.54\pm  0.11$& $ 0.040\pm 0.004$& $  50.2\pm  4.9$ \\
M31N 2002-08a   &$R$ & $ -7.49\pm  0.20$& $ 0.044\pm 0.005$& $  45.8\pm  5.1$ \\
M31N 2004-08b   &$R$ & $ -7.24\pm  0.25$& $ 0.038\pm 0.002$& $  52.0\pm  3.3$ \\
M31N 2004-09a   &$R$ & $ -7.10\pm  0.20$& $ 0.056\pm 0.006$& $  35.5\pm  3.7$ \\
M31N 2004-11a   &$R$ & $ -8.04\pm  0.25$& $ 0.105\pm 0.010$& $  19.1\pm  1.9$ \\
M31N 2004-11b   &$R$ & $ -7.94\pm  0.20$& $ 0.042\pm 0.002$& $  47.2\pm  2.6$ \\
M31N 2005-01a   &$R$ & $ -9.59\pm  0.17$& $ 0.105\pm 0.009$& $  19.0\pm  1.6$ \\
M31N 2005-07a   &$R$ & $ -7.16\pm  0.25$& $ 0.017\pm 0.003$& $ 120.9\pm 23.5$ \\
M31N 2006-06a   &$R$ & $ -7.04\pm  0.11$& $ 0.030\pm 0.006$& $  67.3\pm 13.2$ \\
M31N 2006-09c   &$R$ & $ -7.88\pm  0.14$& $ 0.087\pm 0.006$& $  23.1\pm  1.6$ \\
M31N 2006-10a   &$B$ & $ -7.54\pm  0.11$& $ 0.035\pm 0.002$& $  56.7\pm  3.1$ \\
\dots           &$V$ & $ -7.67\pm  0.11$& $ 0.035\pm 0.002$& $  57.5\pm  3.1$ \\
\dots           &$R$ & $ -6.64\pm  0.20$& $ 0.021\pm 0.005$& $  94.3\pm 20.6$ \\
M31N 2006-10b   &$B$ & $ -7.85\pm  0.11$& $ 0.287\pm 0.018$& $   7.0\pm  0.4$ \\
\dots           &$V$ & $ -7.98\pm  0.11$& $ 0.345\pm 0.018$& $   5.8\pm  0.3$ \\
M31N 2006-11a   &$R$ & $ -8.54\pm  0.20$& $ 0.070\pm 0.006$& $  28.7\pm  2.6$ \\
M31N 2006-12a   &$R$ & $ -7.24\pm  0.25$& $ 0.057\pm 0.007$& $  34.9\pm  4.5$ \\
M31N 2007-02b   &$R$ & $ -7.88\pm  0.20$& $ 0.059\pm 0.006$& $  34.1\pm  3.6$ \\
M31N 2007-07e   &$R$ & $ -6.64\pm  0.35$& $ 0.044\pm 0.010$& $  45.6\pm 10.5$ \\
M31N 2007-08d   &$R$ & $ -6.44\pm  0.11$& $ 0.025\pm 0.003$& $  80.5\pm 10.5$ \\
M31N 2007-10a   &$B$ & $ -8.25\pm  0.11$& $ 0.222\pm 0.013$& $   9.0\pm  0.5$ \\
\dots           &$V$ & $ -8.38\pm  0.11$& $ 0.255\pm 0.013$& $   7.9\pm  0.4$ \\
\dots           &$i'$ & $ -6.51\pm  0.13$& $ 0.225\pm 0.010$& $   8.9\pm  0.4$ \\
M31N 2007-10b   &$B$ & $ -7.60\pm  1.15$& $ 0.508\pm 0.067$& $   3.9\pm  0.5$ \\
\dots           &$V$ & $ -8.14\pm  1.56$& $ 0.690\pm 0.065$& $   2.9\pm  0.3$ \\
\dots           &$R$ & $ -8.18\pm  1.44$& $ 0.639\pm 0.085$& $   3.1\pm  0.4$ \\
M31N 2007-11b   &$B$ & $ -5.65\pm  0.11$& $ 0.055\pm 0.004$& $  36.3\pm  2.6$ \\
\dots           &$V$ & $ -5.78\pm  0.11$& $ 0.044\pm 0.003$& $  45.4\pm  3.5$ \\
\dots           &$R$ & $ -6.36\pm  0.25$& $ 0.046\pm 0.012$& $  43.6\pm 11.0$ \\
\dots           &$i'$ & $ -5.45\pm  0.14$& $ 0.027\pm 0.006$& $  74.4\pm 16.7$ \\
M31N 2007-11c   &$B$ & $ -8.25\pm  0.24$& $ 0.138\pm 0.006$& $  14.5\pm  0.6$ \\
\dots           &$V$ & $ -8.20\pm  0.27$& $ 0.155\pm 0.007$& $  12.9\pm  0.6$ \\
\dots           &$i'$ & $ -7.33\pm  0.18$& $ 0.171\pm 0.013$& $  11.7\pm  0.9$ \\
M31N 2007-11d   &$B$ & $ -9.35\pm  0.11$& $ 0.143\pm 0.009$& $  14.0\pm  0.9$ \\
\dots           &$V$ & $ -9.48\pm  0.11$& $ 0.158\pm 0.009$& $  12.6\pm  0.7$ \\
\dots           &$r'$ & $ -9.64\pm  0.11$& $ 0.153\pm 0.009$& $  13.1\pm  0.8$ \\
\dots           &$i'$ & $ -8.72\pm  0.11$& $ 0.216\pm 0.011$& $   9.2\pm  0.5$ \\
M31N 2007-12a   &$B$ & $ -7.02\pm  0.12$& $ 0.078\pm 0.005$& $  25.6\pm  1.6$ \\
\dots           &$V$ & $ -7.06\pm  0.11$& $ 0.081\pm 0.004$& $  24.8\pm  1.4$ \\
\dots           &$i'$ & $ -7.19\pm  0.11$& $ 0.068\pm 0.004$& $  29.6\pm  2.0$ \\
M31N 2007-12b   &$B$ & $ -8.15\pm  0.11$& $ 0.497\pm 0.034$& $   4.0\pm  0.3$ \\
\dots           &$V$ & $ -8.28\pm  0.11$& $ 0.548\pm 0.033$& $   3.6\pm  0.2$ \\
\dots           &$R$ & $ -8.44\pm  0.11$& $ 0.403\pm 0.043$& $   5.0\pm  0.5$ \\
M31N 2008-05c   &$R$ & $ -7.54\pm  0.11$& $ 0.038\pm 0.002$& $  53.0\pm  2.8$ \\
M31N 2008-06b   &$R$ & $ -8.64\pm  0.11$& $ 0.083\pm 0.005$& $  24.2\pm  1.5$ \\
M31N 2008-07a   &$R$ & $ -6.24\pm  0.11$& $ 0.005\pm 0.001$& $ 410.0\pm 51.3$ \\
M31N 2008-07b   &$R$ & $ -6.14\pm  0.20$& $ 0.049\pm 0.006$& $  40.9\pm  4.7$ \\
M31N 2008-08a   &$R$ & $ -8.04\pm  0.20$& $ 0.094\pm 0.011$& $  21.2\pm  2.5$ \\
M31N 2008-10a   &$B$ & $ -7.15\pm  0.11$& $ 0.053\pm 0.003$& $  37.7\pm  2.2$ \\
\dots           &$V$ & $ -7.28\pm  0.11$& $ 0.052\pm 0.003$& $  38.3\pm  2.2$ \\
\dots           &$r'$ & $ -7.44\pm  0.11$& $ 0.030\pm 0.003$& $  67.5\pm  6.7$ \\
M31N 2008-10b   &$B$ & $ -7.08\pm  0.12$& $ 0.020\pm 0.003$& $  98.4\pm 14.9$ \\
\dots           &$V$ & $ -6.88\pm  0.12$& $ 0.017\pm 0.003$& $ 117.3\pm 21.5$ \\
\dots           &$R$ & $ -6.56\pm  0.25$& $ 0.032\pm 0.015$& $  63.3\pm 30.2$ \\
\dots           &$r'$ & $ -6.71\pm  0.11$& $ 0.027\pm 0.003$& $  73.2\pm  8.1$ \\
M31N 2008-11a   &$B$ & $ -7.75\pm  0.11$& $ 0.522\pm 0.029$& $   3.8\pm  0.2$ \\
\dots           &$V$ & $ -7.88\pm  0.11$& $ 0.457\pm 0.024$& $   4.4\pm  0.2$ \\
\dots           &$R$ & $ -8.04\pm  0.11$& $ 0.378\pm 0.044$& $   5.3\pm  0.6$ \\
\dots           &$r'$ & $ -8.04\pm  0.11$& $ 0.253\pm 0.013$& $   7.9\pm  0.4$ \\
\dots           &$i'$ & $ -6.13\pm  0.13$& $ 0.120\pm 0.006$& $  16.6\pm  0.8$ \\
\dots           &$z'$ & $ -6.61\pm  0.12$& $ 0.237\pm 0.017$& $   8.4\pm  0.6$ \\
M31N 2008-12b   &$B$ & $ -7.48\pm  0.11$& $ 0.055\pm 0.006$& $  36.7\pm  4.1$ \\
\dots           &$V$ & $ -7.61\pm  0.11$& $ 0.065\pm 0.006$& $  30.7\pm  2.9$ \\
\dots           &$r'$ & $ -7.74\pm  0.11$& $ 0.073\pm 0.006$& $  27.6\pm  2.2$ \\
\dots           &$i'$ & $ -7.00\pm  0.15$& $ 0.081\pm 0.012$& $  24.7\pm  3.6$ \\
\dots           &$z'$ & $ -7.73\pm  0.12$& $ 0.044\pm 0.011$& $  45.0\pm 10.9$ \\
M31N 2009-08a   &$B$ & $ -7.05\pm  0.11$& $ 0.006\pm 0.001$& $ 351.1\pm 84.5$ \\
\dots           &$V$ & $ -7.18\pm  0.11$& $ 0.011\pm 0.001$& $ 190.0\pm 24.5$ \\
\dots           &$R$ & $ -7.34\pm  0.11$& $ 0.014\pm 0.001$& $ 142.3\pm  9.0$ \\
\dots           &$r'$ & $ -7.34\pm  0.11$& $ 0.014\pm 0.001$& $ 142.3\pm 13.1$ \\
M31N 2009-08b   &$B$ & $ -7.15\pm  0.11$& $ 0.111\pm 0.005$& $  18.0\pm  0.8$ \\
\dots           &$V$ & $ -7.28\pm  0.11$& $ 0.112\pm 0.005$& $  17.8\pm  0.8$ \\
\dots           &$r'$ & $ -7.44\pm  0.11$& $ 0.086\pm 0.004$& $  23.1\pm  1.2$ \\
\dots           &$i'$ & $ -6.47\pm  0.12$& $ 0.074\pm 0.006$& $  26.9\pm  2.2$ \\
M31N 2009-08d   &$B$ & $ -7.23\pm  0.24$& $ 0.072\pm 0.017$& $  27.9\pm  6.5$ \\
\dots           &$V$ & $ -7.18\pm  0.11$& $ 0.063\pm 0.008$& $  31.7\pm  3.9$ \\
\dots           &$R$ & $ -7.34\pm  0.20$& $ 0.055\pm 0.008$& $  36.2\pm  5.2$ \\
\dots           &$r'$ & $ -7.34\pm  0.11$& $ 0.077\pm 0.019$& $  25.9\pm  6.5$ \\
M31N 2009-08e   &$R$ & $ -6.74\pm  0.11$& $ 0.016\pm 0.001$& $ 121.3\pm  8.0$ \\
M31N 2009-09a   &$R$ & $ -7.04\pm  0.25$& $ 0.012\pm 0.001$& $ 163.7\pm 15.4$ \\
M31N 2009-10a   &$B$ & $ -7.15\pm  0.11$& $ 0.123\pm 0.007$& $  16.2\pm  0.9$ \\
\dots           &$V$ & $ -7.28\pm  0.11$& $ 0.132\pm 0.006$& $  15.2\pm  0.7$ \\
M31N 2009-10b   &$B$ & $ -9.55\pm  0.11$& $ 0.250\pm 0.007$& $   8.0\pm  0.2$ \\
\dots           &$V$ & $ -9.68\pm  0.11$& $ 0.224\pm 0.005$& $   8.9\pm  0.2$ \\
\dots           &$R$ & $ -9.84\pm  0.11$& $ 0.156\pm 0.010$& $  12.8\pm  0.8$ \\
M31N 2009-10c   &$B$ & $ -8.75\pm  0.13$& $ 0.134\pm 0.008$& $  14.9\pm  0.8$ \\
\dots           &$V$ & $ -8.24\pm  0.14$& $ 0.124\pm 0.011$& $  16.1\pm  1.4$ \\
\dots           &$R$ & $ -8.14\pm  0.25$& $ 0.065\pm 0.020$& $  30.6\pm  9.4$ \\
M31N 2009-11a   &$B$ & $ -6.65\pm  0.11$& $ 0.095\pm 0.007$& $  21.1\pm  1.6$ \\
\dots           &$V$ & $ -6.78\pm  0.11$& $ 0.092\pm 0.005$& $  21.7\pm  1.2$ \\
\dots           &$R$ & $ -6.94\pm  0.11$& $ 0.096\pm 0.009$& $  20.8\pm  1.9$ \\
M31N 2009-11b   &$B$ & $ -5.85\pm  0.11$& $ 0.022\pm 0.004$& $  92.5\pm 16.6$ \\
\dots           &$V$ & $ -5.98\pm  0.11$& $ 0.027\pm 0.004$& $  74.8\pm 10.6$ \\
\dots           &$R$ & $ -6.14\pm  0.11$& $ 0.023\pm 0.002$& $  88.0\pm  6.6$ \\
M31N 2009-11c   &$B$ & $ -7.28\pm  0.11$& $ 0.048\pm 0.003$& $  42.0\pm  2.7$ \\
\dots           &$V$ & $ -7.84\pm  0.14$& $ 0.062\pm 0.005$& $  32.5\pm  2.4$ \\
\dots           &$R$ & $ -7.57\pm  0.11$& $ 0.046\pm 0.004$& $  43.1\pm  3.6$ \\
M31N 2009-11d   &$B$ & $ -7.85\pm  0.11$& $ 0.178\pm 0.007$& $  11.2\pm  0.4$ \\
\dots           &$V$ & $ -7.98\pm  0.11$& $ 0.276\pm 0.025$& $   7.2\pm  0.7$ \\
\dots           &$R$ & $ -8.14\pm  0.11$& $ 0.122\pm 0.013$& $  16.3\pm  1.7$ \\
M31N 2009-11e   &$R$ & $ -7.56\pm  0.20$& $ 0.036\pm 0.002$& $  55.7\pm  3.1$ \\
\enddata
\end{planotable}




\end{document}